\documentclass[cmp]{svjour} 

\usepackage{amsmath,a4,amssymb,epsfig}
\usepackage[hypertex]{hyperref}

\allowdisplaybreaks[4]

\makeatletter
\@addtoreset{equation}{section}

\date{}

\makeatother

\newtheorem{lem}{Lemma} 
\newtheorem{dfn}[lem]{Definition}
\newtheorem{thm}[lem]{Theorem}
\newtheorem{prp}[lem]{Proposition}

\newcommand{\di}{\genfrac{}{}{0pt}{}}

\begin{document}

\title{Renormalisation of $\phi^4$-theory on noncommutative
    $\mathbb{R}^4$ in the matrix base}

\author{Harald Grosse\inst{1} \and 
        Raimar Wulkenhaar\inst{2}}

\institute{Institut f\"ur Theoretische Physik, Universit\"at Wien,
Boltzmanngasse 5, A-1090 Wien, Austria,
\email{harald.grosse@univie.ac.at}
\and 
Max-Planck-Institut f\"ur Mathematik in den
  Naturwissenschaften,
Inselstra\ss{}e 22--26, D-04103 Leipzig, Germany, 
\email{raimar.wulkenhaar@mis.mpg.de}}

\maketitle

\begin{abstract}
  We prove that the real four-dimensional Euclidean noncommutative
  $\phi^4$-model is renormalisable to all orders in perturbation
  theory. Compared with the commutative case, the bare action of
  relevant and marginal couplings contains necessarily an additional
  term: an harmonic oscillator potential for the free scalar field
  action. This entails a modified dispersion relation for the free
  theory, which becomes important at large distances
  (UV/IR-entanglement). The renormalisation proof relies on flow
  equations for the expansion coefficients of the effective action
  with respect to scalar fields written in the matrix base of the
  noncommutative $\mathbb{R}^4$. The renormalisation flow depends on
  the topology of ribbon graphs and on the asymptotic and local
  behaviour of the propagator governed by orthogonal Meixner
  polynomials.
\end{abstract}

\section{Introduction}

Noncommutative $\phi^4$-theory is widely believed to be not
renormalisable in four dimensions. To underline the belief one usually
draws the non-planar one-loop two-point function resulting from the
noncommutative $\phi^4$-action. The corresponding integral is finite,
but behaves $\sim (\theta p)^{-2}$ for small momenta $p$ of the
two-point function. The finiteness is important, because the
$p$-dependence of the non-planar graph has no counterpart in the
original $\phi^4$-action, and thus (if divergent) cannot be absorbed
by multiplicative renormalisation. However, if we insert the
non-planar graph declared as finite as a subgraph into a bigger graph,
one easily builds examples (with an arbitrary number of external legs)
where the $\sim p^{-2}$ behaviour leads to non-integrable integrals at
small inner momenta. This is the so-called UV/IR-mixing problem
\cite{Minwalla:1999px}.

The heuristic argumentation can be made exact: Using a sophisticated
mathematical machinery, Chepelev and Roiban have proven a
power-counting theorem \cite{Chepelev:1999tt,Chepelev:2000hm} which
relates the power-counting degree of divergence to the topology of
\emph{ribbon graphs}. The rough summary of the power-counting theorem
is that noncommutative field theories with quadratic divergences
become meaningless beyond a certain loop order\footnote{There exist
  proposals to resum the perturbation series, see
  \cite{Chepelev:2000hm}, but there is no complete proof that this is
  consistent to all orders.}. For example, in the real noncommutative
$\phi^4$-model there exist three-loop graphs which cannot be
integrated.

In this paper we prove that the real noncommutative $\phi^4$-model is
\emph{renormalisable to all orders}. At first sight, this seems to be
a grave contradiction. However, we do not say that the famous papers
\cite{Minwalla:1999px,Chepelev:1999tt,Chepelev:2000hm} are wrong. In
fact, we reconfirm their results, it is only that we take their
message serious. The message of the UV/IR-entanglement is that 
\begin{verse}\emph{noncommutativity relevant at very short distances
    modifies the physics of the model at very large distances.}
  \end{verse}
At large distances we have approximately a free theory. Thus, we have
to alter the free theory, whereas the quasi-local $\phi^4$-interaction 
could hopefully be left unchanged. 

But how to modify the free action? We found the distinguished
modification in the course of a long refinement process of our method.
But knowing the result, it can be made plausible. It was pointed out
by Langmann and Szabo \cite{Langmann:2002cc} that the $\star$-product
interaction is invariant under a duality
transformation between positions and momenta, 
\begin{align}
\hat{\phi}(p) &\leftrightarrow \pi^2\sqrt{|\det \theta|} \phi(x)\;, & 
p_\mu &\leftrightarrow \tilde{x}_\mu:= 2(\theta^{-1})_{\mu\nu} x^\nu\;,
\end{align}
where $ \hat{\phi}(p_a)=\int d^4x_a
\;\mathrm{e}^{(-1)^a \mathrm{i} p_{a,\mu} x_a^\mu} \phi(x_a)$.
Using the definition of the $\star$-product given in (\ref{starprod}) 
and the reality $\phi(x)=\overline{\phi(x)}$ one obtains
\begin{align}
S_{\text{int}} &= \int d^4x\;\frac{\lambda}{4!}
( \phi\star\phi\star\phi\star \phi)(x)
\nonumber
\\*
&= \int \Big(\prod_{a=1}^4 d^4x_a\Big) \,
\phi(x_1)\phi(x_2)\phi(x_3)\phi(x_4)\,V(x_1,x_2,x_3,x_4)
\nonumber
\\*
&= \int \Big(\prod_{a=1}^4 \frac{d^4p_a}{(2\pi)^4} \Big) \,
\hat{\phi}(p_1)\hat{\phi}(p_2)\hat{\phi}(p_3)\hat{\phi}(p_4)\,
\hat{V}(p_1,p_2,p_3,p_4)\;,
\label{Sintdual}
\end{align}
with 
\begin{align}
\hat{V}(p_1,p_2,p_3,p_4) &= \frac{\lambda}{4!} (2\pi)^4 
\delta^4(p_1{-}p_2{-}p_3{+}p_4)  \,
\cos\Big(\tfrac{1}{2} \theta^{\mu\nu}(p_{1,\mu} p_{2,\nu} 
+ p_{3,\mu} p_{4,\nu})\Big)\;, 
\nonumber
\\*
V(x_1,x_2,x_3,x_4) &= \frac{\lambda}{4!} \frac{1}{\pi^4 \det \theta} \,
\delta^4(x_1{-}x_2{+}x_3{-}x_4)  \,
\cos\Big(2 (\theta^{-1})_{\mu\nu}(x_1^\mu x_2^\nu 
+ x_3^\mu x_4^\nu)\Big)\;.
\end{align}
Passing to quantum field theory, $V(x_1,x_2,x_3,x_4)$ and 
$\hat{V}(p_1,p_2,p_3,p_4)$ become the Feynman rules for the vertices
in position space and momentum space, respectively. Multiplicative
renormalisability of the four-point function requires that its
divergent part has to be self-dual, too. This requires an appropriate
Feynman rule for the propagator. Building now two-point functions with
these Feynman rules, it is very plausible that if the two-point
function is divergent in momentum space, also the duality-transformed
two-point function will be divergent. That divergence has to be
absorbed in a multiplicative renormalisation of the initial action.

However, the usual free scalar field action is not invariant under
that duality transformation and therefore cannot absorb the expected
divergence in the two-point function. In order to cure this problem we
have to extend the free scalar field action by a harmonic oscillator
potential:
\begin{align}
S_{\text{free}} = \int d^4x\;\Big( \frac{1}{2} (\partial_\mu \phi) \star
(\partial^\mu \phi) + \frac{\Omega^2}{2} (\tilde{x}_\mu \phi) \star
(\tilde{x}^\mu \phi) + \frac{\mu_0^2}{2} \phi\star\phi\Big)(x)\;,
\label{Sfreedual}
\end{align}
The action $S_{\text{free}}+ S_{\text{int}}$ according to
(\ref{Sfreedual}) and (\ref{Sintdual}) is now preserved by duality
transformation, up to rescaling. From the previous considerations we
can expect that also the renormalisation flow preserves that action
$S_{\text{free}}+ S_{\text{int}}$. We prove in this paper that this is
indeed the case. Thus, the duality-covariance of the action
$S_{\text{free}}+ S_{\text{int}}$ implements precisely the
UV/IR-entanglement.

Of course, we cannot treat the quantum field theory associated with
the action (\ref{Sfreedual}) in momentum space.  Fortunately, there is
a matrix representation of the noncommutative $\mathbb{R}^D$, where the
$\star$-product becomes a simple product of infinite matrices and
where the duality between positions and momenta is manifest. The
matrix representation plays an important r\^ole in the proof that the
noncommutative $\mathbb{R}^D$ is a spectral triple
\cite{Gayral:2003dm}. It is also crucial for the exact solution of
 quantum field theories on noncommutative phase space 
\cite{Langmann:2002ai,Langmann:2003cg,Langmann:2003if}.

Coincidently, the matrix base is also required for another
reason. In the traditional Feynman graph approach, the value of the
integral associated to non-planar graphs is not unique, because one
exchanges the order of integrations in integrals which are not
absolutely convergent. To avoid this problem one should use a
renormalisation scheme where the various limiting processes are better
controlled. The preferred method is the use of flow equations.  The
idea goes back to Wilson \cite{Wilson:1973jj}. It was then used by
Polchinski \cite{Polchinski:1983gv} to give a very efficient
renormalisability proof for commutative $\phi^4$-theory. Several
improvements have been suggested in \cite{Keller:1992ej}. Applying
Polchinski's method to the noncommutative $\phi^4$-model there is,
however, a serious problem in momentum space. We have to
guarantee that planar graphs only appear in the distinguished
interaction coefficients for which we fix the boundary condition at
the renormalisation scale $\Lambda_R$.  Non-planar graphs have phase
factors which involve inner momenta.  Polchinski's method consists in
taking norms of the interaction coefficients, and these norms ignore
possible phase factors. Thus, we would find that boundary conditions
for non-planar graphs at $\Lambda_R$ are required. Since there is an
infinite number of different non-planar structures, the model is not
renormalisable in this way. A more careful examination of the phase
factors is also not possible, because the cut-off integrals prevent the
Gau\ss{}ian integration required for the parametric integral
representation \cite{Chepelev:1999tt,Chepelev:2000hm}.

As we show in this paper, the bare action $S_{\text{free}}+
S_{\text{int}}$ according to (\ref{Sfreedual}) and (\ref{Sintdual})
corresponds to a quantum field theory which is renormalisable to all
orders. Together with the duality argument, this is now the conclusive
indication that the usual noncommutative $\phi^4$-action (with
$\Omega=0$) has to be dismissed in favour of the duality-covariant 
action of (\ref{Sfreedual}) and (\ref{Sintdual}). 

Our proof is very technical. We do not claim that it is the most
efficient one. However, it was for us, for the time being, the only
possible way. We encountered several ``miracles'' without which the
proof had failed. The first is that the propagator is complicated but
numerically accessible. We had thus convinced ourselves that the
propagator has such an asymptotic behaviour that all non-planar graphs
and all graphs with $N>4$ external legs are irrelevant according to
our general power-counting theorem for dynamical matrix models
\cite{Grosse:2003aj}. However, this still leaves an infinite number of
planar two- or four-point functions which would be relevant or
marginal according to \cite{Grosse:2003aj}. In the first versions of
\cite{Grosse:2003aj} we had, therefore, to propose some consistency
relations in order to get a meaningful theory.

Miraculously, all this is not necessary. We have further found
numerically that the propagator has some universal locality properties
suggesting that the infinite number of relevant\,/\,marginal planar
two- or four-point functions can be decomposed into four
relevant\,/\,marginal base interactions and an irrelevant remainder.
Of course, there must exist a reason for such a coincidence, and the
reason are orthogonal polynomials. In our case, it means that the
kinetic matrix corresponding to the free action (\ref{Sfreedual})
written in the matrix base of the noncommutative $\mathbb{R}^D$ is
diagonalised by orthogonal Meixner polynomials
\cite{Meixner:1934}\footnote{In our renormalisation proof
  \cite{Grosse:2003nw} of the two-dimensional noncommutative
  $\phi^4$-model we had originally termed these polynomials ``deformed
  Laguerre polynomials'', which we had only constructed via its
  recursion relation. The closed formula was not known to us.  Thus,
  we are especially grateful to Stefan Schraml who provided us first
  with \cite{Masson:1991}, from which we got the information that we
  were using Meixner polynomials, and then with the encyclopaedia
  \cite{Koekoek:1996} of orthogonal polynomials, which was the key to
  complete the renormalisation proof.} Now, having a closed solution
for the free theory in the preferred base of the interaction, the
desired local and asymptotic behaviour of the propagator can be
derived.

We stress, however, that some of the corresponding estimations of
Section~\ref{bounds} are, so far, verified numerically only. There is
no doubt that the estimations are correct, but for the purists we have
to formulate our result as follows: The quantum field theory
corresponding to the action (\ref{Sfreedual}) and (\ref{Sintdual}) is
renormalisable to all orders provided that the estimations given in
Section~\ref{bounds} hold. Already this weaker result is a
considerable progress, because the elimination of the last possible
doubt amounts to estimate a single integral. This estimation will be
performed in \cite{Rivasseau:2004??}. 
The method, further motivation
and an outlook to constructive applications are already presented in
\cite{Rivasseau:2004az}.

Finally, let us recall that the noncommutative $\phi^4$-theory in two
dimensions is different. We also need the harmonic oscillator
potential of (\ref{Sfreedual}) in all intermediate steps of the
renormalisation proof, but at the end it can be switched off with the
removal of the cut-off \cite{Grosse:2003nw}. This is in agreement with
the common belief that the UV/IR-mixing problem can be cured in models
with only logarithmic divergences. 

\subsection{Strategy of the proof}

\begin{figure}
\normalsize%
\begin{picture}(130,130)
\put(0,120){\fbox{\parbox{37\unitlength}{propagator (\ref{noalpha})\\
            derived in App.~\ref{app:evalprop}}}}
\put(80,122){\fbox{\parbox{10em}{initial interaction (\ref{ct4})}}}
\put(85,105){\fbox{\parbox{13.4em}{Polchinski equation (\ref{polL})\\
             derived in \cite{Grosse:2003aj}}}}
\put(21,95){\fbox{\parbox{10em}{integration procedure\\
            Def.~\ref{defint}}}}
\put(85,89){\fbox{\parbox{8em}{numerical bounds \\ App.~\ref{appB}}}}
\put(0,78){\fbox{\parbox{10em}{composite propagators \\
           Sec.~\ref{seccomposite}, App.~\ref{appcompositeidentity}}}}
\put(0,51){\fbox{\fbox{\parbox{63\unitlength}{Power-counting behaviour \\
            of interaction coefficients, 
            Prop.~\ref{power-counting-prop}}}}}
\put(77,70){\fbox{\parbox{54\unitlength}{general power-counting theorem\\
            for non-local matrix models \\
            proven in \cite{Grosse:2003aj}}}}
\put(15,32){\fbox{\parbox{49\unitlength}{$\Lambda_0$-dependence of \\
            interaction coefficients (\ref{Rlim}) }}}
\put(81,20){\fbox{\parbox{14.6em}{differential equations\\
            -- for $\Lambda_0$-varied funcions (\ref{polV4}) \\
            -- for auxiliary functions (\ref{polB4})  }}}
\put(0,9){\fbox{\parbox{60\unitlength}{Power-counting behaviour \\
           -- of auxiliary functions, Prop.~\ref{prop-H} \\ 
           -- of $\Lambda_0$-varied functions, Prop.~\ref{R-prop} }}}
\put(75,-5){\fbox{\fbox{\parbox{14.5em}{Convergence Theorem, 
           Thm.~\ref{final-theorem} }}}}
\put(80,123){\line(-2,-1){9.5}}
\put(68,117){\line(-2,-1){10.5}}
\put(56,111){\vector(-2,-1){18}}
\put(40,125){\vector(4,-1){50}}
\put(40,120){\line(2,-1){27}}
\put(70,105){\line(2,-1){10.5}}
\put(82,99){\vector(2,-1){7}}
\put(85,110){\vector(-4,-1){33}}
\put(40,90.5){\vector(-2,-1){11.5}}
\put(10,115){\vector(0,-1){30}}
\put(85,105){\line(-2,-3){17.7}}
\put(65.7,76){\vector(-2,-3){11}}
\put(130,100){\vector(0,-1){21}}
\put(105,84){\vector(0,-1){5}}
\put(77,74){\line(-1,-1){5}}
\put(70,67){\vector(-1,-1){8}}
\put(55,90.5){\vector(-1,-4){7.7}}
\put(35,73){\vector(0,-1){13.5}}
\put(60,90.5){\vector(1,-2){30.5}}
\put(67,35){\vector(3,-1){16}}
\put(81,25){\line(-4,-1){13.5}}
\put(66,21.25){\vector(-4,-1){9}}
\put(138,100){\vector(0,-1){70.5}}
\put(7,45){\vector(0,-1){26.5}}
\put(61,27){\vector(1,-1){26}}
\put(63,7.7){\vector(3,-1){20}}
\end{picture}
\vspace*{8mm}
\caption{Relations between the main steps
  of the proof. The central results are the power-counting behaviour
  of Proposition~\ref{power-counting-prop} and the convergence theorem
  (Theorem~\ref{final-theorem}). Note that the numerical estimations
  for the propagator influence the entire chain of the proof.  }
\label{fig:flowchart}
\end{figure}
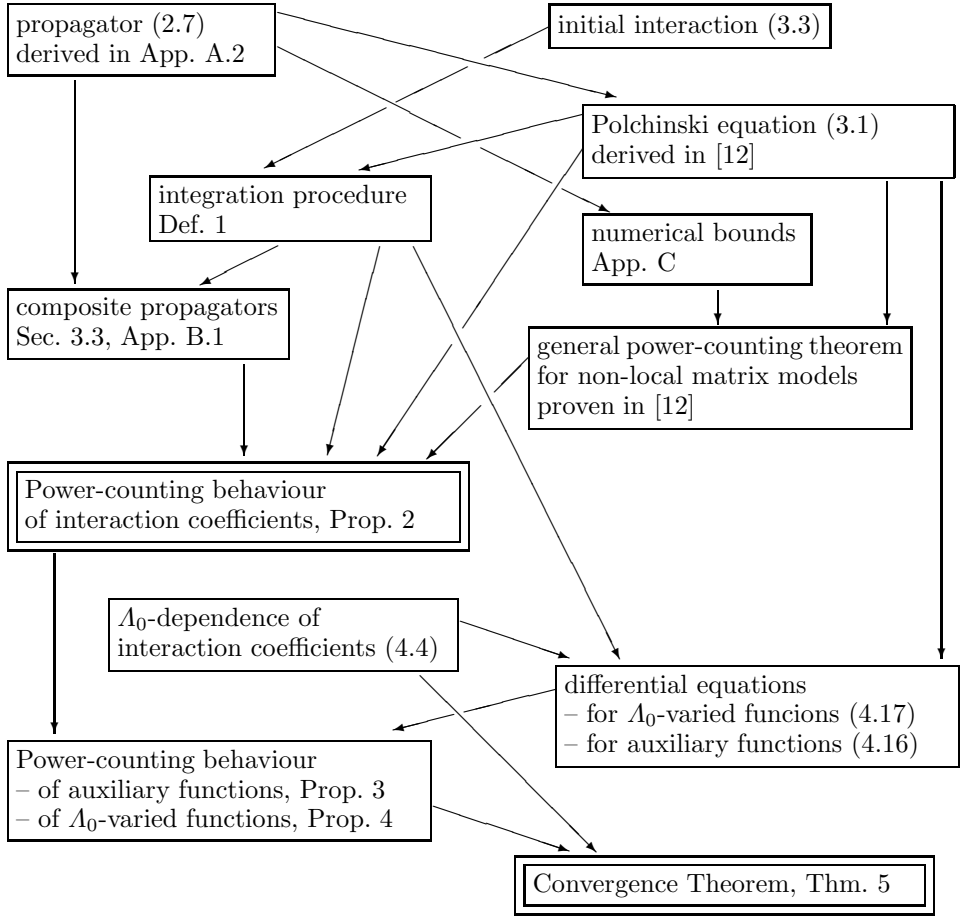

As the renormalisation proof is long and technical, we list here the
most important steps and the main ideas. A flow chart of these steps
is presented in Fig.~\ref{fig:flowchart}. A more detailed introduction
is given in \cite{Grosse:2004ik}.

The first step is to rewrite the $\phi^4$-action
(\ref{Sfreedual})+(\ref{Sintdual}) in the harmonic oscillator base of
the Moyal plane, see (\ref{action}) and (\ref{G4D}). The free theory
is solved by the propagator (\ref{noalpha}), which we compute in
Appendix~\ref{appMeixner} using Meixner polynomials in an essential
way. The propagator is represented by a finite sum which enables a
fast numerical evaluation. Unfortunately, we can offer analytic
estimations only in a few special cases.

The propagator is so complicated that a direct calculation of Feynman
graphs is not practicable. Therefore, we employ the renormalisation
method based on flow equations \cite{Polchinski:1983gv,Keller:1992ej}
which we have previously adapted to non-local (dynamical) matrix
models \cite{Grosse:2003aj}. The modification $K[\Lambda]$ of the
weights of the matrix indices in the kinetic term is undone in the
partition function by a careful adaptation of effective action
$L[\phi,\Lambda]$, which is described by the matrix Polchinski
equation (\ref{polL}). For a modification given by a cut-off function
$K[\Lambda]$, renormalisation of the model amounts to prove that the
matrix Polchinski equation (\ref{polL4}) admits a regular solution
which depends on a finite number of initial data.

In a perturbative expansion, the matrix Polchinski equation is solved
by ribbon graphs drawn on Riemann surfaces. The existence of a
regular solution follows from the general power-counting theorem
proven in \cite{Grosse:2003aj} together with the numerical
determination of the propagator asymptotics in Appendix~\ref{appB}.
However, the general proof involved an infinite number of initial
conditions, which is physically not acceptable.  Therefore, the
challenge is to prove the reduction to a finite number of initial data
for the renormalisation flow.

The answer is the integration procedure given in
Definition~\ref{defint}, Sec.~\ref{intPolEq}, which entails mixed
boundary conditions for certain planar two- and four-point functions.
The idea is to introduce four types of reference graphs with vanishing
external indices and to split the integration of the Polchinski
equation for the distinguished two- and four-point graphs into an
integration of the difference to the reference graphs and a different
integration of the reference graphs themselves. The difference between
original graph and reference graph is further reduced to differences
of propagators, which we call ``composite propagators''. See
Sec.~\ref{seccomposite}.

The proof of the power-countung estimations for the interaction
coefficients (Proposition~\ref{power-counting-prop} in Sec.~\ref{pct})
requires the following extensions of the general case treated in
\cite{Grosse:2003aj}:
\begin{itemize}
\item We have to prove that graphs where the index jumps along the
  trajectory between incoming and outgoing indices are suppressed.
  This leaves 1PI planar four-point functions with constant index
  along the trajectory and 1PI planar two point functions with in
  total at most two index jumps along the trajectories as the only
  graphs which are marginal or relevant. 
  
\item For these types of graphs we have to prove that the leading
  relevant/marginal contribution is captured by reference graphs with
  vanishing external indices, whereas the difference to the reference
  graphs is irrelevant. This is the discrete analogue of the BPHZ
  Taylor subtraction of the expansion coefficients to lowest-order in
  the external momenta.

\end{itemize}

Thus, Proposition~\ref{power-counting-prop} provides bounds for the
interaction coefficients for the effective action at a scale $\Lambda
\in [\Lambda_R,\Lambda_0]$. Here, $\Lambda_R$ is the renormalisation
scale where the four reference graphs are normalised, and $\Lambda_0$
is the initial scale for the integration which has to be sent to
$\infty$ in order to scale away possible initial conditions for the
irrelevant functions. The estimations of
Proposition~\ref{power-counting-prop} are actually independent of
$\Lambda_0$ so that the limit $\Lambda_0 \to \infty$ can be taken.
This already ensures the renormalisation of the model. However, one
would also like to know whether the interaction coefficients converge
in the limit $\Lambda_0 \to \infty$ and if so, with which rate. That
analysis is performed in Section~\ref{sec:convergence} which
culminates in Theorem~\ref{final-theorem}, confirming convergence with
a rate $\Lambda_0^{-2}$.

\section{The duality-covariant noncommutative $\phi^4$-action in 
the matrix base}

The noncommutative $\mathbb{R}^4$ is defined as the algebra
$\mathbb{R}^4_\theta$ which as a vector space is given by the space
$\mathcal{S}(\mathbb{R}^4)$ of (complex-valued) Schwartz class functions of
rapid decay, equipped with the multiplication rule \cite{Gracia-Bondia:1987kw}
\begin{align}
  (a\star b)(x) &= \int \frac{d^4k}{(2\pi)^4} \int d^4 y \; a(x{+}\tfrac{1}{2}
  \theta {\cdot} k)\, b(x{+}y)\, \mathrm{e}^{\mathrm{i} k \cdot y}\;,
\label{starprod}
\\*
& (\theta {\cdot} k)^\mu = \theta^{\mu\nu} k_\nu\;,\quad k{\cdot}y = k_\mu
y^\mu\;,\quad \theta^{\mu\nu}=-\theta^{\nu\mu}\;.  \nonumber
\end{align}
The entries $\theta^{\mu\nu}$ in (\ref{starprod}) have the dimension of an
area. We place ourselves into a coordinate system in which $\theta$ has the
form
\begin{align}
  \theta_{\mu\nu} &=\left(\begin{array}{cccc}
      0 & \theta_1 & 0 & 0 \\
      -\theta_1 & 0 & 0 & 0 \\
      0 & 0 & 0 & \theta_2  \\
      0 & 0 & -\theta_2 & 0
\end{array}\right)\;.
\end{align}
We use an adapted base
\begin{align}
  b_{mn}(x) &= f_{m^1n^1}(x^1,x^2) \,f_{m^2n^2}(x^3,x^4) \;,\qquad
  m=\textstyle{\di{m^1}{m^2}}\in \mathbb{N}^2\,,~
n=\textstyle{\di{n^1}{n^2}} \in \mathbb{N}^2\;,
\label{bbas}
\end{align}
where the base $f_{m^1n^1}(x^1,x^2) \in \mathbb{R}^2_\theta$ is given in
\cite{Grosse:2003nw}. This base satisfies
\begin{align}
  (b_{mn} \star b_{kl})(x) &= \delta_{nk} b_{ml}(x) \;, & \int d^4x\, b_{mn} =
  4\pi^2 \theta_1 \theta_2 \delta_{mn} \;.
\end{align}
More information about the noncommutative $\mathbb{R}^D$ can be found
in \cite{Gayral:2003dm,Gracia-Bondia:1987kw}.

We are going to study a duality-covariant $\phi^4$-theory on
$\mathbb{R}^4_\theta$.  This means that we add a harmonic oscillator
potential to the standard $\phi^4$-action, which breaks translation
invariance but is required for renormalisation. Expanding the scalar
field in the matrix base, $\phi(x)=\sum_{m,n \in \mathbb{N}^2}
\phi_{mn} b_{mn}(x)$, we have
\begin{align}
  S[\phi] &= \int d^4x \Big( \frac{1}{2} g^{\mu\nu} \big(\partial_\mu \phi
  \star \partial_\nu \phi +4 \Omega^2 ((\theta^{-1})_{\mu\rho} x^\rho \phi )
  \star ((\theta^{-1})_{\nu\sigma} x^\sigma \phi) \big) + \frac{1}{2} \mu_0^2
  \,\phi \star \phi \nonumber
  \\
  &\hspace*{10em} + \frac{\lambda}{4!} \phi \star \phi \star \phi \star
  \phi\Big)(x) \nonumber
  \\
  &= 4\pi^2 \theta_1 \theta_2 \sum_{m,n,k,l \in \mathbb{N}^2} \Big(
  \frac{1}{2} G_{mn;kl} \phi_{mn} \phi_{kl} + \frac{\lambda}{4!}  \phi_{mn}
  \phi_{nk} \phi_{kl} \phi_{lm}\Big)\;,
\label{action}
\end{align}
where according to \cite{Grosse:2003nw}
\begin{align}
&  G_{mn;kl} 
\nonumber
\\*
&= \big(\mu_0^2{+} \frac{2}{\theta_1}(1{+}\Omega^2)
  (n^1{+}m^1{+}1) {+} \frac{2}{\theta_2}(1{+}\Omega^2)(n^2{+}m^2{+}1) \big)
  \delta_{n^1k^1} \delta_{m^1l^1} \delta_{n^2k^2} \delta_{m^2l^2} \nonumber
  \\*
  & - \frac{2}{\theta_1}(1{-}\Omega^2) \big(\sqrt{k^1l^1}\,
  \delta_{n^1+1,k^1}\delta_{m^1+1,l^1 } 
+ \sqrt{m^1n^1}\, \delta_{n^1-1,k^1}
  \delta_{m^1-1,l^1} \big) \delta_{n^2k^2} \delta_{m^2l^2} \nonumber
  \\*
  & - \frac{2}{\theta_2}(1{-}\Omega^2) \big(\sqrt{k^2l^2}\,
  \delta_{n^2+1,k^2}\delta_{m^2+1,l^2 } 
+ \sqrt{m^2n^2}\, \delta_{n^2-1,k^2}
  \delta_{m^2-1,l^2} \big) \delta_{n^1k^1} \delta_{m^1l^1}\;.
\label{G4D}
\end{align}
We need, in particular, the inverse of the kinetic matrix $G_{mn;kl}$, the
propagator $\Delta_{nm;kl}$, which solves the partition function of the free
theory ($\lambda=0$) with respect to the preferred base of the interaction. We
present the computation of the propagator in Appendix~\ref{appMeixner}. The
result is
\begin{align}
  &\Delta_{\di{m^1}{m^2}\di{n^1}{n^2}; \di{k^1}{k^2}\di{l^1}{l^2}} 
= \frac{\theta}{2(1{+}\Omega)^2} 
\delta_{m^1+k^1,n^1+l^1}\delta_{m^2+k^2,n^2+l^2}
  \nonumber
  \\*
  &\times \!\!
\sum_{v^1=\frac{|m^1-l^1|}{2}}^{\frac{\min(m^1+l^1,n^1+k^1)}{2}}
  \sum_{v^2=\frac{|m^2-l^2|}{2}}^{\frac{\min(m^2+l^2,n^2+k^2)}{2}} 
\!\!\!\!
B\big(1{+} \tfrac{\mu_0^2\theta}{8\Omega}
  {+}\tfrac{1}{2}(m^1{+}m^2{+}k^1{+}k^2){-}v^1{-}v^2, 1{+}2v^1{+}2v^2 \big)
  \nonumber
  \\*
  &\times {}_2F_1\bigg(\di{1{+} 2v^1{+}2v^2\,,\; \frac{\mu_0^2\theta}{8\Omega}
    {-}\frac{1}{2}(m^1{+}m^2{+}k^1{+}k^2){+}v^1{+}v^2 }{2{+}
    \frac{\mu_0^2\theta}{8\Omega}
    {+}\frac{1}{2}(m^1{+}m^2{+}k^1{+}k^2){+}v^1{+}v^2} \bigg|
  \frac{(1{-}\Omega)^2}{(1{+}\Omega)^2} \bigg) \nonumber
  \\*
  &\times \prod_{i=1}^2 \sqrt{ \binom{n^i}{v^i{+}\frac{n^i-k^i}{2}}
    \binom{k^i}{v^i{+}\frac{k^i-n^i}{2}} \binom{m^i}{v^i{+}\frac{m^i-l^i}{2}}
    \binom{l^i}{v^i{+}\frac{l^i-m^i}{2}}}
  \Big(\frac{1{-}\Omega}{1{+}\Omega} \Big)^{2v^i} \,.
\label{noalpha}
\end{align}
Here, $B(a,b)$ is the Beta-function and
${}_2F_1\big(\di{a,b}{c}\big|z\big)$ the hypergeometric function.

\section{Estimation of the interaction coefficients}
\label{sec:est-int-coeff}

\subsection{The Polchinski equation}

We have developed in \cite{Grosse:2003aj} the Wilson-Polchinski
renormalisation programme
\cite{Wilson:1973jj,Polchinski:1983gv,Keller:1992ej} for non-local
matrix models where the kinetic term (Taylor coefficient matrix of the
part of the action which is bilinear in the fields) is neither
constant nor diagonal. Introducing a cut-off in the measure
$\prod_{m,n} d\phi_{mn}$ of the partition function $Z$, the resulting
effect is undone by adjusting the effective action $L[\phi]$ (and
other terms which are easy to evaluate). If the cut-off function is a
smooth function of the cut-off scale $\Lambda$, the adjustment of
$L[\phi,\Lambda]$ is described by a differential equation,
\begin{align}
 \Lambda \frac{\partial L[\phi,\Lambda]}{\partial \Lambda} &=
  \sum_{m,n,k,l} \frac{1}{2} \Lambda \frac{\partial
    \Delta^K_{nm;lk}(\Lambda)}{\partial \Lambda} \bigg( \frac{\partial
    L[\phi,\Lambda]}{\partial \phi_{mn}}\frac{\partial
    L[\phi,\Lambda]}{\partial \phi_{kl}} - \frac{1}{4\pi^2\theta_1\theta_2}
  \Big[\frac{\partial^2 L[\phi,\Lambda]}{\partial \phi_{mn}\,\partial
    \phi_{kl}}\Big]_\phi \bigg) \;,
\label{polL}
\end{align}
where $\big[F[\phi]\big]_\phi:= F[\phi]-F[0]$ and 
\begin{align}
\Delta^K_{nm;lk}(\Lambda) 
= \prod_{r=1}^2 \Big(\prod_{i^r \in \{m^r,n^r,k^r,l^r\}}
  K\Big( \frac{i^r}{\theta_r\Lambda^2}\Big)\Big)
\Delta_{nm;lk} \;.
\label{GKDK1}  
\end{align}
Here, $K(x)$ is a smooth monotonous cut-off function with $K(x)=1$ for
$x\leq 1$ and $K(x)=0$ for $x\geq 2$. The differential equation
(\ref{polL}) is referred to as the Polchinski equation. 

In \cite{Grosse:2003aj} we have derived a power-counting theorem for
$L[\phi,\Lambda]$ by integrating (\ref{polL}) perturbatively between
the initial scale $\Lambda_0$ and the renormalisation scale $\Lambda_R
\ll \Lambda_0$. The power-counting degree is given by topological data
of ribbon graphs and two scaling exponents of the (summed and
differentiated) cut-off propagator. The power-counting theorem in
\cite{Grosse:2003aj} is model independent, but it relied on boundary
conditions for the integrations which do not correspond to a
physically meaningful model. 

In this paper we will show that the four-dimensional duality-covariant
noncommutative $\phi^4$-theory given by the action (\ref{action})
admits an improved power-counting behaviour which only relies on a
finite number of physical boundary conditions for the integration. The
first step is to extract from the power-counting theorem
\cite{Grosse:2003aj} the set of relevant and marginal interactions,
which on the other hand is used as an input to derive the
power-counting theorem. To say it differently: One has to be lucky to
make the right ansatz for the initial interaction which is then
reconfirmed by the power-counting theorem as the set of relevant and
marginal interactions. We are going to prove that the following ansatz
for the initial interaction is such a lucky choice:
\begin{align}
&L[\phi,\Lambda_0,\Lambda_0,\rho^0] 
\nonumber
\\*
&= \sum_{m^1,m^2,n^1,n^2 \in \mathbb{N}}\frac{1}{2\pi\theta} 
\Big(\frac{1}{2} \big(\rho_1^0 {+} (n^1{+}m^1{+} n^2{+}m^2)\rho_2^0 
  \big)  \phi_{\di{m^1}{m^2}\di{n^1}{n^2}} 
\phi_{\di{n^1}{n^2}\di{m^1}{m^2}}
\nonumber
\\*
& \hspace*{5em} - \rho_3^0 \big( 
\sqrt{n^1m^1}
\phi_{\di{m^1}{m^2}\di{n^1}{n^2}} 
\phi_{\di{n^1-1}{n^2}\di{m^1-1}{m^2}}
+ \sqrt{n^2m^2}
\phi_{\di{m^1}{m^2}\di{n^1}{n^2}} 
\phi_{\di{n^1}{n^2-1}\di{m^1}{m^2-1}}\big)\Big)
\nonumber
\\*
&+ \sum_{m^1,m^2,n^1,n^2,k^1,k^2, l^1,l^2 \in \mathbb{N}}
\frac{1}{4!} \rho_4^0 \,\phi_{\di{m^1}{m^2}\di{n^1}{n^2}} 
\phi_{\di{n^1}{n^2}\di{k^1}{k^2}}
\phi_{\di{k^1}{k^2}\di{l^1}{l^2}}
\phi_{\di{l^1}{l^2}\di{m^1}{m^2}}\;.
\label{ct4}
\end{align}
For simplicity we impose a symmetry between the two components $m^i$
of matrix indices $m=\di{m^1}{m^2}\in \mathbb{N}$, which could be
relaxed by taking different $\rho$-coefficients in front of
$m^i{+}n^i$ and $\sqrt{m^i n^i}$. Accordingly, we choose the same
weights in the noncommutativity matrix,
$\theta_1=\theta_2\equiv\theta$.

The differential equation (\ref{polL}) is non-perturbatively defined.
However, we shall solve it perturbatively as a formal power series in
a coupling constant $\lambda$ which later on will be related to a
normalisation condition at $\Lambda=\Lambda_R$, see (\ref{initrho}).
We thus consider the following expansion:
\begin{align}
&L[\phi,\Lambda,\Lambda_0,\rho^0] 
\nonumber
\\*
&= 
\sum_{V =1}^\infty \lambda^V \sum_{N=2}^{2V+2}
\frac{(2\pi\theta)^{\frac{N}{2}-2}}{N!} \sum_{m_i,n_i}
A^{(V)}_{m_1n_1;\dots;m_Nn_N}[\Lambda,\Lambda_0,\rho^0]
\phi_{m_1n_1}\cdots \phi_{m_Nn_N}\;.
\label{Lg4}
\end{align}
Inserting (\ref{Lg4}) into (\ref{polL}) we obtain
\begin{align}
&\Lambda \frac{\partial}{\partial \Lambda} 
A^{(V)}_{m_1n_1;\dots;m_Nn_N}[\Lambda,\Lambda_0,\rho^0] 
\nonumber
\\*
&= \!
\sum_{N_1=2}^N \sum_{V_1=1}^{V-1}
\sum_{m,n,k,l \in \mathbb{N}^2} \frac{1}{2} Q_{nm;lk}(\Lambda)  
A^{(V_1)}_{m_1n_1;\dots;m_{N_1-1}n_{N_1-1};mn}[\Lambda]
 A^{(V-V_1)}_{m_{N_1}n_{N_1};\dots;m_{N}n_{N};kl}[\Lambda]
\nonumber
\\*[-1ex]
&\hspace*{15em} + \Big(\binom{N}{N_1{-}1} -1\Big) \text{ permutations}
\nonumber
\\*
& - \sum_{m,n,k,l\in \mathbb{N}^2} \frac{1}{2} Q_{nm;lk}(\Lambda) 
A^{(V)}_{m_1n_1;\dots;m_{N}n_{N};mn;kl}[\Lambda]\;,
\label{polL4}
\end{align}
with
\begin{align}
  Q_{nm;lk}(\Lambda) := \frac{1}{2\pi \theta} \Lambda
  \frac{\partial \Delta^K_{nm;lk}(\Lambda)}{\partial \Lambda}\;.
\label{Q4}
\end{align}

\subsection{Integration of the Polchinski equation}
\label{intPolEq}

We are going to compute the functions $A^{(V)}_{m_1n_1;\dots;m_Nm_N}$
by iteratively integrating the Polchinski equation (\ref{polL4})
starting from boundary conditions either at $\Lambda_R$ or at
$\Lambda_0$. The right choice of the integration direction is an art:
The boundary condition influences crucially the estimation, which in
turn justifies or discards the original choice of the boundary
condition. At the end of numerous trial-and-error experiments with the
boundary condition, one convinces oneself that the procedure described
in Definition~\ref{defint} below is---up to finite re-normalisations
discussed later---the unique possibility\footnote{We ``only'' prove
  that the method works, not its uniqueness. The reader who doubts
  uniqueness of the integration procedure is invited to attempt a
  different way.} to renormalise the model. 

First we have to recall \cite{Grosse:2003aj} the graphology resulting
from the Polchinski equation (\ref{polL4}). The Polchinski equation is
solved by \emph{ribbon graphs} drawn on a Riemann surface of uniquely
determined \emph{genus} $g$ and uniquely determined number $B$
of \emph{boundary components} (holes). The ribbons are made of
double-line propagators 
\begin{align}
\parbox{20\unitlength}{\begin{picture}(15,6)
\put(0,2){\epsfig{file=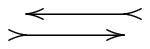,scale=0.9,bb=71 667 101 675}}
       \put(1,0){\mbox{\scriptsize$n$}}
       \put(8,-0.5){\mbox{\scriptsize$k$}}
       \put(2,5.5){\mbox{\scriptsize$m$}}
       \put(9,5.5){\mbox{\scriptsize$l$}}
   \end{picture}}
=Q_{mn;kl}(\Lambda)
\end{align}
attached to vertices
\begin{align}
\parbox{19\unitlength}{\begin{picture}(17,17)
       \put(0,0){\epsfig{file=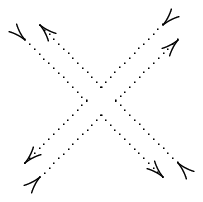,scale=0.9,bb=71 638 117 684}}
       \put(-1,5){\mbox{\scriptsize$m_1$}}
       \put(-2,12){\mbox{\scriptsize$n_4$}}
       \put(3,0){\mbox{\scriptsize$n_1$}}
       \put(8,1){\mbox{\scriptsize$m_2$}}
       \put(16,3){\mbox{\scriptsize$n_2$}}
       \put(15,11){\mbox{\scriptsize$m_3$}}
       \put(10,16){\mbox{\scriptsize$n_3$}}
       \put(4,14){\mbox{\scriptsize$m_4$}}
   \end{picture}} =\delta_{n_1m_2}\delta_{n_2m_3}
\delta_{n_3m_4}\delta_{n_4m_1}\;.
\label{vertex}
\end{align}
Under certain conditions verified by our model, the rough power-counting
behaviour of the ribbon graph is determined by the topology $g,B$ of
the Riemann surface and the number of vertices $V$ and external legs $N$.
However, in order to prove this behaviour we need some auxiliary notation: the
number $V^e$ of external vertices (vertices to which at least one external leg
is attached), a certain segmentation index $\iota$ and a certain summation of
graphs with appropriately varying indices.

We recall in detail the index summation, because we need it for a
refinement of the general proof given in \cite{Grosse:2003aj}. Viewing
the ribbon graph as a set of single-lines, we can distinguish closed
and open lines. The open lines are called \emph{trajectories} starting
at an incoming index $n$, running through a chain of inner indices
$k_j$ and ending at an outgoing index $m$. Each index belongs to
$\mathbb{N}^2$, its components are labelled by superscripts, e.g.\ 
$m_j=\di{m^1_j}{m^2_j}$. We define
$n=\mathfrak{i}[m]=\mathfrak{i}[k_j]$ and
$m=\mathfrak{o}[n]=\mathfrak{o}[k_j]$. There is a conservation of the
total amount of indices, $\sum_{j=1}^N n_j=\sum_{j=1}^N
m_j$ (as vectors in $\mathbb{N}^2$). An index summation
$\sum_{\mathcal{E}^s}$ is a summation over the graphs with outgoing
indices $\mathcal{E}^s =\{ m_1,\dots,m_s\}$ where
$\mathfrak{i}[m_1],\dots,\mathfrak{i}[m_s]$ are kept fixed. The number
of these summations is restricted by $s\leq V^e+ \iota- 1$.  Due to
the symmetry properties of the propagator one could equivalently sum
over $n_j$ where $\mathfrak{o}[n_j]$ are fixed.

A graph $\gamma$ is produced via a certain \emph{history} of
contractions of (in each step) either two smaller subgraphs (with
fewer vertices) or a self-contraction of a subgraph with two
additional external legs. At a given order $V$ of vertices there are
finitely many graphs (distinguished by their topology and the
permutation of external indices) contributing the part
$A^{(V)\gamma}_{m_1n_1;\dots;m_Nm_N}$ to a function
$A^{(V)}_{m_1n_1;\dots;m_Nm_N}$. It is therefore sufficient to prove
estimations for each $A^{(V)\gamma}_{m_1n_1;\dots;m_Nm_N}$ separately. 

A ribbon graph is called one-particle irreducible (1PI) if it remains
connected when removing a single propagator. The first term on the rhs
of the Polchinski equation (\ref{polL4}) leads always to one-particle
reducible graphs, because it is left disconnected when removing the
propagator $Q_{nm;lk}$ in (\ref{polL4}).

According to the detailed properties a graph $\gamma$, which is
possibly a generalisation of the original ribbon graphs as explaned in
Section~\ref{seccomposite} below, we define the following recursive
procedure (starting with the vertex (\ref{vertex}) which does not have
any subgraphs) to integrate the Polchinski equation (\ref{polL4}):
\begin{dfn}
\label{defint}
We consider generalised\/\footnote{This refers to graphs with
  composite propagators as defined in Section~\ref{seccomposite}.}
ribbon graphs $\gamma$ which result via a history of contractions of
subgraphs which at each contraction step have already been integrated
according to the rules given below.

\begin{enumerate}
  
\item \label{defint4} Let $\gamma$ be a planar ($B=1$, $g=0$) one-particle
  irreducible graph with $N=4$ external legs, where the index along each of
  its trajectories is constant (this includes the two external indices of a
  trajectory and the chain of indices at contracting inner vertices in between
  them). Then, the contribution $A^{(V)\gamma}_{\di{m^1}{m^2}\di{n^1}{n^2};
    \di{n^1}{n^2}\di{k^1}{k^2};\di{k^1}{k^2}\di{l^1}{l^2};
    \di{l^1}{l^2}\di{m^1}{m^2}}[\Lambda]$ (using the natural cyclic order of
  legs of a planar graph) of $\gamma$ to the effective action is integrated as
  follows:
\begin{align}
&
A^{(V)\gamma}_{\di{m^1}{m^2}\di{n^1}{n^2};
\di{n^1}{n^2}\di{k^1}{k^2};\di{k^1}{k^2}\di{l^1}{l^2};
\di{l^1}{l^2}\di{m^1}{m^2}}[\Lambda] 
\nonumber
\\*
&
:= -\int_{\Lambda}^{\Lambda_0} \!\frac{d \Lambda'}{\Lambda'} \;
\Bigg\{ \widehat{\Lambda'\frac{\partial}{\partial \Lambda'}  
A^{(V)\gamma}_{\di{m^1}{m^2}\di{n^1}{n^2};
\di{n^1}{n^2}\di{k^1}{k^2};\di{k^1}{k^2}\di{l^1}{l^2};
\di{l^1}{l^2}\di{m^1}{m^2}}[\Lambda'] }
\nonumber
\\*
&\hspace*{10em} - ~
\parbox{17\unitlength}{\begin{picture}(17,17)
       \put(0,0){\epsfig{file=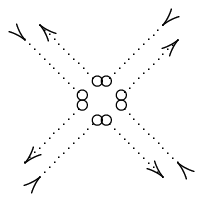,scale=0.9,bb=71 638 117 684}}
       \put(-1,7.5){\mbox{\scriptsize$\di{m^1}{m^2}$}}
       \put(6,0){\mbox{\scriptsize$\di{n^1}{n^2}$}}
       \put(12,7.5){\mbox{\scriptsize$\di{k^1}{k^2}$}}
       \put(6,14){\mbox{\scriptsize$\di{l^1}{l^2}$}}
   \end{picture}}
\cdot\left[\widehat{\Lambda' \frac{\partial}{\partial \Lambda'}  
A^{(V)\gamma}_{\di{0}{0}\di{0}{0};
\di{0}{0}\di{0}{0};\di{0}{0}\di{0}{0};\di{0}{0}\di{0}{0}}
[\Lambda']}\right] \Bigg\}
\nonumber
\\*
&
+ ~~
\parbox{19\unitlength}{\begin{picture}(17,17)
       \put(0,0){\epsfig{file=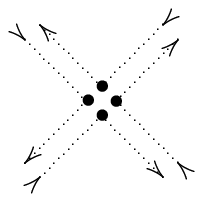,scale=0.9,bb=71 638 117 684}}
       \put(-1,7.5){\mbox{\scriptsize$\di{m^1}{m^2}$}}
       \put(6,0){\mbox{\scriptsize$\di{n^1}{n^2}$}}
       \put(12,7.5){\mbox{\scriptsize$\di{k^1}{k^2}$}}
       \put(6,14){\mbox{\scriptsize$\di{l^1}{l^2}$}}
   \end{picture}}
\left[ 
\int_{\Lambda_R}^\Lambda \!\frac{d \Lambda'}{\Lambda'} \;
\Big(\widehat{\Lambda' \frac{\partial}{\partial \Lambda'}  
A^{(V)\gamma}_{\di{0}{0}\di{0}{0};
\di{0}{0}\di{0}{0};\di{0}{0}\di{0}{0};\di{0}{0}\di{0}{0}}
[\Lambda']}\Big)  + A^{(V)\gamma}_{\di{0}{0}\di{0}{0};
\di{0}{0}\di{0}{0};\di{0}{0}\di{0}{0};\di{0}{0}\di{0}{0}}
[\Lambda_R]\right]\;.
\label{intA4}
\end{align}
Here (and in the sequel), the wide hat over the $\Lambda'$-derivative
of an $A^\gamma$-function indicates that the rhs of the Polchinski
equation (\ref{polL4}) has to be inserted. The two vertices in the
third and fourth lines of (\ref{intA4}) are identical (both are equal
to $1$). The four-leg graph in the third line of (\ref{intA4})
indicates that the graph corresponding to the function in brackets
right of it has to be inserted into the holes. The result is a graph
with the same topology as the function in the second line, but
different indices on inner trajectories. The graph in the fourth line
of (\ref{intA4}) is identical to the original vertex (\ref{vertex}).
The different symbol shall remind us that in the analytic expression
for subgraphs containing the vertex of the last line in (\ref{intA4})
we have to insert the value in brackets right of it.

\vskip 1ex

Remark: We use here (and in all other cases discussed below) the 
convention (its consistency will be shown later) that at
$\Lambda=\Lambda_0$ the contribution to the initial four-point
function is independent of the external matrix indices,
$A^{(V)\gamma}_{\di{0}{0}\di{0}{0};
  \di{0}{0}\di{0}{0};\di{0}{0}\di{0}{0};\di{0}{0}\di{0}{0}}
[\Lambda_0]= A^{(V)\gamma}_{\di{m^1}{m^2}\di{n^1}{n^2};
  \di{n^1}{n^2}\di{k^1}{k^2};\di{k^1}{k^2}\di{l^1}{l^2};
  \di{l^1}{l^2}\di{m^1}{m^2}}[\Lambda_0]$. This is not really
necessary, we could admit $A^{(V)\gamma}_{\di{m^1}{m^2}\di{n^1}{n^2};
  \di{n^1}{n^2}\di{k^1}{k^2};\di{k^1}{k^2}\di{l^1}{l^2};
  \di{l^1}{l^2}\di{m^1}{m^2}}[\Lambda_0]-A^{(V)\gamma}_{\di{0}{0}\di{0}{0};
  \di{0}{0}\di{0}{0};\di{0}{0}\di{0}{0};\di{0}{0}\di{0}{0}}
[\Lambda_0]=C_{\di{m^1}{m^2}\di{n^1}{n^2};
  \di{n^1}{n^2}\di{k^1}{k^2};\di{k^1}{k^2}\di{l^1}{l^2};
  \di{l^1}{l^2}\di{m^1}{m^2}}[\Lambda_0]$ with
$\big|C_{\di{m^1}{m^2}\di{n^1}{n^2};
  \di{n^1}{n^2}\di{k^1}{k^2};\di{k^1}{k^2}\di{l^1}{l^2};
  \di{l^1}{l^2}\di{m^1}{m^2}}[\Lambda_0]\big| \leq
\frac{\mathrm{const}}{\theta \Lambda_0^2}$. 

\vskip 2ex

\item \label{defint0} Let $\gamma$ be a planar ($B=1$, $g=0$) 1PI graph with
  $N=2$ external legs, where either the index is constant along each
  trajectory, or one component of the index jumps\footnote{A jump
    forward and backward means the following: Let $k_1,\dots,k_{a-1}$ be the
    sequence of indices at inner vertices on the considered trajectory
    $\overrightarrow{nm}$, in correct order between $n$ and $m$. Then, for
    either $r=1$ or $r=2$ we require $n^r=k_i^r=m^r$ for all $i \in [1,p{-}1]
    \cup [q,a{-}1]$ and $k_i^r=n^r \pm 1$ (fixed sign) for all $i \in
    [p,q{-}1]$. The cases $p=1$, $q=p{+}1$ and $q=a$ are admitted.  The other
    index component is constant along the trajectory.} once by $\pm 1$ and
  back on one of the trajectories, whereas the index along the possible other
  trajectory remains constant.  Then, the contribution
  $A^{(V)\gamma}_{\di{m^1}{m^2}\di{n^1}{n^2};
    \di{n^1}{n^2}\di{m^1}{m^2}}[\Lambda]$ of $\gamma$ to the effective action
  is integrated as follows:
\begin{align}
&
A^{(V)\gamma}_{\di{m^1}{m^2}\di{n^1}{n^2};
\di{n^1}{n^2}\di{m^1}{m^2}}[\Lambda] 
\nonumber
\\*
& :=  -\int_{\Lambda}^{\Lambda_0} \!\frac{d \Lambda'}{\Lambda'} \;
\left\{ \widehat{\Lambda'\frac{\partial}{\partial \Lambda'}  
A^{(V)\gamma}_{\di{m^1}{m^2}\di{n^1}{n^2};
\di{n^1}{n^2}\di{m^1}{m^2}}[\Lambda']}\right.
\nonumber
\\*
& 
\quad
- ~~\parbox{23\unitlength}{\begin{picture}(20,10)
       \put(0,3){\epsfig{file=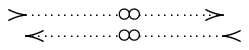,scale=0.9,bb=71 666 130 677}}
       \put(2,0){\mbox{\scriptsize$\di{m^1}{m^2}$}}
       \put(16,0){\mbox{\scriptsize$\di{m^1}{m^2}$}}
       \put(1,8){\mbox{\scriptsize$\di{n^1}{n^2}$}}
       \put(16,8){\mbox{\scriptsize$\di{n^1}{n^2}$}}
   \end{picture}} \cdot
\left[\widehat{\Lambda'\frac{\partial}{\partial \Lambda'}  
A^{(V)\gamma}_{\di{0}{0}\di{0}{0};
\di{0}{0}\di{0}{0}}[\Lambda']} \right.
\nonumber
\\*[-1ex]
& \hspace*{10em} 
+ m^1 \Big(\widehat{\Lambda'\frac{\partial}{\partial \Lambda'}  
A^{(V)\gamma}_{\di{1}{0}\di{0}{0};\di{0}{0}\di{1}{0}}[\Lambda']}
-\widehat{\Lambda'\frac{\partial}{\partial \Lambda'}  
A^{(V)\gamma}_{\di{0}{0}\di{0}{0};\di{0}{0}\di{0}{0}}[\Lambda']}\Big)
\nonumber
\\*
& \hspace*{10em} 
+ m^2 \Big(\widehat{\Lambda'\frac{\partial}{\partial \Lambda'}  
A^{(V)\gamma}_{\di{0}{1}\di{0}{0};\di{0}{0}\di{0}{1}}[\Lambda']}
-\widehat{\Lambda'\frac{\partial}{\partial \Lambda'}  
A^{(V)\gamma}_{\di{0}{0}\di{0}{0};\di{0}{0}\di{0}{0}}[\Lambda']}\Big)
\nonumber
\\*
& \hspace*{10em} 
+ n^1 \Big(\Lambda'\widehat{\frac{\partial}{\partial \Lambda'}  
A^{(V)\gamma}_{\di{0}{0}\di{1}{0};\di{1}{0}\di{0}{0}}[\Lambda']}
-\widehat{\Lambda'\frac{\partial}{\partial \Lambda'}  
A^{(V)\gamma}_{\di{0}{0}\di{0}{0};\di{0}{0}\di{0}{0}}[\Lambda']}\Big)
\nonumber
\\*
& \hspace*{10em} 
+ n^2 \left.\left. \Big(\widehat{\Lambda'\frac{\partial}{\partial \Lambda'}  
A^{(V)\gamma}_{\di{0}{0}\di{0}{1};\di{0}{1}\di{0}{0}}[\Lambda']}
-\widehat{\Lambda'\frac{\partial}{\partial \Lambda'}  
A^{(V)\gamma}_{\di{0}{0}\di{0}{0};\di{0}{0}\di{0}{0}}[\Lambda']}\Big) 
\right]\right\}
\nonumber
\\
& + 
~~\parbox{24\unitlength}{\begin{picture}(20,10)
       \put(0,3){\epsfig{file=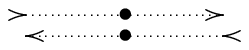,scale=0.9,bb=71 666 130 677}}
       \put(2,0){\mbox{\scriptsize$\di{m^1}{m^2}$}}
       \put(16,0){\mbox{\scriptsize$\di{m^1}{m^2}$}}
       \put(1,8){\mbox{\scriptsize$\di{n^1}{n^2}$}}
       \put(16,8){\mbox{\scriptsize$\di{n^1}{n^2}$}}
   \end{picture}}
\left[
\int_{\Lambda_R}^{\Lambda} \!\frac{d \Lambda'}{\Lambda'} \;
\Big(\widehat{\Lambda'\frac{\partial}{\partial \Lambda'}  
A^{(V)\gamma}_{\di{0}{0}\di{0}{0};\di{0}{0}\di{0}{0}}[\Lambda']} \Big)
+ A^{(V)\gamma}_{\di{0}{0}\di{0}{0};\di{0}{0}\di{0}{0}}[\Lambda_R] 
\right.
\nonumber
\\*
& \hspace*{8em} 
+ m^1 \bigg(\int_{\Lambda_R}^{\Lambda} \!\frac{d \Lambda'}{\Lambda'} \; 
\Big(\widehat{\Lambda'\frac{\partial}{\partial \Lambda'}  
A^{(V)\gamma}_{\di{1}{0}\di{0}{0};\di{0}{0}\di{1}{0}}[\Lambda']}
-\widehat{\Lambda'\frac{\partial}{\partial \Lambda'}  
A^{(V)\gamma}_{\di{0}{0}\di{0}{0};\di{0}{0}\di{0}{0}}[\Lambda']}\Big)
\nonumber
\\*[-0.5ex]
& \hspace*{15em} 
+ A^{(V)\gamma}_{\di{1}{0}\di{0}{0};\di{0}{0}\di{1}{0}}[\Lambda_R]
{-}A^{(V)\gamma}_{\di{0}{0}\di{0}{0};\di{0}{0}\di{0}{0}}[\Lambda_R]\bigg)
\nonumber
\\*
& \hspace*{8em} 
+ m^2 \bigg(\int_{\Lambda_R}^{\Lambda} \!\frac{d \Lambda'}{\Lambda'} \; 
\Big(\widehat{\Lambda'\frac{\partial}{\partial \Lambda'}  
A^{(V)\gamma}_{\di{0}{1}\di{0}{0};\di{0}{0}\di{0}{1}}[\Lambda']}
-\widehat{\Lambda'\frac{\partial}{\partial \Lambda'}  
A^{(V)\gamma}_{\di{0}{0}\di{0}{0};\di{0}{0}\di{0}{0}}[\Lambda']}\Big)
\nonumber
\\*[-0.5ex]
& \hspace*{15em} 
+ A^{(V)\gamma}_{\di{0}{1}\di{0}{0};\di{0}{0}\di{0}{1}}[\Lambda_R]
{-} A^{(V)\gamma}_{\di{0}{0}\di{0}{0};\di{0}{0}\di{0}{0}}[\Lambda_R]\bigg)
\nonumber
\\*
& 
\hspace*{8em} 
+ n^1 \bigg(\int_{\Lambda_R}^{\Lambda} \!\frac{d \Lambda'}{\Lambda'} \; 
\Big(\widehat{\Lambda'\frac{\partial}{\partial \Lambda'}  
A^{(V)\gamma}_{\di{0}{0}\di{1}{0};\di{1}{0}\di{0}{0}}[\Lambda']}
-\widehat{\Lambda'\frac{\partial}{\partial \Lambda'}  
A^{(V)\gamma}_{\di{0}{0}\di{0}{0};\di{0}{0}\di{0}{0}}[\Lambda']}\Big)
\nonumber
\\*[-0.5ex]
& \hspace*{15em} 
+ A^{(V)\gamma}_{\di{0}{0}\di{1}{0};\di{1}{0}\di{0}{0}}[\Lambda_R]
{-}A^{(V)\gamma}_{\di{0}{0}\di{0}{0};\di{0}{0}\di{0}{0}}[\Lambda_R]\bigg)
\nonumber
\\*
& 
\hspace*{8em} + 
n^2 \bigg(\int_{\Lambda_R}^{\Lambda} \!\frac{d \Lambda'}{\Lambda'} \; 
\Big(\widehat{\Lambda'\frac{\partial}{\partial \Lambda'}  
A^{(V)\gamma}_{\di{0}{0}\di{0}{1};\di{0}{1}\di{0}{0}}[\Lambda']}
-\widehat{\Lambda'\frac{\partial}{\partial \Lambda'}  
A^{(V)\gamma}_{\di{0}{0}\di{0}{0};\di{0}{0}\di{0}{0}}[\Lambda']}\Big) 
\nonumber
\\*[-0.5ex]
& \hspace*{15em} 
+ \left.
A^{(V)\gamma}_{\di{0}{0}\di{0}{1};\di{0}{1}\di{0}{0}}[\Lambda_R]
{-}A^{(V)\gamma}_{\di{0}{0}\di{0}{0};\di{0}{0}\di{0}{0}}[\Lambda_R]\bigg) 
\right] .
\label{intA1}
\end{align}

\item \label{defint1} Let $\gamma$ be a planar ($B=1$, $g=0$)
  1PI graph having $N=2$ external legs with external indices
  $m_1n_1;m_2n_2 = \di{m^1 \pm 1}{m^2}\di{n^1 \pm 1}{n^2};
  \di{n^1}{n^2}\di{m^1}{m^2}$ (equal sign) or $m_1n_1;m_2n_2 =
  \di{m^1}{m^2 \pm 1}\di{n^1}{n^2 \pm 1}; \di{n^1}{n^2}\di{m^1}{m^2}$,
  with a single\footnote{For an index arrangement
    $m_1n_1;m_2n_2=\di{m^1 + 1}{m^2}\di{n^1 +1}{n^2};
    \di{n^1}{n^2}\di{m^1}{m^2}$ and sequences $k_1,\dots,k_{a-1}$
    ($l_1,\dots,l_{b-1}$) of indices at inner vertices on the
    trajectory $\overrightarrow{n_1m_2}$ ($\overrightarrow{n_2m_1}$)
    this means that there exist labels $p,q$ with $n^1{+}1=k_i^1$ for
    all $i \in [1,p{-}1]$, $n^1=k_i^1$ for all $i \in [p,a{-}1]$ and
    $n^2=k_i^2$ for all $i \in [1,a{-}1]$ on one trajectory and
    $m^1{+}1=l_j^1$ for all $j \in [q,b{-}1]$, $m^1=k_j^1$ for all $j
    \in [1,q{-}1]$ and $m^2=k_j^2$ for all $j \in [1,b{-}1]$ on the
    other trajectory. The cases $p\in \{1,a\}$ and $q\in\{1,b\}$ are
    admitted.} jump in the index component of each trajectory. Under
  these conditions the contribution of $\gamma$ to the effective
  action is integrated as follows:
\begin{align}
&
A^{(V)\gamma}_{\di{m^1+1}{m^2}\di{n^1+1}{n^2};
\di{n^1}{n^2}\di{m^1}{m^2}}[\Lambda]
\nonumber
\\*
& :=  -\int_{\Lambda}^{\Lambda_0} \!\frac{d \Lambda'}{\Lambda'} \;
\left\{ \widehat{\Lambda' \frac{\partial}{\partial \Lambda'}  
A^{(V)\gamma}_{\di{m^1+1}{m^2}\di{n^1+1}{n^2};
\di{n^1}{n^2}\di{m^1}{m^2}}[\Lambda']} \right.
\nonumber
\\*[-2ex]
&\hspace*{7em} \left.
- \sqrt{(m^1{+}1)(n^1{+}1)} ~~\parbox{24\unitlength}{\begin{picture}(20,10)
       \put(0,3){\epsfig{file=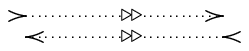,scale=0.9,bb=71 666 130 677}}
       \put(2,0){\mbox{\scriptsize$\di{m^1+1}{m^2}$}}
       \put(16,0){\mbox{\scriptsize$\di{m^1}{m^2}$}}
       \put(1,8){\mbox{\scriptsize$\di{n^1+1}{n^2}$}}
       \put(16,8){\mbox{\scriptsize$\di{n^1}{n^2}$}}
   \end{picture}} \cdot \left[
\widehat{\Lambda'\frac{\partial}{\partial \Lambda'}  
A^{(V)\gamma}_{\di{1}{0}\di{1}{0};\di{0}{0}\di{0}{0}}[\Lambda']} \right]
\right\}
\nonumber
\\*
&  + \sqrt{(m^1{+}1)(n^1{+}1)} ~~\parbox{24\unitlength}{\begin{picture}(20,10)
       \put(0,3){\epsfig{file=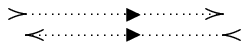,scale=0.9,bb=71 665 130 677}}
       \put(2,0){\mbox{\scriptsize$\di{m^1+1}{m^2}$}}
       \put(16,0){\mbox{\scriptsize$\di{m^1}{m^2}$}}
       \put(1,8){\mbox{\scriptsize$\di{n^1+1}{n^2}$}}
       \put(16,8){\mbox{\scriptsize$\di{n^1}{n^2}$}}
   \end{picture}}
\Bigg[\int_{\Lambda_R}^{\Lambda} \!\frac{d \Lambda'}{\Lambda'} \;
\Big(\widehat{\Lambda' \frac{\partial}{\partial \Lambda'}  
A^{(V)\gamma}_{\di{1}{0}\di{1}{0};\di{0}{0}\di{0}{0}}[\Lambda']} \Big)
\nonumber
\\*
&\hspace*{22em}
+ A^{(V)\gamma}_{\di{1}{0}\di{1}{0};\di{0}{0}\di{0}{0}}[\Lambda_R] 
\Bigg]\,,
\label{intA2}
\\
&
A^{(V)\gamma}_{\di{m^1}{m^2+1}\di{n^1}{n^2+1};
\di{n^1}{n^2}\di{m^1}{m^2}}[\Lambda]
\nonumber
\\*
& :=  -\int_{\Lambda}^{\Lambda_0} \!\frac{d \Lambda'}{\Lambda'} \;
\left\{ \widehat{\Lambda' \frac{\partial}{\partial \Lambda'}  
A^{(V)\gamma}_{\di{m^1}{m^2+1}\di{n^1}{n^2+1};
\di{n^1}{n^2}\di{m^1}{m^2}}[\Lambda']} \right.
\nonumber
\\*[-2ex]
&\hspace*{7em} \left.
- \sqrt{(m^2{+}1)(n^2{+}1)} ~~\parbox{24\unitlength}{\begin{picture}(20,10)
       \put(0,3){\epsfig{file=h03,scale=0.9,bb=71 666 130 677}}
       \put(2,0){\mbox{\scriptsize$\di{m^1}{m^2+1}$}}
       \put(16,0){\mbox{\scriptsize$\di{m^1}{m^2}$}}
       \put(1,8){\mbox{\scriptsize$\di{n^1}{n^2+1}$}}
       \put(16,8){\mbox{\scriptsize$\di{n^1}{n^2}$}}
   \end{picture}} \cdot \left[
\widehat{\Lambda'\frac{\partial}{\partial \Lambda'}  
A^{(V)\gamma}_{\di{0}{1}\di{0}{1};\di{0}{0}\di{0}{0}}[\Lambda']} \right]
\right\}
\nonumber
\\*
&  + \sqrt{(m^2{+}1)(n^2{+}1)} ~~\parbox{24\unitlength}{\begin{picture}(20,10)
       \put(0,3){\epsfig{file=at3b,scale=0.9,bb=71 665 130 677}}
       \put(2,0){\mbox{\scriptsize$\di{m^1}{m^2+1}$}}
       \put(16,0){\mbox{\scriptsize$\di{m^1}{m^2}$}}
       \put(1,8){\mbox{\scriptsize$\di{n^1}{n^2+1}$}}
       \put(16,8){\mbox{\scriptsize$\di{n^1}{n^2}$}}
   \end{picture}}
\Bigg[\int_{\Lambda_R}^{\Lambda} \!\frac{d \Lambda'}{\Lambda'} \;
\Big(\widehat{\Lambda' \frac{\partial}{\partial \Lambda'}  
A^{(V)\gamma}_{\di{0}{1}\di{0}{1};\di{0}{0}\di{0}{0}}[\Lambda']} \Big)
\nonumber
\\*
&\hspace*{22em}
+ A^{(V)\gamma}_{\di{0}{1}\di{0}{1};\di{0}{0}\di{0}{0}}[\Lambda_R] 
\Bigg]\,.
\label{intA3}
\end{align}

\item \label{defintnp}
Let $\gamma$ be any other type of graph. This includes
  non-planar graphs ($B>1$ and/or $g>0$), graphs with $N \geq
  6$ external legs, one-particle reducible graphs, four-point graphs
  with non-constant index along at least one trajectory and two-point
  graphs where the integrated absolute value of the jump along the
  trajectories is bigger than $2$. Then the contribution of $\gamma$
  to the effective action is integrated as follows:
\begin{align}
A^{(V)\gamma}_{m_1n_1;\dots;m_Nn_N} [\Lambda]
& :=  -\int_{\Lambda}^{\Lambda_0} \!\frac{d \Lambda'}{\Lambda'} \;
\Big(\widehat{\Lambda'\frac{\partial}{\partial \Lambda'}  
A^{(V)\gamma}_{m_1n_1;\dots;m_Nn_N}[\Lambda']}\Big) \;.
\end{align}
\end{enumerate}
\end{dfn}

The previous integration procedure identifies the following distinguished
functions $\rho_a[\Lambda,\Lambda_0,\rho^0]$:
\begin{subequations}
\label{rhoa}
\begin{align}
  \rho_1[\Lambda,\Lambda_0,\rho^0] &:= \sum_{\text{$\gamma$ as in
      Def.~\ref{defint}.\ref{defint0}}}
  A^\gamma_{\di{0}{0}\di{0}{0};\di{0}{0}\di{0}{0}}
  [\Lambda,\Lambda_0,\rho^0]\;,
\label{rho1}
\\
\rho_2[\Lambda,\Lambda,\rho^0] &:= \sum_{ \text{$\gamma$ as in 
Def.~\ref{defint}.\ref{defint0}}}
\Big(A^\gamma_{\di{1}{0}\di{0}{0};\di{0}{0}\di{1}{0}}[\Lambda,\Lambda_0,\rho^0]
- A^\gamma_{\di{0}{0}\di{0}{0};\di{0}{0}\di{0}{0}}
[\Lambda,\Lambda_0,\rho^0]\Big)\;,
\label{rho2}
\\
\rho_3[\Lambda,\Lambda_0,\rho^0] 
&:= \sum_{\text{$\gamma$ as in Def.~\ref{defint}.\ref{defint1}}} \Big(-
A^\gamma_{\di{1}{0}\di{1}{0};\di{0}{0}\di{0}{0}} [\Lambda,\Lambda_0,\rho^0]
\Big) \;,
\label{rho3}
\\
\rho_4[\Lambda,\Lambda_0,\rho^0] &:= \sum_{\text{$\gamma$ as in 
Def.~\ref{defint}.\ref{defint4}}}
A^\gamma_{\di{0}{0}\di{0}{0};\di{0}{0}\di{0}{0};
  \di{0}{0}\di{0}{0};\di{0}{0}\di{0}{0}}[\Lambda,\Lambda_0,\rho^0]\;.
\label{rho4}
\end{align}
\end{subequations}
This identification uses the symmetry properties of the $A$-functions
when summed over all contributing graphs. It follows from
Definition~\ref{defint} and (\ref{ct4}) that
\begin{align}
  \rho_a[\Lambda_0,\Lambda_0,\rho^0] \equiv \rho_a^0\;,\qquad a=1,\dots,4\;.
\label{rhoa0}
\end{align}

As part of the renormalisation strategy encoded in
Definition~\ref{defint}, the coefficients (\ref{rhoa}) are kept
constant at $\Lambda=\Lambda_R$. We define
\begin{align}
  \rho_a[\Lambda_R,\Lambda_0,\rho^0] &=0 \qquad \text{for } a=1,2,3\;, &
  \rho_4[\Lambda_R,\Lambda_0,\rho^0] &= \lambda\;.
\label{initrho}
\end{align}
The normalisation (\ref{initrho}) for $\rho_1,\rho_2,\rho_3$
identifies $\Delta^K_{nm;lk}(\Lambda_R)$ as the cut-off propagator
related to the normalised two-point function at $\Lambda_R$. This
entails a normalisation of the mass $\mu_0$, the oscillator frequency
$\Omega$ and the amplitude of the fields $\phi_{mn}$. The
normalisation condition for $\rho_4[\Lambda_R,\Lambda_0,\rho^0]$
defines the coupling constant used in the expansion (\ref{Lg4}).

\subsection{Ribbon graphs with composite propagators}
\label{seccomposite}

It is convenient to write the linear combination of the functions in 
braces $\{~\}$ in (\ref{intA4})--(\ref{intA3}) as a (non-unique)
linear combination of graphs in which we find at least one of the following
composite propagators:
\begin{subequations}
\label{comp-prop}
\begin{align}
&\mathcal{Q}^{(0)}_{\di{m^1}{m^2}\di{n^1}{n^2};\di{n^1}{n^2}\di{m^1}{m^2}}
:= Q_{\di{m^1}{m^2}\di{n^1}{n^2};\di{n^1}{n^2}\di{m^1}{m^2}}
- Q_{\di{0}{0}\di{n^1}{n^2}; \di{n^1}{n^2}\di{0}{0}} 
&= \parbox{23\unitlength}{\begin{picture}(20,8)
       \put(0,2){\epsfig{file=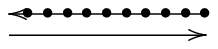,scale=0.9,bb=71 667 132 677}}
       \put(1,-1.5){\mbox{\scriptsize$\di{n^1}{n^2}$}}
       \put(16,-1.5){\mbox{\scriptsize$\di{n^1}{n^2}$}}
       \put(2,7){\mbox{\scriptsize$\di{m^1}{m^2}$}}
       \put(17,7){\mbox{\scriptsize$\di{m^1}{m^2}$}}
\end{picture}}
\label{comp-prop-0}
\\[2ex]
&\mathcal{Q}^{(1)}_{\di{m^1}{m^2}\di{n^1}{n^2};\di{n^1}{n^2}\di{m^1}{m^2}}
\nonumber
\\*
&:= \mathcal{Q}^{(0)}_{\di{m^1}{m^2}\di{n^1}{n^2};\di{n^1}{n^2}\di{m^1}{m^2}}
- m^1 \mathcal{Q}^{(0)}_{\di{1}{0}\di{n^1}{n^2}; \di{n^1}{n^2}\di{1}{0}} 
- m^2 \mathcal{Q}^{(0)}_{\di{0}{1}\di{n^1}{n^2}; \di{n^1}{n^2}\di{0}{1}} 
\hspace*{-1em}
&= \parbox{23\unitlength}{\begin{picture}(20,8)
       \put(0,2){\epsfig{file=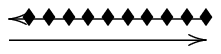,scale=0.9,bb=71 667 131 677}}
       \put(1,-1.5){\mbox{\scriptsize$\di{n^1}{n^2}$}}
       \put(16,-1.5){\mbox{\scriptsize$\di{n^1}{n^2}$}}
       \put(2,7.5){\mbox{\scriptsize$\di{m^1}{m^2}$}}
       \put(17,7.5){\mbox{\scriptsize$\di{m^1}{m^2}$}}
\end{picture}}
\label{comp-prop-1}
\\[2ex]
&\mathcal{Q}^{(+\frac{1}{2})}_{\di{m^1+1}{m^2}\di{n^1+1}{n^2};
\di{n^1}{n^2}\di{m^1}{m^2}}
\nonumber
\\*
&:= Q_{\di{m^1+1}{m^2}\di{n^1+1}{n^2};\di{n^1}{n^2}\di{m^1}{m^2}}
- \sqrt{m^1{+}1} Q_{\di{1}{0}\di{n^1+1}{n^2}; \di{n^1}{n^2}\di{0}{0}} 
&= \parbox{23\unitlength}{\begin{picture}(20,8)
       \put(0,2){\epsfig{file=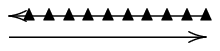,scale=0.9,bb=71 667 132 677}}
       \put(1,-1.5){\mbox{\scriptsize$\di{n^1+1}{n^2}$}}
       \put(16,-1.5){\mbox{\scriptsize$\di{n^1}{n^2}$}}
       \put(2,7){\mbox{\scriptsize$\di{m^1+1}{m^2}$}}
       \put(17,7){\mbox{\scriptsize$\di{m^1}{m^2}$}}
\end{picture}}
\label{comp-prop-p12}
\\[2ex]
&\mathcal{Q}^{(-\frac{1}{2})}_{\di{m^1}{m^2+1}\di{n^1}{n^2+1};
\di{n^1}{n^2}\di{m^1}{m^2}}
\nonumber
\\*[-1ex]
&:= Q_{\di{m^1}{m^2+1}\di{n^1}{n^2+1};\di{n^1}{n^2}\di{m^1}{m^2}}
- \sqrt{m^2{+}1} Q_{\di{0}{1}\di{n^1}{n^2+1}; \di{n^1}{n^2}\di{0}{0}} 
&= \raisebox{1ex}{\parbox{23\unitlength}{\begin{picture}(20,8)
       \put(0,2){\epsfig{file=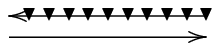,scale=0.9,bb=71 667 132 677}}
       \put(1,-1.5){\mbox{\scriptsize$\di{n^1}{n^2+1}$}}
       \put(16,-1.5){\mbox{\scriptsize$\di{n^1}{n^2}$}}
       \put(2,7){\mbox{\scriptsize$\di{m^1}{m^2+1}$}}
       \put(17,7){\mbox{\scriptsize$\di{m^1}{m^2}$}}
\end{picture}}}
\label{comp-prop-m12}
\end{align}
\end{subequations}
To obtain the linear combination we recall how the graph $\gamma$ under
consideration is produced via a history of contractions and
integrations of subgraphs. For a history $a$-$b$-$\dots$-$n$ ($a$ first) we
have 
\begin{align}
&\widehat{\Lambda \frac{\partial}{\partial \Lambda} A^{(V)\gamma}_{
m_1n_1;\dots;m_Nn_N}}[\Lambda]
\nonumber
\\*
&=\sum_{m_a,n_a,k_a,l_a,\dots,m_n,n_n,k_n,l_n}
\int_{(\Lambda)_A}^{(\Lambda)_B} \!
\frac{d \Lambda_n}{\Lambda_n}
\int_{(\Lambda_{n})_A}^{(\Lambda_{n})_B} \!
\frac{d \Lambda_{n-1}}{\Lambda_{n-1}}
\dots
\int_{(\Lambda_b)_A}^{(\Lambda_b)_B} \!
\frac{d \Lambda_a}{\Lambda_a}
\nonumber
\\*
&\qquad \times 
Q_{m_nn_n;k_nl_n}(\Lambda_n)  \dots Q_{m_bn_b;k_bl_b}(\Lambda_b)  
Q_{m_an_a;k_al_a}(\Lambda_a)\,
V^{m_an_ak_al_a\dots m_nn_nk_nl_n}_{
m_1n_1\dots m_N n_N}\;,
\label{inthist}
\end{align}
where $V^{m_an_ak_al_a\dots m_nn_nk_nl_n}_{ m_1n_1\dots m_N n_N}$ is
the vertex operator and either $(\Lambda_i)_A=\Lambda_i,
(\Lambda_i)_B=\Lambda_0$ or $(\Lambda_i)_A=\Lambda_R,
(\Lambda_i)_B=\Lambda_i$. The graph
$\widehat{\Lambda \frac{\partial}{\partial \Lambda} 
A^{(V)\gamma}_{00;\dots;00}}[\Lambda]$ is obtained via the same
procedure (including the choice of the integration direction), except
that we use the vertex operator $V^{m_an_ak_al_a\dots
  m_nn_nk_nl_n}_{ 00\dots 00}$. This means that all propagator indices
which are not determined by the external indices are the same.
Therefore, we can factor out in the difference of graphs all
completely inner propagators and the integration operations. 

We first consider the difference in (\ref{intA4}). Since $\gamma$ is
one-particle irreducible with constant index on each trajectory, we
get for a certain permutation $\pi$ ensuring the history of integrations
\begin{align}
&\widehat{\Lambda' \frac{\partial}{\partial \Lambda'} 
A^{(V)\gamma}_{mn;nk;kl;lm}[\Lambda']}
-\widehat{\Lambda'\frac{\partial}{\partial \Lambda'} 
A^{(V)\gamma}_{00;00;00;00}[\Lambda']}
\nonumber
\\*
&= \dots \left\{ 
\prod_{i=1}^a Q_{m_{\pi_i}k_{\pi_i};k_{\pi_i}m_{\pi_i}}(\Lambda_{\pi_i}) 
- \prod_{i=1}^a Q_{0k_{\pi_i};k_{\pi_i}0}(\Lambda_{\pi_i}) \right\}
\nonumber
\\*
&= \dots \Bigg\{ \sum_{b=1}^a 
\Big(\prod_{i=1}^{b-1} Q_{0k_{\pi_i};k_{\pi_i}0}(\Lambda_{\pi_i}) \Big)
\mathcal{Q}^{(0)}_{m_{\pi_b}k_{\pi_b};k_{\pi_b}m_{\pi_b}}(\Lambda_{\pi_b}) 
\nonumber
\\*[-1ex]
&\hspace*{10em}  \times  \Big(\prod_{j=b+1}^a Q_{m_{\pi(j)}k_{\pi(j)};
k_{\pi(j)}m_{\pi(j)}}(\Lambda_{\pi(j)}) \Big)
\Bigg\}\;,
\label{decA4}
\end{align}
where $\gamma$ contains $a$ propagators with external indices and
$m_{\pi_i} \in \{m,n,k,l\}$. The parts of the analytic expression
common to both $\widehat{\Lambda'\frac{\partial}{\partial \Lambda'}
A^{(V)\gamma}_{mn;nk;kl;lm}[\Lambda']}$ and $\widehat{\Lambda' 
\frac{\partial}{\partial
  \Lambda'} A^{(V)\gamma}_{00;00;00;00}[\Lambda']}$ are symbolised by the
dots. The $k_{\pi_i}$ are inner indices. We thus learn that the
difference of graphs appearing in the braces in (\ref{intA4}) can be
written as a sum of graphs each one having a composite propagator
(\ref{comp-prop-0}). Of course, the identity (\ref{decA4}) is
nothing but a generalisation of $a^n-b^n=\sum_{k=0}^{n-1} b^k(a{-}b)
a^{n-k-1}$. There are similar identities for the differences appearing
in (\ref{intA1})--(\ref{intA3}). We delegate their derivation to
Appendix~\ref{appcompositeidentity}. In Appendix~\ref{app:graph} we
show how the difference operation works for a concrete example of a
two-leg graph.

\subsection{Bounds for the cut-off propagator}
\label{bounds}

Differentiating the cut-off propagator (\ref{GKDK1}) with respect to
$\Lambda$ and recalling that the cut-off function $K(x)$ is constant
unless $x \in [1,2]$, we notice that for 
our choice $\theta_1=\theta_2\equiv\theta$ the indices are 
restricted as follows:
\begin{align}
 \Lambda \frac{\partial \Delta_{\di{m^1}{m^2}\di{n^1}{n^2};
\di{k^1}{k^2}\di{l^1}{l^2}}^K(\Lambda)}{\partial \Lambda}
& =0 
\nonumber
\\* 
\text{unless}\quad
 \theta\Lambda^2 & \leq \max(m^1,m^2,n^1,n^2,k^1,k^2,l^1,l^2) \leq
  2 \theta\Lambda^2 \;.
\label{Qc4}
\end{align}
In particular, the volume of the support of the differentiated cut-off
propagator (\ref{Qc4}) with respect to a single index $m,n,k,l \in
\mathbb{N}^2$ equals $4\theta^2 \Lambda^4$, which is the correct
normalisation of a four-dimensional model \cite{Grosse:2003aj}.

We compute in Appendix~\ref{appB} the $\Lambda$-dependence of the
maximised propagator $\Delta^{\mathcal{C}}_{mn;kl}$, which is the
application of the sharp cut-off realising the condition (\ref{Qc4})
to the propagator, for selected values of
$\mathcal{C}=\theta\Lambda^2$ and $\Omega$, which is extremely well
reproduced by (\ref{Form4-1}). We thus obtain for the maximum of
(\ref{Q4})
\begin{align}
\max_{m,n,k,l}  \big|Q_{mn;kl}(\Lambda)\big| & \leq
  \frac{1}{2\pi\theta} \,(32 \max_x
  |K'(x)|)\, \max_{m,n,k,l} \big|\Delta^{\mathcal{C}}_{mn;kl}
  \big|_{\mathcal{C} = \Lambda^2\theta }
\nonumber  \\*
  & \leq \left\{ \begin{array}{ll} \displaystyle \frac{C_0}{\Omega
  \Lambda^2 \theta} \, \delta_{m+k,n+l}\qquad & \text{for } \Omega>0\;,
      \\[3ex]
      \displaystyle \frac{C_0}{\sqrt{\Lambda^2\theta}}\, 
\delta_{m+k,n+l} \qquad
  & \text{for } \Omega=0\;,
\end{array}\right.
\label{est4-0}
\end{align}
where $C_0=C_0' \frac{40}{3\pi} \max_x |K'(x)|$.
The constant $C_0' \gtrapprox 1$ corrects the fact that
(\ref{Form4-1}) holds asymptotically only. Next, from (\ref{Form4-2}) we
obtain
\begin{align}
 \max_m \Big(\sum_l \max_{n,k} \big|Q_{mn;kl}(\Lambda)\big| \Big)
& \leq  \frac{1}{2\pi\theta} (32 \max_x
  |K'(x)|)\, \max_m \sum_l \max_{n,k}
 \big|\Delta^{\mathcal{C}}_{mn;kl}
  \big|_{\mathcal{C} =\Lambda^2 \theta}
\nonumber 
  \\*
& \leq \frac{C_1}{\Omega^2 \theta \Lambda^2} \;,
\label{est4-1}
\end{align}
where $C_1=48 C_1' /(7\pi)\, \max_x |K'(x)|$.  The
product of (\ref{est4-0}) by the volume $4 \theta^2 \Lambda^4$ 
of the support of the cut-off propagator with respect to a single
index leads to the following bound:
\begin{align}
\sum_m 
\Big( \max_{n,k,l}  \big|Q_{mn;kl}(\Lambda)\big|\Big)
& \leq \left\{ \begin{array}{ll} \displaystyle  4 C_0
    \frac{\theta \Lambda^2}{\Omega} \qquad 
& \text{for } \Omega>0\;,
      \\[3ex]
      \displaystyle 4C_0  \,\big(\sqrt{\theta}\Lambda\big)^3\, 
\qquad  & \text{for } \Omega=0\;.
\end{array}\right.
\label{est4-2}
\end{align}
According to (\ref{Delta-ss}) there is the
following refinement of the estimation (\ref{est4-0}):
\begin{align}
\Big|Q_{\di{m^1}{m^2}\di{n^1}{n^2};
\di{n^1-a^1}{n^2-a^2}\di{m^1-a^1}{m^2-a^2}}(\Lambda)\Big|_{a^r \geq 0,\, m^r
\leq n^r}
\leq C_{a^1,a^2} 
\Big(\frac{m^1}{\theta \Lambda^2}\Big)^{\frac{a_1}{2}}
\Big(\frac{m^2}{\theta \Lambda^2}\Big)^{\frac{a_2}{2}}
\frac{1}{\Omega \theta \Lambda^2} \;. 
\label{est4-3}
\end{align}
This property will imply that graphs with big total jump along the
trajectories are suppressed, provided that the indices on the
trajectory are ``small''. However, there is a potential danger from
the presence of completely inner vertices, where the index summation
runs over ``large'' indices as well. Fortunately, according to
(\ref{Form4-3}) this case can be controlled by the following property
of the propagator:
\begin{align}
\bigg( \sum_{\mbox{\scriptsize$\begin{array}{c}l \in \mathbb{N}^2 \\
\|m-l\|_1\geq 5\end{array}$}}
\max_{k,n \in \mathbb{N}^2 } 
\big|Q_{\di{m^1}{m^2}\di{n^1}{n^2};
\di{k^1}{k^2}\di{l^1}{l^2}}(\Lambda)\big|\bigg)
\leq C_4 \Big(\frac{\|m\|_\infty{+}1}{\theta \Lambda^2}\Big)^{2}  
\frac{1}{\Omega^2\theta \Lambda^2}\;,
\label{est4-4}
\end{align}
where we have defined the following norms:
\begin{align}
\|m-l\|_1:=\sum_{r=1}^2|m^r-l^r|\;,\qquad
\|m\|_\infty := \max(m^1,m^2)
\qquad \text{if}~
m={\textstyle\di{m^1}{m^2}}\;,~
\quad
l={\textstyle\di{l^1}{l^2}}\;.
\end{align}
Moreover, we define 
\begin{align}
\|m_1n_1;\dots;m_Nn_N\|_\infty := \max_{i=1,\dots,N} 
\big(\|m_i\|_\infty,\|n_i\|_\infty\big)\;.
\end{align}
Finally, we need estimations for the composite propagators (\ref{comp-prop})
and (\ref{comp-prop-rest}):
\begin{align}
\Big| \mathcal{Q}^{(0)}_{\di{m^1}{m^2}\di{n^1}{n^2};
    \di{n^1}{n^2}\di{m^1}{m^2}}(\Lambda) \Big| & 
\leq C_5 \frac{\|m\|_\infty}{\theta \Lambda^2} 
\frac{1}{\Omega \theta \Lambda^2}\;,
\label{Qpl1}
  \\
\Big| \mathcal{Q}^{(1)}_{\di{m^1}{m^2}\di{n^1}{n^2};
    \di{n^1}{n^2}\di{m^1}{m^2}}(\Lambda) \Big| & 
\leq C_6 \Big( \frac{\|m\|_\infty}{\theta \Lambda^2}\Big)^2 
\frac{1}{\Omega \theta  \Lambda^2}\;,
\label{Qpl2}
  \\
\Big| \mathcal{Q}^{(+\frac{1}{2})}_{\di{m^1+1}{m^2}\di{n^1+1}{n^2};
    \di{n^1}{n^2}\di{m^1}{m^2}}(\Lambda) \Big| 
& \leq C_7 \bigg(\frac{\big\|\di{m^1+1}{m^2}\big\|_\infty}{\theta \Lambda^2}
\bigg)^{\frac{3}{2}}
\;\frac{1}{\Omega \theta\Lambda^2}\;.
\label{Qpl3}
\end{align}
These estimations follow from (\ref{Delta-00}) and (\ref{Delta-01}).

\subsection{The power-counting estimation}
\label{pct}

Now we are going to prove a power-counting theorem for the
$\phi^4$-model in the matrix base, generalising the theorem proven in
\cite{Grosse:2003aj}. The generalisation concerns 1PI planar graphs
and their subgraphs. A subgraph of a planar graph has necessarily
genus $g= 0$ and an even number of legs on each boundary component.
We distinguish one boundary component of the subgraph which after a
sequence of contractions will be part of the unique boundary component
of an 1PI planar graph. For a trajectory $\overrightarrow{nm}$ on the
distinguished boundary component, which passes through the indices
$k_1,\dots,k_{a}$ when going from $n$ to $m=\mathfrak{o}[n]$, we
define the total jump as
\begin{align}
\langle\overrightarrow{nm}\rangle:= \|n-k_1\|_1 
+ \Big(\sum_{c=1}^{a-1} \|k_c-k_{c+1}\|_1 \Big) +  \|k_a-m\|_1 \;.
\label{jump}
\end{align}
Clearly, the jump is additive: if we connect two trajectories
$\overrightarrow{nm}$ and $\overrightarrow{mm'}$ to a new trajectory
$\overrightarrow{nm'}$, then $\langle\overrightarrow{nm'}\rangle=
\langle\overrightarrow{nm}\rangle+\langle\overrightarrow{mm'}\rangle$.
We let $T$ be a set of trajectories
$\overrightarrow{n_j\mathfrak{o}[n_j]}$ on the distinguished boundary
component for which we measure the total jump.  By definition, the end
points of a trajectory in $T$ cannot belong to
$\mathcal{E}^s$.

Moreover, we consider a second set $T'$ of $t'$ trajectories
$\overrightarrow{n_j\mathfrak{o}[n_j]}$ of the distinguished boundary
component where one of the end points $m_j$ or $n_j$ is kept fixed and the
other end point is summed over.  However, we require the summation to run over
$\langle \overrightarrow{n_j\mathfrak{o}[n_j]} \rangle \geq 5$ only, see
(\ref{est4-4}). We let $\sum_{\mathcal{E}^{t'}}$ be the corresponding
summation operator.
%Accordingly, we
%have to replace in previous formulae related to Theorem~\ref{thm1} the
%number $s$ of summations by $s{+}t'$.

Additionally, we have to introduce a new notation in order to control
\begin{itemize}
\item the behaviour for large indices and given $\Lambda$,
\item the behaviour for given indices and large $\Lambda$.
\end{itemize}
For this purpose we let $P^a_b\Big[ \frac{m_1n_1;\dots;m_Nn_N}{\theta
  \Lambda^2}\Big]$ denote a function of the indices $m_1,n_1,\dots,$ $m_N,n_N$
and the scale $\Lambda$ which is bounded as follows:
\begin{align}
0 & \leq P^a_b\Big[ \frac{m_1n_1;\dots;m_Nn_N}{\theta \Lambda^2}\Big] \leq  
\left\{ \begin{array}{ll}
C_a M^{a} & \text{for } M \geq 1\;,
\\
C_b M^{b} & \text{for } M \leq 1 \;,
\end{array} \right.
\label{polab}
\\*
M &:=\max_{m_i,n_i \notin \mathcal{E}^s,\mathcal{E}^{t'}}
\Big( \frac{m_1^r+1}{2\theta
      \Lambda^2}, \frac{n_1^r+1}{2\theta \Lambda^2}, \dots,
  \frac{n_N^r+1}{2 \theta \Lambda^2}\Big)  \;,
\nonumber
\end{align}
for some constants $C_a,C_b$. The maximisation over the indices
$m_i^r,n_i^r$ excludes the summation indices $\mathcal{E}^{\prime t}$.
Fixing the indices and varying $\Lambda$ we have
\begin{align}
P^{a-a'}_{b+b'}\Big[ \frac{m_1n_1;\dots;m_Nn_N}{\theta \Lambda^2}\Big]  
\leq P^{a}_{b}\Big[ \frac{m_1n_1;\dots;m_Nn_N}{\theta \Lambda^2}\Big]  \;,
\label{polab-range}
\end{align}
for $0\leq a' \leq a$ and $b'\geq 0$, assuming appropriate $C_a,C_b$.
Moreover,
\begin{align}
P^{a_1}_{b_1}\Big[ \frac{m_1n_1;\dots;m_{N_1}n_{N_2}}{\theta \Lambda^2}\Big]
&P^{a_2}_{b_2}\Big[ \frac{m_{N_1+1}n_{N_1+1};\dots;m_Nn_N}{
\theta \Lambda^2}\Big] 
\nonumber
\\*
& \leq 
P^{a_1+a_2}_{b_1+b_2}\Big[ \frac{m_1n_1;\dots;m_Nn_N}{\theta
  \Lambda^2}\Big] \;.  
\label{Pabring}
\end{align}

We are going to prove:

\begin{prp} \label{power-counting-prop}
  Let $\gamma$ be a ribbon graph having $N$ external legs, $V$
  vertices, $V^e$ external vertices and segmentation index $\iota$,
  which is drawn on a genus-$g$ Riemann surface with $B$ boundary
  components. We require the graph $\gamma$ to be constructed via a
  history of subgraphs and an integration procedure according to
  Definition~\ref{defint}. Then the
  contribution $A^{(V,V^e,B,g,\iota)\gamma}_{m_1n_1;\dots;m_Nn_N}$ of
  $\gamma$ to the expansion coefficient of the effective action
  describing a duality-covariant $\phi^4$-theory on
  $\mathbb{R}^4_\theta$ in the matrix base is bounded as follows:
\begin{enumerate}
\item \label{prp-planar-A4} If $\gamma$ is as in
  Definition~\ref{defint}.\ref{defint4}, we have
\begin{subequations}
\label{prp-A4}
\begin{align}
\Big|   A^{(V,V^e,1,0,0)\gamma}_{
    \di{m^1}{m^2}\di{n^1}{n^2};\di{n^1}{n^2}\di{k^1}{k^2};
    \di{k^1}{k^2}\di{l^1}{l^2};\di{l^1}{l^2}\di{m^1}{m^2}} 
[\Lambda,\Lambda_0,\rho_0]
&- A^{(V,V^e,1,0,0)\gamma}_{\di{0}{0}\di{0}{0};\di{0}{0}\di{0}{0};
    \di{0}{0}\di{0}{0};\di{0}{0}\di{0}{0}}[\Lambda,\Lambda_0,\rho_0]\Big|
\nonumber
\\*
&\hspace*{-13em} \leq 
  P^{4V-N}_1 \bigg[\frac{\di{m^1}{m^2}\di{n^1}{n^2};\di{n^1}{n^2}\di{k^1}{k^2};
    \di{k^1}{k^2}\di{l^1}{l^2};\di{l^1}{l^2}\di{m^1}{m^2}}{ 
\theta \Lambda^2} \bigg] \Big(\frac{1}{\Omega}\Big)^{3V-2-V^e} 
\,P^{2V-2}\Big[\ln\frac{\Lambda}{\Lambda_R}\Big] \;,
\label{hatA4}  
\\
\Big|A^{(V,V^e,1,0,0)\gamma}_{\di{0}{0}\di{0}{0};\di{0}{0}\di{0}{0};
    \di{0}{0}\di{0}{0};\di{0}{0}\di{0}{0}} [\Lambda,\Lambda_0,\rho_0] \Big|
&\leq  \Big(\frac{1}{\Omega}\Big)^{3V-2-V^e} 
P^{2V-2}\Big[\ln\frac{\Lambda}{\Lambda_R}\Big] \;.
\label{hatA4-0}  
\end{align}
\end{subequations}

\item \label{prp-planar-A0}
If $\gamma$ is as in 
  Definition~\ref{defint}.\ref{defint0}, we have 
\begin{subequations}
\label{prp-A0}
\begin{align}
\Big|A^{(V,V^e,1,0,0)\gamma}_{\di{m^1}{m^2}\di{n^1}{n^2};
    \di{n^1}{n^2}\di{m^1}{m^2}} [\Lambda,\Lambda_0,\rho_0] &-
  A^{(V,V^e,1,0,0)\gamma}_{\di{0}{0}\di{0}{0};\di{0}{0}\di{0}{0}} 
[\Lambda,\Lambda_0,\rho_0]
\nonumber
\\* 
&-  m^1 \big(
  A^{(V,V^e,1,0,0)\gamma}_{\di{1}{0}\di{0}{0};\di{0}{0}\di{1}{0}}
[\Lambda,\Lambda_0,\rho_0]  
-A^{(V,V^e,1,0,0)\gamma}_{\di{0}{0}\di{0}{0};\di{0}{0}\di{0}{0}}
[\Lambda,\Lambda_0,\rho_0]  \big) 
\nonumber
\\*
&-  n^1\big(
  A^{(V,V^e,1,0,0)\gamma}_{\di{0}{0}\di{1}{0};\di{1}{0}\di{0}{0}}
[\Lambda,\Lambda_0,\rho_0]  
-A^{(V,V^e,1,0,0)\gamma}_{\di{0}{0}\di{0}{0};\di{0}{0}\di{0}{0}}
[\Lambda,\Lambda_0,\rho_0]  \big) 
\nonumber
\\*
&-  m^2\big(
  A^{(V,V^e,1,0,0)\gamma}_{\di{0}{1}\di{0}{0};\di{0}{0}\di{0}{1}}
[\Lambda,\Lambda_0,\rho_0]  
-A^{(V,V^e,1,0,0)\gamma}_{\di{0}{0}\di{0}{0};\di{0}{0}\di{0}{0}}
[\Lambda,\Lambda_0,\rho_0]  \big) 
\nonumber
\\*
&
- n^2 \big(
  A^{(V,V^e,1,0,0)\gamma}_{\di{0}{0}\di{0}{1};\di{0}{1}\di{0}{0}}
[\Lambda,\Lambda_0,\rho_0]  
-A^{(V,V^e,1,0,0)\gamma}_{\di{0}{0}\di{0}{0};\di{0}{0}\di{0}{0}}
[\Lambda,\Lambda_0,\rho_0]  \big) \Big|
\nonumber
\\*
& \hspace*{-8em}
\leq (\theta \Lambda^2)\, P^{4V-N}_2 \bigg[\frac{\di{m^1}{m^2}\di{n^1}{n^2};
    \di{n^1}{n^2}\di{m^1}{m^2} }{\theta \Lambda^2} \bigg]
\Big(\frac{1}{\Omega}\Big)^{3V-1-V^e} \,
P^{2V-1}\Big[\ln\frac{\Lambda}{\Lambda_R}\Big] \;,
\label{hatA0}
\\
\Big|A^{(V,V^e,1,0,0)\gamma}_{\di{0}{0}\di{0}{0};\di{0}{0}\di{0}{0}}
[\Lambda,\Lambda_0,\rho_0]\Big|
& \leq  (\theta \Lambda^2)\Big(\frac{1}{\Omega}\Big)^{3V-1-V^e} \, 
P^{2V-1}\Big[\ln\frac{\Lambda}{\Lambda_R}\Big] \;,
\label{hatA0-0}
\\
\Big|A^{(V,V^e,1,0,0)\gamma}_{\di{1}{0}\di{0}{0};\di{0}{0}\di{1}{0}}
[\Lambda,\Lambda_0,\rho_0]
&-A^{(V,V^e,1,0,0)\gamma}_{\di{0}{0}\di{0}{0};\di{0}{0}\di{0}{0}}
[\Lambda,\Lambda_0,\rho_0]\Big|
\nonumber
\\*
&\leq \Big(\frac{1}{\Omega}\Big)^{3V-1-V^e} \,
P^{2V-1}\Big[\ln\frac{\Lambda}{\Lambda_R}\Big] \;.
\label{hatA0-1}
\end{align}
\end{subequations}

\item \label{prp-planar-A1} If $\gamma$ is as in
  Definition~\ref{defint}.\ref{defint1}, we have
\begin{subequations}
\label{prp-A1}
\begin{align}
&\Big| A^{(V,V^e,1,0,0)\gamma}_{\di{m^1+1}{m^2}\di{n^1+1}{n^2};
    \di{n^1}{n^2}\di{m^1}{m^2}}[\Lambda,\Lambda_0,\rho_0]  
- \sqrt{(m^1{+}1)(n^1{+}1)}
  A^{(V,V^e,1,0,0)\gamma}_{\di{1}{0}\di{1}{0};\di{0}{0}\di{0}{0} }
[\Lambda,\Lambda_0,\rho_0] \Big|
\nonumber
\\*
&\quad \leq (\theta \Lambda^2)\, 
P^{4V-N}_2 \bigg[\frac{\di{m^1+1}{m^2}\di{n^1+1}{n^2};
    \di{n^1}{n^2}\di{m^1}{m^2} }{\theta \Lambda^2} \bigg]
\Big(\frac{1}{\Omega}\Big)^{3V-1-V^e} \,
P^{2V-1}\Big[\ln\frac{\Lambda}{\Lambda_R}\Big] \;,
\label{hatA1}
\\
& \Big| A^{(V,V^e,1,0,0)\gamma}_{\di{1}{0}\di{1}{0};\di{0}{0}\di{0}{0} }
 [\Lambda,\Lambda_0,\rho_0]\Big| 
\leq  \Big(\frac{1}{\Omega}\Big)^{3V-1-V^e} \,
P^{2V-1}\Big[\ln\frac{\Lambda}{\Lambda_R}\Big] \;.
\label{hatA1-0}
\end{align}
\end{subequations}

\item  \label{prp-sub-planar}
If $\gamma$ is a subgraph of an 1PI planar graph with a selected set $T$ of
  trajectories on one distinguished boundary component and a second set $T'$
  of summed trajectories on that boundary component, we have
\begin{align}
&\sum_{\mathcal{E}^{s}} \sum_{\mathcal{E}^{t'}}
\big|A^{(V,V^e,B,0,\iota)\gamma}_{
    m_1n_1;\dots;m_Nn_N}[\Lambda,\Lambda_0,\rho_0] \big|
\nonumber
  \\*
  &~~
\leq \big(\theta \Lambda^2\big)^{(2-\frac{N}{2})+2(1-B)}
 P^{4V-N}_{\big(2t' + \sum_{\overrightarrow{n_j\mathfrak{o}[n_j]} \in T} 
\min(2, \frac{1}{2} 
\langle \overrightarrow{n_j\mathfrak{o}[n_j]} \rangle )\big)} 
\Big[ \frac{m_1n_1;\dots;m_Nn_N}{\theta \Lambda^2} \Big]
\nonumber
  \\*
  & \qquad \times 
\Big(\frac{1}{\Omega}\Big)^{3V-\frac{N}{2}-1 
+B-V^e-\iota+s+t'}
P^{2V-\frac{N}{2}}\Big[\ln \frac{\Lambda}{\Lambda_R}\Big]\;.
\label{AN4-1plan}
\end{align}

\item \label{prp-non-planar}
If $\gamma$ is a non-planar graph, we have
\begin{align}
&\sum_{\mathcal{E}^{s}} \big|A^{(V,V^e,B,g,\iota)}_{
    m_1n_1;\dots;m_Nn_N}[\Lambda,\Lambda_0,\rho_0] \big| 
\nonumber
  \\*
  &\quad 
\leq \big(\theta \Lambda^2\big)^{(2-\frac{N}{2})+2(1-B-2g)}\,
P^{4V-N}_0
\Big[ \frac{m_1n_1;\dots;m_Nn_N}{\theta \Lambda^2} \Big]
\nonumber
  \\*
  & \qquad \times 
  \Big(\frac{1}{\Omega}\Big)^{3V-\frac{N}{2}-1+B+2g-V^e-\iota+s}\, 
  \,P^{2V-\frac{N}{2}}\Big[\ln \frac{\Lambda}{\Lambda_R}\Big]\;.
\label{AN4-1}
\end{align}

\end{enumerate}

\end{prp}
\textit{Proof.} We prove the Proposition by induction upward in the
vertex order $V$ and for given $V$ downward in the number $N$ of
external legs. 

\begin{enumerate}

\item[\ref{prp-non-planar}.] We start with with the proof for
  non-planar graphs, noticing that due to (\ref{polab-range}) the
  estimations (\ref{prp-A4}), (\ref{prp-A0}), (\ref{prp-A1}) and
  (\ref{AN4-1plan}) can be further bounded by (\ref{AN4-1}). The proof
  of (\ref{AN4-1}) reduces to the proof of the general power-counting
  theorem given in \cite{Grosse:2003aj}, where we have to take for
  $\big(\frac{\mu}{\Lambda}\big)^{\delta_0}$
  $\big(\frac{\mu}{\Lambda}\big)^{\delta_1}$ and
  $\big(\frac{\Lambda}{\mu}\big)^{\delta_2}$ the estimations
  (\ref{est4-0}), (\ref{est4-1}) and (\ref{est4-2}) with both their
  $\Lambda$- and $\Omega$-dependence. Independent of the factor
  (\ref{polab}), the non-planarity of the graph guarantees the
  irrelevance of the corresponding function so that the integration
  according to Definition~\ref{defint} agrees with the procedure used
  in \cite{Grosse:2003aj}. The dependence on
  $\frac{m_i^r}{\theta\Lambda^2},\frac{n_i^r}{\theta\Lambda^2}$
  through (\ref{polab}) is preserved in its
  structure, because for $\omega>0$ we have
\begin{subequations}
\label{CCb}
\begin{align}
    \int^{\Lambda_0}_\Lambda \frac{d\Lambda'}{\Lambda'} 
\frac{C}{\Lambda^{\prime \omega}} P^a_b\Big[\frac{m}{\theta
  \Lambda^{\prime 2}} \Big] \leq 
 \int^{\Lambda_0}_\Lambda \frac{d\Lambda'}{\Lambda'} 
\frac{C}{\Lambda^{\prime \omega}} C_b \Big(\frac{m{+}1}{2\theta
  \Lambda^{\prime 2}}\Big)^b 
\leq \frac{1}{\omega{+}2b} \frac{C}{\Lambda^\omega}  C_b 
\Big(\frac{m{+}1}{2\theta \Lambda^2}\Big)^b
\end{align}
for $m{+}1\leq 2\theta \Lambda^2$ and
 \begin{align}
&    \int^{\Lambda_0}_\Lambda \frac{d\Lambda'}{\Lambda'} 
\frac{C}{\Lambda^{\prime \omega}} P^a_b\Big[\frac{m}{\theta
  \Lambda^{\prime 2}} \Big] 
\nonumber
\\*
&\leq 
\int^{\Lambda_0}_{\sqrt{\frac{m+1}{2\theta}}} \frac{d\Lambda'}{\Lambda'} 
\frac{C}{\Lambda^{\prime \omega}} C_b \Big(\frac{m{+}1}{2\theta
  \Lambda^{\prime 2}}\Big)^b 
+  \int^{\sqrt{\frac{m+1}{2\theta}}}_\Lambda \frac{d\Lambda'}{\Lambda'} 
\frac{C}{\Lambda^{\prime \omega}} C_a \Big(\frac{m{+}1}{2\theta
  \Lambda^{\prime 2}}\Big)^a 
\nonumber
\\*
&\leq \frac{1}{\omega{+}2a} \frac{C}{\Lambda^\omega}  C_a 
\Big(\frac{m{+}1}{2\theta \Lambda^2}\Big)^a 
+ \frac{(\omega{+}2a) C_b{-}(\omega{+}2b) C_a}{(\omega{+}2a)(\omega{+}2b)}
\frac{C}{\big(\frac{m+1}{2\theta}\big)^{\frac{\omega}{2}}} 
\label{CCbb}
  \end{align}
\end{subequations}
for $m{+}1 \geq 2\theta \Lambda^2$. For $(\omega{+}2b) C_a >
(\omega{+}2a) C_b$ we can omit the last term in the second line of
(\ref{CCbb}), and for $(\omega{+}2b) C_a < (\omega{+}2a) C_b$ we
estimate it by $\frac{(\omega{+}2a) C_b{-}(\omega{+}2b)
  C_a}{C_a(\omega+2b)}$ times the first term.  Taking a polynomial in
$\ln \frac{\Lambda}{\Lambda_R}$ into account, the spirit of
(\ref{CCb}) is unchanged according to \cite{Grosse:2003aj}.

The general power-counting theorem in \cite{Grosse:2003aj} uses
analogues of the bounds (\ref{est4-0}) and (\ref{est4-1}) of the
propagator, which do not add factors $\frac{m}{\theta \Lambda^2}$.
Since two legs of the subgraph(s) are contracted, the total $a$-degree
of (\ref{polab}) becomes $4V-N-2$, which due to (\ref{polab-range})
can be regarded as degree $4V-N$, too.

\vskip 1ex 
  
\item[\ref{prp-sub-planar}.] The proof of (\ref{AN4-1plan}) is
  essentially a repetition of the proof of (\ref{AN4-1}), with
  particular care when contracting trajectories on the distinguished
  boundary component.  The verification of the exponents of $(\theta
  \Lambda^2)$, $\frac{1}{\Omega}$ and $\ln \frac{\Lambda}{\Lambda_R}$
  in (\ref{AN4-1plan}) is identical to the proof of (\ref{AN4-1}). 
%  We can thus restrict ourselves in verifying 
  It remains to verify
  the $a,b$-degrees of the factor (\ref{polab}).
  
 We first consider the contraction of two smaller graphs $\gamma_1$
  (left subgraph) and $\gamma_2$ (right subgraph) to the total graph
  $\gamma$.
  \renewcommand\theenumi{\ref{prp-sub-planar}}

\begin{enumerate}
\item We first assume additionally that all indices of the contracting
  propagator are determined (this is the case for $V_1^e{+} V_2^e=V^e$
  and $\iota_1{+} \iota_2 = \iota$), e.g.\
\begin{align}
  \parbox{47\unitlength}{\begin{picture}(40,28)
\put(0,0){\epsfig{file=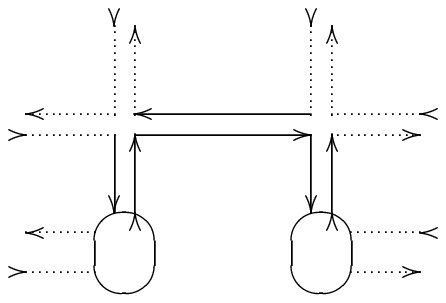,scale=0.9,bb=71 605 187 684}}
\put(6,26){\mbox{\scriptsize$n_1$}}
\put(12.5,24){\mbox{\scriptsize$m_1$}}
\put(0,14.5){\mbox{\scriptsize$n_2$}}
\put(2,20){\mbox{\scriptsize$m_2$}}
\put(0,0.5){\mbox{\scriptsize$\sigma n$}}
\put(2,8){\mbox{\scriptsize$\sigma m$}}
\put(26.5,26){\mbox{\scriptsize$l_1$}}
\put(32.5,24){\mbox{\scriptsize$k_1$}}
\put(36,14){\mbox{\scriptsize$k_2$}}
\put(38,20){\mbox{\scriptsize$l_2$}}
\put(36,0){\mbox{\scriptsize$\sigma k$}}
\put(38,8){\mbox{\scriptsize$\sigma l$}}
\put(12.5,14){\mbox{\scriptsize$m$}}
\put(26,14){\mbox{\scriptsize$l$}}
\end{picture}}
\qquad
\parbox{77\unitlength}{\begin{picture}(70,28)
\put(0,0){\epsfig{file=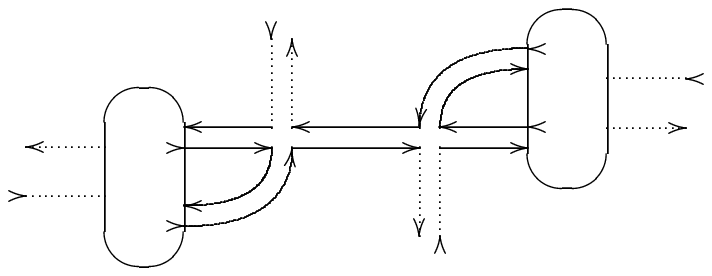,scale=0.9,bb=71 608 263 684}}
\put(22,22){\mbox{\scriptsize$n_1$}}
\put(28,20){\mbox{\scriptsize$m_1$}}
\put(0,5.5){\mbox{\scriptsize$\sigma n$}}
\put(2,13.5){\mbox{\scriptsize$\sigma m$}}
\put(43,4){\mbox{\scriptsize$l_1$}}
\put(36.5,6){\mbox{\scriptsize$k_1$}}
\put(62,12){\mbox{\scriptsize$\sigma k$}}
\put(64,20.5){\mbox{\scriptsize$\sigma l$}}
\put(28.5,10){\mbox{\scriptsize$m$}}
\put(37,16){\mbox{\scriptsize$k$}}
\end{picture}}
\label{graph-aa1}
\raisetag{3ex}
\end{align}
As a subgraph of an 1PI planar graph, at most one side $ml$ or
$m_1l_1$ ($mk_1$ or $m_1k$) of the contracting propagator
$Q_{m_1m;ll_1}$ ($Q_{m_1m;k_1k}$) can belong to a trajectory in $T$. 
In the left graph of (\ref{graph-aa1}) let us assume that the side
$\overrightarrow{ml}$ connects two trajectories
$\overrightarrow{\mathfrak{i}[m]m}\in T_1$ and
$\overrightarrow{l\mathfrak{o}[l]} \in T_2$ to a new trajectory 
$\overrightarrow{\mathfrak{i}[m]\mathfrak{o}[l]} \in T$. The proof for
the small-$\Lambda$ degree $a=4V-N$ in (\ref{AN4-1plan}) is immediate,
because the contraction reduces the number of external legs by $2$ and
we are free to estimate the contracting propagator by its global
maximisation (\ref{est4-0}). Concerning the large-$\Lambda$ degree
$b$, there is nothing to prove if already
$\langle\overrightarrow{\mathfrak{i}[m]m}\rangle +
\langle\overrightarrow{l\mathfrak{o}[l]}\rangle\geq 4$. For
$\langle\overrightarrow{\mathfrak{i}[m]m}\rangle +
\langle\overrightarrow{l\mathfrak{o}[l]}\rangle < 4$ we use the
refined estimation (\ref{est4-3}) for the contracting propagator,
which gives a relative factor $M^{\frac{1}{2}\langle
  \overrightarrow{lm}\rangle}$ compared with (\ref{est4-0}), where
$M=\max\big(\frac{\|m\|+1}{2\theta \Lambda^2}, \frac{\|l\|+1}{2\theta
  \Lambda^2}\big)$. Now, the result follows from (\ref{jump}). Because
of $\langle\overrightarrow{\mathfrak{i}[m]m}\rangle +
\langle\overrightarrow{l\mathfrak{o}[l]}\rangle < 4$, the indices
$m^r,l^r$ from the propagator can be estimated by $\mathfrak{i}[m]^r$
and $\mathfrak{o}[l]^r$. If the resulting jump leads to $\frac{1}{2}
\langle\overrightarrow{\mathfrak{i}[m] \mathfrak{o}[l]}\rangle > 2$,
we use (\ref{polab-range}) to reduce it to $2$. In this way we can
guarantee that the $b$-degree does not exceed
the $a$-degree. Alternatively, if $\langle
\overrightarrow{lm}\rangle \geq 5$ we can avoid a huge
$\langle\overrightarrow{\mathfrak{i}[m]\mathfrak{o}[l]}\rangle$ by
estimating the contracting propagator via (\ref{est4-4}) instead
of\footnote{In this case there is an additional factor
  $\frac{1}{\Omega}$ in (\ref{est4-1}) compared with (\ref{est4-0}).
  It is plausible that this is due to the
  summation, which we do not need here. However, we do not prove a
  corresponding formula without summation. In order to be on a safe
  side, one could replace $\Omega$ in the final estimation
  (\ref{limA4}) by $\Omega^2$. Since $\Omega$ is finite anyway, there
  is no change of the final result. We therefore ignore the
  discrepancy in $\frac{1}{\Omega}$.}  (\ref{est4-3}), because a
single propagator $|Q_{mm_1;l_1l}|$ is clearly smaller than the entire
sum over $\|m-l\|_1 \geq 5$.

If $\overrightarrow{\mathfrak{i}[m]\mathfrak{o}[l]} \in T'$, then the
sum over $\mathfrak{o}[l]$ with $\langle
\overrightarrow{\mathfrak{i}[m]\mathfrak{o}[l]}\rangle \geq 5$ can be
estimated by the combined sum
\begin{itemize}
\item[--] over finitely many combinations of $m,l$ with
  $\max(\langle \overrightarrow{\mathfrak{i}[m]m}\rangle,\langle
  \overrightarrow{ml} \rangle,\langle
  \overrightarrow{l\mathfrak{o}[l]}\rangle)\leq 4$, which via
  (\ref{est4-3}) and the induction hypotheses relative to 
  $T_1,T_2$  contributes a factor $M^{\langle
    \overrightarrow{\mathfrak{i}[m]\mathfrak{o}[l]}\rangle}$ to
  the rhs of (\ref{AN4-1plan}), where
  $M=\max\Big(\frac{\mathfrak{o}[l]^r+1}{\theta \Lambda^2},
  \frac{\mathfrak{i}[m]^r+1}{\theta \Lambda^2},\frac{l^r+1}{\theta
    \Lambda^2}, \frac{m^r+1}{\theta \Lambda^2}\Big)$. We use
  (\ref{polab-range}) to reduce the $b$-degree from $\frac{1}{2}\langle
    \overrightarrow{\mathfrak{i}[m]\mathfrak{o}[l]}\rangle $ to $2$. 

\item[--] over $m$ via the
  induction hypothesis relative to $\overrightarrow{\mathfrak{i}[m]m} 
  \in T_1'$, combined with the usual maximisation (\ref{est4-0}) of the
  contracting propagator and an estimation of $\gamma_2$ where 
  $\overrightarrow{l\mathfrak{o}[l]} \notin T_2,T_2'$,
  
\item[--] over $l$ for fixed $m \approx \mathfrak{i}[m]$,
  via (\ref{est4-4}), taking  $\overrightarrow{\mathfrak{i}[m]m}
  \notin T_1,T_1'$ and $\overrightarrow{l\mathfrak{o}[l]} \notin T_2,T_2'$.
  
\item[--] over $\mathfrak{o}[l]$ for fixed $m$ and $l$, with
  $\mathfrak{i}[m] \approx m \approx l$, via the induction hypothesis
  relative to $\overrightarrow{l\mathfrak{o}[l]} \in T_2'$, the bound
  (\ref{est4-0}) of the propagator and $\overrightarrow{\mathfrak{i}[m]m}
  \notin T_1,T_1'$.
\end{itemize}
A summation over $\mathfrak{i}[m]$ with given $\mathfrak{o}[l]$ is
analogous.

In conclusion, we have proven that the integrand for the graph
$\gamma$ is bounded by (\ref{AN4-1plan}). Since we are dealing with a
$N\geq 6$-point function, the total $\Lambda$-exponent is negative.
Using (\ref{CCb}) we thus obtain the same bound
(\ref{AN4-1plan}) after integration from $\Lambda_0$ down to
$\Lambda$. If $\overrightarrow{m_1l_1} \in T$ or
$\overrightarrow{m_1l_1} \in T'$ we get (\ref{AN4-1plan}) directly
from (\ref{est4-3}) or (\ref{est4-4}).

The discussion of the right graph in (\ref{graph-aa1}) is similar, showing
that the integrand is bounded by (\ref{AN4-1plan}). As long as the integrand
is irrelevant (i.e.\ the total $\Lambda$-exponent is negative), we get
(\ref{AN4-1plan}) after $\Lambda$-integration, too.  However, $\gamma$ might
have two legs only with $\langle\overrightarrow{i(m)m}\rangle +
\langle\overrightarrow{lo(l)}\rangle \leq 2$. In this case the integrand is
marginal or relevant, but according to Definition~\ref{defint}.\ref{defintnp}
we nonetheless integrate from $\Lambda_0$ down to $\Lambda$. We have to take into account that the
cut-off propagator at the scale $\Lambda$ vanishes for $\Lambda^2 \geq
\|m_1m;k_1k\|_\infty /\theta$. Assuming two relevant two-leg subgraphs
$\gamma_1,\gamma_2$ bounded by $\theta \Lambda^{\prime 2}$ times a
polynomial in $\ln \frac{\Lambda'}{\Lambda_R}$ each, we have
\begin{align}
\int_\Lambda^{\Lambda_0} & \frac{d \Lambda'}{\Lambda'} 
\big|Q_{m_1m;k_1k}(\Lambda') A^{\gamma_1}(\Lambda') 
A^{\gamma_2}(\Lambda')\big| 
\nonumber
\\*[-1ex]
&\leq \frac{C_0}{\Omega}
\int_\Lambda^{\sqrt{\|m_1m;k_1k\|_\infty/\theta}} 
\frac{d \Lambda'}{\Lambda'} (\theta \Lambda^{\prime 2})
P^{2V-2}\Big[\ln\frac{\Lambda'}{\Lambda_R}\Big] 
\nonumber
\\*
&\leq \frac{C_0}{\Omega} \|m_1m;k_1k\|_\infty 
P^{2V-2}\bigg[\ln \Big(\frac{\|m_1m;k_1k\|_\infty}{\theta\Lambda_R^2}
\Big)^{\frac{1}{2}}\bigg] 
\nonumber
\\*
& \leq \frac{C_0}{\Omega} 
(\theta \Lambda^2) P^2_1\Big[\frac{m_1m;k_1k}{2\theta\Lambda^2}\Big]
P^{2V-2}\Big[\frac{\Lambda}{\Lambda_R}\Big]\;.
\label{1PR}
\end{align}
Here, I have inserted the estimation (\ref{est4-0}) for the
propagator, restricted to its support. In the logarithm we expanded
$\ln \sqrt{\frac{m}{\theta\Lambda_R^2}}= \ln
\sqrt{\frac{m}{\theta\Lambda^2}}+\ln\frac{\Lambda}{\Lambda_R}$ and
estimated $\big(\ln \sqrt{\frac{m}{\theta\Lambda^2}}\big)^q< c
\frac{m}{2\theta\Lambda^2}$. Thus, the small-$\Lambda$ degree $a$ of
the total graph is increased by 2 over the sum of the small-$\Lambda$
degrees of the subgraphs (taken $=0$ here), in agreement with
(\ref{AN4-1plan}). The estimation for the logarithm is not necessary
for the large-$\Lambda$ degree $b$ in (\ref{polab}). Using
(\ref{polab-range}) we could reduce that degree to $b=0$. We would
like to underline that the integration of 1PR graphs is one of the
sources for the factor (\ref{polab}) in the power-counting theorem.
Taking the factors (\ref{polab}) in the bounds for the subgraphs
$\gamma_i$ into account, the formula modifies accordingly.  We confirm
(\ref{AN4-1plan}) in any case.

It is clear that all other possibilities with determined propagator
indices as discussed in \cite{Grosse:2003aj} are treated
similarly.

\item 
Next, let one index of the contracting propagator be an undetermined summation
index, e.g.\
\begin{align}
\parbox{80\unitlength}{\begin{picture}(70,27)
\put(0,0){\epsfig{file=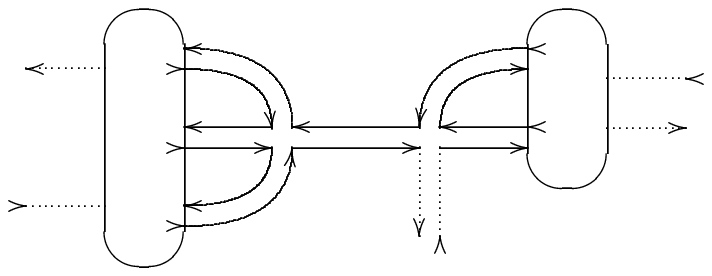,scale=0.9,bb=71 608 263 684}}
\put(0,4){\mbox{\scriptsize$\sigma n$}}
\put(2,22){\mbox{\scriptsize$\sigma m$}}
\put(43,4){\mbox{\scriptsize$l_1$}}
\put(36.5,6){\mbox{\scriptsize$k_1$}}
\put(62,12.5){\mbox{\scriptsize$\sigma k$}}
\put(64,20.5){\mbox{\scriptsize$\sigma l$}}
\put(28.5,10){\mbox{\scriptsize$m$}}
\put(28,16){\mbox{\scriptsize$n$}}
\put(37,16){\mbox{\scriptsize$k$}}
\end{picture}}
\label{aa3-4}
\end{align}

\noindent
Let $\overrightarrow{\mathfrak{i}[k]\mathfrak{o}[n]} \in T$. Then $k$
is determined by the external indices of $\gamma_2$. There is nothing
to prove for $\langle \overrightarrow{\mathfrak{i}[k]k}\rangle \geq
4$. For $\langle \overrightarrow{\mathfrak{i}[k]k}\rangle <4$ we
partitionate the sum over $n$ into $\langle \overrightarrow{nk}\rangle
\leq 4$, where each term yields the integrand (\ref{AN4-1plan}) as
before in the case of determined indices (\ref{graph-aa1}), and the
sum over $\langle \overrightarrow{nk} \rangle \geq 5$, which yields
the desired factor in (\ref{AN4-1plan}) via (\ref{est4-4}) and the
similarity $k \approx \mathfrak{i}[k]$ of indices. As a subgraph of a
planar graph, $m \neq \mathfrak{o}[n]$ in $\gamma_1$, so that a
possible $k_1$-summation can be transferred to $m$.  If
$\overrightarrow{\mathfrak{i}[k]\mathfrak{o}[n]} \in T'$ then in the
same way as for (\ref{graph-aa1}) the summation splits into the four
possibilities related to the pieces
$\overrightarrow{n\mathfrak{o}[n]}$, $\overrightarrow{k n}$ and
$\overrightarrow{\mathfrak{i}[k]k}$, which yield the integrand
(\ref{AN4-1plan}) via the induction hypotheses for the subgraphs and
via (\ref{est4-3}) or (\ref{est4-4}). The $\Lambda$-integration yields
(\ref{AN4-1plan}) via (\ref{CCb}) if the integrand is irrelevant, whereas
we have to perform similar considerations as in (\ref{1PR}) if the
integrand is relevant or marginal.

\item The discussion of graphs with two summation indices on the contracting
propagator, such as in
\begin{align}
  \parbox{82\unitlength}{\begin{picture}(82,29)
\put(0,0){\epsfig{file=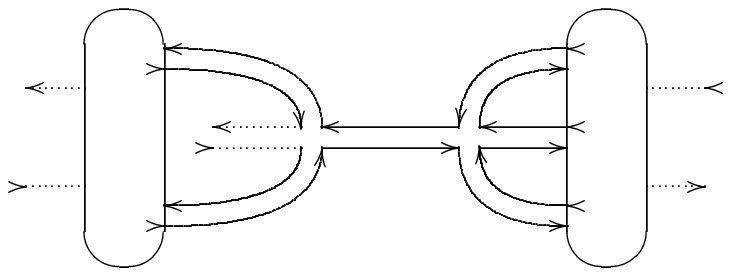,scale=0.9,bb=71 608 269 684}}
\put(1,7){\mbox{\scriptsize$\sigma n$}}
\put(1.5,19.5){\mbox{\scriptsize$\sigma m$}}
\put(19.5,10){\mbox{\scriptsize$n_1$}}
\put(21.5,16){\mbox{\scriptsize$m_1$}}
\put(65.5,6.5){\mbox{\scriptsize$\sigma k$}}
\put(67,19.5){\mbox{\scriptsize$\sigma l$}}
\put(32,10.5){\mbox{\scriptsize$m$}}
\put(32,16){\mbox{\scriptsize$n$}}
\put(41.5,15.5){\mbox{\scriptsize$k$}}
\put(41.5,10){\mbox{\scriptsize$l$}}
\end{picture}}
\end{align}
is similar. Note that the planarity requirement implies $m\neq
\mathfrak{o}[n]$ and $l \neq \mathfrak{i}[k]$.

\item Next, we look at self-contractions of the same vertex of a
  graph. Among the examples discussed in \cite{Grosse:2003aj}
  there are only two possibilities which can appear in subgraphs of
  planar graphs:
\begin{align}
  \parbox{50\unitlength}{\begin{picture}(50,20)
\put(0,0){\epsfig{file=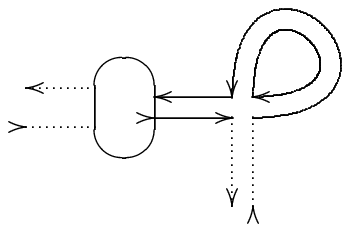,scale=0.9,bb=71 625 164 68}}
\put(0,6.5){\mbox{\scriptsize$\sigma n$}}
\put(2,13.5){\mbox{\scriptsize$\sigma m$}}
\put(17,12.5){\mbox{\scriptsize$n_1$}}
\put(24,13){\mbox{\scriptsize$l$}}
\put(17,3){\mbox{\scriptsize$m_1$}}
\put(24,1){\mbox{\scriptsize$n_1$}}
\end{picture}}
\qquad
\parbox{50\unitlength}{\begin{picture}(50,23)
\put(0,0){\epsfig{file=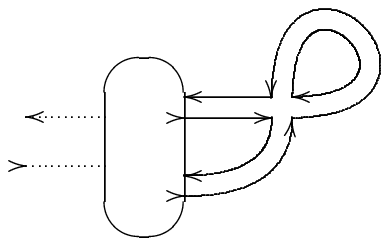,scale=0.9,bb=71 616 175 684}}
\put(0,6){\mbox{\scriptsize$\sigma n$}}
\put(2,13.5){\mbox{\scriptsize$\sigma m$}}
\put(22,15.5){\mbox{\scriptsize$n$}}
\put(28,10){\mbox{\scriptsize$n$}}
\put(28,15.5){\mbox{\scriptsize$l$}}
\end{picture}}
\label{graph-al2}
\end{align}
There is nothing to prove for the left graph in (\ref{graph-al2}). To
verify the large-$\Lambda$ degree $b$ relative to the right graph, we
partitionate the sum over $n$ into
\begin{itemize}
\item $\langle \overrightarrow{\mathfrak{i}[n]n}\rangle \leq 4$ and $\langle
\overrightarrow{n\mathfrak{o}[n]}\rangle \leq 4$, where each term yields
(\ref{AN4-1plan}) via the induction hypothesis for the trajectories
$\overrightarrow{\mathfrak{i}[n]n} \in T_1$ and 
$\overrightarrow{n\mathfrak{o}[n]}\in T_1$ of the
subgraph (in the same way as for the examples with determined propagator
indices), 

\item $\langle \overrightarrow{\mathfrak{i}[n]n}\rangle \leq 4$ and
  $\langle \overrightarrow{n\mathfrak{o}[n]}\rangle \geq 5$, for which
  the induction hypothesis for
  $\overrightarrow{\mathfrak{i}[n]n}\notin T_1,T_1'$ and
  $\overrightarrow{n\mathfrak{o}[n]}\in T_1'$, together with
  $\mathfrak{i}[n]\approx n$, gives a contribution of $2$ to the
  $b$-degree in (\ref{polab}), and 
  
\item $\langle \overrightarrow{\mathfrak{i}[n]n}\rangle \geq 5$, 
  which via the induction hypothesis for
  $\overrightarrow{\mathfrak{i}[n]n}\in T_1'$ and
  $\overrightarrow{n\mathfrak{o}[n]}\notin T_1,T_1'$ gives a
  contribution of $2$ to the $b$-degree in (\ref{polab}).
\end{itemize}
The case $\overrightarrow{\mathfrak{i}[n]\mathfrak{o}[n]} \in T'$ is
similar to discuss. At the end we always arrive at the integrand
(\ref{AN4-1plan}). If it is irrelevant the integration from
$\Lambda_0$ down to $\Lambda$ yields (\ref{AN4-1plan}) according to
(\ref{CCb}). If the integrand is marginal/relevant and $\gamma$ is
one-particle reducible, then the indices of the propagator contracting
1PI subgraphs are of the same order as the incoming and outgoing
indices of the trajectories through the propagator (otherwise the 1PI
subgraphs are irrelevant). Now a procedure similar to (\ref{1PR})
yields (\ref{AN4-1plan}) after integration from $\Lambda_0$ down to
$\Lambda$, too. If $\gamma$ is 1PI and marginal or relevant, it is
actually of the type \ref{defint4}--\ref{defint1} of
Definition~\ref{defint} and will be discussed below.

\addtocounter{enumii}{-1}
\renewcommand\theenumii{.\alph{enumii}}
\refstepcounter{enumii}
\label{contsamevertex}
\renewcommand\theenumii{\alph{enumii}}

\item 
Finally, there will be self-contractions of different vertices of a subgraph,
such as in 
\begin{align}
  \parbox{85\unitlength}{\begin{picture}(55,28)
      \put(0,0){\epsfig{file=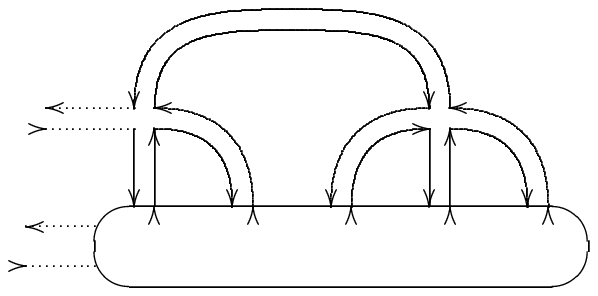,scale=0.9,bb=71 605 235 687}}
      \put(2.5,14){\mbox{\scriptsize$n_1$}}
      \put(4.5,20){\mbox{\scriptsize$m_1$}}
      \put(2.5,8){\mbox{\scriptsize$\sigma m$}}
      \put(1,1){\mbox{\scriptsize$\sigma n$}}
      \put(15,20){\mbox{\scriptsize$m$}}
      \put(45,20){\mbox{\scriptsize$k$}}
      \put(38,20){\mbox{\scriptsize$l$}}
\end{picture}}
\label{graph-ab5}
\end{align}
The vertices to contract have to be situated on the same (distinguished)
boundary component, because the contraction of different boundary components
increases the genus and for contractions of other boundary components the
proof is immediate. Only the large-$\Lambda$ degree $b$ is questionable.

Let $\overrightarrow{\mathfrak{i}[m]\mathfrak{o}[l]}\in T$, with
$\mathfrak{i}[m]\neq \mathfrak{o}[l]$ due to planarity. Here, $m$ is
regarded as a summation index. As before we split that sum over $m$
into a piece with $\langle \overrightarrow{\mathfrak{i}[m]m}\rangle
\leq 4$, which yields the $b$-degree of the integrand
(\ref{AN4-1plan}) term by term via the induction hypothesis relative
to $\overrightarrow{\mathfrak{i}[m]m},
\overrightarrow{l\mathfrak{o}[l]}\in T_1$ and (\ref{est4-3}) for the
contracting propagator, and a piece with $\langle
\overrightarrow{\mathfrak{i}[m]m}\rangle \geq 5$, which gives
(\ref{AN4-1plan}) via the induction hypothesis relative to
$\overrightarrow{\mathfrak{i}[m]m}\in T_1'$ and
$\overrightarrow{l\mathfrak{o}[l]}\notin T_1,T_1'$.  If
$\overrightarrow{\mathfrak{i}[m]\mathfrak{o}[l]}\in T'$ the sum over
$\mathfrak{o}[l]$ with
$\overrightarrow{\mathfrak{i}[m]\mathfrak{o}[l]} \geq 5$ is estimated
by a finite number of combinations of $m,l$ with $\max( \langle
\overrightarrow{\mathfrak{i}[m]m}\rangle, \langle
\overrightarrow{ml}\rangle, \langle
\overrightarrow{l\mathfrak{o}[l]}\rangle)\leq 4$, which yields the
integrand (\ref{AN4-1plan}) via the induction hypothesis for $T_1$ and
(\ref{est4-3}), and the sum over index combinations
\begin{itemize}
\item $\langle \overrightarrow{\mathfrak{i}[m]m}\rangle \leq 4$, 
$\langle \overrightarrow{ml}\rangle \leq 4$, $\langle
\overrightarrow{l\mathfrak{o}[l]}\rangle \geq 5$ 

\item $\langle \overrightarrow{\mathfrak{i}[m]m}\rangle \leq 4$, 
$\langle \overrightarrow{ml}\rangle \geq 5$

\item $\langle \overrightarrow{\mathfrak{i}[m]m}\rangle \geq 5$ 
\end{itemize}
which is controlled by the induction hypothesis relative to $T_1'$ or
(\ref{est4-4}), together with the similarity of trajectory indices at
those parts where the jumps is bounded by $4$. The case where
$\overrightarrow{\mathfrak{i}[k]m_1} \in T$ or
$\overrightarrow{\mathfrak{i}[k]m_1} \in T'$ is easier to treat. We
thus arrive in any case at the estimation (\ref{AN4-1plan}) for the
integrand of $\gamma$, which leads to the same estimation
(\ref{AN4-1plan}) for $\gamma$ itself according to the considerations
at the end of \ref{contsamevertex}. If $\gamma$ is of type
\ref{defint4}--\ref{defint1} of Definition~\ref{defint} we will treat
it below.

The discussion of all other possible self-contractions as listed in
\cite{Grosse:2003aj} is similar.

\end{enumerate}
This finishes the part \ref{prp-sub-planar} of the proof of 
Proposition~\ref{power-counting-prop}.

\vskip 1ex 

\item[\ref{prp-planar-A4}.] Now we consider 1PI planar 4-leg graphs
  $\gamma$ with constant index on each trajectory. If the external
  indices are zero, we get (\ref{hatA4-0}) directly from the general
  power-counting theorem \cite{Grosse:2003aj}, because the integration
  direction used there agrees with
  Definition~\ref{defint}.\ref{defint4}.
  
  For non-zero external indices we decompose the difference
  (\ref{hatA4}) according to (\ref{decA4}) into graphs with composite
  propagators (\ref{comp-prop-0}) bounded by (\ref{Qpl1}). The
  composite propagators appear on one of the trajectories of $\gamma$,
  and as such already on the trajectory of a sequence of subgraphs of
  $\gamma$, starting with some minimal subgraph $\gamma_0$. The
  composite propagator is the contracting propagator for $\gamma_0$.
  Now, the integrand of the minimal subgraph $\gamma_0$ with composite
  propagator is bounded by a factor $C_5' \frac{\|m\|}{\Lambda^2
    \theta}$ times the integrand of the would-be graph $\gamma_0$ with
  ordinary propagator, where $m$ is the index at the trajectory under
  consideration. 

If $\gamma_0$ is irrelevant, the factor $C_5'
  \frac{\|m\|}{\Lambda^2 \theta}$ of the integrand survives according
  to (\ref{CCb}) to the subgraph $\gamma_0$ itself. Otherwise, if
  $\gamma_0$ is relevant or marginal, it is decomposed according to
  \ref{defint4}--\ref{defint1} of Definition~\ref{defint}. Here, the
  last lines of (\ref{intA4})--(\ref{intA3}) are independent of the
  external index $m$ so that in the difference relative to the
  composite propagator these last lines of
  (\ref{intA4})--(\ref{intA3}) cancel identically. There remains the
  first part of (\ref{intA4})--(\ref{intA3}), which is integrated from
  $\Lambda_0$ downward and which is irrelevant by induction. Thus,
  (\ref{CCb}) applies in this case, too, saving the factor $C_5'
  \frac{\|m\|}{\Lambda^2 \theta}$ to $\gamma_0$ in any case.  This
  factor thus appears in the integrand of the subgraph of $\gamma$
  next larger than $\gamma_0$. By iteration of the procedure we obtain
  the additional factor $C_5' \frac{\|m\|}{\Lambda^2 \theta}$ in the
  integrand of the total graph $\gamma$ with composite propagators,
  the $\Lambda$-degree of which being thus reduced by $2$ compared
  with the original graph $\gamma$. Since $\gamma$ itself is a
  marginal graph according to the general power-counting behaviour
  (\ref{AN4-1}), the graph with composite propagator is irrelevant and
  according to Definition~\ref{defint}.\ref{defint4} to be integrated
  from $\Lambda_0$ down to $\Lambda$. This explains (\ref{hatA4}).
  
\vskip 1ex plus 0.4ex 

\item[\ref{prp-planar-A1}.] Similarly, we conclude from the proof of
  (\ref{AN4-1plan}) that the integrands of graphs $\gamma$ according to
  Definition~\ref{defint}.\ref{defint1} are marginal. In particular, we
  immediately confirm (\ref{hatA1-0}).  For non-zero external indices we
  decompose the difference (\ref{hatA1}) according to (\ref{decA2-all}) into
  graphs either with composite propagators (\ref{comp-prop-0}) bounded by
  (\ref{Qpl1}) or with composite propagators
  (\ref{comp-prop-p12})/(\ref{comp-prop-m12}) bounded by (\ref{Qpl3}). In such
  a graph there are---apart from usual propagators with bound
  (\ref{est4-0})/(\ref{est4-1})---two propagators with $a^1+a^2=1$ in
  (\ref{est4-3}) and a composite propagator with bound (\ref{Qpl1}), or one
  propagator with $a^1+a^2=1$ in (\ref{est4-3}) and one composite propagator
  with bound (\ref{Qpl3}). In both cases we get a total factor
  $\frac{\|\di{m^1+1}{m^2}\di{n^1+1}{n^2}\|_\infty^2}{(\theta \Lambda^2)^2}$
  compared with a general planar two-point graph (\ref{AN4-1}).  The detailed
  discussion of the subgraphs is similar as under \ref{prp-planar-A4}.
  
\vskip 1ex plus 0.4ex

\item[\ref{prp-planar-A0}.] Finally, we have to discuss graphs $\gamma$
  according to Definition~\ref{defint}.\ref{defint0}. We first consider the
  case that $\gamma$ has constant index on each trajectory.  It is then clear
  from the proof of (\ref{AN4-1}) that (in particular) at vanishing indices
  the graph $\gamma$ is relevant, which is expressed by (\ref{hatA0-0}). Next,
  the difference (\ref{hatA0-1}) of graphs can as in (\ref{decA4}) be written
  as a sum of graphs with one composite propagator (\ref{comp-prop-0}), the
  bound of which is given by (\ref{Qpl1}). After the treatment of subgraphs as
  described under \ref{prp-planar-A4}, the integrand of each term in the
  linear combination is marginal. According to
  Definition~\ref{defint}.\ref{defint0} we have to integrate these terms from
  $\Lambda_R$ up to $\Lambda$ which agrees with the procedure in
  \cite{Grosse:2003aj} and leads to (\ref{hatA0-1}). Finally, according
  to (\ref{decA1-all}) and (\ref{decA1-b-all}), the linear combination
  constituting the lhs of (\ref{hatA0}) results in a linear combination of
  graphs with either one propagator (\ref{comp-prop-1}) with bound
  (\ref{Qpl2}), or with two propagators (\ref{comp-prop-0}) with bound
  (\ref{Qpl1}). A similar discussion as under \ref{prp-planar-A4} then leads
  to (\ref{hatA0}).
  
  The second case is when one index component jumps once on a trajectory and
  back. According to the proof of (\ref{AN4-1plan}) the integrand of $\gamma$
  at vanishing external indices is marginal.  We regard it nevertheless as
  relevant using the inequality $1 \leq (\theta \Lambda^2) (\theta
  \Lambda_R^2)^{-1}$, where $(\theta \Lambda_R^2)^{-1}$ is some number kept
  constant in our renormalisation procedure. We now obtain (\ref{hatA0-0}).
  Similarly, the integrand relative to the difference (\ref{hatA0-1}) would be
  irrelevant, but is considered as marginal via the same trick. Finally, the
  linear combination constituting the lhs of (\ref{hatA0}) is according to
  (\ref{decA1-all}) and (\ref{decA1-c-all})--(\ref{decA1-c-d}) a linear
  combination of graphs having either two propagators with $a^1+a^2=1$ in
  (\ref{est4-3}) and a composite propagator with bound (\ref{Qpl1}), or one
  propagator with $a^1+a^2=1$ in (\ref{est4-3}) and one composite propagator
  (\ref{comp-prop-p12})/(\ref{comp-prop-m12}) with bound (\ref{Qpl3}). The
  discussion as before would lead to an increased large-$\Lambda$ degrees
  $P^{4V-N}_3$ instead of $P^{4V-N}_2$ in (\ref{hatA0}), which can be reduced
  to $P^{4V-N}_2$ according to (\ref{polab-range}).

\end{enumerate}

\noindent
This finishes the proof of Proposition~\ref{power-counting-prop}.
\hfill $\square$% 

\bigskip

It is now important to realise \cite{Keller:1992ej} that the
estimations (\ref{prp-A4})--(\ref{AN4-1}) of
Proposition~\ref{power-counting-prop} do not make any reference to the
initial scale $\Lambda_0$. Therefore, the estimations
(\ref{prp-A4})--(\ref{AN4-1}), which give finite bounds for the
interaction coefficients with finite external indices, also hold in
the limit $\Lambda_0 \to \infty$. This is the renormalisation of the
duality-covariant noncommutative $\phi^4$-model.

In numerical computations the limit $\Lambda_0 \to \infty$ is
difficult to realise. Taking instead a large but finite $\Lambda_0$,
it is then important to estimate the error and the rate of convergence
as $\Lambda_0$ approaches $\infty$. This type of estimations is the
subject of the next section.

\smallskip

We finish this section with a remark on the freedom of
normalisation conditions. One
of the most important steps in the proof is the integration procedure
for the Polchinski equation given in Definition~\ref{defint}. For
presentational reasons we have chosen the smallest possible set of
graphs to be integrated from $\Lambda_R$ upward. This can easily be
generalised. We could admit in (\ref{intA4}) any planar 1PI four-point
graphs for which the incoming index of each trajectory is equal to the
outgoing index on that trajectory, but with arbitrary jump along the
trajectory.  There is no change of the estimation (\ref{hatA4}),
because (according to
Proposition~\ref{power-counting-prop}.\ref{prp-sub-planar})
$\widehat{\Lambda'\frac{\partial}{\partial\Lambda'}
  A^{(V,V^e,1,0,0)\gamma}_{\di{m^1}{m^2}\di{n^1}{n^2};
    \di{n^1}{n^2}\di{k^1}{k^2};\di{k^1}{k^2}\di{l^1}{l^2};
    \di{l^1}{l^2}\di{m^1}{m^2}}[\Lambda']}$ is already irrelevant for
these graphs, so is the difference in braces in (\ref{intA4}).
Moreover, integrating such an irrelevant graph according to the last
line of (\ref{intA4}) from $\Lambda_R$ upward we obtain a bound
$\frac{1}{(\Lambda_R^2\theta)} P^{2V-2}[\ln
\frac{\Lambda}{\Lambda_R}]$, which agrees with (\ref{hatA4-0}),
because $\Lambda_R^2\theta$ is finite.  Similarly, we can relax the
conditions on the jump along the trajectory in
(\ref{intA1})--(\ref{intA3}). We would then define the
$\rho_a[\Lambda,\Lambda_0,\rho^0]$-functions in (\ref{rhoa}) for that
enlarged set of graphs $\gamma$.

In a second generalisation we could admit one-particle reducible
graphs in \ref{defint4}--\ref{defint1} of Definition~\ref{defint} and
even non-planar graphs with the same
condition on the external indices as in \ref{defint4}--\ref{defint1}
of Definition~\ref{defint}. Since there is no difference in the
power-counting behaviour between non-planar graphs and 
planar graphs with large jump, the discussion is as before. However,
the convergence theorem developed in the next section cannot be
adapted in an easy way to normalisation conditions involving
non-planar graphs. 

In summary, the proposed generalisations constitute different
normalisation conditions of the same duality-covariant $\phi^4$-model.
Passing from one normalisation to another one is a finite
re-normalisation. The invariant characterisation of our model is its
definition via four independent normalisation conditions for the
$\rho$-functions so that at large scales the effective action
approaches (\ref{ct4}).

\section{The convergence theorem}
\label{sec:convergence}

In this section we prove the \emph{convergence} of the coefficients of
the effective action in the limit $\Lambda_0 \to \infty$, relative to
the integration procedure given in Definition~\ref{defint}. This is a
stronger result than the power-counting estimation of
Proposition~\ref{power-counting-prop}, which e.g.\ would be compatible
with bounded oscillations. Additionally, we identify the rate of
convergence of the interaction coefficients.

\subsection{The $\Lambda_0$-dependence of the effective action}

We have to control the $\Lambda_0$-dependence which enters the
effective action via the integration procedure of
Definition~\ref{defint}. There is an explicit dependence via the
integration domain of irrelevant graphs and an implicit dependence
through the normalisation (\ref{initrho}), which requires a carefully
adapted $\Lambda_0$-dependence of $\rho^0_a$. For fixed
$\Lambda=\Lambda_R$ but variable $\Lambda_0$ we consider the identity
\begin{align}
  &L[\phi,\Lambda_R,\Lambda_0',\rho^0[\Lambda_0']]
  -L[\phi,\Lambda_R,\Lambda_0'',\rho^0[\Lambda_0'']] \equiv
  \int_{\Lambda_0''}^{\Lambda_0'} \frac{d\Lambda_0}{\Lambda_0} \,
  \Big(\Lambda_0 \frac{d}{d\Lambda_0}
  L[\phi,\Lambda_R,\Lambda_0,\rho^0[\Lambda_0]]\Big) \nonumber
  \\*
  &\qquad\qquad = \int_{\Lambda_0''}^{\Lambda_0'} \frac{d\Lambda_0}{\Lambda_0}
  \; \Big(\Lambda_0 \frac{\partial
    L[\phi,\Lambda_R,\Lambda_0,\rho^0]}{\partial\Lambda_0} +\sum_{a=1}^4
  \Lambda_0 \frac{d \rho^0_a}{d \Lambda_0} \frac{\partial
    L[\phi,\Lambda_R,\Lambda_0,\rho^0]}{\partial\rho^0_a} \Big)\;.
\label{taut}
\end{align}
The model is defined by fixing the boundary condition for $\rho_b$ at
$\Lambda_R$, i.e.\ by keeping $\rho_b[\Lambda_R,\Lambda_0,\rho^0]
=\text{constant}$:
\begin{align}
  0 = d \rho_b[\Lambda_R,\Lambda_0,\rho^0] &= \frac{\partial
    \rho_b[\Lambda_R,\Lambda_0,\rho^0]}{ \partial \Lambda_0} d\Lambda_0 +
  \sum_{a=1}^4 \frac{\partial \rho_b[\Lambda_R,\Lambda_0,\rho^0]}{ \partial
    \rho^0_a} \frac{d\rho^0_a}{d\Lambda_0} d\Lambda_0\;.
\end{align}
Assuming that we can invert the matrix $\frac{\partial
  \rho_b[\Lambda_R,\Lambda_0,\rho^0]}{ \partial \rho^0_a}$, which is
possible in perturbation theory, we get
\begin{align}
  \frac{d \rho^0_a}{d \Lambda_0} &= - \sum_{b=1}^4 \frac{\partial \rho^0_a}{
    \partial \rho_b[\Lambda_R,\Lambda_0,\rho^0]} \frac{\partial
    \rho_b[\Lambda_R,\Lambda_0,\rho^0]}{\partial \Lambda_0} \;.
\label{rhoLam}
\end{align}
Inserting (\ref{rhoLam}) into (\ref{taut}) we obtain
\begin{align}
  &L[\phi,\Lambda_R,\Lambda_0',\rho^0[\Lambda_0']] -
  L[\phi,\Lambda_R,\Lambda_0'',\rho^0[\Lambda_0'']] =
  \int_{\Lambda_0''}^{\Lambda_0'} \frac{d\Lambda_0}{\Lambda_0} \;
  R[\phi,\Lambda_R,\Lambda_0,\rho^0[\Lambda_0]]\;,
\label{Rlim}
\end{align}
with
\begin{align}
  R[\phi,\Lambda,\Lambda_0,\rho^0] &:= \Lambda_0 \frac{\partial
    L[\phi,\Lambda,\Lambda_0,\rho^0]}{ \partial \Lambda_0} \nonumber
  \\*
  &- \sum_{a,b=1}^4 \frac{\partial L[\phi,\Lambda,\Lambda_0,\rho^0]}{ \partial
    \rho_a^0} \frac{\partial \rho_a^0}{\partial
    \rho_b[\Lambda,\Lambda_0,\rho^0]} \Lambda_0 \frac{\partial
    \rho_b[\Lambda,\Lambda_0,\rho^0]}{ \partial \Lambda_0}\;.
\label{Rdef}
\end{align}

Following \cite{Polchinski:1983gv} we differentiate (\ref{Rdef}) with
respect to $\Lambda$:
\begin{align}
  \Lambda \frac{\partial R}{\partial \Lambda} &= \Lambda_0
  \frac{\partial}{\partial \Lambda_0} \Big( \Lambda \frac{\partial L}{\partial
    \Lambda}\Big) - \sum_{a,b=1}^4 \frac{\partial}{\partial \rho_{a}^0}
  \Big(\Lambda \frac{\partial L}{\partial \Lambda}\Big) \frac{\partial
    \rho_{a}^0}{\partial \rho_{b}} \Lambda_0 \frac{\partial \rho_{b}}{\partial
    \Lambda_0} \nonumber
  \\*
  & + \sum_{a,b,c,d=1}^4 \frac{\partial L}{\partial \rho_{a}^0} \frac{\partial
    \rho_{a}^0}{\partial \rho_{b}} \frac{\partial }{\partial \rho_{c}^0}
  \Big(\Lambda \frac{\partial \rho_{b}}{\partial \Lambda}\Big) \frac{\partial
    \rho_{c}^0}{\partial \rho_{d}} \Lambda_0 \frac{\partial \rho_{d}}{\partial
    \Lambda_0} - \sum_{a,b=1}^4 \frac{\partial L}{\partial \rho_{a}^0}
  \frac{\partial \rho_{a}^0}{\partial \rho_{b}} \Lambda_0
  \frac{\partial}{\partial \Lambda_0} \Big( \Lambda \frac{\partial
    \rho_{b}}{\partial \Lambda}\Big)\;.
\label{VV}
\end{align}
We have omitted the dependencies for simplicity and made use of the
fact that the derivatives with respect to $\Lambda,\Lambda_0,\rho^0$
commute.  Using (\ref{polL}), with $\theta_1=\theta_2\equiv \theta$, we
compute the terms on the rhs of (\ref{VV}):
\begin{align}
  \Lambda_0 \frac{\partial}{\partial \Lambda_0 }
& \Big( \Lambda \frac{\partial
    L[\phi,\Lambda,\Lambda_0,\rho^0]}{\partial \Lambda} \Big) \nonumber
  \\*
  &= \sum_{m,n,k,l} \frac{1}{2} \Lambda \frac{\partial
    \Delta^K_{nm;lk}(\Lambda)}{\partial \Lambda} \bigg( 2 \frac{\partial
    L[\phi,\Lambda,\Lambda_0,\rho^0]}{ \partial \phi_{mn}}
  \frac{\partial}{\partial \phi_{kl}} \Big( \Lambda_0 \frac{\partial}{\partial
    \Lambda_0} L[\phi,\Lambda,\Lambda_0,\rho^0]\Big) \nonumber
  \\*
  &\hspace*{6em} - \frac{1}{(2\pi\theta)^2} \Big[\frac{\partial^2}{ \partial
    \phi_{mn}\,\partial \phi_{kl}}\Big( \Lambda_0 \frac{\partial}{\partial
    \Lambda_0} L[\phi,\Lambda,\Lambda_0,\rho^0]\Big)\Big]_\phi \bigg)
\nonumber
  \\*
  &
  \equiv M\Big[L,\Lambda_0 \frac{\partial L}{\partial \Lambda_0} \Big] \;.
\label{V2}
\end{align}
Similarly, we have
\begin{align}
  \frac{\partial}{\partial \rho_a^0}\Big( \Lambda \frac{\partial
    L[\phi,\Lambda,\Lambda_0,\rho^0]}{\partial \Lambda} \Big) &=
  M\Big[L,\frac{\partial L}{\partial \rho_a^0}\Big]\;.
\label{V2a}
\end{align}
For (\ref{VV}) we also need the function $\Lambda_0 \frac{\partial}{\partial
  \Lambda_0} \big( \Lambda \frac{\partial \rho_{b}}{\partial \Lambda}\big)$,
which is obtained from (\ref{V2}) by first expanding $L$ on the lhs according
to (\ref{Lg4}) and by further choosing the indices at the $A$-coefficients
according to (\ref{rhoa}).  Applying these operations to the rhs of
(\ref{V2}), we obtain for $U\mapsto \Lambda_0 \frac{\partial L}{\partial
  \Lambda_0}$ or $U \mapsto \frac{\partial L}{\partial \rho_a^0}$ the
expansions
\begin{align}
  M[L, U]
  &= \sum_{N=2}^\infty \sum_{m_i,n_i\in \mathbb{N}^2} \frac{1}{N!}
  M_{m_1n_1;\dots;m_Nn_N}[L,U] \phi_{m_1n_1}\cdots \phi_{m_Nn_N}
\label{MU}
\end{align}
and the projections
\begin{subequations}
\label{MUa}
\begin{align}
  M_1[L,U]&:= \sum_{\text{$\gamma$ as in 
      Def.~\ref{defint}.\ref{defint0}}}
  M^\gamma_{\di{0}{0}\di{0}{0};\di{0}{0}\di{0}{0}}[L,U]\;,
  \\*
  M_2[L,U] &:= \sum_{\text{$\gamma$ as in 
      Def.~\ref{defint}.\ref{defint0}}}
  \big(M^\gamma_{\di{1}{0}\di{0}{0};\di{0}{0}\di{1}{0}}[L,U] -
  M^\gamma_{\di{0}{0}\di{0}{0};\di{0}{0}\di{0}{0}}[L,U]\big)\;,
  \\
  M_3[M,U] &:= \sum_{\text{$\gamma$ as in 
      Def.~\ref{defint}.\ref{defint1}}} \big(-
  M^\gamma_{\di{1}{0}\di{1}{0};\di{0}{0}\di{0}{0}}[L,U]\big)\;,
  \\*
  M_4[L,U] &:= \sum_{\text{$\gamma$ as in
      Def.~\ref{defint}.\ref{defint4}}}
  M^\gamma_{\di{0}{0}\di{0}{0};\di{0}{0}\di{0}{0};
    \di{0}{0}\di{0}{0};\di{0}{0}\di{0}{0}}[L,U]\;.
\end{align}
\end{subequations}
Since the graphs $\gamma$ in (\ref{MUa}) are one-particle irreducible,
only the third line of (\ref{V2}) can contribute\footnote{If
  one-particle reducible graphs are included in the normalisation
  conditions as discussed at the end of
  Section~\ref{sec:est-int-coeff}, also the second line of (\ref{V2})
  must be taken into account.} to $M_a$.  Using (\ref{V2}),
(\ref{V2a}) and (\ref{MUa}) as well as the linearity of $M[L,U]$ in
the second argument we can rewrite (\ref{VV}) as
\begin{align}
  \Lambda \frac{\partial R}{\partial \Lambda} = M[L,R] - \sum_{a=1}^4
  \frac{\partial L}{\partial \rho_a} M_a[L,R] \;,
\label{VVV}
\end{align}
where the function
\begin{align}
  \frac{\partial L}{\partial \rho_a} [\Lambda,\Lambda_0,\rho^0] &:=
  \sum_{b=1}^4 \frac{\partial L[\Lambda,\Lambda_0,\rho^0]}{ \partial \rho_b^0}
  \frac{\partial \rho_b^0}{\partial \rho_a[\Lambda,\Lambda_0,\rho^0]}
\label{Lr}
\end{align}
scales according to
\begin{align}
  \Lambda \frac{\partial}{\partial \Lambda} \Big( \frac{\partial L}{\partial
    \rho_a} \Big) &= M\Big[L, \frac{\partial L}{\partial \rho_a}\Big] -
  \sum_{b=1}^4 \frac{\partial L}{\partial \rho_b} M_b\Big[L, \frac{\partial
    L}{\partial \rho_a}\Big]\;,
\label{LLr}
\end{align}
as a similar calculation shows.

Next, we also expand (\ref{Rdef}) and (\ref{Lr}) as power series in the
coupling constant:
\begin{align}
&  \frac{\partial L}{\partial \rho_a} [\phi,\Lambda,\Lambda_0,\rho^0] 
\nonumber
\\*
&=
  \sum_{V =0}^\infty \lambda^V \sum_{N=2}^{2V+4}
  \frac{(2\pi\theta)^{\frac{N}{2}-2}}{N!}  \sum_{m_i,n_i}
  H^{a(V)}_{m_1n_1;\dots;m_Nn_N}[\Lambda,\Lambda_0,\rho^0] \phi_{m_1n_1}\cdots
  \phi_{m_Nn_N}\;,
  \\
&  R[\phi,\Lambda,\Lambda_0,\rho^0] 
\nonumber
\\*
&= \sum_{V =1}^\infty \lambda^V
  \sum_{N=2}^{2V+2} \frac{(2\pi\theta)^{\frac{N}{2}-2}}{N!} \sum_{m_i,n_i}
  R^{(V)}_{m_1n_1;\dots;m_Nn_N}[\Lambda,\Lambda_0,\rho^0] \phi_{m_1n_1}\cdots
  \phi_{m_Nn_N}\;.
\end{align}
The differential equations (\ref{LLr}) and (\ref{VVV}) can now with (\ref{MU})
and (\ref{MUa}) be written as
\begin{align}
&  \Lambda \frac{\partial}{\partial \Lambda} H^{a(V)}_{m_1n_1;\dots;m_Nn_N}[
  \Lambda,\Lambda_0,\rho^0] \nonumber
  \\*
  &= \bigg\{\sum_{N_1=2}^N \sum_{V_1=1}^{V} \sum_{m,n,k,l} Q_{nm;lk}(\Lambda)
  A^{(V_1)}_{m_1n_1;\dots;m_{N_1-1}n_{N_1-1};mn}[\Lambda]
  H^{a(V-V_1)}_{m_{N_1}n_{N_1};\dots;m_{N}n_{N};kl}[\Lambda] \nonumber
  \\*
  &\quad + \Big(\binom{N}{N_1{-}1} -1\Big) \text{ permutations } \bigg\} -
  \sum_{m,n,k,l} \frac{1}{2} Q_{nm;lk}(\Lambda)
  H^{a(V)}_{m_1n_1;\dots;m_{N}n_{N};mn;kl}[\Lambda] \nonumber
  \\
  & - \sum_{V_1=0}^{V} H^{1(V-V_1)}_{m_1n_1;\dots;m_{N}n_{N}}[\Lambda] \bigg\{
  -\frac{1}{2} \sum_{m,n,k,l} Q_{nm;lk}(\Lambda)
  H^{a(V_1)}_{\di{0}{0}\di{0}{0};\di{0}{0}\di{0}{0};mn;kl}[\Lambda]
  \bigg\}_{[\text{Def.~\ref{defint}.\ref{defint0}}]} \nonumber
  \\
  & - \sum_{V_1=0}^{V} H^{2(V-V_1)}_{m_1n_1;\dots;m_{N}n_{N}}[\Lambda] 
\nonumber
\\*[-1ex]
& \qquad\qquad \times 
\bigg\{
  {-}\frac{1}{2} \sum_{m,n,k,l}\!\! Q_{nm;lk}(\Lambda)
  \Big(H^{a(V_1)}_{\di{1}{0}\di{0}{0};\di{0}{0}\di{1}{0};mn;kl}[\Lambda]
  -H^{a(V_1)}_{\di{0}{0}\di{0}{0};\di{0}{0}\di{0}{0};mn;kl}[\Lambda]\Big)
  \!\bigg\}_{[\text{Def.~\ref{defint}.\ref{defint0}}]} \nonumber
  \\
  & + \sum_{V_1=0}^{V} H^{3(V-V_1)}_{m_1n_1;\dots;m_{N}n_{N}}[\Lambda] \bigg\{
  -\frac{1}{2} \sum_{m,n,k,l} Q_{nm;lk}(\Lambda)
  H^{a(V_1)}_{\di{1}{0}\di{1}{0};\di{0}{0}\di{0}{0};mn;kl}[\Lambda]
  \bigg\}_{[\text{Def.~\ref{defint}.\ref{defint1}}]} \nonumber
  \\
  & - \sum_{V_1=1}^{V} H^{4(V-V_1)}_{m_1n_1;\dots;m_{N}n_{N}}[\Lambda] \bigg\{
  -\frac{1}{2} \sum_{m,n,k,l} Q_{nm;lk}(\Lambda)
  H^{a(V_1)}_{\di{0}{0}\di{0}{0};\di{0}{0}\di{0}{0};
    \di{0}{0}\di{0}{0};\di{0}{0}\di{0}{0};mn;kl}[\Lambda]
  \bigg\}_{[\text{Def.~\ref{defint}.\ref{defint4}}]} \;,
\label{polB4}
\\
&\Lambda \frac{\partial}{\partial \Lambda}  R^{(V)}_{m_1n_1;\dots;m_Nn_N}[
\Lambda,\Lambda_0,\rho^0] \nonumber
\\*
&= \bigg\{\sum_{N_1=2}^N \sum_{V_1=1}^{V-1} \sum_{m,n,k,l} Q_{nm;lk}(\Lambda)
A^{(V_1)}_{m_1n_1;\dots;m_{N_1-1}n_{N_1-1};mn}[\Lambda]
R^{(V-V_1)}_{m_{N_1}n_{N_1};\dots;m_{N}n_{N};kl}[\Lambda] \nonumber
\\*
&\quad + \Big(\binom{N}{N_1{-}1} -1\Big) \text{ permutations } \bigg\} -
\sum_{m,n,k,l} \frac{1}{2} Q_{nm;lk}(\Lambda)
R^{(V)}_{m_1n_1;\dots;m_{N}n_{N};mn;kl}[\Lambda] \nonumber
\\
& - \sum_{V_1=1}^{V} H^{1(V-V_1)}_{m_1n_1;\dots;m_{N}n_{N}}[\Lambda] \bigg\{
-\frac{1}{2} \sum_{m,n,k,l} Q_{nm;lk}(\Lambda)
R^{(V_1)}_{\di{0}{0}\di{0}{0};\di{0}{0}\di{0}{0};mn;kl}[\Lambda]
\bigg\}_{[\text{Def.~\ref{defint}.\ref{defint0}}]} \nonumber
\\
& - \sum_{V_1=1}^{V} H^{2(V-V_1)}_{m_1n_1;\dots;m_{N}n_{N}}[\Lambda] 
\nonumber
\\*[-1ex]
& \qquad\qquad \times 
\bigg\{
{-}\frac{1}{2} \sum_{m,n,k,l}\! Q_{nm;lk}(\Lambda)
\Big(R^{(V_1)}_{\di{1}{0}\di{0}{0};\di{0}{0}\di{1}{0};mn;kl}[\Lambda]
-R^{(V_1)}_{\di{0}{0}\di{0}{0};\di{0}{0}\di{0}{0};mn;kl}[\Lambda]\Big)
\bigg\}_{[\text{Def.~\ref{defint}.\ref{defint0}}]} \nonumber
\\
& + \sum_{V_1=1}^{V} H^{3(V-V_1)}_{m_1n_1;\dots;m_{N}n_{N}}[\Lambda] \bigg\{
-\frac{1}{2} \sum_{m,n,k,l} Q_{nm;lk}(\Lambda)
R^{(V_1)}_{\di{1}{0}\di{1}{0};\di{0}{0}\di{0}{0};mn;kl}[\Lambda]
\bigg\}_{[\text{Def.~\ref{defint}.\ref{defint1}}]} \nonumber
\\
& - \sum_{V_1=1}^{V} H^{4(V-V_1)}_{m_1n_1;\dots;m_{N}n_{N}}[\Lambda] \bigg\{
-\frac{1}{2} \sum_{m,n,k,l} Q_{nm;lk}(\Lambda)
R^{(V_1)}_{\di{0}{0}\di{0}{0};\di{0}{0}\di{0}{0};
  \di{0}{0}\di{0}{0};\di{0}{0}\di{0}{0};mn;kl}[\Lambda]
\bigg\}_{[\text{Def.~\ref{defint}.\ref{defint4}}]} \;.
\label{polV4}
\end{align}
We have used several times symmetry properties of the expansion
coefficients and of the propagator and the fact that to the 1PI
projections (\ref{MUa}) only the last line of (\ref{MU}) can
contribute. By $\{\dots\}_{[\text{Def.~\ref{defint}.?.}]}$ we
understand the restriction to $H$-graphs and $R$-graphs, respectively,
which satisfy the index criteria on the trajectories as given in
Definition~\ref{defint}. The $H$-graphs will be constructed later in
Section~\ref{secHinit}. The $R$-graphs are in their structure identical
to the previously considered graphs for the $A$-functions, but have a 
different meaning. See Section~\ref{graphR}.

\subsection{Initial data and graphs for the auxiliary functions}
\label{secHinit}

Next, we derive the bounds for the $H$-functions. Inserting
(\ref{ct4}) into the definition (\ref{Lr}) we obtain immediately the initial
condition at $\Lambda=\Lambda_0$:
\begin{align}
&  H^{1(V)}_{m_1n_1;\dots;m_Nn_N}[\Lambda_0,\Lambda_0,\rho^0] = \delta_{N2}
  \delta^{V0} \delta_{n_1m_2}\delta_{n_2m_1}\;,
\label{initrho1}
\\
&H^{2(V)}_{m_1n_1;\dots;m_Nn_N}[\Lambda_0,\Lambda_0,\rho^0] = \delta_{N2}
\delta^{V0} (m_1^1{+}n_1^1{+}m_1^2{+}n_1^2) \delta_{n_1m_2}\delta_{n_2m_1}\;,
\label{initrho2}
\\
&H^{3(V)}_{m_1n_1;\dots;m_Nn_N}[\Lambda_0,\Lambda_0,\rho^0] 
\nonumber
\\*
& = -\delta_{N2}
\delta^{V0} \Big( \big(\sqrt{n_1^1m_1^1}
\delta_{n_1^1,m_2^1+1}\delta_{n_2^1+1,m_1^1} 
+\sqrt{n_2^1m_2^1}
\delta_{n_1^1,m_2^1-1}\delta_{n_2^1-1,m_1^1}\big) 
\delta_{n_1^2,m_2^2}\delta_{n_2^2,m_1^2} \nonumber
\\
& + (\sqrt{n_1^2m_1^2}
\delta_{n_1^2,m_2^2+1}\delta_{n_2^2+1,m_1^2} 
+\sqrt{n_2^2m_2^2}
\delta_{n_1^2,m_2^2-1}\delta_{n_2^2-1,m_1^2}) 
\delta_{n_1^1,m_2^1}\delta_{n_2^1,m_1^1}\Big)\;,
\label{initrho3}
\\
&H^{4(V)}_{m_1n_1;\dots;m_Nn_N}[\Lambda_0,\Lambda_0,\rho^0] = \delta_{N4}
\delta^{V0} \Big( \frac{1}{6}
\delta_{n_1m_2}\delta_{n_2m_3}\delta_{n_3m_4}\delta_{n_4m_1} 
+ 5\;\text{permutations}\,\Big)\,.
\label{initrho4}
\end{align}

We first compute $H^{a(0)}_{m_1n_1;\dots;m_4n_4}[\Lambda]$ for
$a\in\{1,2,3\}$.  Since there is no 6-point function at order 0 in $V$, the
differential equation (\ref{polB4}) reduces to
\begin{align}
  &\Lambda \frac{\partial H^{a(0)}_{m_1n_1;\dots;m_4n_4} [\Lambda]}{\partial
    \Lambda} \nonumber
  \\*
  &= \sum_{b=1}^3 \sum_{m,n,k,l,m',n',k',l'} C_b^{m'n';k'l'}
  H^{b(0)}_{m_1n_1;\dots;m_4n_4} [\Lambda]\Big( Q_{nm;lk}(\Lambda)
  H^{a(0)}_{m'n';k'l';mn;kl}[\Lambda] \Big)\;,
\end{align}
for certain coefficients $ C_b^{m'n';k'l'}$. The solution is due to
(\ref{Q4}) given by
\begin{align}
  &H^{a(0)}_{m_1n_1;\dots;m_4n_4} [\Lambda] 
= H^{a(0)}_{m_1n_1;\dots;m_4n_4} [\Lambda_0] \nonumber
  \\*
  &+  \sum_{b=1}^3 \sum_{m,n,k,l,m',n',k',l'} C_b^{m'n';k'l'}
  H^{b(0)}_{m_1n_1;\dots;m_4n_4}[\Lambda_0] \nonumber
  \\*
  &\quad \times \Big( \big(\Delta^K_{nm;lk}(\Lambda) -
  \Delta^K_{nm;lk}(\Lambda_0) \big) H^{a(0)}_{m'n';k'l';mn;kl}[\Lambda_0]\Big)
  \nonumber
  \\*
  &+ \sum_{b=1}^3 \sum_{m,n,k,l,m',n',k',l'} C_b^{m'n';k'l'}
  H^{b(0)}_{m_1n_1;\dots;m_4n_4}[\Lambda_0] \nonumber
  \\*
  &\quad \times \Big( \big(\Delta^K_{nm;lk}(\Lambda) -
  \Delta^K_{nm;lk}(\Lambda_0) \big) 
\nonumber
\\*
&\qquad\qquad \times \sum_{b'=1}^3
  \sum_{m'',n'',k'',l'',m''',n''',k''',l'''} C_{b'}^{m'''n''';k'''l'''}
  H^{b'(0)}_{m'n';k'l';mn;kl}[\Lambda_0]\Big) \nonumber
  \\*
  &\quad\quad \times \Big( \big(\Delta^K_{n''m'';l''k''}(\Lambda) -
  \Delta^K_{n''m'';l''k''}(\Lambda_0) \big)
  H^{a(0)}_{m'''n''';k'''l''';m''n'';k''l''}[\Lambda_0]\Big) \nonumber
  \\*
  &\quad + \dots\;.
\end{align}
With the initial conditions (\ref{initrho1})--(\ref{initrho3}) we get
\begin{align}
  H^{a(0)}_{m_1n_1;\dots;m_4n_4} [\Lambda,\Lambda_0,\rho^0] \equiv 0 \qquad
  \text{for } a \in \{1,2,3\}\;.
\label{H0-24}
\end{align}
Inserting (\ref{H0-24}) into the rhs of (\ref{polB4}) we see that
$H^{a(0)}_{m_1n_1;m_2n_2}$ for $ a \in \{1,2,3\}$ and
$H^{4(0)}_{m_1n_1;\dots;m_4n_4}$ are constant, which means that the relations
(\ref{initrho1})--(\ref{initrho4}) hold actually at any value $\Lambda$ and
not only at $\Lambda=\Lambda_0$.

We need a graphical notation for the $H$-functions. We represent the
base functions (\ref{initrho1})--(\ref{initrho4}), valid for any
$\Lambda$, as follows:
\begin{align}
  H^{1(0)}_{\di{m^1}{m^2}\di{n^1}{n^2};\di{n^1}{n^2}\di{m^1}{m^2}} [\Lambda]
  &= \quad \parbox{24\unitlength}{\begin{picture}(20,10)
      \put(0,3){\epsfig{file=h01,scale=0.9,bb=71 666 130 677}}
      \put(2,0){\mbox{\scriptsize$\di{m^1}{m^2}$}}
      \put(16,0){\mbox{\scriptsize$\di{m^1}{m^2}$}}
      \put(1,8){\mbox{\scriptsize$\di{n^1}{n^2}$}}
      \put(16,8){\mbox{\scriptsize$\di{n^1}{n^2}$}}
      \put(10,8){\mbox{\scriptsize$1$}}
   \end{picture}}
\label{h01}
\\[2ex]
H^{2(0)}_{\di{m^1}{m^2}\di{n^1}{n^2};\di{n^1}{n^2}\di{m^1}{m^2}} [\Lambda] &=
\quad \parbox{24\unitlength}{\begin{picture}(20,10) \put(0,3){\epsfig{file=h01,scale=0.9,bb=71
        666 130 677}} \put(2,0){\mbox{\scriptsize$\di{m^1}{m^2}$}}
    \put(16,0){\mbox{\scriptsize$\di{m^1}{m^2}$}}
    \put(1,8){\mbox{\scriptsize$\di{n^1}{n^2}$}}
    \put(16,8){\mbox{\scriptsize$\di{n^1}{n^2}$}}
    \put(10,8){\mbox{\scriptsize$2$}}
   \end{picture}}
\label{h02}
\\[2ex]
H^{3(0)}_{\di{m^1+1}{m^2}\di{n^1+1}{n^2};\di{n^1}{n^2}\di{m^1}{m^2}} [\Lambda]
&= \quad \parbox{24\unitlength}{\begin{picture}(20,10) \put(0,3){\epsfig{file=h03,scale=0.9,bb=71
        666 130 677}} \put(2,0){\mbox{\scriptsize$\di{m^1+1}{m^2}$}}
    \put(16,0){\mbox{\scriptsize$\di{m^1}{m^2}$}}
    \put(1,8){\mbox{\scriptsize$\di{n^1+1}{n^2}$}}
    \put(16,8){\mbox{\scriptsize$\di{n^1}{n^2}$}}
    \put(10,8){\mbox{\scriptsize$3^1$}}
   \end{picture}}
\label{h03}
\\[1ex]
H^{4(0)}_{\di{m^1}{m^2}\di{n^1}{n^2};\di{n^1}{n^2}\di{k^1}{k^2};
  \di{k^1}{k^2}\di{l^1}{l^2};\di{l^1}{l^2}\di{m^1}{m^2}} [\Lambda] &=
\frac{1}{6} \quad \parbox{19\unitlength}{\begin{picture}(17,17)
    \put(0,0){\epsfig{file=h04,scale=0.9,bb=71 638 117 684}}
    \put(-1,7.5){\mbox{\scriptsize$\di{m^1}{m^2}$}}
    \put(6,0){\mbox{\scriptsize$\di{n^1}{n^2}$}}
    \put(12,7.5){\mbox{\scriptsize$\di{k^1}{k^2}$}}
    \put(6,14){\mbox{\scriptsize$\di{l^1}{l^2}$}}
   \end{picture}}
+ \text{ 5 permutations}  \;.
 \label{h04}
\end{align}
The special vertices stand for some sort of hole into which we can
insert planar two- or four-point functions at vanishing external
indices. However, the graph remains connected at these holes, in
particular, there is index conservation at the hole ${\circ}\!{\circ}$
and a jump by $\di{1}{0}$ or $\di{0}{1}$ at the hole
\mbox{\small${\triangleright}\!\!\:{\triangleright}$}.

By repeated contraction with $A$-graphs and self-contractions we build
out of (\ref{h01})--(\ref{h04}) more complicated graphs with holes.
We use this method to compute $H^{4(0)}_{m_1n_1;m_2n_2}[\Lambda]$. 
In this case, we need the planar and non-planar self-contractions of
(\ref{h04}):
\begin{align}
&  \sum_{m,n,k,l} Q_{mn;kl} H^{4(0)}_{m_1n_1;m_2n_2;mn;kl} 
\nonumber
\\*
&= \sum_l \left(\frac{1}{3} ~~
  \parbox{18\unitlength}{\begin{picture}(18,25) 
      \put(0,0){\epsfig{file=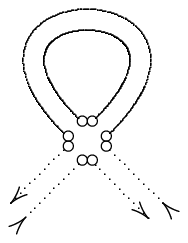,scale=0.9,bb=96 613 143 675}} 
      \put(-2,5){\mbox{\scriptsize$m_1$}}
      \put(3,-1){\mbox{\scriptsize$n_1$}} 
      \put(9,0){\mbox{\scriptsize$m_2$}}
      \put(16,3){\mbox{\scriptsize$n_2$}} 
      \put(7,13){\mbox{\scriptsize$l$}}
   \end{picture}}
+ \frac{1}{3} ~~ \parbox{20\unitlength}{\begin{picture}(20,25)
     \put(0,0){\epsfig{file=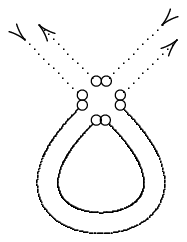,scale=0.9,bb=96 622 143 684}}
     \put(-1,17){\mbox{\scriptsize$n_1$}} 
     \put(5,20){\mbox{\scriptsize$m_1$}}
     \put(10,21){\mbox{\scriptsize$n_2$}} 
     \put(16,17){\mbox{\scriptsize$m_2$}}
     \put(7,8.5){\mbox{\scriptsize$l$}}
   \end{picture}} \right)
+ \frac{1}{3} ~~ \parbox{20\unitlength}{\begin{picture}(20,25)
     \put(0,0){\epsfig{file=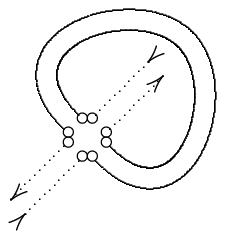,scale=0.9,bb=79 625 141 686}}
     \put(-2,5){\mbox{\scriptsize$m_1$}} \put(3,-1){\mbox{\scriptsize$n_1$}}
     \put(13,9){\mbox{\scriptsize$m_2$}} \put(10,16){\mbox{\scriptsize$n_2$}}
   \end{picture}}
\label{h041}
\end{align}
These contractions correspond (with a factor $-\frac{1}{2}$) to last
term in the third line of (\ref{polB4}), for $a=4$ and $N=2$. We also
have to subtract (again up to the factor $-\frac{1}{2}$) the fourth to
last lines of (\ref{polB4}). For instance, the fourth line amounts to
insert the \emph{planar graphs} of (\ref{h041}) with
$m_1=n_1=m_2=n_2=\di{0}{0}$ into (\ref{h01}). The total contribution
to the rhs of (\ref{polB4}) corresponding to the first graph in
(\ref{h041}), with $m_1=n_2=\di{m^1}{m^2}$ and
$n_1=m_2=\di{n^1}{n^2}$, reads
\begin{align}
 \sum_l & \left\{ \frac{1}{3} ~~ \parbox{20\unitlength}{\begin{picture}(20,25)
      \put(0,0){\epsfig{file=h041,scale=0.9,bb=96 613 143 675}}
      \put(-3,5){\mbox{\scriptsize$\di{m^1}{m^2}$}}
      \put(6,0){\mbox{\scriptsize$\di{n^1}{n^2}$}}
      \put(16,4){\mbox{\scriptsize$\di{m^1}{m^2}$}}
      \put(7,13){\mbox{\scriptsize$\di{l^1}{l^2}$}}
   \end{picture}}
 - \frac{1}{3} \quad \parbox{20\unitlength}{\begin{picture}(20,25)
     \put(0,0){\epsfig{file=h041,scale=0.9,bb=96 613 143 675}}
     \put(-3,5){\mbox{\scriptsize$\di{m^1}{m^2}$}}
     \put(0,11){\mbox{\scriptsize$\di{0}{0}$}}
     \put(14,10){\mbox{\scriptsize$\di{0}{0}$}}
     \put(6,0){\mbox{\scriptsize$\di{n^1}{n^2}$}}
     \put(16,4){\mbox{\scriptsize$\di{m^1}{m^2}$}}
     \put(7,13){\mbox{\scriptsize$\di{l^1}{l^2}$}}
   \end{picture}} \right.
- \frac{1}{3} m^1 \left(~~ \parbox{20\unitlength}{\begin{picture}(20,25)
       \put(0,0){\epsfig{file=h041,scale=0.9,bb=96 613 143 675}}
       \put(-3,5){\mbox{\scriptsize$\di{m^1}{m^2}$}}
       \put(0,11){\mbox{\scriptsize$\di{1}{0}$}}
       \put(14,10){\mbox{\scriptsize$\di{1}{0}$}}
       \put(6,0){\mbox{\scriptsize$\di{n^1}{n^2}$}}
       \put(16,4){\mbox{\scriptsize$\di{m^1}{m^2}$}}
       \put(7,13){\mbox{\scriptsize$\di{l^1}{l^2}$}}
   \end{picture}} 
 - \quad \parbox{20\unitlength}{\begin{picture}(20,25)
     \put(0,0){\epsfig{file=h041,scale=0.9,bb=96 613 143 675}}
     \put(-3,5){\mbox{\scriptsize$\di{m^1}{m^2}$}}
     \put(0,11){\mbox{\scriptsize$\di{0}{0}$}}
     \put(14,10){\mbox{\scriptsize$\di{0}{0}$}}
     \put(6,0){\mbox{\scriptsize$\di{n^1}{n^2}$}}
     \put(16,4){\mbox{\scriptsize$\di{m^1}{m^2}$}}
     \put(7,13){\mbox{\scriptsize$\di{l^1}{l^2}$}}
   \end{picture}}\right)
 \nonumber
 \\*
&\left. 
- \frac{1}{3} m^2 \left(~~ \parbox{20\unitlength}{\begin{picture}(20,25)
      \put(0,0){\epsfig{file=h041,scale=0.9,bb=96 613 143 675}}
      \put(-3,5){\mbox{\scriptsize$\di{m^1}{m^2}$}}
      \put(0,11){\mbox{\scriptsize$\di{0}{1}$}}
      \put(14,10){\mbox{\scriptsize$\di{0}{1}$}}
      \put(6,0){\mbox{\scriptsize$\di{n^1}{n^2}$}}
      \put(16,4){\mbox{\scriptsize$\di{m^1}{m^2}$}}
      \put(7,13){\mbox{\scriptsize$\di{l^1}{l^2}$}}
   \end{picture}} 
 - \quad \parbox{20\unitlength}{\begin{picture}(20,25)
     \put(0,0){\epsfig{file=h041,scale=0.9,bb=96 613 143 675}}
     \put(-3,5){\mbox{\scriptsize$\di{m^1}{m^2}$}}
     \put(0,11){\mbox{\scriptsize$\di{0}{0}$}}
     \put(14,10){\mbox{\scriptsize$\di{0}{0}$}}
     \put(6,0){\mbox{\scriptsize$\di{n^1}{n^2}$}}
     \put(16,4){\mbox{\scriptsize$\di{m^1}{m^2}$}}
     \put(7,13){\mbox{\scriptsize$\di{l^1}{l^2}$}}
   \end{picture}}\right) \right\}
\nonumber
\\*
&= \parbox{20\unitlength}{\begin{picture}(20,25) \put(0,0){\epsfig{file=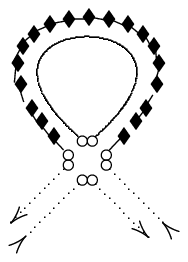,scale=0.9,bb=96 608
        144 676}} \put(-3,5){\mbox{\scriptsize$\di{m^1}{m^2}$}}
    \put(6,0){\mbox{\scriptsize$\di{n^1}{n^2}$}}
    \put(16,4){\mbox{\scriptsize$\di{m^1}{m^2}$}}
    \put(7,13){\mbox{\scriptsize$\di{l^1}{l^2}$}}
   \end{picture}}
\label{h044}
\end{align}
The second graph in the first line of (\ref{h044}) corresponds to the
fourth line of (\ref{polB4}). The second line of (\ref{h044})
represents the fifth/sixth lines of (\ref{polB4}), undoing the
symmetry properties of the upper and lower component used in
(\ref{polB4}). The difference of graphs corresponding to the
$\di{n^1}{n^2}$ component vanishes, because the value of the graph is
independent of $\di{n^1}{n^2}$. There is no planar contribution from
the last two lines of (\ref{polB4}). In total, we get the projection
(\ref{Qpl2}) to the irrelevant part of the graph. The same procedure
leads to the irrelevant part of the second graph in (\ref{h041}).

With these considerations, the differential equation (\ref{polB4})
takes for $N=2$ and $a=4$ the form
\begin{align}
  \Lambda \frac{\partial}{\partial \Lambda} H^{4(0)}_{m_1n_1;m_2n_2}[\Lambda]
  &= -\frac{1}{6} \delta_{n_1m_2}\delta_{n_2m_1} \Big(\sum_{l\in \mathbb{N}^2
    } \mathcal{Q}^{(1)}_{n_1l;ln_1} (\Lambda) + \sum_{l\in \mathbb{N}^2 }
  \mathcal{Q}^{(1)}_{n_2l;ln_2} (\Lambda) \Big) 
\nonumber
\\*
& - \frac{1}{6} Q_{m_1n_1;m_2n_2}(\Lambda)\;.
\label{H0-42}
\end{align}
The first line comes from the planar graphs in (\ref{h041}) and the
subtraction terms according to (\ref{h044}), whereas the second line
of (\ref{H0-42}) is obtained from the last (non-planar) graph in
(\ref{h041}). Using the initial condition (\ref{initrho4}) at
$\Lambda=\Lambda_0$, the bounds (\ref{est4-0}) and (\ref{Qpl2})
combined with the volume factor $(C_2 \theta \Lambda^2)^2$ for the
$l$-summation we get
\begin{align}
  |H^{4(0)}_{m_1n_1;m_2n_2}[\Lambda]| \leq \frac{C
    \big(\|m_1\|_\infty^2+\|n_1\|_\infty^2\big)}{ \Omega \theta \Lambda^2 }\,
  \delta_{n_1m_2}\delta_{n_2m_1} +
\frac{C_0}{ \Omega \theta \Lambda^2 }\,
  \delta_{n_1m_2}\delta_{n_2m_1} \;.
\label{H0-42N}
\end{align}
It is extremely important here that the irrelevant projection
$\mathcal{Q}^{(1)}_{nl;ln}$ and not the propagator $Q_{nl;ln}$ itself
appears in the first line of (\ref{H0-42}).

\subsection{The power-counting behaviour of the auxiliary functions}

The example suggests that similar cancellations of relevant and marginal parts
appear in general, too. Thus, we expect all $H$-functions to be irrelevant.
This is indeed the case:
\begin{prp} 
\label{prop-H}
Let $\gamma$ be a ribbon graph with holes having $N$ external legs,
$V$ vertices, $V^e$ external vertices and segmentation index $\iota$,
which is drawn on a genus-$g$ Riemann surface with $B$ boundary
components.  Then, the contribution
$H^{a(V,V^e,B,g,\iota)\gamma}_{m_1n_1;\dots;m_Nn_N}$ of $\gamma$ to
the expansion coefficient of the auxiliary function of a
duality-covariant $\phi^4$-theory on $\mathbb{R}^4_\theta$ in the
matrix base is bounded as follows:
\begin{enumerate}
 \item \label{H-planar-H4}
For $\gamma$ according to Definition~\ref{defint}.\ref{defint4} we have 
 \begin{align}
& \Big| \sum_{\gamma \textup{ as in Def.~\ref{defint}.\ref{defint4}}}  
H^{a(V,V^e,1,0,0)\gamma}_{\di{m^1}{m^2}\di{n^1}{n^2};
\di{n^1}{n^2}\di{k^1}{k^2}; \di{k^1}{k^2}\di{l^1}{l^2};
\di{l^1}{l^2}\di{m^1}{m^2}} [\Lambda,\Lambda_0,\rho_0] \Big|
 \nonumber
 \\*
 &\leq \big(\theta \Lambda^2\big)^{-\delta^{1a}}
P_{1-\delta^{V0}}^{4V-2+2\delta^{a4}}
\bigg[\frac{\di{m^1}{m^2}\di{n^1}{n^2};\di{n^1}{n^2}\di{k^1}{k^2};
     \di{k^1}{k^2}\di{l^1}{l^2};\di{l^1}{l^2}\di{m^1}{m^2}}{ 
 \theta \Lambda^2} \bigg]\Big(\frac{1}{\Omega}\Big)^{3V-1+\delta^{a4}-V^e}
 \nonumber
 \\*
&\qquad \times P^{2V-1+\delta^{a4}}\Big[\ln\frac{\Lambda}{\Lambda_R}\Big] \;,
 \label{H-H4}  
\end{align}
where all vertices on the trajectories contribute to $V^e$.

\item \label{H-planar-H0}
For $\gamma$ according to Definition~\ref{defint}.\ref{defint0} we have 
\begin{align}
& \Big| \sum_{\gamma \textup{ as in 
    Def.~\ref{defint}.\ref{defint0}}}  H^{a(V,V^e,1,0,0)\gamma}_{
\di{m^1}{m^2}\di{n^1}{n^2};\di{n^1}{n^2}\di{m^1}{m^2}}
 [\Lambda,\Lambda_0,\rho_0] \Big|
\nonumber
\\*
&\leq  \big(\theta \Lambda^2\big)^{1-\delta^{1a}}
P^{4V+2\delta^{a4}}_{2-\delta_{V0}(2\delta^{a1}+\delta^{a2})} 
\bigg[\frac{\di{m^1}{m^2}\di{n^1}{n^2};
     \di{n^1}{n^2}\di{m^1}{m^2} }{\theta \Lambda^2}\bigg]
\Big(\frac{1}{\Omega}\Big)^{3V+\delta^{a4}-V^e}\,
P^{2V+\delta^{a4}}\Big[\ln\frac{\Lambda}{\Lambda_R}\Big] \;,
 \label{H-H0}
 \end{align}
where all vertices on the trajectories contribute to $V^e$.

\item \label{H-planar-H1} 
For $\gamma$ according to Definition~\ref{defint}.\ref{defint1} we have
\begin{align}
&\Big|\sum_{\gamma \textup{ as in Def.~\ref{defint}.\ref{defint1}}}  
H^{a(V,V^e,1,0,0)}_{
\di{m^1+1}{m^2}\di{n^1+1}{n^2};\di{n^1}{n^2}\di{m^1}{m^2}}
[\Lambda,\Lambda_0,\rho_0] \Big|
\nonumber
\\*
&\leq  \big(\theta \Lambda^2\big)^{1-\delta^{1a}}
P^{4V+2\delta^{a4}}_{2-\delta_{V0}} \bigg[
\frac{\di{m^1+1}{m^2}\di{n^1+1}{n^2};
    \di{n^1}{n^2}\di{m^1}{m^2} }{\theta \Lambda^2}\bigg]
\Big(\frac{1}{\Omega}\Big)^{3V+\delta^{a4}-V^e}
\, P^{2V+\delta^{a4}}\Big[\ln\frac{\Lambda}{\Lambda_R}\Big] \;,
\label{H-H1}
\end{align}
where all vertices on the trajectories contribute to $V^e$. 

\item  \label{H-sub-planar}
If $\gamma$ is a subgraph of an 1PI planar graph with a selected set $T$ of
  trajectories on one distinguished boundary component and a second set $T'$
  of summed trajectories on that boundary component, we have
\begin{align}
&\sum_{\mathcal{E}^{s}} \sum_{\mathcal{E}^{t'}}
\big|H^{a(V,V^e,B,0,\iota)\gamma}_{
    m_1n_1;\dots;m_Nn_N}[\Lambda,\Lambda_0,\rho_0] \big|
\nonumber
  \\*
  &~~
\leq \big(\theta \Lambda^2\big)^{(2-\delta^{1a}-\frac{N}{2})+2(1-B)}
P^{4V+2+2\delta^{a4}-N}_{\big(2t' 
+ \sum_{\overrightarrow{n_j\mathfrak{o}[n_j]} \in T} 
\min(2,\frac{1}{2}\langle \overrightarrow{n_j\mathfrak{o}[n_j]} 
\rangle )\big)}
\bigg[ \frac{ m_1n_1;\dots;m_Nn_N}{\theta \Lambda^2} \bigg]
\nonumber
  \\*
  & \qquad \times 
\Big(\frac{1}{\Omega}\Big)^{3V-\frac{N}{2}+\delta^{a4} 
+B-V^e-\iota+s+t'}
P^{2V+1+\delta^{a4}-\frac{N}{2}}
\Big[\ln \frac{\Lambda}{\Lambda_R}\Big]\;.
\label{HN4-1plan}
\end{align}
The number of summations is now restricted by $s+t'\leq V^e{+}\iota$.
  
\item \label{H-non-planar} If $\gamma$ is a non-planar graph or a graph with
  $N>4$ external legs, we have
\begin{align}
  &\sum_{\mathcal{E}^{s}} \big|H^{a(V,V^e,B,g,\iota)}_{
    m_1n_1;\dots;m_Nn_N}[\Lambda,\Lambda_0,\rho_0] \big| \nonumber
  \\*
  &\quad \leq \big(\theta \Lambda^2\big)^{(2-\delta^{a1}-\frac{N}{2})
    +2(1-B-2g)}
 P^{4V+2+2\delta^{a4}-N}_0
\bigg[ \frac{ m_1n_1;\dots;m_Nn_N}{\theta \Lambda^2} \bigg]
\nonumber
  \\*
  & \qquad \times 
 \Big(\frac{1}{\Omega}\Big)^{3V-\frac{N}{2}+\delta^{a4}
    +B+2g-V^e-\iota+s} \,P^{2V+1+\delta^{a4} -\frac{N}{2}}\Big[\ln
  \frac{\Lambda}{\Lambda_R}\Big]\;.
\label{HN4-1}
\end{align}
The number of summations is now restricted by $s\leq V^e{+}\iota$.

\end{enumerate}
\end{prp}
\textit{Proof.} The Proposition will be proven by induction upward in the
number $V$ of vertices and for given $V$ downward in the number $N$ of
external legs. 

\begin{enumerate}
\item[\ref{H-non-planar}.] Taking (\ref{polab-range}) into account, the
  estimations (\ref{H-H4})--(\ref{HN4-1plan}) are further bound by
  (\ref{HN4-1}). In particular, the inequality (\ref{HN4-1}) correctly
  reproduces the bounds for $V=0$ derived in Section~\ref{secHinit}.
  By comparison with (\ref{AN4-1}), the estimation (\ref{HN4-1})
  follows immediately for the $H$-\emph{linear} parts on the rhs of
  (\ref{polB4}) which contribute to the integrand of
  $H^{a(V,V^e,B,g,\iota)}_{ m_1n_1;\dots;m_Nn_N}[\Lambda]$. Since
  planar two- and four-point functions are preliminarily excluded, the
  $\Lambda$-integration (from $\Lambda_0$ down to $\Lambda$) confirms
  (\ref{HN4-1}) for those contributions which arise from $H$-linear
  terms on the rhs of (\ref{polB4}) that are non-planar or have $N>4$
  external legs.
  
  We now consider in the $H$-bilinear part on the rhs of (\ref{polB4})
  the contributions of non-planar graphs or graphs with $N>4$ external
  legs. We start with the fourth line in (\ref{polB4}), with the first
  term being a non-planar $H$-function which (apart from the number of
  vertices and the hole label $a$) has the same topological data as
  the total $H$-graph to estimate. From the induction
  hypothesis it is clear that the term in braces $\{~\}$ is bounded by
  the planar unsummed version ($B_1=1, g_1=0,\iota_1=0,s_1=0$)
  of (\ref{HN4-1plan}), with $N_1=2$ and $T=T'=\emptyset$, and with a
  reduction of the degree of the polynomial in 
$\ln \frac{\Lambda}{\Lambda_R}$ by $1$:
\begin{subequations}
\begin{align}
  &\bigg| \Big\{ -\frac{1}{2} \sum_{m,n,k,l} Q_{nm;lk}(\Lambda)
  H^{a(V_1)}_{\di{0}{0}\di{0}{0};\di{0}{0}\di{0}{0};mn;kl}[\Lambda] 
\Big\}_{\text{Def.~\ref{defint}.\ref{defint0}}} \bigg| \nonumber
  \\*
  &\qquad \leq \big(\theta \Lambda^2\big)^{(1-\delta^{a1})} 
\Big(\frac{1}{\Omega}\Big)^{3V_1+\delta^{a4}-V^e_1} 
\,P^{2V_1-1+\delta^{a4}}\Big[\ln \frac{\Lambda}{\Lambda_R}\Big]\;,
\label{HN4-2a}
  \\
  &\sum_{\mathcal{E}^s} \big|H^{1(V-V_1,V^e,B,g,\iota)}_{
m_1n_1;\dots;m_{N}n_{N}}[\Lambda]\big| \nonumber
  \\*
  &\quad \leq \big(\theta \Lambda^2\big)^{(1-\frac{N}{2})
    +2(1-B-2g)} 
P^{4(V-V_1)+2-N}_0 \Big[\frac{m_1n_1;\dots;m_{N}n_{N}}{\theta
  \Lambda^2} \Big] \nonumber
\\*
& \qquad \times \Big(\frac{1}{\Omega}\Big)^{3(V-V_1)-\frac{N}{2}
    +B+2g-V^e-\iota+s} \,P^{2(V-V_1)+1-\frac{N}{2}}\Big[\ln
  \frac{\Lambda}{\Lambda_R}\Big]\;.
\label{HN4-2b}
\end{align}
\end{subequations}
We can ignore the term $P^a_b[~]$, see (\ref{polab}), in
(\ref{HN4-2a}) because the external indices of that part are zero.
In the first step we exclude $a=4$ so that the sum over $V_1$ in
(\ref{polB4}) starts due to (\ref{initrho1})--(\ref{initrho3}) at
$V_1=1$. For $V_1=V$ there is a contribution to (\ref{HN4-2b}) with
$N=2$ only, where (\ref{HN4-2a}) can be regarded as known by
induction. Since the factor $(\frac{1}{\Omega})^{-V^e_1}$ can safely
be absorbed in the polynomial $P[\ln\frac{\Lambda}{\Lambda_R}]$, the
product of (\ref{HN4-2a}) and (\ref{HN4-2b}) confirms the bound
(\ref{HN4-1}) for the integrand under consideration, preliminarily for
$a\neq 4$. In the next step we repeat the argumentation for $a=4$,
where (\ref{HN4-2b}), with $V_1=0$, is known from the first step.

Second, we consider the fifth/sixth lines in (\ref{polB4}). The
difference of functions in braces $\{~\}$ involves graphs with
constant index along the trajectories. We have seen in
Section~\ref{seccomposite} that such a difference can be written as a
sum of graphs each having a composite propagator (\ref{Qpl1}) at a
trajectory. As such the $(\theta\Lambda^2)$-degree of the part in
braces $\{~\}$ is reduced\footnote{The origin of the reduction is the
  term $P^a_b[~]$ introduced in (\ref{polab}), with $b=1$ in presence
  of a composite propagator (\ref{Qpl1}). The argument in the brackets
  of $P^a_b[~]$ is the ratio of the maximal external index to the
  reference scale $\theta\Lambda^2$. Since the maximal index along the
  trajectory is $1$, we can globally estimate in this case $P^a_1[~]$
  by a constant times $(\theta\Lambda^2)^{-1}$. } by $1$ compared
with planar analogues of (\ref{HN4-1}) for $N=2$. The difference of
functions in braces $\{~\}$ involves also graphs where the index along
one of the trajectories jumps once by $\di{1}{0}$ or $\di{0}{1}$ and
back. For these graphs we conclude from (\ref{est4-3}) (and the fact
that the maximal index along the trajectory is $2$) that the
$(\theta\Lambda^2)$-degree of the part in braces $\{~\}$ is also
reduced by $1$:
\begin{subequations}
\begin{align}
  &\bigg|  
\Big\{
  -\frac{1}{2} \sum_{m,n,k,l} Q_{nm;lk}(\Lambda)
  \Big(H^{a(V_1)}_{\di{1}{0}\di{0}{0};\di{0}{0}\di{1}{0};mn;kl}[\Lambda]
  -H^{a(V_1)}_{\di{0}{0}\di{0}{0};\di{0}{0}\di{0}{0};mn;kl}[\Lambda]\Big)
  \Big\}_{[\text{Def.~\ref{defint}.\ref{defint0}}]} 
\bigg| \nonumber
  \\*
  &\qquad \leq \big(\theta \Lambda^2\big)^{(-\delta^{a1})} 
\Big(\frac{1}{\Omega}\Big)^{3V_1+\delta^{a4}
    (1-V^e_1)} \,P^{2V_1-1+\delta^{a4}}\Big[\ln
  \frac{\Lambda}{\Lambda_R}\Big]\;,
\label{HN4-3a}
  \\
  &\sum_{\mathcal{E}^s} 
\big|H^{2(V-V_1,V^e,B,g,\iota)}_{m_1n_1;\dots;m_{N}n_{N}}
[\Lambda] \big|
\nonumber
  \\*
  &\quad \leq \big(\theta \Lambda^2\big)^{(2-\frac{N}{2})
    +2(1-B-2g)} \,
P^{4(V-V_1)+2-N}_0 \Big[\frac{m_1n_1;\dots;m_{N}n_{N}}{\theta
  \Lambda^2} \Big] \nonumber
\\*
& \qquad \times \Big(\frac{1}{\Omega}\Big)^{3(V-V_1)-\frac{N}{2}+1
    +B+2g-V^e-\iota+s} \,P^{2(V-V_1)+1-\frac{N}{2}}\Big[\ln
  \frac{\Lambda}{\Lambda_R}\Big]\;.
\label{HN4-3b}
\end{align}
\end{subequations}
Again we have to exclude $a=4$ in the first step, which then confirms the
bound (\ref{HN4-1}) for the integrand under consideration. In the second step
we repeat the argumentation for $a=4$. 

Third, the discussion of the seventh line of (\ref{polB4}) is completely
similar, because there the index on each trajectory jumps once by $\di{1}{0}$
or $\di{0}{1}$. This leads again to a reduction by $1$ of the
$(\theta\Lambda^2)$-degree of the part in braces $\{~\}$ compared with planar
analogues of (\ref{HN4-1}) for $N=2$.

Finally, the part in braces in the last line of (\ref{polB4}) can be
estimated by a planar $N=4$ version of (\ref{HN4-1}), again with a reduction
by $1$ of the degree of $P[\ln\frac{\Lambda}{\Lambda_R}]$:
\begin{subequations}
\begin{align}
 & \bigg| \Big\{
  -\frac{1}{2} \sum_{m,n,k,l} Q_{nm;lk}(\Lambda)
  H^{a(V_1)}_{\di{0}{0}\di{0}{0};\di{0}{0}\di{0}{0};
    \di{0}{0}\di{0}{0};\di{0}{0}\di{0}{0};mn;kl}[\Lambda]
  \Big\}_{[\text{Def.~\ref{defint}.\ref{defint4}}]}
\bigg| \nonumber
  \\*
  & \leq \big(\theta \Lambda^2\big)^{-\delta^{a1}}
  \Big(\frac{1}{\Omega}\Big)^{3V_1-1+\delta^{a4}-V^e_1} \,
P^{2V_1-2+\delta^{a4}}\Big[\ln \frac{\Lambda}{\Lambda_R}\Big]\;,
\label{HN4-4a}  
\\
  &\sum_{\mathcal{E}^s} \big|H^{4(V-V_1,B,g,V^e,\iota)}_{
m_1n_1;\dots;m_{N}n_{N}}[\Lambda] \big|\nonumber
  \\*
  &\leq \big(\theta \Lambda^2\big)^{(2-\frac{N}{2}) +2(1-B-2g)} \,
P^{4(V-V_1)+4-N}_0 \Big[\frac{m_1n_1;\dots;m_{N}n_{N}}{\theta
  \Lambda^2} \Big] \nonumber
\\*
& \qquad \times 
  \Big(\frac{1}{\Omega}\Big)^{3(V-V_1)-\frac{N}{2}+1
    +B+2g-V^e-\iota+s} \,P^{2(V-V_1)+2 -\frac{N}{2}}\Big[\ln
  \frac{\Lambda}{\Lambda_R}\Big]\;.
\label{HN4-4b}
\end{align}
\end{subequations}
We confirm again the bound (\ref{HN4-1}) for the integrand
under consideration. 

Since we have assumed that the total $H$-graph is non-planar or has
$N>4$ external legs, the integrand (\ref{HN4-1}) is irrelevant so that
we obtain after integration from $\Lambda_0$ down to $\Lambda$ (and
use of the initial conditions (\ref{initrho1})--(\ref{initrho4})) the
same bound (\ref{HN4-1}) for the graph, too.

\vskip 1ex

\item[\ref{H-sub-planar}.] According to Section~\ref{secHinit}, the
  inequality (\ref{HN4-1plan}) is correct for $V=0$. By comparison
  with (\ref{AN4-1plan}), the estimation (\ref{HN4-1plan}) follows
  immediately for the $H$-\emph{linear} parts on the rhs of
  (\ref{polB4}) which contribute to the integrand of
  $H^{a(V,V^e,B,g,\iota)}_{ m_1n_1;\dots;m_Nn_N}[\Lambda]$. Excluding
  planar two- and four-point functions with constant index on the
  trajectory or with limited jump according to
  \ref{defint4}--\ref{defint1} of Definition~\ref{defint}, the
  $\Lambda$-integration confirms (\ref{HN4-1plan}) for those
  contributions which arise from $H$-linear terms on the rhs of
  (\ref{polB4}) that correspond to subgraphs of planar graphs (subject
  to the above restrictions). The proof of (\ref{HN4-1plan}) for the
  $H$-bilinear terms in (\ref{polB4}) is completely analogous to the
  non-planar case. We only have to replace (\ref{HN4-2b}),
  (\ref{HN4-3b}) and (\ref{HN4-4b}) by the adapted version of
  (\ref{HN4-1plan}). In particular, the distinguished trajectory with
  its subsets $T,T'$ of indices comes exclusively from the
  (\ref{HN4-1plan})-analogues of (\ref{HN4-2b}), (\ref{HN4-3b}) and
  (\ref{HN4-4b}) and not from the terms in braces in (\ref{polB4}).

\vskip 1ex
  
\item[\ref{H-planar-H4}.] We first consider $a\neq 4$. Then, according
  to (\ref{initrho1})--(\ref{initrho3}) we need $V\geq 1$ in order to
  have a non-vanishing contribution to (\ref{H-H4}). Since according
  to Definition~\ref{defint}.\ref{defint4} the index
  along each trajectory of the (planar) graph $\gamma$ is constant, we have 
\begin{align}
H^{a(V,V^e,1,0,0)\gamma}_{m_1n_1;m_2n_2;m_3n_3;m_4n_4}[\Lambda]
= \frac{1}{6} H^{a(V,V^e,1,0,0)}_{m_1m_2;m_2m_3;m_3m_4;m_4n_1}[\Lambda]
+ \text{5 permutations}\;.
\end{align}
Then, using (\ref{initrho1})--(\ref{initrho4}) and the fact that
$\gamma$ is 1PI, the differential equation (\ref{polB4}) reduces to
\begin{align}
&\Lambda \frac{\partial}{\partial \Lambda} \bigg(
\sum_{\gamma \text{ as in Def.~\ref{defint}.\ref{defint4}}} 
H^{a(V,V^e,1,0,0)\gamma}_{m_1n_1;m_2n_2;m_3n_3;m_4n_4}
[\Lambda,\Lambda_0,\rho^0] \bigg)_{a\neq 4}\nonumber
\\*
&= \bigg({-}\frac{1}{12} \bigg\{
  \sum_{m,n,k,l}\! Q_{nm;lk}(\Lambda) \Big(
  H^{a(V,V^e,B,0,\iota)}_{m_1m_2;m_2m_3;m_3m_4;m_4n_1;mn;kl}[\Lambda] 
\nonumber
\\*
&\hspace*{10em} 
-  H^{a(V,V^e,B,0,\iota)}_{00;00;00;00;mn;kl}[\Lambda] \Big)
  \bigg\}_{\textup{[Def.~\ref{defint}.\ref{defint4}]}}
+ \text{5 permutations} \bigg)
\nonumber
\\*
&
+ \text{the $4^{\text{th}}$ to last lines of (\ref{polB4}) with
  $\displaystyle \sum_{V_1=0}^{V} \mapsto \sum_{V_1=1}^{V-1}$}\;.
\label{HH4-a}
\end{align}
Here, the term $H^{a(V,V^e,B,0,\iota)}_{00;00;00;00;mn;kl}[\Lambda]$
in the second line of (\ref{HH4-a}) comes from the
$(V_1=V)$-contribution of the last line in (\ref{polB4}), together
with (\ref{initrho4}). In the same way as in
Section~\ref{seccomposite} we conclude that the second line of
(\ref{HH4-a}) can be written as a linear combination of graphs having
a composite propagator (\ref{comp-prop-0}) on one of the trajectories.
As such we have to replace the bound (\ref{est4-1}) relative to the
contribution of an ordinary propagator by (\ref{Qpl1}). For the total
graph this amounts to multiply the corresponding estimation
(\ref{HN4-1plan}) of ordinary $H$-graphs with $N=4$ by a factor $\frac{\max
  \|m_i\|}{\theta \Lambda^2}$, which yields the subscript $1$ of the
part $P_1^{4V-2+2\delta^{a4}}[~]$ of the integrand (\ref{H-H4}), for
the time being restricted to the second line of (\ref{HH4-a}). Since
the resulting integrand is irrelevant, we also obtain (\ref{H-H4})
after $\Lambda$-integration from $\Lambda_0$ down to $\Lambda$.
Clearly, this is the only contribution for $V=1$ so that (\ref{H-H4})
is proven for $V=1$ and $a\neq 4$.

In the second step we use this result to extend the proof to $V=1$ and
$a=4$. Now the differential equation (\ref{polB4}) reduces to the
second line of (\ref{HH4-a}), with $a=4$, and the fourth to sixth
lines of (\ref{polB4}) with $V=1$ and $V_1=0$. There is no
contribution from the seventh line of (\ref{polB4}) for $V_1=0$, because
the part in braces would be non-planar, which is excluded in
Definition~\ref{defint}.\ref{defint1}. Inserting (\ref{initrho4}) we
obtain the composite propagator (\ref{comp-prop-1}) in the part in
braces $\{~\}$ of the fifth line of (\ref{polB4}).  Together with
(\ref{H-H4}) for $V=1$ and $a\neq 4$ already proven we verify the
integrand (\ref{H-H4}) for $V=1$ and $a=4$. After
$\Lambda$-integration we thus obtain (\ref{H-H4}) for $V=1$ and any
$a$.

This allows us to use (\ref{H-H4}) as induction hypothesis for the
remaining contributions in the last line of (\ref{HH4-a}). This is
similar to the procedure in \ref{H-non-planar}, we only have to
replace (\ref{HN4-2b}), (\ref{HN4-3b}) and (\ref{HN4-4b}) by the
according parametrisation of (\ref{H-H4}). We thus prove (\ref{H-H4})
to all orders. 

\vskip 1ex

\item[\ref{H-planar-H0}.] We first consider $a\neq 4$. Then, according
  to (\ref{initrho1})--(\ref{initrho3}) only terms with $V_1 \geq 1$
  contribute to (\ref{polB4}). Using
  (\ref{initrho1})--(\ref{initrho4}) and the fact that $\gamma$ is
  1PI, the differential equation (\ref{polB4}) reduces to
\begin{subequations}
\label{HH0}
\begin{align}
&\Lambda \frac{\partial}{\partial \Lambda} 
\Big(\sum_{\gamma \text{ as in Def.~\ref{defint}.\ref{defint0}}} 
H^{a(V,V^e,1,0,0)\gamma}_{\di{m^1}{m^2}\di{n^1}{n^2};
\di{n^1}{n^2}\di{m^1}{m^2}}
[\Lambda,\Lambda_0,\rho^0] \Big)_{a\neq 4}\nonumber
\\*
&=  - \frac{1}{2} \bigg\{
\sum_{m,n,k,l}  Q_{nm;lk}(\Lambda) \Big(
  H^{a(V,V^e,B,0,\iota)}_{\di{m^1}{m^2}\di{n^1}{n^2};
\di{n^1}{n^2}\di{m^1}{m^2};mn;kl}[\Lambda] 
-  H^{a(V,V^e,B,0,\iota)}_{\di{0}{0}\di{0}{0};
\di{0}{0}\di{0}{0};mn;kl}[\Lambda]
\nonumber
\\
& \qquad\qquad 
- m^1   \big(H^{a(V,V^e,B,0,\iota)}_{
\di{1}{0}\di{0}{0};\di{0}{0}\di{1}{0};mn;kl}[\Lambda]
  -H^{a(V,V^e,B,0,\iota)}_{
\di{0}{0}\di{0}{0};\di{0}{0}\di{0}{0};mn;kl}[\Lambda]\big)
\nonumber
\\*
& \qquad\qquad 
- n^1 \big(H^{a(V,V^e,B,0,\iota)}_{
\di{0}{0}\di{1}{0};\di{1}{0}\di{0}{0};mn;kl}[\Lambda]
  -H^{a(V,V^e,B,0,\iota)}_{
\di{0}{0}\di{0}{0};\di{0}{0}\di{0}{0};mn;kl}[\Lambda]\big)
\nonumber
\\
& \qquad\qquad 
-m^2   \big(H^{a(V,V^e,B,0,\iota)}_{
\di{0}{1}\di{0}{0};\di{0}{0}\di{0}{1};mn;kl}[\Lambda]
  -H^{a(V,V^e,B,0,\iota)}_{
\di{0}{0}\di{0}{0};\di{0}{0}\di{0}{0};mn;kl}[\Lambda]\big)
\nonumber
\\*
& \qquad\qquad 
- n^2   \big(H^{a(V,V^e,B,0,\iota)}_{
\di{0}{0}\di{0}{1};\di{0}{1}\di{0}{0};mn;kl}[\Lambda]
  -H^{a(V,V^e,B,0,\iota)}_{
\di{0}{0}\di{0}{0};\di{0}{0}\di{0}{0};mn;kl}[\Lambda]\big)
\Big)
\bigg\}_{[\text{Def.~\ref{defint}.\ref{defint0}}]} 
\label{HH0-a}
\\
& - H^{4(0)}_{\di{m^1}{m^2}\di{n^1}{n^2};
\di{n^1}{n^2}\di{m^1}{m^2}}[\Lambda] \bigg\{
  -\frac{1}{2} \sum_{m,n,k,l} Q_{nm;lk}(\Lambda)
  H^{a(V)}_{\di{0}{0}\di{0}{0};\di{0}{0}\di{0}{0};
    \di{0}{0}\di{0}{0};\di{0}{0}\di{0}{0};mn;kl}[\Lambda]
  \bigg\}_{[\text{Def.~\ref{defint}.\ref{defint4}}]}
\label{HH0-b}
\\*
& + \text{the $4^{\text{th}}$ to last lines of (\ref{polB4}) with
  $\displaystyle \sum_{V_1=0}^{V} \mapsto \sum_{V_1=1}^{V-1}$}\;.
\label{HH0-c}
\end{align}
\end{subequations}
If the graphs have constant indices along the trajectories, we
conclude in the same way as in Appendix~\ref{appcompositeidentity}
that the part (\ref{HH0-a}) can be written as a linear combination of
graphs having either a composite propagator (\ref{comp-prop-1}) or two
composite propagators (\ref{comp-prop-0}) on the trajectories.  As
such we have to replace the bound (\ref{est4-1}) relative to the
contribution of an ordinary propagator by (\ref{Qpl2}) or twice
(\ref{est4-1}) by (\ref{Qpl1}). For the total graph this amounts to
multiply the corresponding estimation (\ref{HN4-1plan}) of ordinary
$H$-graphs with $N=2$ by a factor $\big(\frac{\max (m^r,n^r)}{\theta
  \Lambda^2}\big)^2$, which yields the subscript $2$ of the part
$P_2^{4V+2\delta^{a4}}[~]$ of the integrand (\ref{H-H0}), for the time
being restricted to the part (\ref{HH0-a}). For graphs with index jump
in Definition~\ref{defint}.\ref{defint0} we obtain according to
Appendix~\ref{appcompositeidentity} the same improvement by
$\big(\frac{\max (m^r,n^r)}{\theta \Lambda^2}\big)^2$. Next, the product of
(\ref{H0-42N}) with (\ref{HN4-4a}) gives for (\ref{HH0-b}) the same
bound (\ref{H-H0}) for the integrand. Since the resulting integrand is
irrelevant, we also obtain (\ref{H-H0}) after $\Lambda$-integration.
Clearly, this is the only contribution for $V=1$ so that (\ref{H-H0})
is proven for $V=1$ and $a\neq 4$.

In the second step we use this result to extend the proof to $V=1$ and
$a=4$. Now the differential equation (\ref{polB4}) reduces to the sum
of (\ref{HH0-a}) and (\ref{HH0-b}), with $a=4$, and the fourth to 
sixth lines of (\ref{polB4}) with $V=1$ and $V_1=0$. There is again no
contribution of the seventh line of (\ref{polB4}) for $V_1=0$. Inserting
(\ref{initrho4}) we obtain the composite propagators
(\ref{comp-prop-1}) in the fifth/sixth lines of (\ref{polB4}), which together
with (\ref{H-H0}) for $V=1$ and $a\neq 4$ already proven verifies the
integrand (\ref{H-H0}) for $V=1$ and $a=4$. After
$\Lambda$-integration we thus obtain (\ref{H-H0}) for $V=1$ and any
$a$.

This allows us to use (\ref{H-H0}) as induction hypothesis for the
remaining contributions (\ref{HH0-c}) for $V>1$. This is similar to
the procedure in \ref{H-non-planar}, we only have to replace
(\ref{HN4-2b}), (\ref{HN4-3b}) and (\ref{HN4-4b}) by the according
parametrisation of (\ref{H-H0}). We thus prove (\ref{H-H0}) to all
orders.

\vskip 1ex

\item[\ref{H-planar-H1}.] The proof of (\ref{H-H1}) is performed along
  the same lines as the proof of (\ref{H-H4}) and (\ref{H-H0}). There
  is one factor $\frac{\max(m^r,n^r)}{\theta \Lambda^2}$ from $\langle
  \overrightarrow{n_1\mathfrak{o}[n_1]}\rangle + \langle
  \overrightarrow{n_2\mathfrak{o}[n_2]}\rangle =2$ in
  (\ref{HN4-1plan}) and a second factor from the composite propagator
  (\ref{Qpl1}) or (\ref{Qpl3}) appearing according to
  Appendix~\ref{appcompositeidentity} in the $(V_1=V)$-contribution to
  (\ref{polB4}).

\end{enumerate}
This finishes the proof of Proposition~\ref{prop-H}. \hfill $\square$%
\bigskip

\subsection{The power-counting behaviour of the $\Lambda_0$-varied functions} 

\label{graphR}

The estimations in Propositions~\ref{power-counting-prop} and
\ref{prop-H} allow us to estimate the $R$-functions by integrating the
differential equation (\ref{polV4}). Again, the $R$-functions are
expanded in terms of ribbon graphs. Let us look at $R$-ribbon graphs
of the type described in Definition~\ref{defint}.\ref{defint4}. Since
$\sum_{\gamma \text{ as in Def.~\ref{defint}.\ref{defint4}}}
A^{(V)\gamma}_{ \di{0}{0}\di{0}{0};\di{0}{0}\di{0}{0};
  \di{0}{0}\di{0}{0};\di{0}{0}\di{0}{0}}[\Lambda,\Lambda_0,\rho^0]\equiv
\rho_4[\Lambda,\Lambda_0,\rho^0]$, we can rewrite the expansion
coefficients of (\ref{Rdef}) as follows:
\begin{align}
&\sum_{\gamma \text{ as in Def.~\ref{defint}.\ref{defint4}}} 
R^{(V)\gamma}_{\di{m^1}{m^2}\di{n^1}{n^2};\di{n^1}{n^2}\di{k^1}{k^2};
\di{k^1}{k^2}\di{l^1}{l^2};\di{l^1}{l^2}\di{m^1}{m^2}}
[\Lambda,\Lambda_0,\rho^0]\nonumber
\\*
& = \Lambda_0 \frac{\partial}{\partial \Lambda_0} 
\sum_{\gamma \text{ as in Def.~\ref{defint}.\ref{defint4}}}
\Big(A^{(V)\gamma}_{\di{m^1}{m^2}\di{n^1}{n^2};\di{n^1}{n^2}\di{k^1}{k^2};
\di{k^1}{k^2}\di{l^1}{l^2};\di{l^1}{l^2}\di{m^1}{m^2}}
[\Lambda,\Lambda_0,\rho^0]
\nonumber
\\*[-1ex]
&\hspace*{11em}
- A^{(V)\gamma}_{\di{0}{0}\di{0}{0};\di{0}{0}\di{0}{0};
\di{0}{0}\di{0}{0};\di{0}{0}\di{0}{0}}[\Lambda,\Lambda_0,\rho^0]\Big)
\nonumber
\\*
& - \sum_{a,b=1}^4 \frac{\partial}{\partial \rho^0_a} 
\sum_{\gamma \text{ as in Def.~\ref{defint}.\ref{defint4}}}
\Big( A^{(V)\gamma}_{\di{m^1}{m^2}\di{n^1}{n^2};\di{n^1}{n^2}\di{k^1}{k^2};
\di{k^1}{k^2}\di{l^1}{l^2};\di{l^1}{l^2}\di{m^1}{m^2}}
[\Lambda,\Lambda_0,\rho^0]
\nonumber
\\*[-1ex]
&\hspace*{11em}
- A^{(V)\gamma}_{\di{0}{0}\di{0}{0};\di{0}{0}\di{0}{0};
\di{0}{0}\di{0}{0};\di{0}{0}\di{0}{0}}[\Lambda,\Lambda_0,\rho^0]\Big)
\nonumber
\\*
&\qquad\qquad \times 
\frac{\partial \rho^0_a}{\partial \rho_b[\Lambda,\Lambda_0,\rho^0]}
\Lambda_0 \frac{\partial}{\partial \Lambda_0} 
\rho_b[\Lambda,\Lambda_0,\rho^0]\;.
\label{Rproj}
\end{align}
This means that (by construction) only the
$(\Lambda_0,\rho^0)$-derivatives of the projection to the irrelevant
part (\ref{hatA4}) of the planar four-point function contributes to
$R$. Similarly, only the $(\Lambda_0,\rho^0)$-derivatives of the
irrelevant parts (\ref{hatA0}) and (\ref{hatA1}) of the planar
two-point function contribute to $R$. According to the initial
condition (\ref{ct4}), these projections and the other functions given
in Definition~\ref{defint}.\ref{defintnp} vanish at
$\Lambda=\Lambda_0$ \emph{independently} of $\Lambda_0$ or $\rho^0_a$:
\begin{subequations}
\label{initR}
\begin{align}
0&=  \Lambda_0 \frac{\partial}{\partial \Lambda_0} \Big(
A^{(V,V^e,1,0,0)\gamma}_{
\di{m^1}{m^2}\di{n^1}{n^2};\di{n^1}{n^2}\di{k^1}{k^2};
\di{k^1}{k^2}\di{l^1}{l^2};\di{l^1}{l^2}\di{m^1}{m^2}}
[\Lambda_0,\Lambda_0,\rho^0] 
\nonumber
\\*
&\hspace*{7em}
- A^{(V,V^e,1,0,0)\gamma}_{\di{0}{0}\di{0}{0};\di{0}{0}\di{0}{0};
\di{0}{0}\di{0}{0};\di{0}{0}\di{0}{0}}[\Lambda_0,\Lambda_0,\rho^0]\Big)
\Big|_{\gamma \text{ as in Def.~\ref{defint}.\ref{defint4}}} 
\nonumber
\\*
&=  \frac{\partial}{\partial \rho^0_a} \Big(
A^{(V,V^e,1,0,0)\gamma}_{
\di{m^1}{m^2}\di{n^1}{n^2};\di{n^1}{n^2}\di{k^1}{k^2};
\di{k^1}{k^2}\di{l^1}{l^2};\di{l^1}{l^2}\di{m^1}{m^2}}
[\Lambda_0,\Lambda_0,\rho^0] 
\nonumber
\\*
&\hspace*{7em}
- A^{(V,V^e,1,0,0)\gamma}_{\di{0}{0}\di{0}{0};\di{0}{0}\di{0}{0};
\di{0}{0}\di{0}{0};\di{0}{0}\di{0}{0}}[\Lambda_0,\Lambda_0,\rho^0]\Big)
\Big|_{\gamma \text{ as in Def.~\ref{defint}.\ref{defint4}}} \;,
\label{initR4}
\\
0&=  \Lambda_0 \frac{\partial}{\partial \Lambda_0} \Big(
A^{(V,V^e,1,0,0)\gamma}_{
\di{m^1}{m^2}\di{n^1}{n^2};\di{n^1}{n^2}\di{m^1}{m^2}}
[\Lambda_0,\Lambda_0,\rho^0] 
- A^{(V,V^e,1,0,0)\gamma}_{\di{0}{0}\di{0}{0};\di{0}{0}\di{0}{0}}
[\Lambda_0,\Lambda_0,\rho^0]
\nonumber
\\*
&\qquad - m^1 \big( A^{(V,V^e,1,0,0)\gamma}_{
\di{1}{0}\di{0}{0};\di{0}{0}\di{1}{0}}
[\Lambda_0,\Lambda_0,\rho^0] 
- A^{(V,V^e,1,0,0)\gamma}_{\di{0}{0}\di{0}{0};\di{0}{0}\di{0}{0}}
[\Lambda_0,\Lambda_0,\rho^0]\big)
\nonumber
\\*
& \qquad - n^1 \big( A^{(V,V^e,1,0,0)\gamma}_{
\di{0}{0}\di{1}{0};\di{1}{0}\di{0}{0}}
[\Lambda_0,\Lambda_0,\rho^0] 
- A^{(V,V^e,1,0,0)\gamma}_{\di{0}{0}\di{0}{0};\di{0}{0}\di{0}{0}}
[\Lambda_0,\Lambda_0,\rho^0]\big)
\nonumber
\\*
&\qquad - m^2 \big( A^{(V,V^e,1,0,0)\gamma}_{
\di{0}{1}\di{0}{0};\di{0}{0}\di{0}{1}}
[\Lambda_0,\Lambda_0,\rho^0] 
- A^{(V,V^e,1,0,0)\gamma}_{\di{0}{0}\di{0}{0};\di{0}{0}\di{0}{0}}
[\Lambda_0,\Lambda_0,\rho^0]\big)
\nonumber
\\*
&\qquad - n^2 \big( A^{(V,V^e,1,0,0)\gamma}_{
\di{0}{0}\di{0}{1};\di{0}{1}\di{0}{0}}
[\Lambda_0,\Lambda_0,\rho^0] 
- A^{(V,V^e,1,0,0)\gamma}_{\di{0}{0}\di{0}{0};\di{0}{0}\di{0}{0}}
[\Lambda_0,\Lambda_0,\rho^0]\big)\Big)
\Big|_{\gamma \text{ as in Def.~\ref{defint}.\ref{defint0} }} 
\nonumber
\\
&=  \frac{\partial}{\partial \rho^0_a} \Big(
A^{(V,V^e,1,0,0)\gamma}_{
\di{m^1}{m^2}\di{n^1}{n^2};\di{n^1}{n^2}\di{m^1}{m^2}}
[\Lambda_0,\Lambda_0,\rho^0] 
- A^{(V,V^e,1,0,0)\gamma}_{\di{0}{0}\di{0}{0};\di{0}{0}\di{0}{0}}
[\Lambda_0,\Lambda_0,\rho^0]
\nonumber
\\*
&\qquad - m^1 \big( A^{(V,V^e,1,0,0)\gamma}_{
\di{1}{0}\di{0}{0};\di{0}{0}\di{1}{0}}
[\Lambda_0,\Lambda_0,\rho^0] 
- A^{(V,V^e,1,0,0)\gamma}_{\di{0}{0}\di{0}{0};\di{0}{0}\di{0}{0}}
[\Lambda_0,\Lambda_0,\rho^0]\big)
\nonumber
\\*
&\qquad - n^1 \big( A^{(V,V^e,1,0,0)\gamma}_{
\di{0}{0}\di{1}{0};\di{1}{0}\di{0}{0}}
[\Lambda_0,\Lambda_0,\rho^0] 
- A^{(V,V^e,1,0,0)\gamma}_{\di{0}{0}\di{0}{0};\di{0}{0}\di{0}{0}}
[\Lambda_0,\Lambda_0,\rho^0]\big)
\nonumber
\\
&\qquad - m^2 \big( A^{(V,V^e,1,0,0)\gamma}_{
\di{0}{1}\di{0}{0};\di{0}{0}\di{0}{1}}
[\Lambda_0,\Lambda_0,\rho^0] 
- A^{(V,V^e,1,0,0)\gamma}_{\di{0}{0}\di{0}{0};\di{0}{0}\di{0}{0}}
[\Lambda_0,\Lambda_0,\rho^0]\big)
\nonumber
\\*
&\qquad - n^2 \big( A^{(V,V^e,1,0,0)\gamma}_{
\di{0}{0}\di{0}{1};\di{0}{1}\di{0}{0}}
[\Lambda_0,\Lambda_0,\rho^0] 
- A^{(V,V^e,1,0,0)\gamma}_{\di{0}{0}\di{0}{0};\di{0}{0}\di{0}{0}}
[\Lambda_0,\Lambda_0,\rho^0]\big)\Big)
\Big|_{\gamma \text{ as in Def.~\ref{defint}.\ref{defint0}}} \;,
\label{initR0}
\\
0&=  \Lambda_0 \frac{\partial}{\partial \Lambda_0} \Big(
A^{(V,V^e,1,0,0)\gamma}_{
\di{m^1+1}{m^2}\di{n^1+1}{n^2};\di{n^1}{n^2}\di{m^1}{m^2}}
[\Lambda_0,\Lambda_0,\rho^0] 
\nonumber
\\*
&\qquad - \sqrt{(m^1{+}1)(n^1{+}1)} 
A^{(V,V^e,1,0,0)\gamma}_{\di{1}{0}\di{1}{0};\di{0}{0}\di{0}{0}}
[\Lambda_0,\Lambda_0,\rho^0]\Big)
\Big|_{\gamma \text{ as in Def.~\ref{defint}.\ref{defint1}}}
\nonumber
\\*
&= \frac{\partial}{\partial \rho^0_a}
 \Big(
A^{(V,V^e,1,0,0)\gamma}_{
\di{m^1+1}{m^2}\di{n^1+1}{n^2};\di{n^1}{n^2}\di{m^1}{m^2}}
[\Lambda_0,\Lambda_0,\rho^0] 
\nonumber
\\*
&\qquad - \sqrt{(m^1{+}1)(n^1{+}1)} 
A^{(V,V^e,1,0,0)\gamma}_{\di{1}{0}\di{1}{0};\di{0}{0}\di{0}{0}}
[\Lambda_0,\Lambda_0,\rho^0]\Big)
\Big|_{\gamma \text{ as in Def.~\ref{defint}.\ref{defint1}}}\;,
\label{initR1}
\\
0&=  \Lambda_0 \frac{\partial}{\partial \Lambda_0} 
A^{(V,V^e,B,g,\iota)\gamma}_{m_1n_1;\dots;m_Nn_N}
[\Lambda_0,\Lambda_0,\rho^0] 
\Big|_{\gamma \text{ as in Def.~\ref{defint}.\ref{defintnp}}} 
\nonumber
\\*
&=  \frac{\partial}{\partial \rho^0_a} 
A^{(V,V^e,B,g,\iota)\gamma}_{m_1n_1;\dots;m_Nn_N}
[\Lambda_0,\Lambda_0,\rho^0]
\Big|_{\gamma \text{ as in Def.~\ref{defint}.\ref{defintnp}}}\;.
\label{initRnp}
\end{align}
\end{subequations}
The $\Lambda_0$-derivative at $\Lambda=\Lambda_0$ 
has to be considered with care:
\begin{align}
0&=  \Lambda_0 \frac{\partial}{\partial \Lambda_0} 
A^{(V,V^e,B,g,\iota)\gamma}_{m_1n_1;\dots;m_Nn_N}
[\Lambda_0,\Lambda_0,\rho^0] 
\nonumber
\\*
&=  
\Big(\Lambda \frac{\partial}{\partial \Lambda} 
A^{(V,V^e,B,g,\iota)\gamma}_{m_1n_1;\dots;m_Nn_N}
[\Lambda,\Lambda_0,\rho^0] \Big)_{\Lambda = \Lambda_0} \!\!
{+} \Big(\Lambda_0 \frac{\partial}{\partial \Lambda_0} 
A^{(V,V^e,B,g,\iota)\gamma}_{m_1n_1;\dots;m_Nn_N}
[\Lambda,\Lambda_0,\rho^0] \Big)_{\Lambda = \Lambda_0},
\label{initRnp0}
\end{align}
and similarly for (\ref{initR4})--(\ref{initR1}). Inserting
(\ref{Rproj}), (\ref{initR}), (\ref{initRnp0}) and according formulae
into the Taylor expansion of (\ref{Rdef}) we thus have
\begin{subequations}
\label{RN4}
\begin{align}
&\sum_{\gamma \text{ as in Def.~\ref{defint}.\ref{defint4}}} 
R^{(V,V^e,1,0,0)\gamma}_{
\di{m^1}{m^2}\di{n^1}{n^2};\di{n^1}{n^2}\di{k^1}{k^2};
\di{k^1}{k^2}\di{l^1}{l^2};\di{l^1}{l^2}\di{m^1}{m^2}} 
[\Lambda_0,\Lambda_0,\rho^0]
\nonumber
\\*
& = - \sum_{\gamma \text{ as in Def.~\ref{defint}.\ref{defint4}}} 
\bigg(\Lambda\frac{\partial}{\partial \Lambda}\Big( 
A^{(V,V^e,1,0,0)\gamma}_{
\di{m^1}{m^2}\di{n^1}{n^2};\di{n^1}{n^2}\di{k^1}{k^2};
\di{k^1}{k^2}\di{l^1}{l^2};\di{l^1}{l^2}\di{m^1}{m^2}}
[\Lambda,\Lambda_0,\rho^0] 
\nonumber
\\*
& \hspace*{10em}
- A^{(V,V^e,1,0,0)\gamma}_{\di{0}{0}\di{0}{0};\di{0}{0}\di{0}{0};
\di{0}{0}\di{0}{0};\di{0}{0}\di{0}{0}}[\Lambda,\Lambda_0,\rho^0]\Big)
\bigg)_{\Lambda=\Lambda_0}\,,
\label{RN4-40}
\\
&\sum_{\gamma \text{ as in Def.~\ref{defint}.\ref{defint0}}} 
R^{(V,V^e,1,0,0)\gamma}_{
\di{m^1}{m^2}\di{n^1}{n^2};\di{n^1}{n^2}\di{m^1}{m^2}}
[\Lambda_0,\Lambda_0,\rho^0]
\nonumber
\\*
& =- \sum_{\gamma \text{ as in Def.~\ref{defint}.\ref{defint0}}}
\bigg(\Lambda\frac{\partial}{\partial \Lambda}\Big( 
A^{(V,V^e,1,0,0)\gamma}_{
\di{m^1}{m^2}\di{n^1}{n^2};\di{n^1}{n^2}\di{m^1}{m^2}}
[\Lambda,\Lambda_0,\rho^0] 
- A^{(V,V^e,1,0,0)\gamma}_{\di{0}{0}\di{0}{0};\di{0}{0}\di{0}{0}}
[\Lambda,\Lambda_0,\rho^0]
\nonumber
\\*
&\qquad - m^1 \big( A^{(V,V^e,1,0,0)\gamma}_{
\di{1}{0}\di{0}{0};\di{0}{0}\di{1}{0}}
[\Lambda,\Lambda_0,\rho^0] 
- A^{(V,V^e,1,0,0)\gamma}_{\di{0}{0}\di{0}{0};\di{0}{0}\di{0}{0}}
[\Lambda,\Lambda_0,\rho^0]\big)
\nonumber
\\*
&\qquad - n^1 \big( A^{(V,V^e,1,0,0)\gamma}_{
\di{0}{0}\di{1}{0};\di{1}{0}\di{0}{0}}
[\Lambda,\Lambda_0,\rho^0] 
- A^{(V,V^e,1,0,0)\gamma}_{\di{0}{0}\di{0}{0};\di{0}{0}\di{0}{0}}
[\Lambda,\Lambda_0,\rho^0]\big)
\nonumber
\\*
&\qquad - m^2 \big( A^{(V,V^e,1,0,0)\gamma}_{
\di{0}{1}\di{0}{0};\di{0}{0}\di{0}{1}}
[\Lambda,\Lambda_0,\rho^0] 
- A^{(V,V^e,1,0,0)\gamma}_{\di{0}{0}\di{0}{0};\di{0}{0}\di{0}{0}}
[\Lambda,\Lambda_0,\rho^0]\big)
\nonumber
\\*
&\qquad - n^2 \big( A^{(V,V^e,1,0,0)\gamma}_{
\di{0}{0}\di{0}{1};\di{0}{1}\di{0}{0}}
[\Lambda,\Lambda_0,\rho^0] 
- A^{(V,V^e,1,0,0)\gamma}_{\di{0}{0}\di{0}{0};\di{0}{0}\di{0}{0}}
[\Lambda,\Lambda_0,\rho^0]\big)\Big)
\bigg)_{\Lambda=\Lambda_0}\;,
\\
&\sum_{\gamma \text{ as in Def.~\ref{defint}.\ref{defint1}}}
R^{(V,2,1,0,0)\gamma}_{
\di{m^1+1}{m^2}\di{n^1+1}{n^2};\di{n^1}{n^2}\di{m^1}{m^2}}
[\Lambda_0,\Lambda_0,\rho^0]
\nonumber
\\*
& = -\sum_{\gamma \text{ as in Def.~\ref{defint}.\ref{defint1}}} 
\bigg(\Lambda\frac{\partial}{\partial \Lambda}\Big( 
A^{(V,V^e,1,0,0)\gamma}_{
\di{m^1+1}{m^2}\di{n^1+1}{n^2};\di{n^1}{n^2}\di{m^1}{m^2}}
[\Lambda,\Lambda_0,\rho^0] 
\nonumber
\\*
&\qquad - \sqrt{(m^1{+}1)(n^1{+}1)} 
A^{(V,V^e,1,0,0)\gamma}_{\di{1}{0}\di{1}{0};\di{0}{0}\di{0}{0}}
[\Lambda,\Lambda_0,\rho^0]\Big)\bigg)_{\Lambda=\Lambda_0}\;,
\\
& R^{(V,V^e,B,g,\iota)\gamma}_{m_1n_1;\dots;m_Nn_N}
[\Lambda_0,\Lambda_0,\rho^0]
\Big|_{\gamma \text{ as in Def.~\ref{defint}.\ref{defintnp}}} 
\nonumber
\\*
& = -\bigg(\Lambda\frac{\partial}{\partial \Lambda} 
A^{(V,V^e,B,g,\iota)\gamma}_{m_1n_1;\dots;m_Nn_N}
[\Lambda,\Lambda_0,\rho^0] 
\Big|_{\gamma \text{ as in Def.~\ref{defint}.\ref{defintnp}}}
\bigg)_{\Lambda=\Lambda_0}\;.
\end{align}
\end{subequations}
In particular,
\begin{align}
 R^{(1,1,1,0,0)}_{m_1n_1;\dots;m_4n_4}
[\Lambda,\Lambda_0,\rho^0] \equiv 0\;.
\label{RN4-40-0}
\end{align}
We first get (\ref{RN4-40-0}) at $\Lambda=\Lambda_0$ from
(\ref{RN4-40}). Since the rhs of (\ref{polV4}) vanishes for $V=1$ and
$N=4$, we conclude (\ref{RN4-40-0}) for any $\Lambda$. 

\begin{prp} \label{R-prop}
  Let $\gamma$ be an $R$-ribbon graph having $N$
  external legs, $V$ vertices, $V^e$ external vertices and
  segmentation index $\iota$, which is drawn on a genus-$g$ Riemann
  surface with $B$ boundary components. Then the contribution
  $R^{(V,V^e,B,g,\iota)\gamma}_{m_1n_1;\dots;m_Nn_N}$ of $\gamma$ to
  the expansion coefficient of the $\Lambda_0$-varied effective action
  describing a duality-covariant $\phi^4$-theory on
  $\mathbb{R}^4_\theta$ in the matrix base is bounded as follows:
\begin{enumerate}
\item \label{R-planar-R4}
If $\gamma$ is of the type described under \ref{defint4}--\ref{defint1} of
  Definition~\ref{defint}, we have 
\begin{align}
&\sum_{\gamma \text{ as in Def.~\ref{defint}.\ref{defint4}}} 
\Big| R^{(V,V^e,1,0,0)\gamma}_{
    \di{m^1}{m^2}\di{n^1}{n^2};\di{n^1}{n^2}\di{k^1}{k^2};
    \di{k^1}{k^2}\di{l^1}{l^2};\di{l^1}{l^2}\di{m^1}{m^2}} 
[\Lambda,\Lambda_0,\rho_0]\Big| 
\nonumber
\\*
&\quad \qquad
\leq \Big(\frac{\Lambda^2}{\Lambda_0^2}\Big) 
P^{4V-4}_1 \bigg[ \frac{\di{m^1}{m^2}\di{n^1}{n^2};\di{n^1}{n^2}\di{k^1}{k^2};
    \di{k^1}{k^2}\di{l^1}{l^2};\di{l^1}{l^2}\di{m^1}{m^2}}{
\theta \Lambda^2} \bigg]
\Big(\frac{1}{\Omega}\Big)^{3V-2-V^e}
\,P^{2V-2}\Big[\ln\frac{\Lambda_0}{\Lambda_R}\Big] \;,
\label{R4}  
\\
&\sum_{\gamma \text{ as in Def.~\ref{defint}.\ref{defint0}}} 
\Big|R^{(V,V^e,1,0,0)\gamma}_{\di{m^1}{m^2}\di{n^1}{n^2};
    \di{n^1}{n^2}\di{m^1}{m^2}} [\Lambda,\Lambda_0,\rho_0] \Big|
\nonumber
\\*
&\quad \qquad
\leq \Big(\frac{\Lambda^2}{\Lambda_0^2}\Big) (\theta \Lambda^2)
P^{4V-2}_2 \bigg[
\frac{\di{m^1}{m^2}\di{n^1}{n^2};\di{n^1}{n^2}\di{m^1}{m^2}}{
\theta \Lambda^2} \bigg]
\Big(\frac{1}{\Omega}\Big)^{3V-1-V^e}
\,P^{2V-1}\Big[\ln\frac{\Lambda_0}{\Lambda_R}\Big] \;,
\label{R0}
\\
&\sum_{\gamma \text{ as in Def.~\ref{defint}.\ref{defint1}}} 
\Big| R^{(V,V^e,1,0,0)}_{\di{m^1+1}{m^2}\di{n^1+1}{n^2};
    \di{n^1}{n^2}\di{m^1}{m^2}}[\Lambda,\Lambda_0,\rho_0]  \Big|
\nonumber
\\*
&\quad \qquad
\leq \Big(\frac{\Lambda^2}{\Lambda_0^2}\Big) (\theta \Lambda^2)
P^{4V-2}_2 \bigg[
\frac{\di{m^1+1}{m^2}\di{n^1+1}{n^2};\di{n^1}{n^2}\di{m^1}{m^2}}{
\theta \Lambda^2} \bigg]
\Big(\frac{1}{\Omega}\Big)^{3V-1-V^e}
\,P^{2V-1}\Big[\ln\frac{\Lambda_0}{\Lambda_R}\Big] \;,
\label{R1}
\end{align}

\item  \label{R-sub-planar}
If $\gamma$ is a subgraph of an 1PI planar graph with a selected set $T$ of
  trajectories on one distinguished boundary component and a second set $T'$
  of summed trajectories on that boundary component, we have
\begin{align}
&\sum_{\mathcal{E}^{s}} \sum_{\mathcal{E}^{t'}}
\big|R^{(V,V^e,B,0,\iota)\gamma}_{
    m_1n_1;\dots;m_Nn_N}[\Lambda,\Lambda_0,\rho_0] \big|
\nonumber
\\*
& \quad \leq \Big(\frac{\Lambda^2}{\Lambda_0^2}\Big) 
\big(\theta \Lambda^2\big)^{(2-\frac{N}{2})+2(1-B)}
\Big(\frac{1}{\Omega}\Big)^{3V-\frac{N}{2}-1 +B+2g-V^e-\iota+s+t'}
\nonumber
\\*
&\qquad \times  
 P^{4V-N}_{\big(2t'+ \sum_{\overrightarrow{n_j\mathfrak{o}[n_j]} \in T}
\min(2,\frac{1}{2}\langle \overrightarrow{n_j\mathfrak{o}[n_j]} \rangle)\big)}
 \bigg[ \frac{ m_1n_1;\dots;m_Nn_N}{\theta \Lambda^2} \bigg]
P^{2V-\frac{N}{2}}\Big[\ln \frac{\Lambda_0}{\Lambda_R}\Big]\;.
\label{RN4-sub}
\end{align}

\item \label{R-non-planar}
If $\gamma$ is a non-planar graph, we have
\begin{align}
&\sum_{\mathcal{E}^{s}} \big|R^{(V,V^e,B,g,\iota)}_{
    m_1n_1;\dots;m_Nn_N}[\Lambda,\Lambda_0,\rho_0] \big| 
\nonumber
  \\*
  &\quad 
\leq 
\Big(\frac{\Lambda^2}{\Lambda_0^2}\Big) 
\big(\theta \Lambda^2\big)^{(2-\frac{N}{2})+2(1-B-2g)}
 P^{4V-N}_0 \bigg[ \frac{ m_1n_1;\dots;m_Nn_N}{\theta \Lambda^2} \bigg]
\nonumber
\\*
&\qquad \times  
\Big(\frac{1}{\Omega}\Big)^{3V-\frac{N}{2}-1 
+B+2g-V^e-\iota+s}
  \,P^{2V-\frac{N}{2}}\Big[\ln \frac{\Lambda_0}{\Lambda_R}\Big]\;.
\label{RN4-np}
\end{align}

\end{enumerate}

\noindent
We have $R^{(V,V^e,B,g,\iota)}_{ m_1n_1;\dots;m_Nn_N}\equiv 0$ for
$N>2V{+}2$ or $\sum_{i=1}^N (m_i{-}n_i)\neq 0$.
\end{prp} 
\textit{Proof.}  Inserting the estimations of
Proposition~\ref{power-counting-prop} into (\ref{RN4}) we confirm
Proposition~\ref{R-prop} for $\Lambda=\Lambda_0$, which serves as
initial condition for the $\Lambda$-integration of (\ref{polV4}). This
entails the polynomial in $\ln \frac{\Lambda_0}{\Lambda_R}$ instead of
$\ln \frac{\Lambda}{\Lambda_R}$ appearing in
Propositions~\ref{power-counting-prop} and \ref{prop-H}. Accordingly,
when using Propositions~\ref{power-counting-prop} and \ref{prop-H} as
the input for (\ref{polV4}), we will further bound these estimations
by replacing $\ln \frac{\Lambda}{\Lambda_R}$ by $\ln
\frac{\Lambda_0}{\Lambda_R}$.

Due to (\ref{RN4-40-0}) the rhs of (\ref{polV4}) vanishes for
$N=2,V=1$ and for $N=6,V=2$. This means that the corresponding
$R$-functions are constant in $\Lambda$ so that the Proposition holds
for $R^{(1,1,1,0,0)}_{m_1n_1;m_2n_2}[\Lambda]$, 
$R^{(1,1,2,0,1)}_{m_1n_1;m_2n_2}[\Lambda]$ and
$R^{(2,2,1,0,0)}_{m_1n_1;\dots;m_2n_2}[\Lambda]$.  Since (\ref{polV4})
is a linear differential equation, the factor
$\frac{\Lambda^2}{\Lambda_0^2}$ relative to the estimation of the
$A$-functions of Proposition~\ref{power-counting-prop}, first appearing
in $R^{(1,1,1,0,0)}_{m_1n_1;m_2n_2}[\Lambda]$, 
$R^{(1,1,2,0,1)}_{m_1n_1;m_2n_2}[\Lambda]$ and
$R^{(2,2,1,0,0)}_{m_1n_1;\dots;m_2n_2}[\Lambda]$, survives to more
complicated graphs, provided that none of the $R$-functions is
relevant in $\Lambda$. 

For graphs according to Definition~\ref{defint}.\ref{defintnp}, the
first two lines on the rhs of (\ref{polV4}) yield in the same way as
in the proof of (\ref{AN4-1}) the integrand
(\ref{RN4-np}), with the degree of the polynomial in $\ln
\frac{\Lambda_0}{\Lambda_R}$ lowered by $1$. Since under the given
conditions an $A$-graph would be irrelevant, an $R$-graph with the
additional factor $\frac{\Lambda^2}{\Lambda_0^2}$ is relevant or
marginal. Thus, the $\Lambda$-integration of the first two lines on
the rhs of (\ref{polV4}) can be estimated by the integrand and a
factor $P^1[\ln \frac{\Lambda_0}{\Lambda_R}]$, in agreement with
(\ref{RN4-np}). In the same way we verify (\ref{RN4-sub}) for the
first two lines on the rhs of (\ref{polV4}).

In the remaining lines of (\ref{polV4}) we get by induction the
following estimation:
\begin{subequations}
\label{RN4-int}
\begin{align}
&\bigg|\bigg\{
\sum_{m,n,k,l} Q_{nm;lk}(\Lambda)
R^{(V_1)}_{\di{0}{0}\di{0}{0};\di{0}{0}\di{0}{0};mn;kl}[\Lambda]
\bigg\}_{[\text{Def.~\ref{defint}.\ref{defint0}}]}\bigg|
\nonumber
\\*
& \qquad \leq \Big(\frac{\Lambda^2}{\Lambda_0^2}\Big) 
\big(\theta \Lambda^2\big)
  \Big(\frac{1}{\Omega}\Big)^{3V-1-V^e}
  \,P^{2V-2}\Big[\ln \frac{\Lambda_0}{\Lambda_R}\Big]\;,
\label{RN4-int-a}
\\
&\bigg|\bigg\{
\sum_{m,n,k,l} Q_{nm;lk}(\Lambda)
\Big(R^{(V_1)}_{\di{1}{0}\di{0}{0};\di{0}{0}\di{1}{0};mn;kl}[\Lambda]
-R^{(V_1)}_{\di{0}{0}\di{0}{0};\di{0}{0}\di{0}{0};mn;kl}[\Lambda]\Big)
\bigg\}_{[\text{Def.~\ref{defint}.\ref{defint0}}]} \bigg|
\nonumber
\\*
& \qquad \leq \Big(\frac{\Lambda^2}{\Lambda_0^2}\Big) 
  \Big(\frac{1}{\Omega}\Big)^{3V-1-V^e}
  \,P^{2V-2}\Big[\ln \frac{\Lambda_0}{\Lambda_R}\Big]\;,
\label{RN4-int-b}
\\
& \bigg|\bigg\{
\sum_{m,n,k,l} Q_{nm;lk}(\Lambda)
R^{(V_1)}_{\di{1}{0}\di{1}{0};\di{0}{0}\di{0}{0};mn;kl}[\Lambda]
\bigg\}_{[\text{Def.~\ref{defint}.\ref{defint1}}]}\bigg| 
\nonumber
\\*
& \qquad \leq \Big(\frac{\Lambda^2}{\Lambda_0^2}\Big) 
  \Big(\frac{1}{\Omega}\Big)^{3V-1-V^e}
  \,P^{2V-2}\Big[\ln \frac{\Lambda_0}{\Lambda_R}\Big]\;,
\label{RN4-int-c}
\\
& \bigg|\bigg\{
\sum_{m,n,k,l} Q_{nm;lk}(\Lambda)
R^{(V_1)}_{\di{0}{0}\di{0}{0};\di{0}{0}\di{0}{0};
  \di{0}{0}\di{0}{0};\di{0}{0}\di{0}{0};mn;kl}[\Lambda]
\bigg\}_{[\text{Def.~\ref{defint}.\ref{defint4}}]}\bigg|
\nonumber
\\*
& \qquad \leq \Big(\frac{\Lambda^2}{\Lambda_0^2}\Big) 
  \Big(\frac{1}{\Omega}\Big)^{3V-2-V^e}
  \,P^{2V-3}\Big[\ln \frac{\Lambda_0}{\Lambda_R}\Big]\;.
\label{RN4-int-d}
\end{align}
\end{subequations}
These estimations are obtained in a similar way as (\ref{HN4-2a}),
(\ref{HN4-3a}) and (\ref{HN4-4a}). In particular, the improvement by
$(\theta \Lambda^2)^{-1}$ in (\ref{RN4-int-b}) is due to the
difference of graphs which according to Section~\ref{seccomposite}
yield a composite propagator (\ref{comp-prop-0}). To obtain
(\ref{RN4-int-c}) we have to use (\ref{RN4-sub}) with $\langle
\overrightarrow{n_1o(n_1)} \rangle + \langle
\overrightarrow{n_2o(n_2)} \rangle=2$, which for the graphs under
consideration is known by induction.

Multiplying (\ref{RN4-int}) by versions of Proposition~\ref{prop-H}
according to (\ref{polV4}), for $V_1<V$, we obtain again
(\ref{RN4-np}) or (\ref{RN4-sub}), with the degree of the polynomial
in $\ln \frac{\Lambda_0}{\Lambda_R}$ lowered by $1$, for the
integrand. Then the  $\Lambda$-integration proves (\ref{RN4-np}) and 
(\ref{RN4-sub}).

For graphs as in \ref{defint4}--\ref{defint1} of
Definition~\ref{defint} one shows in the same way as in the proof of
\ref{H-planar-H4}--\ref{H-planar-H1} of Proposition~\ref{prop-H} that
the last term in the third line of (\ref{polV4}) and the
$(V_1=V)$-terms in the remaining lines project to the irrelevant part
of these $R$-functions, i.e.\ lead to (\ref{R4})--(\ref{R1}). This was
already clear from (\ref{Rproj}). For the remaining $(V_1<V)$-terms in the
fourth to last lines of (\ref{polV4}) we obtain (\ref{R4})--(\ref{R1})
from (\ref{RN4-int}) and (\ref{H-H4})--(\ref{H-H1}). This finishes the
proof.\hfill $\square$%

\subsection{Finishing the convergence and renormalisation
  theorem}

We return now to the starting point of the entire estimation
procedure---the identity (\ref{Rlim}). We put $\Lambda=\Lambda_R$ in
Proposition~\ref{R-prop} and perform the $\Lambda_0$-integration in
(\ref{Rlim}): 
\begin{thm}
\label{final-theorem}
  The $\phi^4$-model on $\mathbb{R}^4_\theta$ is (order by order in
  the coupling constant) renormalisable in the matrix base by
  adjusting the coefficients $\rho^0_a[\Lambda_0]$ defined in
  (\ref{rhoa0}) and (\ref{rhoa}) of the initial interaction
  (\ref{ct4}) to give (\ref{initrho}) and by integrating the 
  the Polchinski equation according to Definition~\ref{defint}.

  The limit
  $A^{(V,V^e,B,g,\iota)}_{m_1n_1;\dots;m_Nn_N}[\Lambda_R,\infty]
  :=\lim_{\Lambda_0 \to \infty}
  A^{(V,V^e,B,g,\iota)}_{m_1n_1;\dots;m_Nn_N}
  [\Lambda_R,\Lambda_0,\rho^0[\Lambda_0]]$ of the
  expansion coefficients of the effective action
  $L[\phi,\Lambda_R,\Lambda_0,\rho^0[\Lambda_0]]$,
  see (\ref{Lg4}), exists and satisfies
\begin{align}
  & \Big| (2\pi\theta)^{\frac{N}{2}-2}
  A^{(V,V^e,B,g,\iota)}_{m_1n_1;\dots;m_Nn_N}[\Lambda_R,\infty] -
  (2\pi\theta)^{\frac{N}{2}-2}
  A^{(V,V^e,B,g,\iota)}_{m_1n_1;\dots;m_Nn_N}
[\Lambda_R,\Lambda_0,\rho^0]
  \Big|
\nonumber
  \\*
  &\quad \leq \frac{\Lambda_R^{6-N}}{\Lambda_0^2} 
\Big(\frac{1}{\Omega\theta\Lambda_R^2}\Big)^{2(B+2g-1)} 
\nonumber
  \\*
&\qquad\times 
P^{4V-N}_0 \bigg[ \frac{ m_1n_1;\dots;m_Nn_N}{\theta \Lambda_R^2} \bigg]
\Big(\frac{1}{\Omega}\Big)^{3V-\frac{N}{2}-V^e-\iota}
  \,P^{2V-\frac{N}{2}}\Big[\ln \frac{\Lambda_0}{\Lambda_R}\Big]\;.
\label{limA4}
\end{align}
\end{thm}
\textit{Proof.} We insert Proposition~\ref{R-prop}, taken at
$\Lambda=\Lambda_R$, into (\ref{Rlim}).  We also use
(\ref{polab-range}) in Proposition~\ref{R-prop}.\ref{R-planar-R4}.
Now, the existence of the limit and its property (\ref{limA4}) are a
consequence of Cauchy's criterion.  Note that $\int
\frac{dx}{x^3}\,P^q[\ln x] = \frac{1}{x^2}
P^{\prime q}[\ln x]$. \hfill $\square$% 
\bigskip

\section{Conclusion}

In this paper we have proven that the real $\phi^4$-model on
(Euclidean) noncommutative $\mathbb{R}^4$ is renormalisable to all
orders in perturbation theory. The bare action of relevant and
marginal couplings of the model is parametrised by four (divergent)
quantities which require normalisation to the experimental data at a
physical renormalisation scale. The corresponding physical parameters
which determine the model are the mass, the field amplitude (to be
normalised to $1$), the coupling constant and (in addition to the
commutative version) the frequency of an harmonic oscillator
potential. The appearance of the oscillator potential is not a bad
trick but a true physical effect. It is the self-consistent solution
of the UV/IR-mixing problem found in the traditional noncommutative
$\phi^4$-model in momentum space. It implements the duality (see also
\cite{Langmann:2002cc}) that \emph{noncommutativity relevant at short
distances goes hand in hand with a modified structure of space
relevant at large distances}.

Such a modified structure of space at very large distances seems to be
in contradiction with experimental data. But this is not true. Neither
position space nor momentum space are the adapted frames to interpret
the model. An invariant characterisation of the model is the spectrum
of the Laplace-like operator which defines the free theory. Due to the
link to Meixner polynomials, the spectrum is discrete. Comparing
(\ref{B1}) with (\ref{B10}) and (\ref{B11}) we see that the spectrum
of the squared momentum variable has an equidistant spacing of $\frac{4
  \Omega}{\theta}$. Thus, $\sqrt{\frac{4 \Omega}{\theta}}$ is the
minimal (non-vanishing) momentum of the scalar field which is allowed
in the noncommutative universe. We can thus identify the parameter
$\sqrt{\Omega}$ with the ratio of the Planck length to the size of the
(finite!) universe. Thus, for typical momenta on earth, the
discretisation is not visible. However, there should be an observable
effect at extremely huge scales. Indeed, there is some evidence of
discrete momenta in the spectrum of the cosmic microwave background
\cite{Luminet:2003dx}\footnote{According to the main purpose of
\cite{Luminet:2003dx} one should also discuss other topologies than
the noncommutative $\mathbb{R}^D$.}.

Of course, when we pass to a frame where the propagator becomes
$\frac{1}{\mu_0^2+p^2}$, with $p$ now being discrete, we also have to
transform the interactions. We thus have to shift the unitary matrices
$U^{(\alpha)}_m$ appearing in (\ref{B1}) from the kinetic matrix or
the propagator into the vertex. The properties of that dressed
(physical) vertex will be studied elsewhere. 

Another interesting exercise is the evaluation of the $\beta$-function
of the duality-covariant $\phi^4$-model \cite{Grosse:2004by}.  It
turns out that the one-loop $\beta$-function for the coupling constant
remains non-negative and and vanishes for the self-dual case
$\Omega=1$.  Moreover, the limit $\Omega \to 0$ exists at the one-loop
level. This is related to the fact that the UV/IR-mixing in momentum
space becomes problematic only at higher loop order.

Of particular interest would be the limit $\theta \to 0$.  In the
developed approach, $\theta$ defines the reference size of an
elementary cell in the Moyal plane.  All dimensionful quantities, in
particular the energy scale $\Lambda$, are measured in units of
(appropriate powers of) $\theta$. In the final result of
Theorem~\ref{final-theorem}, these mass dimensions are restored. Then,
we learn from (\ref{limA4}) that a finite $\theta$ regularises the
non-planar graphs. This means that for given $\Lambda_0$ and
$\Lambda_R$ the limit $\theta \to 0$ cannot be taken.

On the other hand, there could be a chance to let $\theta$ depend on
$\Lambda_0$ in the same way as in the two-dimensional case
\cite{Grosse:2003nw} the oscillator frequency $\Omega$ was switched
off with the limit $\Lambda_0 \to \infty$. However, this does not
work. The point is that taking in (\ref{V2}) instead of the
$\Lambda_0$-derivative the $\theta$-derivative, there is now a
contribution from the $\theta$-dependence of the propagator.  This
leads in the analogue of the differential equation (\ref{VVV}) to a
term bilinear in $L$. Looking at the proof of
Proposition~\ref{R-prop}, we see that this $L$-bilinear term will
remove the factor $\Lambda_0^{-2}$.

Thus, the limit $\theta \to 0$ is singular. This is not surprising.
In the limit $\theta \to 0$ the distinction between planar and
non-planar graphs disappears (which is immediately clear in momentum
space). Then, non-planar two- and four-point functions should yield
the same divergent values as their planar analogues. Whereas the bare
divergences in the planar sector are avoided by the mixed boundary
conditions in \ref{defint4}-\ref{defint1} of Definition~\ref{defint},
the na\"{\i}ve initial condition in
Definition~\ref{defint}.\ref{defintnp} for non-planar graphs leaves the
bare divergences in the limit $\theta \to 0$.

The next goal must be to generalise the renormalisation proof to gauge
theories. This requires probably a gauge-invariant extension of the
harmonic osillator potential. The result should be compared with
string theory, because gauge theory on the Moyal plane arises in the
zero-slope limit of string theory in presence of a Neveu-Schwarz
$B$-field \cite{Seiberg:1999vs}. As renormalisation requires an
appropriate structure of the space at very large distances, the
question arises whether the oscillator potential has a counterpart in
string theory. In this respect, it is tempting\footnote{We would like
  to thank G.~Bonelli for this interesting remark.} to relate the
oscillator potential to the maximally supersymmetric pp-wave
background metric of type IIB string theory found in
\cite{Blau:2001ne},
  \begin{align}
    ds^2 = 2 dx^+ dx^- - 4\lambda^2 \sum_{i=1}^8 (x^i)^2 (dx^-)^2
    +  \sum_{i=1}^8 (dx^i)^2\;, 
  \end{align}
for $dx^{\pm} = \frac{1}{\sqrt{2}}(dx^9\pm dx^{10})$,
which solves Einstein's equations for an energy-momentum tensor
relative to the 5-form field strength
\begin{align}
F_5 = \lambda dx^-\big( dx^1\wedge dx^2 \wedge dx^3 \wedge dx^4
+ dx^5\wedge dx^6 \wedge dx^7 \wedge dx^8\big)\;.
\end{align}

\begin{acknowledgement}

We are indebted to Stefan Schraml for providing us with the references
to orthogonal polynomials, without which the completion of the proof
would have been impossible. We had stimulating discussions with Edwin
Langmann, Vincent Rivasseau and Harold Steinacker. We are
grateful to Christoph Kopper for indicating to us a way to
reduce in our original power-counting estimation the polynomial in
$\big(\ln \frac{\Lambda_0}{\Lambda_R}\big)$ to a polynomial in
$\big(\ln \frac{\Lambda}{\Lambda_R}\big)$, thus permitting immediately
the limit $\Lambda_0 \to \infty$. We would like to
thank the Max-Planck-Institute for Mathematics in the Sciences
(especially Eberhard Zeidler), the Erwin-Schr\"odinger-Institute and 
the Institute for Theoretical Physics of the University of Vienna for
the generous support of our collaboration.

\end{acknowledgement}

\begin{appendix}

\section{Evaluation of the propagator}
\label{appMeixner}

\subsection{Diagonalisation of the kinetic matrix via Meixner polynomials}

Our goal is to diagonalise the (four-dimensional) kinetic matrix
$G_{\di{m^1}{m^2}\di{n^1}{n^2};\di{k^1}{k^2}\di{l^1}{l^2}}$ given in
(\ref{G4D}), making use of the angular momentum conservation
$\alpha^r=n^r-m^r=k^r-l^r$ (which is due to the $SO(2)\times
SO(2)$-symmetry of the action.  For $\alpha^r\geq 0$ we thus look for
a representation
\begin{align}
  G_{\di{m^1}{m^2}\di{m^1+\alpha^1}{m^2+\alpha^2};
    \di{l^1+\alpha^1}{l^2+\alpha^2}\di{l^1}{l^2}} &= \sum_{i^1,i^2}
  U^{(\alpha^1)}_{m^1i^1}U^{(\alpha^2)}_{m^2i^2} \big(\tfrac{2}{\theta_1}
  v_{i^1} +\tfrac{2}{\theta_2} v_{i^2}+\mu_0^2\big)
  U^{(\alpha^1)}_{i^1l^1}U^{(\alpha^2)}_{i^2l^2}\;,
\label{B1}
  \\
  \delta_{ml} &= \sum_{i} U^{(\alpha)}_{mi}U^{(\alpha)}_{il} \;. 
\label{B1a}
\end{align}
The sum over $i^1,i^2$ would be an integration for continuous eigenvalues
$v_{i^r}$. Comparing this ansatz with (\ref{G4D}) we obtain, eliminating $i$
in favour of $v$, the recurrence relation
\begin{align}
  (1{-}\Omega^2) \sqrt{m(\alpha{+}m)} U^{(\alpha)}_{m-1}(v) &+
  \big(v-(1{+}\Omega^2)(\alpha{+}1{+}2m)\big)U^{(\alpha)}_m(v) \nonumber
  \\*
  &+ (1{-}\Omega^2)\sqrt{(m{+}1)(\alpha{+}m{+}1)} U^{(\alpha)}_{m+1}(v)=0
\label{B2}
\end{align}
to determine $U^{(\alpha)}_{m}(v)$ and $v$.  We are interested in the
case $\Omega>0$. In order to make contact with standard formulae we
put
\begin{align}
  U^{(\alpha)}_m(v)
=f^{(\alpha)}(v)\, \frac{1}{\tau^m} \sqrt{\frac{(\alpha{+}m)! }{m!}}
V^{(\alpha)}_m(v)\;, \qquad v=  \nu x + \rho \;.
\label{B-UV}
\end{align}
We obtain after division by $f^{(\alpha)}(v)$
\begin{align}
  0&=\frac{(1{-}\Omega^2)}{\tau^{m-1}} \sqrt{\frac{m^2(\alpha{+}m)!}{m!}}
  V^{(\alpha)}_{m-1}(\nu x{+}\rho) \nonumber
  \\*
  &- \frac{1}{\tau^{m}}\big((1{+}\Omega^2) (\alpha{+}1{+}2m)-\rho-\nu
  x\big) \sqrt{\frac{(\alpha{+}m)!}{m!}} 
V^{(\alpha)}_{m}(\nu x{+}\rho) \nonumber
  \\*
  & + \frac{(1{-}\Omega^2)}{\tau^{m+1}}\sqrt{\frac{ (\alpha{+}m{+}1)^2
      (\alpha{+}m)!}{m!}}  V^{(\alpha)}_{m+1}(\nu x{+}\rho) \;,
 \end{align}
 i.e.\ 
\begin{align}
  -\frac{\nu}{\tau(1{-}\Omega^2)} x 
V^{(\alpha)}_{m}(\nu x{+}\rho) 
&= m  V^{(\alpha)}_{m-1}(\nu x{+}\rho) -
  \frac{(1{+}\Omega^2)(\alpha{+}1{+}2m)-\rho}{\tau(1{-}\Omega^2)}
  V^{(\alpha)}_{m}(\nu x{+}\rho) \nonumber
  \\*
  &+ \frac{1}{\tau^2} (\alpha{+}m{+}1) 
V^{(\alpha)}_{m+1}(\nu x{+}\rho) \;,
 \end{align}
 Now we put
\begin{align}
  1{+}\alpha &=\beta \;, & \frac{1}{\tau^2}&=c\;, &
  \frac{2(1{+}\Omega^2)}{\tau(1{-}\Omega^2)} &=1{+}c \;, &
  \frac{(1{+}\Omega^2)\beta-\rho}{\tau(1{-}\Omega^2)}&=\beta c\;, &
  \frac{\nu}{(1{-}\Omega^2)\tau} &= 1{-}c
\label{lagmeix}
\end{align}
and
\begin{align}
  V^{(\alpha)}_n(\nu x{+}\rho)= M_n (x;\beta,c)\;,
\end{align}
which yields the recursion relation for the Meixner polynomials
\cite{Koekoek:1996}: 
\begin{align}
  (c{-}1)xM_m(x;\beta,c) &= c(m{+}\beta) M_{m+1}(x;\beta,c) \nonumber
  \\*
  &- (m+(m{+}\beta)c) M_m(x;\beta,c)+m M_{m-1}(x;\beta,c) \;.
\end{align}
The solution of (\ref{lagmeix}) is
\begin{align}
  \tau &= \frac{(1{\pm} \Omega)^2}{1{-}\Omega^2} \equiv \frac{1{\pm}
    \Omega}{1{\mp}\Omega} \;, & c&=\frac{(1{\mp} \Omega)^2}{(1{\pm}
    \Omega)^2}\;, & \nu &= \pm 4 \Omega\;, & \rho = \pm
  2\Omega(1{+}\alpha)\;.
\end{align}
We have to chose the upper sign, because the eigenvalues $v$ are positive. We
thus obtain
\begin{align}
  U^{(\alpha)}_m(v_x)&= f^{(\alpha)}(x) 
\sqrt{\frac{(\alpha{+}n)!}{n!}}
  \Big(\frac{1{-}\Omega}{1{+}\Omega}\Big)^m M_m\Big(x;1{+}\alpha, \frac{(1{-}
    \Omega)^2}{(1{+}\Omega)^2} \Big) \;, 
\nonumber
\\*
v_x &= 2\Omega (2 x {+} \alpha{+}1)  \;.
\label{B10}
\end{align}
The function $f^{(\alpha)}(x)$ is identified by comparison of
(\ref{B1a}) with the orthogonality relation of Meixner polynomials
\cite{Koekoek:1996},
\begin{align}
  \sum_{x=0}^\infty \frac{\Gamma(\beta{+}x) c^x}{\Gamma(\beta) x!}
  M_m(x;\beta,c) M_n(x;\beta,c) &= \frac{c^{-n} n!
    \Gamma(\beta)}{\Gamma(\beta{+}n)(1{-}c)^\beta} \delta_{mn}\;.
\label{B11}
\end{align}
The result is
\begin{align}
  U^{(\alpha)}_m(v_x)&= \sqrt{ \binom{\alpha{+}m}{m}\binom{\alpha{+}x}{x}}
  \Big(\frac{2\sqrt{\Omega}}{1{+}\Omega}\Big)^{\alpha+1}
  \Big(\frac{1{-}\Omega}{1{+}\Omega}\Big)^{m+x}
  M_m\Big(x;1{+}\alpha,\frac{(1{-} \Omega)^2}{(1{+}\Omega)^2} \Big)\;.
\end{align}
The Meixner polynomials can be represented by hypergeometric functions
\cite{Koekoek:1996}
\begin{align}
  M_m\Big(x;1{+}\alpha,\frac{(1{-} \Omega)^2}{(1{+}\Omega)^2}\Big) =
  {}_2F_1\Big(\genfrac{}{}{0pt}{}{-m,-x}{1{+}\alpha}\Big| {-} \frac{4
    \Omega}{(1{-} \Omega)^2}\Big)\;.
\end{align}
This shows that the matrices $U_{ml}^{(\alpha)}$ in (\ref{B1}) and
(\ref{B1a}) are symmetric in the lower indices.

\subsection{Evaluation of the propagator}
\label{app:evalprop}

Now we return to the computation of the propagator, which is obtained
by sandwiching the inverse eigenvalues $(\frac{2}{\theta_1} v_{i^1} +
\frac{2}{\theta_2} v_{i^2} + \mu_0^2)$ between the unitary matrices
$U^{(\alpha)}$. With (\ref{B10}) and the use of Schwinger's trick
$\frac{1}{A}=\int_0^\infty dt \,\mathrm{e}^{-tA}$ we have for
$\theta_1=\theta_2=\theta$
\begin{align}
  &\Delta_{\di{m^1}{m^2}\di{m^1+\alpha^1}{m^2+\alpha^2};
    \di{l^1+\alpha^1}{l^2+\alpha^2}\di{l^1}{l^2}} \nonumber
  \\*
  &= \frac{\theta}{8\Omega} \int_0^\infty \!\! dt\! 
  \sum_{x^1,x^2=0}^\infty\!\! 
  \mathrm{e}^{-\frac{t}{4\Omega}(v_{x^1}+v_{x^2}+\theta \mu_0^2/2)}
  U^{(\alpha^1)}_{m^1}(v_{x^1}) U^{(\alpha^2)}_{m^2}(v_{x^2})
  U^{(\alpha^1)}_{l^1}(v_{x^1}) U^{(\alpha^2)}_{l^2}(v_{x^2}) \nonumber
  \\*
  &= \frac{\theta}{8\Omega} \int_0^\infty \!\! dt \;
  \mathrm{e}^{-t(1+\frac{\mu_0^2\theta}{8\Omega}
    +\frac{1}{2}(\alpha^1+\alpha^2)) } 
\nonumber
\\*
& \qquad \times 
\prod_{i=1}^2 \bigg\{
  \sqrt{\dbinom{\alpha^i{+}m^i}{m^i}\dbinom{\alpha^i{+}l^i}{l^i}}
  \Big(\frac{4\Omega}{(1{+}\Omega)^2}\Big)^{\alpha^i+1}
  \Big(\frac{1{-}\Omega}{1{+}\Omega}\Big)^{m^i+l^i} 
\nonumber
  \\*
  &\qquad\qquad \times 
\sum_{x^i=0}^\infty \bigg\{
\frac{(\alpha^i{+}x^i)!}{x^i!\alpha^i!}
  \Big(\frac{\mathrm{e}^{-t}(1{-}\Omega)^2}{(1{+}\Omega)^2}\Big)^{x^i}
\nonumber
  \\*
  &\qquad\qquad\quad \times 
  {}_2F_1\Big(\di{-m^i,-x^i}{1{+}\alpha^i}\Big| {-} \frac{4 \Omega}{(1{-}
    \Omega)^2}\Big) \,{}_2F_1\Big(\di{-l^i,-x^i}{1{+}\alpha^i}\Big| {-}
  \frac{4 \Omega}{(1{-} \Omega)^2}\Big) \bigg\}\bigg\}\;.
\label{Schwinger}
\end{align}
We use the following identity for hypergeometric functions,
\begin{align}
  \sum_{x=0}^\infty & \frac{(\alpha{+}x)!}{x!\alpha!}  a^x\;
  {}_2F_1\Big(\di{-m,-x}{1{+}\alpha}\Big| b\Big)
  \,{}_2F_1\Big(\di{-l,-x}{1{+}\alpha}\Big| b\Big) \nonumber
  \\*
  &= \frac{(1{-}(1{-}b)a)^{m+l}}{(1{-}a)^{\alpha+m+l+1}}
  {}_2F_1\Big(\di{-m\,,\;-l}{1{+}\alpha}\Big| \frac{a
    b^2}{(1{-}(1{-}b)a)^2}\Big)\;,\qquad |a|<1\;.
\label{FFF}
\end{align}
The identity (\ref{FFF}) is probably known, but because it is crucial
for the solution of the free theory, we provide the proof in
Section~\ref{id-hypergeom}. We insert the rhs of (\ref{FFF}), expanded
as a finite sum, into (\ref{Schwinger}), where we also put
$z=\mathrm{e}^{-t}$:
\begin{align}
  &\Delta_{\di{m^1}{m^2}\di{m^1+\alpha^1}{m^2+\alpha^2};
    \di{l^1+\alpha^1}{l^2+\alpha^2}\di{l^1}{l^2}} \nonumber
  \\*
  &= \frac{\theta}{8\Omega} \sum_{u^1=0}^{\min(m^1,l^1)}
  \sum_{u^2=0}^{\min(m^2,l^2)} \int_0^1 \!\! dz \;
  \frac{z^{\frac{\mu_0^2\theta}{8\Omega} +\frac{1}{2}(\alpha^1+\alpha^2)
      +u^1+u^2} (1-z)^{m^1+m^2+l^1+l^2-2u^1-2u^2}}{
    \big(1-\frac{(1{-}\Omega)^2}{(1{+}\Omega)^2}
    z\big)^{\alpha^1+\alpha^2+m^1+m^2+l^1+l^2+2}}
\nonumber  
\\*
  &\times \prod_{i=1}^2 \bigg\{
  \Big(\frac{4\Omega}{(1{+}\Omega)^2}\Big)^{\alpha^i+2u^i+1}
  \Big(\frac{1{-}\Omega}{1{+}\Omega}\Big)^{m^i+l^i-2u^i} 
  \frac{\sqrt{m^i!(\alpha^i{+}m^i)!l^i! (\alpha^i{+}l^i)!}}{
    (m^i{-}u^i)!(l^i{-}u^i)! (\alpha^i{+}u^i)!u^i!}  \bigg\}\;.
\label{Schwinger1}
\end{align}
This formula tells us the important property
\begin{align}
  0 \leq \Delta_{\di{m^1}{m^2}\di{m^1+\alpha^1}{m^2+\alpha^2};
    \di{l^1+\alpha^1}{l^2+\alpha^2}\di{l^1}{l^2}} \leq
  \Delta_{\di{m^1}{m^2}\di{m^1+\alpha^1}{m^2+\alpha^2};
    \di{l^1+\alpha^1}{l^2+\alpha^2}\di{l^1}{l^2}}\Big|_{\mu_0^2=0}\;,
\end{align}
i.e.\ all matrix elements of the propagator are positive and majorised
by the massless matrix elements. The representation (\ref{Schwinger1})
seems to be the most convenient one for analytical estimations of the
propagator. The strategy\footnote{We are grateful to Vincent Rivasseau
  for this idea.} would be to divide the integration domain into
slices and to maximise the individual $z$-dependent terms of the
integrand over the slice, followed by resummation
\cite{Rivasseau:2004az,Rivasseau:2004??}. 

The $z$-integration in (\ref{Schwinger1}) leads
according to \cite[\S 9.111]{GR} again to a hypergeometric function:
\begin{align}
  &\Delta_{\di{m^1}{m^2}\di{m^1+\alpha^1}{m^2+\alpha^2};
    \di{l^1+\alpha^1}{l^2+\alpha^2}\di{l^1}{l^2}} =
    \frac{\theta}{8\Omega}
\nonumber
  \\*
  & \times \sum_{u^1=0}^{\min(m^1,l^1)}
  \sum_{u^2=0}^{\min(m^2,l^2)} \frac{\Gamma\big(1{+}
    \frac{\mu_0^2\theta}{8\Omega}
    {+}\frac{1}{2}(\alpha^1{+}\alpha^2){+}u^1{+}u^2\big)
    (m^1{+}m^2{+}l^1{+}l^2{-}2u^1{-}2u^2)!}{ \Gamma\big(2{+}
    \frac{\mu_0^2\theta}{8\Omega} {+}\frac{1}{2}(\alpha^1{+}\alpha^2)
    {+}m^1{+}m^2{+}l^1{+}l^2{-}u^1{-}u^2\big)} \nonumber
  \\*
  &\times {}_2F_1\bigg(\di{1{+}\frac{\mu_0^2\theta}{8\Omega}
    {+}\frac{1}{2}(\alpha^1{+}\alpha^2){+}u^1{+}u^2\,,\;
    2{+}m^1{+}m^2{+}l^1{+}l^2{+}\alpha^1{+}\alpha^2 }{ 2{+}
    \frac{\mu_0^2\theta}{8\Omega} {+}\frac{1}{2}(\alpha^1{+}\alpha^2)
    {+}m^1{+}m^2{+}l^1{+}l^2{-}u^1{-}u^2} \bigg|
  \frac{(1{-}\Omega)^2}{(1{+}\Omega)^2} \bigg) \nonumber
  \\*
  &\times \prod_{i=1}^2 \bigg\{
  \Big(\frac{4\Omega}{(1{+}\Omega)^2}\Big)^{\alpha^i+2u^i+1}
  \Big(\frac{1{-}\Omega}{1{+}\Omega}\Big)^{m^i+l^i-2u^i} \nonumber
  \frac{\sqrt{m^i!(\alpha^i{+}m^i)!l^i! (\alpha^i{+}l^i)!}}{
    (m^i{-}u^i)!(l^i{-}u^i)! (\alpha^i{+}u^i)!u^i!}  \bigg\} \nonumber
  \\
  &= \frac{\theta}{2(1{+}\Omega)^2} \sum_{u^1=0}^{\min(m^1,l^1)}
  \sum_{u^2=0}^{\min(m^2,l^2)} \prod_{i=1}^2 \sqrt{
    \binom{\alpha^i{+}m^i}{\alpha^i{+}u^i}
    \binom{\alpha^i{+}l^i}{\alpha^i{+}u^i} \binom{m^i}{u^i}\binom{l^i}{u^i}}
\nonumber
  \\*
  &\times \Big(\frac{1{-}\Omega}{1{+}\Omega}\Big)^{m^i+l^i-2u^i} 
\,B\big(1{+} \tfrac{\mu_0^2\theta}{8\Omega}
  {+}\tfrac{1}{2}(\alpha^1{+}\alpha^2){+}u^1{+}u^2,
  1{+}m^1{+}m^2{+}l^1{+}l^2{-}2u^1{-}2u^2 \big) \nonumber
  \\*
  &\times {}_2F_1\bigg(\di{1{+} m^1{+}m^2{+}l^1{+}l^2{-}2u^1{-}2u^2\,,\;
    \frac{\mu_0^2\theta}{8\Omega}
    {-}\frac{1}{2}(\alpha^1{+}\alpha^2){-}u^1{-}u^2 }{ 2{+}
    \frac{\mu_0^2\theta}{8\Omega} {+}\frac{1}{2}(\alpha^1{+}\alpha^2)
    {+}m^1{+}m^2{+}l^1{+}l^2{-}u^1{-}u^2} \bigg|
  \frac{(1{-}\Omega)^2}{(1{+}\Omega)^2} \bigg)\;.
\label{propalpha}
\end{align}
We have used \cite[\S 9.131.1]{GR} to obtain the last line. The form
(\ref{propalpha}) will be useful for the evaluation of special cases and of
the asymptotic behaviour. In the main part, for presentational purposes, 
$\alpha^i$ is eliminated in favour of $k^i,n^i$ and the summation variable
$v^i:=m^i+l^i-2u^i$ is used. The final result is given in
(\ref{noalpha}).

For $\mu_0=0$ we can in few cases evaluate the sum over $u^i$ exactly. First,
for $l^i=0$ we also have $u^i=0$. If additionally $\alpha^i=0$ we get
\begin{align}
\Delta_{\di{m^1}{m^2}\di{m^1}{m^2}; \di{0}{0}\di{0}{0}} \Big|_{\mu_0=0}
 &= \frac{\theta}{2(1{+}\Omega)^2(1{+}m^1{+}m^2)} 
\Big(\frac{1{-}\Omega}{1{+}\Omega}\Big)^{m^1+m^2}\;.
\label{propalpha0l0}
\end{align}
One should notice here the exponential decay for $\Omega>0$. It can be seen
numerically that this is a general feature of the propagator: Given $m^i$ and
$\alpha^i$, the maximum of the propagator is attained at $l^i=m^i$. Moreover, 
the decay with $|l^i-m^i|$ is exponentially so that the sum 
\begin{align}
\sum_{l^1,l^2}  \Delta_{\di{m^1}{m^2}\di{m^1+\alpha^1}{m^2+\alpha^2};
    \di{l^1+\alpha^1}{l^2+\alpha^2}\di{l^1}{l^2}} 
\end{align}
converges. We confirm this argumentation numerically in
(\ref{Form4-2}).

It turns out numerically that the
maximum of the propagator for indices restricted by $\mathcal{C}
\leq \max(m^1,m^2,n^1,n^2,k^1,k^2,l^1,l^2) \leq 2\mathcal{C}$ 
is found in the subclass $
\Delta_{\di{m^1}{0}\di{n^1}{0}; \di{n^1}{0}\di{m^1}{0}} $ of
propagators.  Coincidently, the computation in case of
$m^2=l^2=\alpha^2=0$ simplifies considerably. If additionally
$m^1=n^1$ we obtain a closed result:
\begin{align}
  \Delta_{\di{m}{0}\di{m}{0}; \di{m}{0}\di{m}{0}} 
&= \frac{\theta}{2(1{+}\Omega)^2} 
\sum_{u=0}^{m} \frac{(m!)^2(2u)!}{(m{-}u)!(u!)^2
(1{+}m{+}u)!} 
\nonumber
\\*
&\qquad\qquad \times 
{}_2F_1\bigg(\di{1{+} 2u\,,\; u{-}m
    }{2{+}m{+}u} \bigg| 
   \frac{(1{-}\Omega)^2}{(1{+}\Omega)^2} \bigg) 
\bigg(\frac{1{-}\Omega}{1{+}\Omega} \bigg)^{\!2u} 
\nonumber
\\
&=  \frac{\theta}{2(1{+}\Omega)^2} \sum_{u=0}^{m}\sum_{s=0}^{m-u} (-1)^s 
\frac{(m!)^2(2u{+}s)!}{(m{-}u{-}s)!(u!)^2
(1{+}m{+}u{+}s)!s!} 
\bigg(\frac{(1{-}\Omega)^2}{(1{+}\Omega)^2} \bigg)^{u+s} 
\nonumber
\\
&=  \frac{\theta}{2(1{+}\Omega)^2} \sum_{u=0}^{m}\sum_{r=u}^{m} (-1)^{u+r} 
\frac{(m!)^2(r{+}u)!}{(m{-}r)!(u!)^2
(1{+}m{+}r)!(r{-}u)!} 
\bigg(\frac{(1{-}\Omega)^2}{(1{+}\Omega)^2} \bigg)^{r} 
\nonumber
\\
&=  \frac{\theta}{2(1{+}\Omega)^2} \sum_{r=0}^{m}  (-1)^{r} 
\frac{(m!)^2}{(m{-}r)!(1{+}m{+}r)!} \;
{}_2F_1\Big(\di{r{+}1\,,\;-r}{1}\Big| 1\Big)
\bigg(\frac{(1{-}\Omega)^2}{(1{+}\Omega)^2} \bigg)^{r} 
\nonumber
\\
&=  \frac{\theta}{2(1{+}\Omega)^2} \sum_{r=0}^{m}  
\frac{(m!)^2}{(m{-}r)!(1{+}m{+}r)!} 
\bigg(\frac{(1{-}\Omega)^2}{(1{+}\Omega)^2} \bigg)^{r} 
\nonumber
\\
&=  \frac{\theta}{2(1{+}\Omega)^2(m{+}1)} 
\;{}_2F_1\Big(\di{1\,,\;-m}{m{+}2}\Big| 
-\frac{(1{-}\Omega)^2}{(1{+}\Omega)^2} \Big)
\nonumber
\\*
& \sim \left\{ \begin{array}{cc} 
\dfrac{\theta}{8\Omega(m{+}1)}  \qquad & \text{for } \Omega>0\;,~m \gg
1\;,
\\[1ex]
\dfrac{\sqrt{\pi}\theta}{4 \sqrt{m{+}\frac{3}{4}}}
\qquad & \text{for } \Omega=0\;,~m \gg 1\;.
\end{array}\right.
\label{propmmmm}
\end{align}
We see a crucial difference in the asymptotic behaviour for $\Omega>0$
versus $\Omega=0$. The slow decay with $m^{-\frac{1}{2}}$ of the
propagator is responsible for the non-renormalisability of the
$\phi^4$-model in case of $\Omega=0$. The numerical result 
(\ref{Form4-1}) shows that 
the maximum of the propagator for indices
restricted by $\mathcal{C} \leq \max(m^1,m^2,n^1,n^2,k^1,k^2,l^1,l^2)
\leq 2\mathcal{C}$ is very close to the result (\ref{propmmmm}), for
$m=\mathcal{C}$. For $\Omega=0$ the maximum is exactly given by
the $6^{\text{th}}$ line of (\ref{propmmmm}).

\subsection{Asymptotic behaviour of the propagator for large $\alpha^i$}

We consider various limiting cases of the propagator, making use of the
asymptotic expansion (Stirling's formula) of the $\Gamma$-function,
\begin{align}
\Gamma(n{+}1) \sim \Big(\frac{n}{\mathrm{e}}\Big)^n \sqrt{2\pi
    (n+\tfrac{1}{6})} + \mathcal{O}(n^{-2})\;.
\label{Stirling}
    \end{align}
This implies
\begin{align}
  \frac{\Gamma(n{+}1{+}a)}{\Gamma(n{+}1{+}b)} 
  &\sim n^{a-b} \Big(1 + \frac{(a{-}b)(a{+}b{+}1)}{2n} +
  \mathcal{O}(n^{-2})\Big)\;.
\end{align}
We rewrite the propagator (\ref{propalpha}) in a manner where the
large-$\alpha^i$ behaviour is easier to discuss:
\begin{align}
  &\Delta_{\di{m^1}{m^2}\di{m^1+\alpha^1}{m^2+\alpha^2};
    \di{l^1+\alpha^1}{l^2+\alpha^2}\di{l^1}{l^2}} \nonumber
  \\*
  &= \sum_{u^1=0}^{\min(m^1,l^1)} \sum_{u^2=0}^{\min(m^2,l^2)}
  \frac{\theta}{2(1{+}\Omega)^2 ( 1{+} \frac{\mu_0^2\theta}{8\Omega}
    {+}\frac{1}{2}(\alpha^1{+}\alpha^2) {+}m^1{+}m^2{+}l^1{+}l^2{-}u^1{-}u^2)}
  \nonumber
  \\*
  &\times \frac{(m^1{+}m^2{+}l^1{+}l^2{-}2u^1{-}2u^2)!\sqrt{m^1!l^1!m^2!l^2!}
    }{(m^1{-}u^1)!(l^1{-}u^1)!u^1!  (m^2{-}u^2)!(l^2{-}u^2)!u^2!}
  \Big(\frac{1{-} \Omega}{1{+}\Omega}
  \Big)^{m^1+m^2+l^1+l^2-2u^1-2u^2} \nonumber
  \\
  &\times \bigg\{ \frac{\Gamma\big(1{+} \frac{\mu_0^2\theta}{8\Omega} 
{+}\frac{1}{2}(\alpha^1{+}\alpha^2){+}u^1{+}u^2\big)}{
    \Gamma\big(1{+} \frac{\mu_0^2\theta}{8\Omega}
    {+}\frac{1}{2}(\alpha^1{+}\alpha^2)
    {+}m^1{+}m^2{+}l^1{+}l^2{-}u^1{-}u^2\big)} \nonumber
  \\*
  &\qquad\qquad\qquad 
\times \sqrt{ \frac{(\alpha^1{+}m^1)!(\alpha^1{+}l^1)!}{
      (\alpha^1{+}u^1)!(\alpha^1{+}u^1)!}}  \sqrt{
    \frac{(\alpha^2{+}m^2)!(\alpha^2{+}l^2)!}{
      (\alpha^2{+}u^2)!(\alpha^2{+}u^2)!}} \bigg\} \nonumber
  \\*
  &\times {}_2F_1\bigg(\di{1{+} m^1{+}m^2{+}l^1{+}l^2{-}2u^1{-}2u^2\,,\;
    \frac{\mu_0^2\theta}{8\Omega} 
    {-}\frac{1}{2}(\alpha^1{+}\alpha^2){-}u^1{-}u^2 }{ 2{+}
    \frac{\mu_0^2 \theta}{8\Omega}
    {+}\frac{1}{2}(\alpha^1{+}\alpha^2) {+}m^1{+}m^2{+}l^1{+}l^2{-}u^1{-}u^2}
  \bigg| \frac{(1{-}\Omega)^2}{(1{+}\Omega)^2} \bigg)\;.
\label{Delta-as}
\end{align}
We assume $\frac{1}{2}(\alpha_1+\alpha_2) \geq
\max(\frac{\mu_0^2\theta}{8\Omega},m,l)$. The term in braces $\{~\}$ in
(\ref{Delta-as}) behaves like
\begin{align}
  \big\{\dots\big\} &\sim
  \big(\tfrac{1}{2}(\alpha^1{+}\alpha^2)\big)^{2u^1+2u^2-m^1-m^2-l^2-l^2}
  (\alpha^1)^{\frac{1}{2}(m^1+l^1-2u^1)} (\alpha^2)^{\frac{1}{2}(m^2+l^2-2u^1)}
  \nonumber
  \\*
  & \times \Big(1+ \frac{(2u^1{+}2u^2{-}m^1{-}m^2{-}l^2{-}l^2)
    (m^1{+}m^2{+}l^1{+}l^2{+}\frac{\mu_0^2\theta}{4\Omega}
{+}1)}{(\alpha^1+\alpha^2)} 
\nonumber
\\*[-1ex]
& \qquad +
  \mathcal{O}\big((\alpha^1{+}\alpha^2)^{-2}\big) \Big) \nonumber
  \\*
  & \times \Big(1+ \frac{(m^1{-}u^1)(m^1{+}u^1{+}1)}{4\alpha^1} +
  \frac{(l^1{-}u^1)(l^1{+}u^1{+}1)}{4\alpha^1} +
  \mathcal{O}\big((\alpha^1)^{-2}\big) \Big) \nonumber
  \\*
  & \times \Big(1+ \frac{(m^2{-}u^2)(m^2{+}u^2{+}1)}{4\alpha^2} +
  \frac{(l^2{-}u^2)(l^2{+}u^2{+}1)}{4\alpha^1} +
  \mathcal{O}\big((\alpha^2)^{-2}\big) \Big)\;.
\label{est-brace-1}
\end{align}
We look for the maximum of the propagator under the condition
$\mathcal{C} \leq \max(\alpha^1,\alpha^2) \leq 2\mathcal{C}$. Defining
$s^i=m^i+l^i-2u^i$ and $s=s^1+s^2$, the dominating term in
(\ref{est-brace-1}) is
\begin{align}
\frac{ (\alpha^1)^{\frac{s^1}{2}}
(\alpha^2)^{\frac{s^2}{2}}}{
\big(\tfrac{1}{2}(\alpha^1{+}\alpha^2)\big)^{s}}
\bigg|_{\mathcal{C}
\leq \max(\alpha^1,\alpha^2) \leq 2\mathcal{C}}
\leq \frac{\max\Big(
\dfrac{(\frac{s^1}{s^1+2s^2})^{\frac{s^1}{2}}}{
(\frac{s^1+s^2}{s^1+2s^2})^{s^1+s^2}},
\dfrac{(\frac{s^2}{s^2+2s^1})^{\frac{s^2}{2}}}{
(\frac{s^1+s^2}{s^2+2s^1})^{s^1+s^2}}\Big)}{
\mathcal{C}^{\frac{s}{2}} }\;.
\label{est-brace}
\end{align}
The maximum is attained at $(\alpha^1,\alpha^2)=(
\frac{s^1\mathcal{C}}{s^1+2s^2},\mathcal{C})$ for $s^1\leq s^2$ and at
$(\alpha^1,\alpha^2)=(\mathcal{C}, \frac{s^2\mathcal{C}}{s^2+2s^1})$ for
$s^1\geq s^2$.  Thus, the leading contribution to the propagator 
will come from the summation index $u^i=\min(m^i,l^i)$.

Next we evaluate the leading contribution of the hypergeometric function:
\begin{align}
&  {}_2F_1\Big(\di{1{+} m^1{+}m^2{+}l^1{+}l^2{-}2u^1{-}2u^2\,,\;
    \frac{\mu_0^2\theta}{8\Omega}
    {-}\frac{1}{2}(\alpha^1{+}\alpha^2){-}u^1{-}u^2 }{ 2{+}
    \frac{\mu_0^2\theta}{8\Omega} {+}\frac{1}{2}(\alpha^1{+}\alpha^2)
    {+}m^1{+}m^2{+}l^1{+}l^2{-}u^1{-}u^2}\Big| 
\frac{(1{-}\Omega)^2}{(1{+}\Omega)^2} \Big) 
\nonumber
\\*
& \sim \sum_{k=0}^\infty
  \frac{(m^1{+}m^2{+}l^1{+}l^2{-}2u^1{-}2u^2{+}k)!}{
    (m^1{+}m^2{+}l^1{+}l^2{-}2u^1{-}2u^2)!} \frac{\big(
- \frac{(1{-}\Omega)^2}{(1{+}\Omega)^2} \big)^k}{k!}  
\Big(1+\frac{k(2u^1{+}2u^2
{-}\frac{\mu_0^2\theta}{4\Omega}{+}1{-}k)}{\alpha^1{+}\alpha^2}
\nonumber
  \\*
  & \qquad\qquad 
  -\frac{k(3{+}2m^1{+}2m^2{+}2l^1{+}2l^2{-}2u^1{-}2u^2 
{+}\frac{\mu_0^2\theta}{4\Omega}{+}k)}
  {\alpha^1{+}\alpha^2} +
  \mathcal{O}\big((\alpha^1{+}\alpha^2)^{-2}\big)\Big) \nonumber
  \\*
  &= \sum_{k=0}^\infty \frac{(s{+}k)!}{s!}  \Big(1-\frac{2
    k(1{+}s{+}\frac{\mu_0^2\theta}{4\Omega}{+}k)}{
\alpha^1{+}\alpha^2}\Big) 
\frac{\big(-\frac{(1{-}\Omega)^2}{(1{+}\Omega)^2} \big)^k}{k!}
+ \mathcal{O}\big((\alpha^1{+}\alpha^2)^{-2}\big)
\nonumber
 \\*
 & = \Big(\frac{(1{+}\Omega)^2}{2(1{+}\Omega^2)}\Big)^{1+s} 
\bigg(1+
  \frac{\frac{(1{-}\Omega)^2}{1{+}\Omega^2}(1{+}s)}{
(\alpha^1{+}\alpha^2)}
\Big(1{+}\tfrac{\mu_0^2\theta}{4\Omega}{+}\tfrac{s}{2}
{+}(s{+}2) \tfrac{\Omega}{(1{+}\Omega^2)}  
\Big) \bigg)
\nonumber
 \\*
&+ \mathcal{O}\big((\alpha^1{+}\alpha^2)^{-2}\big)\;.
\label{est-F}
\end{align}

Assuming $s^1\leq s^2$, we obtain from (\ref{Stirling}), (\ref{est-brace}) and
(\ref{est-F}) the following leading contribution to the propagator
(\ref{Delta-as})
\begin{align}
  &\Delta_{\di{m^1}{m^2}\di{m^1+\alpha^1}{m^2+\alpha^2};
    \di{l^1+\alpha^1}{l^2+\alpha^2}\di{l^1}{l^2}} 
\bigg|_{\max(m^1,m^2,l^1,l^2) \ll \mathcal{C}
\leq \max(\alpha^1,\alpha^2) \leq 2\mathcal{C}}
\nonumber
  \\*
&= \frac{\theta \big(\max(m^1,l^1)\big)^{\frac{s^1}{2}}
\big(\max(m^2,l^2)\big)^{\frac{s^2}{2}}}{(1{+}\Omega)^2 
\mathcal{C}^{1+\frac{s^1+s^2}{2}}}
\Big(\frac{1{-}\Omega}{1{+}\Omega}\Big)^{s^1+s^2}
\Big(\frac{(1{+}\Omega)^2}{2(1{+}\Omega^2)}\Big)^{1+s^1+s^2}
\nonumber
\\
& \qquad\qquad \times 
\frac{(s^1{+}s^2)^{s^1+s^2} \sqrt{2\pi (s^1{+}s^2)}}{
(s^1)^{s^1} (s^2)^{s^2} 2\pi \sqrt{s^1s^2}}
\frac{(\frac{s^1}{s^1+2s^2})^{\frac{s^1}{2}}}{
(\frac{s^1+s^2}{s^1+2s^2})^{1+s^1+s^2}}
\Big(1+\mathcal{O}(\mathcal{C}^{-1})\Big)
\bigg|_{s^i:=|m^i-l^i|}\;.
\label{Delta-ss}
\end{align}
The numerator comes from $\sqrt{\frac{m!}{l!}} \leq
m^{\frac{m-l}{2}}$ for $m\geq l$. The estimation (\ref{Delta-ss}) is the
explanation of (\ref{est4-3}).

Let us now look at propagators with $m^i=l^i$ and $m^i \ll \mathcal{C}
\leq \max(\alpha^1,\alpha^2) \leq 2\mathcal{C}$: 
\begin{align}
  &\Delta_{\di{m^1}{m^2}\di{m^1+\alpha^1}{m^2+\alpha^2};
    \di{m^1+\alpha^1}{m^2+\alpha^2}\di{m^1}{m^2}} \nonumber
  \\*
  &= \frac{\theta}{2(1{+}\Omega)^2 \big( 1{+} \frac{\mu_0^2\theta}{8\Omega}
    {+}\frac{1}{2}(\alpha^1{+}\alpha^2) {+}m^1{+}m^2\big)} 
\nonumber
\\*
& \qquad \times \bigg(\frac{(1{+}\Omega)^2}{2(1{+}\Omega^2)} 
+  \frac{\big(\frac{1{-}\Omega^2}{1{+}\Omega^2}\big)^2}{
2(\alpha_1{+}\alpha_2)}
\Big(1{+}\tfrac{\mu_0^2\theta}{4\Omega}
{+} \tfrac{2\Omega}{(1{+}\Omega^2)}  
\Big) \bigg)
\nonumber
\\*
&+\frac{\theta}{2(1{+}\Omega)^2 \big( 1{+} \frac{\mu_0^2\theta}{8\Omega}
    {+}\frac{1}{2}(\alpha^1{+}\alpha^2) {+}m^1{+}m^2{+}1\big)} 
\; \Big(\frac{1{-}\Omega^2}{1{+}\Omega^2}\Big)^2 \;
\frac{(1{+}\Omega)^2}{1{+}\Omega^2}\;
\frac{m^1 \alpha^1+m^2\alpha^2 }{(\alpha^1{+}\alpha^2)^2}
\nonumber
\\*
&+  \mathcal{O}\big((\alpha^1{+}\alpha^2)^{-3}\big)\;.
\end{align}
This means
\begin{align}
  &\Delta_{\di{m_1}{m_2}\di{m_1+\alpha_1}{m_2+\alpha_2};
    \di{m_1+\alpha_1}{m_2+\alpha_2}\di{m_1}{m_2}}
  -\Delta_{\di{0}{0}\di{m^1+\alpha_1}{m^2+\alpha_2};
    \di{m^1+\alpha_1}{m^2+\alpha_2}\di{0}{0}}\nonumber
  \\*
  &= -(m_1{+}m_2) \frac{\theta}{8(1{+}\Omega^2) 
\big( 1{+} \frac{\mu_0^2\theta}{8\Omega}
 {+}\frac{1}{2}(\alpha^1{+}\alpha^2{+}m^1{+}m^2)\big)^2}
\nonumber
\\*
&+\frac{\theta}{2(1{+}\Omega^2) \big( 1{+} \frac{\mu_0^2\theta}{8\Omega}
    {+}\frac{1}{2}(\alpha^1{+}\alpha^2{+}m^1{+}m^2)\big)} 
\; \Big(\frac{1{-}\Omega^2}{1{+}\Omega^2}\Big)^2 \;
\frac{m^1 (\alpha^1{+}m^1)+m^2(\alpha^2{+}m^2) }{
(\alpha^1{+}\alpha^2{+}m^1{+}m^2)^2}
\nonumber
\\*
&+  \mathcal{O}\Big(\frac{1}{(\alpha^1{+}\alpha^2{+}m^1{+}m^2)} 
\frac{(m^1{+}m^2)^2}{(\alpha^1{+}\alpha^2{+}m^1{+}m^2)^2} \Big) 
\nonumber
\\
&= m_1 \Big(\Delta_{\di{1}{0}\di{m_1+\alpha_1}{m_2+\alpha_2};
    \di{m_1+\alpha_1}{m_2+\alpha_2}\di{1}{0}}
  -\Delta_{\di{0}{0}\di{m_1+\alpha_1}{m_2+\alpha_2};
    \di{m_1+\alpha_1}{m_2+\alpha_2}\di{0}{0}}\Big) 
\nonumber
\\
&+ m_2
  \Big(\Delta_{\di{0}{1}\di{m_1+\alpha_1}{m_2+\alpha_2};
    \di{m_1+\alpha_1}{m_2+\alpha_2}\di{0}{1}}
  -\Delta_{\di{0}{0}\di{m_1+\alpha_1}{m_2+\alpha_2};
    \di{m_1+\alpha_1}{m_2+\alpha_2}\di{0}{0}}\Big) 
\nonumber
\\*
&+  \mathcal{O}\Big(\frac{1}{(\alpha^1{+}\alpha^2{+}m^1{+}m^2)} 
\frac{(m^1{+}m^2)^2}{(\alpha^1{+}\alpha^2{+}m^1{+}m^2)^2} \Big)\;.
\label{Delta-00}
\end{align}
The second and third line of (\ref{Delta-00}) explains the estimation
(\ref{Qpl1}). Clearly, the next term in the expansion
is of the order $\frac{(m^1{+}m^2)^2}{(\alpha^1{+}\alpha^2{+}m^1{+}m^2)^3}$,
which explains the estimation (\ref{Qpl2}).

For $m_1=l_1+1$ and $m_2=l_2$ we have
\begin{align}
\Delta_{\di{l_1{+}1}{l_2}\di{l_1+1+\alpha_1}{l_2+\alpha_2};
    \di{l_1+\alpha_1}{l_2+\alpha_2}\di{l_1}{l_2}} 
  &=  \frac{\theta}{2(1{+}\Omega)^2 \big( \frac{\mu_0^2\theta}{8\Omega}
    {+}\frac{1}{2}(\alpha^1{+}\alpha^2) {+}l^1{+}l^2{+}2\big)} 
\nonumber
\\*
& \times 
\frac{1{-}\Omega^2}{1{+}\Omega^2} 
\frac{\sqrt{(l^1{+}1)(l^1{+}\alpha^1{+}1)}}{\alpha^1{+}\alpha^2}
\Big(1+\mathcal{O}\big((\alpha_1{+}\alpha_2)^{-1}\Big)\;.
\end{align}
This yields
\begin{align}
&\Delta_{\di{l_1{+}1}{l_2}\di{l_1+1+\alpha_1}{l_2+\alpha_2};
    \di{l_1+\alpha_1}{l_2+\alpha_2}\di{l_1}{l_2}} - \sqrt{l_1{+}1}
  \Delta_{\di{1}{0}\di{l_1+1+\alpha_1}{l_2+\alpha_2};
    \di{l_1+\alpha_1}{l_2+\alpha_2}\di{0}{0}} 
\nonumber
\\*
  &=  \mathcal{O}\bigg(\frac{1}{(\alpha^1{+}\alpha^2{+}l^1{+}l^2)}
\sqrt{\frac{l^1{+}1}{\alpha^1{+}\alpha^2{+}l^1{+}l^2}}
\frac{(l^1{+}1)}{(\alpha^1{+}\alpha^2{+}l^1{+}l^2)} \bigg)\;,
\label{Delta-01}
\end{align}
which explains the estimation (\ref{Qpl3}). Similarly, we have 
\begin{align}
\Delta_{\di{1}{1}\di{1+\alpha_1}{1+\alpha_2};
    \di{\alpha_1}{1+\alpha_2}\di{0}{1}} - 
  \Delta_{\di{1}{0}\di{1+\alpha_1}{1+\alpha_2};
    \di{\alpha_1}{1+\alpha_2}\di{0}{0}} 
  &=  \mathcal{O}\bigg(
\frac{\theta\sqrt{\alpha^1{+}1}}{
(\alpha^1{+}\alpha^2{+}1)^3}\bigg)\;,
\label{Delta-01a}
\end{align}
which shows that the norm of (\ref{comp-prop-rest}) is of the same
order as (\ref{Qpl3}).

\subsection{An identity for hypergeometric functions}
\label{id-hypergeom}

For terminating hypergeometric series ($m,l \in \mathbb{N}$) we compute the
sum in the last line of (\ref{Schwinger}):
\begin{align}
  & \sum_{x=0}^\infty \frac{(\alpha{+}x)!}{x!\alpha!}
  a^x
  {}_2F_1\Big(\genfrac{}{}{0pt}{}{-m,-x}{1{+}\alpha}\Big| b\Big)
  \,{}_2F_1\Big(\genfrac{}{}{0pt}{}{-l,-x}{1{+}\alpha}\Big| b\Big)
\nonumber
  \\*
  &= \sum_{x=0}^\infty \sum_{r=0}^{\min(x,m)}\sum_{s=0}^{\min(x,l)}
  \frac{(\alpha{+}x)!}{x!\alpha!} a^x 
\frac{m! x! \alpha!}{(m{-}r)!(x{-}r)!(\alpha{+}r)!r!}  b^r 
\frac{l! x!  \alpha!}{(m{-}s)!(x{-}s)!(\alpha{+}s)!s!} b^s
\nonumber
  \\
  &= \sum_{r=0}^{m}\sum_{s=0}^{l} \sum_{x=\max(r,s)}^\infty (\alpha{+}x)!
  x!\alpha!m!l!  a^x
\frac{b^{r+s}}{(m{-}r)!(x{-}r)!(\alpha{+}r)!r!
    (l{-}s)!(x{-}s)!(\alpha{+}s)!s!}  
\nonumber
  \\
  &= \sum_{r=0}^{m}\sum_{s=0}^{l} \frac{\alpha!m!l!}{(m{-}r)!(\alpha{+}r)!r!
    (m{-}s)!(\alpha{+}s)!s!}
a^{\max(r,s)} b^{r+s} \nonumber
  \\*
  & \qquad \times \sum_{y=0}^\infty \frac{(\alpha{+}y{+}\max(r,s))!
    (y{+}\max(r,s))!}{(y{+}|r{-}s|)!y!} a^y
\nonumber
  \\
  &= \sum_{r=0}^{m}\sum_{s=0}^{l} \frac{\alpha!m!l!}{(m{-}r)!(\alpha{+}r)!r!
    (l{-}s)!(\alpha{+}s)!s!} a^{\max(r,s)} b^{r+s}
\nonumber
  \\*
  & \qquad \times \frac{(\alpha{+}\max(r,s))!(\max(r,s))!} {(|r{-}s|)!}
  \,{}_2F_1\Big(\di{\alpha{+}\max(r,s){+}1\,,\; \max(r,s){+}1}{|r{-}s|{+}1}
  \Big| a \Big) 
\nonumber
  \\
  &=^* \sum_{r=0}^{m}\sum_{s=0}^{l} \frac{\alpha!m!l!}{(m{-}r)!(\alpha{+}r)!r!
    (l{-}s)!(\alpha{+}s)!s!} 
  \frac{a^{\max(r,s)}\,b^{r+s}}{(1- a)^{\alpha+r+s+1}}
  \nonumber
  \\*
  & \qquad \times \frac{(\alpha{+}\max(r,s))!(\max(r,s))!} {(|r{-}s|)!}
  \,{}_2F_1\Big(\di{-\min(\alpha{+}r,\alpha{+}s)\,,\; -\min(r,s)}{|r{-}s|{+}1}
  \Big| a \Big)
\nonumber
  \\
  &= \sum_{r=0}^{m}\sum_{s=0}^{l} \frac{\alpha!m!l!}{(m{-}r)!(\alpha{+}r)!r!
    (l{-}s)!(\alpha{+}s)!s!}
  \frac{a^{\max(r,s)}\,b^{r+s}}{(1- a)^{\alpha+r+s+1}}
  \nonumber
  \\*
  & \qquad \times \sum_{u'=0}^{\min(r,s)}
\frac{(\alpha{+}\max(r,s))!(\max(r,s))!
(\alpha{+}\min(r,s))!(\min(r,s))!} {(|r{-}s|{+}u')!
(\min(r,s){-}u')!(\alpha{+}\min(r,s){-}u')!u'!} a^{u'}
\nonumber
  \\
  &= \sum_{r=0}^{m}\sum_{s=0}^{l} \sum_{u=0}^{\min(r,s)}
\frac{\alpha!m!l!}{(m{-}r)!(r{-}u)!
    (l{-}s)!(s{-}u)!(\alpha{+}u)!u!}
\frac{a^{r+s-u}\,b^{r+s}}{(1- a)^{\alpha+r+s+1}}
\nonumber
  \\
  &= \sum_{u=0}^{\min(m,l)}\sum_{r=u}^{m}\sum_{s=u}^{l} 
\frac{\alpha!m!l!}{(m{-}r)!(r{-}u)!
    (l{-}s)!(s{-}u)!(\alpha{+}u)!u!}
\frac{a^{r+s-u}\,b^{r+s}}{(1- a)^{\alpha+r+s+1}}
\nonumber
  \\
  &= \sum_{u=0}^{\min(m,l)}
\frac{\alpha!m!l!}{(m{-}u)! (l{-}u)!(\alpha{+}u)!u!}
\Big(\frac{ab}{1{-} a}\Big)^{2u} 
\Big(1+\frac{ab}{1{-} a}\Big)^{m+l-2u} 
\frac{1}{a^u (1{-}a)^{\alpha+1}}
\nonumber
  \\
  &= \frac{(1{-}a{+}ab)^{m+l}}{(1{-}a)^{m+l+\alpha+1}} 
\,{}_2F_1\Big(\di{-m\,,\;-l}{\alpha{+}1}\Big| 
\frac{ab^2}{(1{-}a{+}ab)^2}\Big)\;.
\end{align}
In the step denoted by $=^*$ we have used \cite[\S 9.131.1]{GR}. All other
transformations should be self-explaining.

\section{On composite propagators}
\label{appcomposite}

\subsection{Identities for differences of ribbon graphs}
\label{appcompositeidentity}

We continue here the discussion of Section~\ref{seccomposite} on
composite propagators generated by differences of interaction
coefficients. 

After having derived (\ref{decA4}), we now have a look at
(\ref{intA2}). Since $\gamma$ is one-particle irreducible, we get for
a certain permutation $\pi$ ensuring the history of integrations the
following linear combination:
\begin{subequations}
\label{decA2-all}
\begin{align}
&\widehat{\Lambda'\frac{\partial}{\partial \Lambda'} A^{(V)\gamma}_{
\di{m^1+1}{m^2}\di{n^1+1}{n^2};\di{n^1}{n^2}\di{m^1}{m^2}}[\Lambda']}
-\sqrt{(m^1{+}1)(n^1{+}1)}
\widehat{\Lambda'\frac{\partial}{\partial \Lambda'} A^{(V)\gamma}_{
\di{1}{0}\di{1}{0};\di{0}{0}\di{0}{0}}[\Lambda']}
\nonumber
\\*
&= \dots \left\{ 
\prod_{i=1}^{p-1} Q_{\di{n^1+1}{n^2}k_{\pi_i};k_{\pi_i}\di{n^1+1}{n^2}}
(\Lambda_{\pi_i}) 
 Q_{\di{n^1+1}{n^2}(k_{\pi_p}+1^+);k_{\pi_p} \di{n^1}{n^2}}
(\Lambda_{\pi_p}) 
\right.
\nonumber
\\*
& \qquad \qquad \times 
\prod_{i={p+1}}^{a} Q_{\di{n^1}{n^2}k_{\pi_i};k_{\pi_i}\di{n^1}{n^2}}
(\Lambda_{\pi_i}) 
\prod_{j=1}^{q-1} Q_{\di{m^1}{m^2}l_{\pi_j};l_{\pi_j}\di{m^1}{m^2}}
(\Lambda_{\pi_j}) 
\nonumber
\\*
& \qquad \qquad \times 
 Q_{\di{m^1}{m^2}l_{\pi_q}; (l_{\pi_q}+1^+)\di{m^1+1}{m^2}}(\Lambda_{\pi_q}) 
\prod_{j={q+1}}^{b} Q_{\di{m^1+1}{m^2}l_{\pi_j};l_{\pi_j}\di{m^1+1}{m^2}}
(\Lambda_{\pi_j}) 
\nonumber
\\*
&
- \sqrt{(m^1{+}1)( n^1{+}1)} 
\prod_{i=1}^{p-1} Q_{\di{1}{0}k_{\pi_i};k_{\pi_i}\di{1}{0}}
(\Lambda_{\pi_i}) 
 Q_{\di{1}{0}(k_{\pi_p}+1^+);k_{\pi_p} \di{0}{0}}
(\Lambda_{\pi_p}) \!
\prod_{i={p+1}}^{a} \!\! Q_{\di{0}{0}k_{\pi_i};k_{\pi_i}\di{0}{0}}
(\Lambda_{\pi_i}) 
\nonumber
\\*
& \qquad \qquad \times 
\left. \prod_{j=1}^{q} Q_{\di{0}{0}l_{\pi_j};l_{\pi_j}\di{0}{0}}
(\Lambda_{\pi_j}) 
 Q_{\di{0}{0}l_{\pi_q}; (l_{\pi_q}+1^+)\di{1}{0}}
(\Lambda_{\pi_q}) 
\prod_{j={q+1}}^{b} Q_{\di{1}{0}l_{\pi_j};l_{\pi_j}\di{1}{0}}
(\Lambda_{\pi_j}) 
\right\}
\nonumber
\\
&= \dots \left\{ 
\bigg(\prod_{i=1}^{p-1} Q_{\di{n^1+1}{n^2}k_{\pi_i};k_{\pi_i}\di{n^1+1}{n^2}}
(\Lambda_{\pi_i}) \right.
\nonumber
\\*[-2ex]
&\qquad\qquad\qquad \times 
 Q_{\di{n^1+1}{n^2}(k_{\pi_p}+1^+);k_{\pi_p} \di{n^1}{n^2}}
(\Lambda_{\pi_p}) 
\prod_{i={p+1}}^{a} Q_{\di{n^1}{n^2}k_{\pi_i};k_{\pi_i}\di{n^1}{n^2}}
(\Lambda_{\pi_i}) 
\nonumber
\\*[-1ex]
& \qquad \quad - \sqrt{n^1{+}1}
 \prod_{i=1}^{p-1} Q_{\di{1}{0}k_{\pi_i};k_{\pi_i}\di{1}{0}}
(\Lambda_{\pi_i}) 
 Q_{\di{1}{0}(k_{\pi_p}+1^+);k_{\pi_p} \di{0}{0}}
(\Lambda_{\pi_p}) 
\prod_{i={p+1}}^{a} Q_{\di{0}{0}k_{\pi_i};k_{\pi_i}\di{0}{0}}
(\Lambda_{\pi_i}) \bigg)
\nonumber
\\*[-1ex]
& \qquad \times 
\prod_{j=1}^{q-1} Q_{\di{m^1}{m^2}l_{\pi_j};l_{\pi_j}\di{m^1}{m^2}}
(\Lambda_{\pi_j}) 
 Q_{\di{m^1}{m^2}l_{\pi_q}; (l_{\pi_q}+1^+)\di{m^1+1}{m^2}}(\Lambda_{\pi_q}) 
\! \prod_{j={q+1}}^{b} \!\! 
Q_{\di{m^1+1}{m^2}l_{\pi_j};l_{\pi_j}\di{m^1+1}{m^2}}
(\Lambda_{\pi_j}) 
\label{decA2-a}
\\[-1ex]
&
+  \sqrt{n^1{+}1}
 \prod_{i=1}^{p-1} Q_{\di{1}{0}k_{\pi_i};k_{\pi_i}\di{1}{0}}
(\Lambda_{\pi_i}) 
 Q_{\di{1}{0}(k_{\pi_p}+1^+);k_{\pi_p} \di{0}{0}}
(\Lambda_{\pi_p}) 
\prod_{i={p+1}}^{a} Q_{\di{0}{0}k_{\pi_i};k_{\pi_i}\di{0}{0}}
(\Lambda_{\pi_i})
\nonumber
\\*
& \quad \times \bigg(
\prod_{j=1}^{q-1} Q_{\di{m^1}{m^2}l_{\pi_j};l_{\pi_j}\di{m^1}{m^2}}
(\Lambda_{\pi_j}) 
 Q_{\di{m^1}{m^2}l_{\pi_q}; (l_{\pi_q}+1^+)\di{m^1+1}{m^2}}(\Lambda_{\pi_q}) 
\!\prod_{i={q+1}}^{b} \!\! 
Q_{\di{m^1+1}{m^2}l_{\pi_j};l_{\pi_j}\di{m^1+1}{m^2}}
(\Lambda_{\pi_j}) 
\nonumber
\\*
& \qquad -\sqrt{(m^1{+}1)}
\left. \prod_{j=1}^{q} Q_{\di{0}{0}l_{\pi_j};l_{\pi_j}\di{0}{0}}
(\Lambda_{\pi_j}) 
 Q_{\di{0}{0}l_{\pi_q}; (l_{\pi_q}+1^+)\di{1}{0}}
(\Lambda_{\pi_q}) 
\!\!\prod_{j={q+1}}^{b} \!\!
Q_{\di{1}{0}l_{\pi_j};l_{\pi_j}\di{1}{0}} (\Lambda_{\pi_j}) \bigg)
\right\},
\label{decA2-b}
\end{align}
\end{subequations}
with $1^+:=\di{1}{0}$. We further analyse the 
the first three lines of (\ref{decA2-a}): 
\begin{subequations}
\begin{align}
&\bigg(\prod_{i=1}^{p-1} Q_{\di{n^1+1}{n^2}k_{\pi_i};k_{\pi_i}\di{n^1+1}{n^2}}
(\Lambda_{\pi_i}) 
 Q_{\di{n^1+1}{n^2}(k_{\pi_p}+1^+);k_{\pi_p} \di{n^1}{n^2}}
(\Lambda_{\pi_p}) 
\prod_{i={p+1}}^{a} Q_{\di{n^1}{n^2}k_{\pi_i};k_{\pi_i}\di{n^1}{n^2}}
(\Lambda_{\pi_i}) 
\nonumber
\\*[-1ex]
& \quad \quad - \sqrt{n^1{+}1}
 \prod_{i=1}^{p-1} Q_{\di{1}{0}k_{\pi_i};k_{\pi_i}\di{1}{0}}
(\Lambda_{\pi_i}) 
 Q_{\di{1}{0}(k_{\pi_p}+1^+);k_{\pi_p} \di{0}{0}}
(\Lambda_{\pi_p}) 
\prod_{i={p+1}}^{a} Q_{\di{0}{0}k_{\pi_i};k_{\pi_i}\di{0}{0}}
(\Lambda_{\pi_i}) \bigg)
\nonumber
\\[-1ex]
&= \bigg(
\Big(\prod_{i=1}^{p-1} Q_{\di{n^1+1}{n^2}k_{\pi_i};k_{\pi_i}\di{n^1+1}{n^2}}
(\Lambda_{\pi_i}) 
- \prod_{i=1}^{p-1} Q_{\di{1}{0}k_{\pi_i};k_{\pi_i}\di{1}{0}}
(\Lambda_{\pi_i}) \Big)
\nonumber
\\*[-1ex]
&\qquad\qquad \times 
Q_{\di{n^1+1}{n^2}(k_{\pi_p}+1^+);k_{\pi_p} \di{n^1}{n^2}}
(\Lambda_{\pi_p}) 
\prod_{i={p+1}}^{a} Q_{\di{n^1}{n^2}k_{\pi_i};k_{\pi_i}\di{n^1}{n^2}}
(\Lambda_{\pi_i}) \bigg)
\label{decA2-a-a}
\\*[-1ex]
& + \prod_{i=1}^{p-1} \! Q_{\di{1}{0}k_{\pi_i};k_{\pi_i}\di{1}{0}}
(\Lambda_{\pi_i}) 
\Big( \mathcal{Q}^{(+\frac{1}{2})}_{\di{n^1+1}{n^2}(k_{\pi_p}+1^+);
k_{\pi_p} \di{n^1}{n^2}} (\Lambda_{\pi_p}) \Big)
\! \prod_{i={p+1}}^{a} \!\! Q_{\di{n^1}{n^2}k_{\pi_i};k_{\pi_i}\di{n^1}{n^2}}
(\Lambda_{\pi_i}) 
\label{decA2-a-b}
\\*
&
+ \sqrt{n^1{+}1} \bigg(
\prod_{i=1}^{p-1} Q_{\di{1}{0}k_{\pi_i};k_{\pi_i}\di{1}{0}}
(\Lambda_{\pi_i}) 
 Q_{\di{1}{0}(k_{\pi_p}+1^+);k_{\pi_p} \di{0}{0}}
(\Lambda_{\pi_p}) 
\nonumber
\\*
&\qquad\qquad \times 
\Big(\prod_{i={p+1}}^{a} Q_{\di{n^1}{n^2}k_{\pi_i};k_{\pi_i}\di{n^1}{n^2}}
(\Lambda_{\pi_i}) - 
\prod_{i={p+1}}^{a} Q_{\di{0}{0}k_{\pi_i};k_{\pi_i}\di{0}{0}}
(\Lambda_{\pi_i}) \Big)\bigg)\;.
\label{decA2-a-c}
\end{align}
\end{subequations}
According to (\ref{decA4}), the two lines (\ref{decA2-a-a}) and
(\ref{decA2-a-c}) yield graphs having one composite propagator
(\ref{comp-prop-0}), whereas the line (\ref{decA2-a-b}) yields a graph
having one composite propagator\footnote{Note that the   estimation 
  (\ref{est4-3}) yields $\sqrt{n^1{+}1}
  |Q_{\di{1}{0}(k_{\pi_p}+1^+);k_{\pi_p} \di{0}{0}} (\Lambda_{\pi_p})|
  \leq \frac{C}{\Omega\theta\Lambda^2} \big(\frac{n^1+1}{\theta
    \Lambda^2}\big)^{\frac{1}{2}}$. Therefore, the prefactor
  $\sqrt{n^1{+}1}$ in (\ref{decA2-a-c}) combines actually to the ratio
  $\big(\frac{n^1+1}{\theta \Lambda^2}\big)^{\frac{1}{2}}$ which is
  required for the (\ref{polab})-term in
  Proposition~\ref{power-counting-prop}.\ref{prp-planar-A1}.\label{fn-pre}}
(\ref{comp-prop-p12}). In total, we get from (\ref{decA2-all}) $a+b$
graphs with composite propagator (\ref{comp-prop-0}) or
(\ref{comp-prop-p12}).  The treatment of (\ref{intA3}) is similar.

Second, we treat that contribution to (\ref{intA1}) which consists of
graphs with constant index along the trajectories:
\begin{subequations}
\label{decA1-all}
\begin{align}
&\widehat{\Lambda'\frac{\partial}{\partial \Lambda'} A^{(V)\gamma}_{
\di{m^1}{m^2}\di{n^1}{n^2};\di{n^1}{n^2}\di{m^1}{m^2}}[\Lambda']}
\nonumber
\\*
&\qquad 
-\widehat{\Lambda'\frac{\partial}{\partial \Lambda'} A^{(V)\gamma}_{
\di{0}{0}\di{0}{0};\di{0}{0}\di{0}{0}}[\Lambda']}
-m^1\Big(\widehat{\Lambda'\frac{\partial}{\partial \Lambda'} A^{(V)\gamma}_{
\di{1}{0}\di{0}{0};\di{0}{0}\di{1}{0}}[\Lambda']}
-\widehat{\Lambda'\frac{\partial}{\partial \Lambda'} A^{(V)\gamma}_{
\di{0}{0}\di{0}{0};\di{0}{0}\di{0}{0}}[\Lambda']}\Big)
\nonumber
\\*
&\qquad 
-m^2\Big(\widehat{\Lambda'\frac{\partial}{\partial \Lambda'} A^{(V)\gamma}_{
\di{0}{1}\di{0}{0};\di{0}{0}\di{0}{1}}[\Lambda']}
-\widehat{\Lambda'\frac{\partial}{\partial \Lambda'} A^{(V)\gamma}_{
\di{0}{0}\di{0}{0};\di{0}{0}\di{0}{0}}[\Lambda']}\Big)
-n^1\Big(\widehat{\Lambda'\frac{\partial}{\partial \Lambda'} A^{(V)\gamma}_{
\di{0}{0}\di{1}{0};\di{1}{0}\di{0}{0}}[\Lambda']}
\nonumber
\\*
&
\qquad -\widehat{\Lambda'\frac{\partial}{\partial \Lambda'} A^{(V)\gamma}_{
\di{0}{0}\di{0}{0};\di{0}{0}\di{0}{0}}[\Lambda']}\Big)
-n^2\Big(\widehat{\Lambda'\frac{\partial}{\partial \Lambda'} A^{(V)\gamma}_{
\di{0}{0}\di{0}{1};\di{0}{1}\di{0}{0}}[\Lambda']}
-\widehat{\Lambda'\frac{\partial}{\partial \Lambda'} A^{(V)\gamma}_{
\di{0}{0}\di{0}{0};\di{0}{0}\di{0}{0}}[\Lambda']}\Big)
\nonumber
\\
&= \dots \Bigg\{ 
\prod_{i=1}^a Q_{\di{n^1}{n^2}k_{\pi_i};k_{\pi_i}\di{n^1}{n^2}}
(\Lambda_{\pi_i})
\prod_{j=1}^b Q_{\di{m^1}{m^2}l_{\pi_j};l_{\pi_j}\di{m^1}{m^2}}
(\Lambda_{\pi_j}) 
\nonumber
\\*[-1.5ex]
&\qquad\qquad\qquad - 
\prod_{i=1}^a Q_{\di{0}{0}k_{\pi_i};k_{\pi_i}\di{0}{0}}
(\Lambda_{\pi_i})
\prod_{j=1}^b Q_{\di{0}{0}l_{\pi_j};l_{\pi_j}\di{0}{0}} 
(\Lambda_{\pi_j})
\nonumber
\\*[-0.5ex]
&\quad -\prod_{i=1}^a Q_{\di{0}{0}k_{\pi_i};k_{\pi_i}\di{0}{0}}
(\Lambda_{\pi_i})
\bigg(m^1 \Big(\prod_{j=1}^b Q_{\di{1}{0}l_{\pi_j};l_{\pi_j}\di{1}{0}}
(\Lambda_{\pi_j})
- \prod_{j=1}^b Q_{\di{0}{0}l_{\pi_j};l_{\pi_j}\di{0}{0}}
(\Lambda_{\pi_j})\Big)
\nonumber
\\*[-1ex]
&
\qquad\qquad 
+ m^2 \Big(\prod_{j=1}^b Q_{\di{0}{1}l_{\pi_j};l_{\pi_j}\di{0}{1}}
(\Lambda_{\pi_j})
- \prod_{j=1}^b Q_{\di{0}{0}l_{\pi_j};l_{\pi_j}\di{0}{0}}
(\Lambda_{\pi_j})\Big)\bigg)
\nonumber
\\*[-0.5ex]
&\quad -\bigg(n^1 \Big(\prod_{i=1}^a Q_{\di{1}{0}k_{\pi_i};k_{\pi_i}\di{1}{0}}
(\Lambda_{\pi_i})
- \prod_{i=1}^a Q_{\di{0}{0}k_{\pi_i};k_{\pi_i}\di{0}{0}}
(\Lambda_{\pi_i})\Big)
\nonumber
\\*[-1ex]
&
\qquad\qquad 
+ n^2 \Big(\prod_{i=1}^a Q_{\di{0}{1}k_{\pi_i};k_{\pi_i}\di{0}{1}}
(\Lambda_{\pi_i})
- \prod_{i=1}^a Q_{\di{0}{0}k_{\pi_i};k_{\pi_i}\di{0}{0}}
(\Lambda_{\pi_i})\Big)\bigg)
\prod_{j=1}^b Q_{\di{0}{0}l_{\pi_j};l_{\pi_j}\di{0}{0}}
(\Lambda_{\pi_j}) \Bigg\}
\nonumber
\\
& = \dots \Bigg\{ 
\bigg(\Big(\prod_{i=1}^a Q_{\di{n^1}{n^2}k_{\pi_i};k_{\pi_i}\di{n^1}{n^2}}
(\Lambda_{\pi_i})
-\prod_{i=1}^a Q_{\di{0}{0}k_{\pi_i};k_{\pi_i}\di{0}{0}}
(\Lambda_{\pi_i})\Big)  
\nonumber
\\*
&
\qquad\qquad \times
\Big(\prod_{j=1}^b Q_{\di{m^1}{m^2}l_{\pi_j};l_{\pi_j}\di{m^1}{m^2}}
(\Lambda_{\pi_j})
- \prod_{j=1}^b Q_{\di{0}{0}l_{\pi_j};l_{\pi_j}\di{0}{0}} 
(\Lambda_{\pi_j})\Big)\bigg)
\label{decA1-a}
\\[-0.5ex]
&+\prod_{i=1}^a Q_{\di{0}{0}k_{\pi_i};k_{\pi_i}\di{0}{0}}
(\Lambda_{\pi_i})
\bigg(\Big(\prod_{j=1}^b Q_{\di{m^1}{m^2}l_{\pi_j};l_{\pi_j}\di{m^1}{m^2}}
(\Lambda_{\pi_i})
- \prod_{j=1}^b Q_{\di{0}{0}l_{\pi_j};l_{\pi_j}\di{0}{0}} 
(\Lambda_{\pi_j})\Big)
\nonumber
\\*[-1.3ex]
&
\qquad\qquad
- m^1 \Big(\prod_{j=1}^b Q_{\di{1}{0}l_{\pi_j};l_{\pi_j}\di{1}{0}}
(\Lambda_{\pi_j})
- \prod_{j=1}^b Q_{\di{0}{0}l_{\pi_j};l_{\pi_j}\di{0}{0}}
(\Lambda_{\pi_j})\Big)
\nonumber
\\*[-0.5ex]
&
\qquad\qquad 
- m^2 \Big(\prod_{j=1}^b Q_{\di{0}{1}l_{\pi_j};l_{\pi_j}\di{0}{1}}
(\Lambda_{\pi_j})
- \prod_{j=1}^b Q_{\di{0}{0}l_{\pi_j};l_{\pi_j}\di{0}{0}}
(\Lambda_{\pi_j})\Big)\bigg)
\label{decA1-b}
\\[-1ex]
&+\bigg(
\Big(\prod_{i=1}^a Q_{\di{n^1}{n^2}k_{\pi_i};k_{\pi_i}\di{n^1}{n^2}}
(\Lambda_{\pi_i})
-\prod_{i=1}^a Q_{\di{0}{0}k_{\pi_i};k_{\pi_i}\di{0}{0}}
(\Lambda_{\pi_i})\Big) 
\nonumber
\\*[-1ex]
&
\qquad\qquad 
-n^1 \Big(\prod_{i=1}^a Q_{\di{1}{0}k_{\pi_i};k_{\pi_i}\di{1}{0}}
(\Lambda_{\pi_i})
- \prod_{i=1}^a Q_{\di{0}{0}k_{\pi_i};k_{\pi_i}\di{0}{0}}
(\Lambda_{\pi_i})\Big)
\nonumber
\\*[-1ex]
&
\qquad\qquad 
- n^2 \Big(\prod_{i=1}^a Q_{\di{0}{1}k_{\pi_i};k_{\pi_i}\di{0}{1}}
(\Lambda_{\pi_i})
- \prod_{i=1}^a Q_{\di{0}{0}k_{\pi_i};k_{\pi_i}\di{0}{0}}
(\Lambda_{\pi_i})\Big)\bigg)
\prod_{j=1}^b Q_{\di{0}{0}l_{\pi_j};l_{\pi_j}\di{0}{0}}
(\Lambda_{\pi_j}) \Bigg\}.
\label{decA1-c}
\end{align}
\end{subequations}
It is clear from (\ref{decA4}) that the part corresponding to 
(\ref{decA1-a}) can be written as a sum of graphs containing (at
different trajectories) two composite propagators
$\mathcal{Q}^{(0)}_{\di{n^1}{n^2}k_{\pi_i};k_{\pi_i}\di{n^1}{n^2}}
(\Lambda_{\pi_i})$ and
$\mathcal{Q}^{(0)}_{\di{m^1}{m^2}l_{\pi_j};l_{\pi_j}\di{m^1}{m^2}}$ of type
(\ref{comp-prop-0}). We further analyse (\ref{decA1-b}):
\begin{subequations}
\label{decA1-b-all}
\begin{align}
& \prod_{j=1}^b Q_{\di{m^1}{m^2}l_{\pi_j};l_{\pi_j}\di{m^1}{m^2}}
(\Lambda_{\pi_j})
- \prod_{j=1}^b Q_{\di{0}{0}l_{\pi_j};l_{\pi_j}\di{0}{0}} 
(\Lambda_{\pi_j})
- m^1 \Big(\prod_{j=1}^b Q_{\di{1}{0}l_{\pi_j};l_{\pi_j}\di{1}{0}}
(\Lambda_{\pi_j})
\nonumber
\\*[-1ex]
&
\qquad
- \prod_{j=1}^b Q_{\di{0}{0}l_{\pi_j};l_{\pi_j}\di{0}{0}}
(\Lambda_{\pi_j})\Big)
- m^2 \Big(\prod_{j=1}^b Q_{\di{0}{1}l_{\pi_j};l_{\pi_j}\di{0}{1}}
(\Lambda_{\pi_j})
- \prod_{j=1}^b Q_{\di{0}{0}l_{\pi_j};l_{\pi_j}\di{0}{0}}
(\Lambda_{\pi_j})\Big) 
\nonumber
\\[-1ex]
&= 
\mathcal{Q}^{(1)}_{\di{m^1}{m^2}l_{\pi_1};l_{\pi_1}\di{m^1}{m^2}} 
(\Lambda_{\pi_1}) 
\prod_{j=2}^b Q_{\di{0}{0}l_{\pi_j};l_{\pi_j}\di{0}{0}} 
(\Lambda_{\pi_j})
\label{decA1-b-a}
\\*[-1ex]
& + 
\bigg( \mathcal{Q}^{(0)}_{\di{m^1}{m^2}l_{\pi_1};l_{\pi_1}\di{m^1}{m^2}}
(\Lambda_{\pi_1})
 \Big(\prod_{j=2}^b Q_{\di{m^1}{m^2}l_{\pi_j};l_{\pi_j}\di{m^1}{m^2}}
(\Lambda_{\pi_j})
- \prod_{j=2}^b Q_{\di{0}{0}l_{\pi_j};l_{\pi_j}\di{0}{0}} 
(\Lambda_{\pi_j})\Big)
\nonumber
\\*[-0.5ex]
&
\qquad
- m^1 \mathcal{Q}^{(0)}_{\di{1}{0}l_{\pi_1};l_{\pi_1}\di{1}{0}}
(\Lambda_{\pi_1}) 
\Big(
\prod_{j=2}^b Q_{\di{1}{0}l_{\pi_j};l_{\pi_j}\di{1}{0}}
(\Lambda_{\pi_j})
-  \prod_{j=2}^b Q_{\di{0}{0}l_{\pi_j};l_{\pi_j}\di{0}{0}}
(\Lambda_{\pi_j})\Big)
\nonumber
\\*[-0.5ex]
&
\qquad
- m^2 \mathcal{Q}^{(0)}_{\di{0}{1}l_{\pi_1};l_{\pi_1}\di{0}{1}}
(\Lambda_{\pi_1}) 
\Big(
\prod_{j=2}^b Q_{\di{0}{1}l_{\pi_j};l_{\pi_j}\di{0}{1}}
(\Lambda_{\pi_j})
-  \prod_{j=2}^b Q_{\di{0}{0}l_{\pi_j};l_{\pi_j}\di{0}{0}}
(\Lambda_{\pi_j})\Big)\bigg)
\label{decA1-b-b}
\\[-1ex]
& 
+  Q_{\di{0}{0}l_{\pi_1};l_{\pi_1}\di{0}{0}} (\Lambda_{\pi_1}) 
\bigg( \Big(\prod_{j=2}^b Q_{\di{m^1}{m^2}l_{\pi_j};l_{\pi_j}\di{m^1}{m^2}}
(\Lambda_{\pi_j})
- \prod_{j=2}^b Q_{\di{0}{0}l_{\pi_j};l_{\pi_j}\di{0}{0}} 
(\Lambda_{\pi_j})\Big)
\nonumber
\\*[-1ex]
&
\qquad
- m^1 \Big(\prod_{j=2}^b Q_{\di{1}{0}l_{\pi_j};l_{\pi_j}\di{1}{0}}
(\Lambda_{\pi_j})
- \prod_{j=2}^b Q_{\di{0}{0}l_{\pi_j};l_{\pi_j}\di{0}{0}}
(\Lambda_{\pi_j})\Big)
\nonumber
\\*
&
\qquad
- m^2 \Big(\prod_{j=2}^b Q_{\di{0}{1}l_{\pi_j};l_{\pi_j}\di{0}{1}}
(\Lambda_{\pi_j})
- \prod_{j=2}^b Q_{\di{0}{0}l_{\pi_j};l_{\pi_j}\di{0}{0}}
(\Lambda_{\pi_j})\Big) \bigg)\;.
\label{decA1-b-c}
\end{align}
\end{subequations}
The part (\ref{decA1-b-a}) gives rise to graphs with one propagator
(\ref{comp-prop-1}). Due to (\ref{decA4}) the part (\ref{decA1-b-b})
yields graphs with two propagators\footnote{Note that the product 
  $m^1\mathcal{Q}^{(0)}_{\di{1}{0}l_{\pi_1};l_{\pi_1}\di{1}{0}}
(\Lambda_{\pi_1})$ is according to (\ref{Qpl1}) bounded by
  $\frac{C}{\Omega\theta\Lambda^2}
  \big(\frac{m^1}{\theta\Lambda^2}\big)$. This means that the
  prefactors $m^1,m^2$ in (\ref{decA1-b-b}) combine actually to the
  ratio $\frac{m^r}{\theta \Lambda^2}$ which is required for the
  (\ref{polab})-term in
  Proposition~\ref{power-counting-prop}.\ref{prp-planar-A0}.}
(\ref{comp-prop-0}) appearing on the same trajectory. Finally, the
part (\ref{decA1-b-c}) has the same structure as the lhs of the
equation, now starting with $j=2$. After iteration we obtain further
graphs of the type (\ref{decA1-b-a}) and (\ref{decA1-b-b}).

Finally, we look at that contribution to (\ref{intA1}) which consists of
graphs where one index component jumps forward and backward in the
$n^1$-component. We can directly use the decomposition derived in
(\ref{decA1-all}) regarding, if the $n^1$-index jumps up,
\begin{align}
&\prod_{i=1}^{a} Q_{\di{n^1}{n^2}k_{\pi_i};k_{\pi_i}\di{n^1}{n^2}}
(\Lambda_{\pi_i})
\nonumber
\\*[-1ex]
& \quad \mapsto 
\prod_{i=1}^{p-1} Q_{\di{n^1}{n^2}k_{\pi_i};k_{\pi_i}\di{n^1}{n^2}}
(\Lambda_{\pi_i})
Q_{\di{n^1}{n^2}k_{\pi_p};(k_{\pi_p}+1)\di{n^1+1}{n^2}}
(\Lambda_{\pi_p})
\prod_{i=p+1}^{q-1} Q_{\di{n^1+1}{n^2}k_{\pi_i};
k_{\pi_i}\di{n^1+1}{n^2}}(\Lambda_{\pi_i})
\nonumber
\\*[-1.5ex]
& \qquad\qquad \times 
Q_{\di{n^1+ 1}{n^2}(k_{\pi_q}+1^+);k_{\pi_q}\di{n^1}{n^2}}
(\Lambda_{\pi_q})
\prod_{i=q+1}^{a} Q_{\di{n^1}{n^2}k_{\pi_i};k_{\pi_i}\di{n^1}{n^2}}
(\Lambda_{\pi_i})\;. 
\label{decA1-n}
\end{align}
This requires to process (\ref{decA1-all}) slightly differently. The
two parts (\ref{decA1-a}) and (\ref{decA1-b}) need no further
discussion, as they lead to graphs having a composite propagator
(\ref{comp-prop-0}) on the $m$-trajectory. We write (\ref{decA1-c}) as
follows:
\begin{subequations}
\label{decA1-c-all}
\begin{align}
(\textup{\ref{decA1-c}}) &=  
\bigg(
\Big(\prod_{i=1}^a Q_{\di{n^1}{n^2}k_{\pi_i};k_{\pi_i}\di{n^1}{n^2}}
(\Lambda_{\pi_i})
-(n^1{+}1)\prod_{i=1}^a Q_{\di{0}{0}k_{\pi_i};k_{\pi_i}\di{0}{0}}
(\Lambda_{\pi_i})\Big) 
\label{decA1-c-a}
\\*
&
\qquad
-n^1 \Big(\prod_{i=1}^a Q_{\di{1}{0}k_{\pi_i};k_{\pi_i}\di{1}{0}}
(\Lambda_{\pi_i})
- 2 \prod_{i=1}^a Q_{\di{0}{0}k_{\pi_i};k_{\pi_i}\di{0}{0}}
(\Lambda_{\pi_i})\Big)
\label{decA1-c-b}
\\*
&
\qquad
- n^2 \Big(\prod_{i=1}^a Q_{\di{0}{1}k_{\pi_i};k_{\pi_i}\di{0}{1}}
(\Lambda_{\pi_i})
- \prod_{i=1}^a Q_{\di{0}{0}k_{\pi_i};k_{\pi_i}\di{0}{0}}
(\Lambda_{\pi_i})\Big)\bigg)
\prod_{j=1}^b Q_{\di{0}{0}l_{\pi_j};l_{\pi_j}\di{0}{0}}
(\Lambda_{\pi_j})\;.
\label{decA1-c-c}
\end{align}
\end{subequations}
The part (\ref{decA1-c-c}) leads according to (\ref{decA4}) and
(\ref{decA1-n}) to graphs either with composite propagators 
(\ref{comp-prop-0}) or with propagators
\begin{align}
 Q_{\di{1}{1}\di{l^1+1}{l^2};\di{l^1}{l^2}\di{0}{1}} 
- Q_{\di{1}{0}\di{l^1+1}{l^2};\di{l^1}{l^2}\di{0}{0}} \;.
\label{comp-prop-rest}
\end{align}
Inserting (\ref{decA1-n}) into (\ref{decA1-c-a}) we have
\begin{subequations}
\label{decA1-c-a-all}
\begin{align}
&  \Big(\prod_{i=1}^a Q_{\di{n^1}{n^2}k_{\pi_i};k_{\pi_i}\di{n^1}{n^2}}
(\Lambda_{\pi_i})
-(n^1{+}1)\prod_{i=1}^a Q_{\di{0}{0}k_{\pi_i};k_{\pi_i}\di{0}{0}}
(\Lambda_{\pi_i})\Big)
\nonumber
\\*
& \stackrel{(\textup{\ref{decA1-n}})} {\longmapsto}
\Big(\prod_{i=1}^{p-1} Q_{\di{n^1}{n^2}k_{\pi_i};k_{\pi_i}\di{n^1}{n^2}}
(\Lambda_{\pi_i})
-\prod_{i=1}^{p-1} Q_{\di{0}{0}k_{\pi_i};k_{\pi_i}\di{0}{0}}
(\Lambda_{\pi_i})\Big)
Q_{\di{n^1}{n^2}k_{\pi_p};(k_{\pi_p}+1)\di{n^1+1}{n^2}}
(\Lambda_{\pi_p})
\nonumber
\\*
& \qquad \times \!
\prod_{i=p+1}^{q-1} \!\! Q_{\di{n^1+1}{n^2}k_{\pi_i};
k_{\pi_i}\di{n^1+1}{n^2}}(\Lambda_{\pi_i})
Q_{\di{n^1+ 1}{n^2}(k_{\pi_q}+1^+);k_{\pi_q}\di{n^1}{n^2}}
(\Lambda_{\pi_q}) \!
\prod_{i=q+1}^{a} \!\! Q_{\di{n^1}{n^2}k_{\pi_i};k_{\pi_i}\di{n^1}{n^2}}
(\Lambda_{\pi_i})
\label{decA1-c-a-a}
\\
&+ \prod_{i=1}^{p-1} Q_{\di{0}{0}k_{\pi_i};k_{\pi_i}\di{0}{0}}
(\Lambda_{\pi_i})
\mathcal{Q}^{(+\frac{1}{2})}_{\di{n^1}{n^2}k_{\pi_p};
(k_{\pi_p}+1)\di{n^1+1}{n^2}}(\Lambda_{\pi_p})
\prod_{i=p+1}^{q-1} Q_{\di{n^1+1}{n^2}k_{\pi_i};
k_{\pi_i}\di{n^1+1}{n^2}}(\Lambda_{\pi_i})
\nonumber
\\*
& \qquad\qquad \times 
Q_{\di{n^1+ 1}{n^2}(k_{\pi_q}+1^+);k_{\pi_q}\di{n^1}{n^2}}
(\Lambda_{\pi_q})
\prod_{i=q+1}^{a} Q_{\di{n^1}{n^2}k_{\pi_i};k_{\pi_i}\di{n^1}{n^2}}
(\Lambda_{\pi_i})
\label{decA1-c-a-b}
\\
&
+ \sqrt{n^1{+}1} \prod_{i=1}^{p-1} Q_{\di{0}{0}k_{\pi_i};k_{\pi_i}\di{0}{0}}
(\Lambda_{\pi_i})
Q_{\di{0}{0}k_{\pi_p};(k_{\pi_p}+1)\di{1}{0}}
(\Lambda_{\pi_p})
\nonumber
\\*
& \qquad\qquad 
\bigg(\Big(
\prod_{i=p+1}^{q-1} Q_{\di{n^1+1}{n^2}k_{\pi_i};
k_{\pi_i}\di{n^1+1}{n^2}}(\Lambda_{\pi_i})
-\prod_{i=p+1}^{q-1} Q_{\di{0}{0}k_{\pi_i};
k_{\pi_i}\di{0}{0}}(\Lambda_{\pi_i})\Big)
\nonumber
\\*
& \qquad\qquad \qquad
- \Big(
\prod_{i=p+1}^{q-1} Q_{\di{1}{0}k_{\pi_i};
k_{\pi_i}\di{1}{0}}(\Lambda_{\pi_i})
-\prod_{i=p+1}^{q-1} Q_{\di{0}{0}k_{\pi_i};
k_{\pi_i}\di{0}{0}}(\Lambda_{\pi_i})\Big)
\bigg)
\nonumber
\\*
& \qquad\qquad \qquad \times 
Q_{\di{n^1+ 1}{n^2}(k_{\pi_q}+1^+);k_{\pi_q}\di{n^1}{n^2}}
(\Lambda_{\pi_q})
\prod_{i=q+1}^{a} Q_{\di{n^1}{n^2}k_{\pi_i};k_{\pi_i}\di{n^1}{n^2}}
(\Lambda_{\pi_i})
\label{decA1-c-a-c}
\\
& + \sqrt{n^1{+}1} \prod_{i=1}^{p-1} Q_{\di{0}{0}k_{\pi_i};k_{\pi_i}\di{0}{0}}
(\Lambda_{\pi_i}) 
Q_{\di{0}{0}k_{\pi_p};(k_{\pi_p}+1)\di{1}{0}}
(\Lambda_{\pi_p})
\prod_{i=p+1}^{q-1} Q_{\di{1}{0}k_{\pi_i};
k_{\pi_i}\di{1}{0}}(\Lambda_{\pi_i})
\nonumber
\\*
& \qquad\qquad \times 
\mathcal{Q}^{(+\frac{1}{2})}_{\di{n^1}{n^2}(k_{\pi_q}+1^+);
k_{\pi_q}\di{n^1}{n^2}} (\Lambda_{\pi_q})
\prod_{i=q+1}^{a} Q_{\di{n^1}{n^2}k_{\pi_i};k_{\pi_i}\di{n^1}{n^2}}
(\Lambda_{\pi_i})
\label{decA1-c-a-d}
\\
& + (n^1{+}1) \prod_{i=1}^{p-1} Q_{\di{0}{0}k_{\pi_i};k_{\pi_i}\di{0}{0}}
(\Lambda_{\pi_i}) 
Q_{\di{0}{0}k_{\pi_p};(k_{\pi_p}+1)\di{1}{0}}
(\Lambda_{\pi_p})
\nonumber
\\*
& \qquad\qquad \times 
\prod_{i=p+1}^{q-1} Q_{\di{1}{0}k_{\pi_i};
k_{\pi_i}\di{1}{0}}(\Lambda_{\pi_i})
\,
Q_{\di{1}{0}(k_{\pi_q}+1^+);k_{\pi_q}\di{0}{0}}
(\Lambda_{\pi_q})
\nonumber
\\*
& \qquad\qquad \times 
\Big(
\prod_{i=q+1}^{a} Q_{\di{n^1}{n^2}k_{\pi_i};k_{\pi_i}\di{n^1}{n^2}}
(\Lambda_{\pi_i})
- 
\prod_{i=q+1}^{a} Q_{\di{0}{0}k_{\pi_i};k_{\pi_i}\di{0}{0}}
(\Lambda_{\pi_i}) \Big)\;.
\label{decA1-c-a-e}
\end{align}
\end{subequations}
Thus, we obtain (recall also (\ref{decA4})) a linear combination of
graphs either with composite propagator (\ref{comp-prop-0}) or with
composite propagator (\ref{comp-prop-p12}). In power-counting
estimations, the prefactors $\sqrt{n^1{+}1}$ combine according to
footnote~\ref{fn-pre} to the required ratio with the scale $\theta
\Lambda^2$. The part (\ref{decA1-c-b}) is nothing but
(\ref{decA1-c-a}) with $n^1=1$ and $n^2=0$.

If the index jumps down from $n^1$ to $n^1-1$, then the graph with
$n^1=0$ does not exist. There is no change of the discussion of
(\ref{decA1-a}) and (\ref{decA1-b}), but now (\ref{decA1-c}) becomes
\begin{align}
(\textup{\ref{decA1-c}}) &=  
\Big(\prod_{i=1}^a Q_{\di{n^1}{n^2}k_{\pi_i};k_{\pi_i}\di{n^1}{n^2}}
(\Lambda_{\pi_i})
-n^1\prod_{i=1}^a Q_{\di{1}{0}k_{\pi_i};k_{\pi_i}\di{1}{0}}
(\Lambda_{\pi_j})\Big) 
\prod_{j=1}^b Q_{\di{0}{0}l_{\pi_j};l_{\pi_j}\di{0}{0}}
(\Lambda_{\pi_j})\;.
\label{decA1-c-d}
\end{align}
Using the same steps as in (\ref{decA1-c-a-all}) we obtain the desired
representation through graphs either with composite propagator
(\ref{comp-prop-0}) or with composite propagator (\ref{comp-prop-p12}).

We show in Appendix~\ref{app:graph} how the decomposition works in a
concrete example.

\subsection{Example of a difference operation for ribbon graphs}
\label{app:graph}

To make the considerations in Section~\ref{seccomposite} and
Appendix~\ref{appcompositeidentity} about
differences of graphs and composite propagators understandable, we look
at the following example of a planar two-leg graph:
\begin{align}
   \parbox{85\unitlength}{\begin{picture}(55,27)
       \put(0,0){\epsfig{file=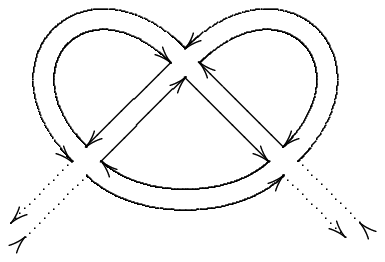,scale=0.9,bb=71 593 191 669}}
       \put(-1,6){\mbox{\scriptsize$m_1$}}
       \put(3,0){\mbox{\scriptsize$n_1$}}
       \put(41,3){\mbox{\scriptsize$n_2$}}
       \put(34,2){\mbox{\scriptsize$m_2$}}
       \put(20,25){\mbox{\scriptsize$k$}}
       \put(4,17){\mbox{\scriptsize$a$}}
       \put(14,17){\mbox{\scriptsize$b$}}
       \put(20,4){\mbox{\scriptsize$c$}}
       \put(24,13){\mbox{\scriptsize$d$}}
       \put(37,17){\mbox{\scriptsize$e$}}
\end{picture}}
\label{graph-at2}
\end{align}
According to Proposition~\ref{power-counting-prop} it depends on the indices
$m_1,n_1,m_2,n_2,k$ whether this graph is irrelevant, marginal, or
relevant. It depends on the history of contraction of subgraphs whether there
are marginal subgraphs or not. 

Let us consider $m_1=k=\di{m^1+1}{m^2}$, $n_2=\di{m^1}{m^2}$,
$n_1=\di{n^1+1}{n^2}$ and $m_2=\di{n^1}{n^2}$ and the history
$a$-$c$-$d$-$e$-$b$ of contraction. Then, all resulting subgraphs are
irrelevant and the total graph is marginal, which leads us to consider the
following difference of graphs:
\begin{align}
&\parbox{45\unitlength}{\begin{picture}(45,28)
       \put(0,0){\epsfig{file=at2,scale=0.9,bb=71 593 191 669}}
       \put(-2,6){\mbox{\scriptsize$\di{m^1+1}{m^2}$}}
       \put(4,-1){\mbox{\scriptsize$\di{n^1+1}{n^2}$}}
       \put(41,4){\mbox{\scriptsize$\di{m^1}{m^2}$}}
       \put(32,2){\mbox{\scriptsize$\di{n^1}{n^2}$}}
       \put(18,26){\mbox{\scriptsize$\di{m^1+1}{m^2}$}}
   \end{picture}}
-\sqrt{(m^1{+}1)(n^1{+}1)} \parbox{45\unitlength}{\begin{picture}(45,28)
       \put(0,0){\epsfig{file=at2,scale=0.9,bb=71 593 191 669}}
       \put(-2,6){\mbox{\scriptsize$\di{m^1+1}{m^2}$}}
       \put(4,-1){\mbox{\scriptsize$\di{n^1+1}{n^2}$}}
       \put(41,4){\mbox{\scriptsize$\di{m^1}{m^2}$}}
       \put(32,2){\mbox{\scriptsize$\di{n^1}{n^2}$}}
       \put(4,14){\mbox{\scriptsize$\di{1}{0}$}}
       \put(12,4){\mbox{\scriptsize$\di{1}{0}$}}
       \put(28,4){\mbox{\scriptsize$\di{0}{0}$}}
       \put(36,14){\mbox{\scriptsize$\di{0}{0}$}}
       \put(20,26){\mbox{\scriptsize$\di{1}{0}$}}
     \end{picture}}
\nonumber
\\*
& = \parbox{37\unitlength}{\begin{picture}(35,28)
       \put(0,0){\epsfig{file=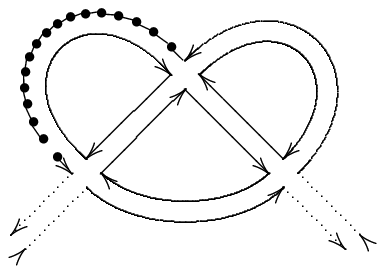,scale=0.9,bb=79 601 182 673}}
       \put(-2,6){\mbox{\scriptsize$\di{m^1+1}{m^2}$}}
       \put(4,-1){\mbox{\scriptsize$\di{n^1+1}{n^2}$}}
       \put(35,4){\mbox{\scriptsize$\di{m^1}{m^2}$}}
       \put(27,1){\mbox{\scriptsize$\di{n^1}{n^2}$}}
       \put(15,24){\mbox{\scriptsize$\di{m^1+1}{m^2}$}}
   \end{picture}} \quad
-~~
\parbox{37\unitlength}{\begin{picture}(35,28)
       \put(0,0){\epsfig{file=at2d,scale=0.9,bb=79 601 182 673}}
       \put(-2,6){\mbox{\scriptsize$\di{m^1+1}{m^2}$}}
       \put(4,-1){\mbox{\scriptsize$\di{n^1+1}{n^2}$}}
       \put(35,4){\mbox{\scriptsize$\di{m^1}{m^2}$}}
       \put(27,1){\mbox{\scriptsize$\di{n^1}{n^2}$}}
       \put(-1,13){\mbox{\scriptsize$\di{1}{0}$}}
       \put(13,26){\mbox{\scriptsize$\di{1}{0}$}}
       \put(16,25){\mbox{\scriptsize$\di{m^1+1}{m^2}$}}
   \end{picture}}
\nonumber
\\*
& + ~~\parbox{37\unitlength}{\begin{picture}(35,28)
       \put(0,0){\epsfig{file=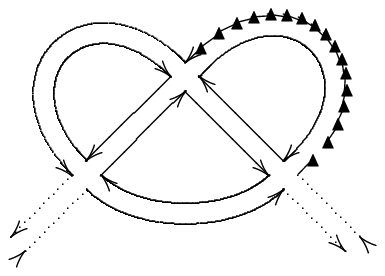,scale=0.9,bb=79 601 182 673}}
       \put(-2,6){\mbox{\scriptsize$\di{m^1+1}{m^2}$}}
       \put(4,-1){\mbox{\scriptsize$\di{n^1+1}{n^2}$}}
       \put(35,4){\mbox{\scriptsize$\di{m^1}{m^2}$}}
       \put(27,1){\mbox{\scriptsize$\di{n^1}{n^2}$}}
       \put(0,13){\mbox{\scriptsize$\di{1}{0}$}}
       \put(13,26){\mbox{\scriptsize$\di{1}{0}$}}
       \put(16,25){\mbox{\scriptsize$\di{m^1+1}{m^2}$}}
   \end{picture}}
+ \sqrt{m^1{+}1} ~~\parbox{37\unitlength}{\begin{picture}(35,28)
       \put(0,0){\epsfig{file=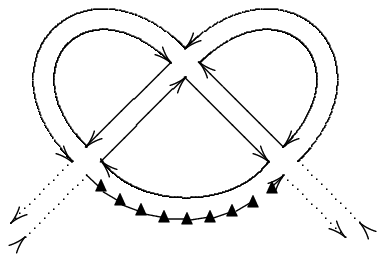,scale=0.9,bb=79 601 182 669}}
       \put(-2,6){\mbox{\scriptsize$\di{m^1+1}{m^2}$}}
       \put(4,-1){\mbox{\scriptsize$\di{n^1+1}{n^2}$}}
       \put(35,4){\mbox{\scriptsize$\di{m^1}{m^2}$}}
       \put(27,1){\mbox{\scriptsize$\di{n^1}{n^2}$}}
       \put(0,13){\mbox{\scriptsize$\di{1}{0}$}}
       \put(34,13){\mbox{\scriptsize$\di{0}{0}$}}
       \put(18,24){\mbox{\scriptsize$\di{1}{0}$}}
   \end{picture}}
\label{brezel3}
\end{align}
It is important to understand that according to (\ref{intA2}) the
indices at the external lines of the reference graph (with
zero-indices) are adjusted to the external indices of the original
(leftmost) graph:
\begin{align}
\parbox{45\unitlength}{\begin{picture}(45,28)
       \put(0,0){\epsfig{file=at2,scale=0.9,bb=71 593 191 669}}
       \put(-2,6){\mbox{\scriptsize$\di{m^1+1}{m^2}$}}
       \put(4,-1){\mbox{\scriptsize$\di{n^1+1}{n^2}$}}
       \put(41,4){\mbox{\scriptsize$\di{m^1}{m^2}$}}
       \put(32,2){\mbox{\scriptsize$\di{n^1}{n^2}$}}
       \put(4,14){\mbox{\scriptsize$\di{1}{0}$}}
       \put(12,4){\mbox{\scriptsize$\di{1}{0}$}}
       \put(28,4){\mbox{\scriptsize$\di{0}{0}$}}
       \put(36,14){\mbox{\scriptsize$\di{0}{0}$}}
       \put(20,26){\mbox{\scriptsize$\di{1}{0}$}}
     \end{picture}}
\equiv \quad
\parbox{24\unitlength}{\begin{picture}(20,10)
       \put(0,3){\reflectbox{\epsfig{file=h03,scale=0.9,bb=71 666 130 677}}}
       \put(0,0){\mbox{\scriptsize$\di{n^1+1}{n^2}$}}
       \put(15,0){\mbox{\scriptsize$\di{n^1}{n^2}$}}
       \put(1,8){\mbox{\scriptsize$\di{m^1+1}{m^2}$}}
       \put(16,8){\mbox{\scriptsize$\di{m^1}{m^2}$}}
   \end{picture}}
\cdot \left[\parbox{45\unitlength}{\begin{picture}(45,28)
       \put(0,0){\epsfig{file=at2,scale=0.9,bb=71 593 191 669}}
       \put(2,8){\mbox{\scriptsize$\di{1}{0}$}}
       \put(7,2){\mbox{\scriptsize$\di{1}{0}$}}
       \put(37,8){\mbox{\scriptsize$\di{0}{0}$}}
       \put(32,2){\mbox{\scriptsize$\di{0}{0}$}}
       \put(20,26){\mbox{\scriptsize$\di{1}{0}$}}
     \end{picture}}\right]\;.
\end{align}
Thus, all graphs with composite propagators have the same index
structure at the external legs. When further contracting these graphs,
the contracting propagator matches the external indices of the
original graph. The argumentation in the proof of
Proposition~\ref{power-counting-prop}.\ref{prp-planar-A1} should be
transparent now. In particular, it becomes understandable why the
difference (\ref{brezel3}) is irrelevant and can be integrated from
$\Lambda_0$ down to $\Lambda$. On the other hand, the reference graph
to be integrated from $\Lambda_R$ up to $\Lambda$ becomes
\begin{align}
\sqrt{(m^1{+}1)(n^1{+}1)} \quad
\parbox{24\unitlength}{\begin{picture}(20,10)
       \put(0,3){\reflectbox{\epsfig{file=at3b,scale=0.9,bb=71 666 130 677}}}
       \put(0,0){\mbox{\scriptsize$\di{n^1+1}{n^2}$}}
       \put(15,0){\mbox{\scriptsize$\di{n^1}{n^2}$}}
       \put(1,8){\mbox{\scriptsize$\di{m^1+1}{m^2}$}}
       \put(16,8){\mbox{\scriptsize$\di{m^1}{m^2}$}}
   \end{picture}}
\quad
\left( \parbox{45\unitlength}{\begin{picture}(45,27)
       \put(0,0){\epsfig{file=at2,scale=0.9,bb=71 593 191 669}}
       \put(2,8){\mbox{\scriptsize$\di{1}{0}$}}
       \put(7,2){\mbox{\scriptsize$\di{1}{0}$}}
       \put(37,8){\mbox{\scriptsize$\di{0}{0}$}}
       \put(32,2){\mbox{\scriptsize$\di{0}{0}$}}
       \put(20,26){\mbox{\scriptsize$\di{1}{0}$}}
     \end{picture}} \right)\;.
\end{align}

We cannot use the same procedure for the history $a$-$b$-$c$-$d$-$e$
of contractions in (\ref{graph-at2}), because we end up with a
marginal subgraph after the $a$-$b$ contractions. According to
Definition~\ref{defint}.\ref{defint4} we have to decompose the $a$-$b$
subgraph into an irrelevant (according to
Proposition~\ref{power-counting-prop}.\ref{prp-planar-A4}) difference
and a marginal reference graph:
\begin{align}
\parbox{28\unitlength}{\begin{picture}(25,28)
       \put(0,0){\epsfig{file=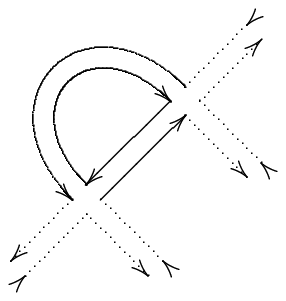,scale=0.9,bb=79 605 154 680}}
       \put(-3,6){\mbox{\scriptsize$\di{m^1+1}{m^2}$}}
       \put(4,-1){\mbox{\scriptsize$\di{n^1+1}{n^2}$}}
       \put(13,7){\mbox{\scriptsize$\di{k^1}{k^2}$}}
       \put(17,12){\mbox{\scriptsize$\di{k^1}{k^2}$}}
       \put(15,24){\mbox{\scriptsize$\di{m^1+1}{m^2}$}}
       \put(23,17){\mbox{\scriptsize$\di{l^1}{l^2}$}}
   \end{picture}}
&= \left\{~~ \parbox{28\unitlength}{\begin{picture}(25,28)
       \put(0,0){\epsfig{file=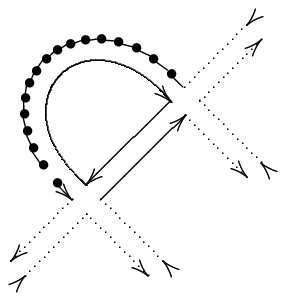,scale=0.9,bb=79 605 154 680}}
       \put(-3,6){\mbox{\scriptsize$\di{m^1+1}{m^2}$}}
       \put(4,-1){\mbox{\scriptsize$\di{n^1+1}{n^2}$}}
       \put(13,7){\mbox{\scriptsize$\di{k^1}{k^2}$}}
       \put(17,12){\mbox{\scriptsize$\di{k^1}{k^2}$}}
       \put(15,24){\mbox{\scriptsize$\di{m^1+1}{m^2}$}}
       \put(23,17){\mbox{\scriptsize$\di{l^1}{l^2}$}}
   \end{picture}}
+ \parbox{28\unitlength}{\begin{picture}(25,28)
       \put(0,0){\epsfig{file=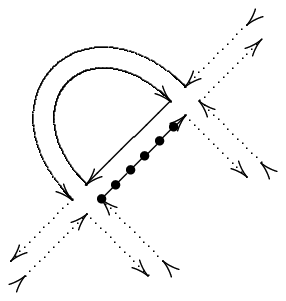,scale=0.9,bb=79 605 154 680}}
       \put(0,13){\mbox{\scriptsize$\di{0}{0}$}}
       \put(-3,6){\mbox{\scriptsize$\di{m^1+1}{m^2}$}}
       \put(4,-1){\mbox{\scriptsize$\di{n^1+1}{n^2}$}}
       \put(13,7){\mbox{\scriptsize$\di{k^1}{k^2}$}}
       \put(17,12){\mbox{\scriptsize$\di{k^1}{k^2}$}}
       \put(16,25){\mbox{\scriptsize$\di{m^1+1}{m^2}$}}
       \put(12,26){\mbox{\scriptsize$\di{0}{0}$}}
       \put(23,17){\mbox{\scriptsize$\di{l^1}{l^2}$}}
   \end{picture}} \right\} 
\nonumber
\\*[-1ex]
&\qquad\qquad + ~~\parbox{14\unitlength}{\begin{picture}(14,14)
       \put(0,0){\epsfig{file=vert1,scale=0.9,bb=71 638 117 684}}
       \put(-3,7){\mbox{\scriptsize$\di{m^1+1}{m^2}$}}
       \put(5,0){\mbox{\scriptsize$\di{n^1+1}{n^2}$}}
       \put(12,7){\mbox{\scriptsize$\di{k^1}{k^2}$}}
       \put(6,14){\mbox{\scriptsize$\di{l^1}{l^2}$}}
   \end{picture}}~~~
 \parbox{28\unitlength}{\begin{picture}(25,28)
       \put(0,0){\epsfig{file=at2f,scale=0.9,bb=79 605 154 680}}
       \put(1,7){\mbox{\scriptsize$\di{0}{0}$}}
       \put(4,-1){\mbox{\scriptsize$\di{0}{0}$}}
       \put(13,7){\mbox{\scriptsize$\di{0}{0}$}}
       \put(17,12){\mbox{\scriptsize$\di{0}{0}$}}
       \put(17,24){\mbox{\scriptsize$\di{0}{0}$}}
       \put(23,17){\mbox{\scriptsize$\di{0}{0}$}}
   \end{picture}}
\label{brezel-a-b}
\end{align}
The two graphs in braces $\{~\}$ are irrelevant and 
integrated from $\Lambda_0$ down to $\Lambda_c$. The remaining piece
can be written as the original $\phi^4$-vertex times a graph with
vanishing external indices, which is integrated from $\Lambda_R$ up to
$\Lambda_c$ and can be bounded by $C \ln \frac{\Lambda_c}{\Lambda_R}$.
Inserting the decomposition (\ref{brezel-a-b}) into
(\ref{graph-at2}) we obtain the following decomposition valid for the
history $a$-$b$-$c$-$d$-$e$:
\begin{subequations}
\label{brezel4}
\begin{align}
&\parbox{45\unitlength}{\begin{picture}(45,28)
       \put(0,0){\epsfig{file=at2,scale=0.9,bb=71 593 191 669}}
       \put(-2,7){\mbox{\scriptsize$\di{m^1+1}{m^2}$}}
       \put(4,-1){\mbox{\scriptsize$\di{n^1+1}{n^2}$}}
       \put(41,4){\mbox{\scriptsize$\di{m^1}{m^2}$}}
       \put(32,2){\mbox{\scriptsize$\di{n^1}{n^2}$}}
       \put(18,26){\mbox{\scriptsize$\di{m^1+1}{m^2}$}}
   \end{picture}}
-\sqrt{(m^1{+}1)(n^1{+}1)}~~ \parbox{45\unitlength}{\begin{picture}(45,28)
       \put(0,0){\epsfig{file=at2,scale=0.9,bb=71 593 191 669}}
       \put(-2,7){\mbox{\scriptsize$\di{m^1+1}{m^2}$}}
       \put(4,-1){\mbox{\scriptsize$\di{n^1+1}{n^2}$}}
       \put(41,4){\mbox{\scriptsize$\di{m^1}{m^2}$}}
       \put(32,2){\mbox{\scriptsize$\di{n^1}{n^2}$}}
       \put(4,14){\mbox{\scriptsize$\di{1}{0}$}}
       \put(12,4){\mbox{\scriptsize$\di{1}{0}$}}
       \put(28,4){\mbox{\scriptsize$\di{0}{0}$}}
       \put(36,14){\mbox{\scriptsize$\di{0}{0}$}}
       \put(20,26){\mbox{\scriptsize$\di{1}{0}$}}
     \end{picture}}
\nonumber
\\*
& 
= \parbox{37\unitlength}{\begin{picture}(35,28)
       \put(0,0){\epsfig{file=at2d,scale=0.9,bb=79 601 182 673}}
       \put(-2,6){\mbox{\scriptsize$\di{m^1+1}{m^2}$}}
       \put(4,-1){\mbox{\scriptsize$\di{n^1+1}{n^2}$}}
       \put(35,4){\mbox{\scriptsize$\di{m^1}{m^2}$}}
       \put(27,1){\mbox{\scriptsize$\di{n^1}{n^2}$}}
       \put(15,24){\mbox{\scriptsize$\di{m^1+1}{m^2}$}}
   \end{picture}} \quad
- \sqrt{(m^1{+}1)(n^1{+}1)} ~~~\parbox{37\unitlength}{\begin{picture}(35,28)
       \put(0,0){\epsfig{file=at2d,scale=0.9,bb=79 601 182 673}}
       \put(-2,6){\mbox{\scriptsize$\di{m^1+1}{m^2}$}}
       \put(4,-1){\mbox{\scriptsize$\di{n^1+1}{n^2}$}}
       \put(35,4){\mbox{\scriptsize$\di{m^1}{m^2}$}}
       \put(27,1){\mbox{\scriptsize$\di{n^1}{n^2}$}}
       \put(-1,13){\mbox{\scriptsize$\di{1}{0}$}}
       \put(34,13){\mbox{\scriptsize$\di{0}{0}$}}
       \put(18,24){\mbox{\scriptsize$\di{1}{0}$}}
   \end{picture}} 
\label{brezel4a}
\\
&  
+~  \parbox{37\unitlength}{\begin{picture}(35,28)
       \put(0,0){\epsfig{file=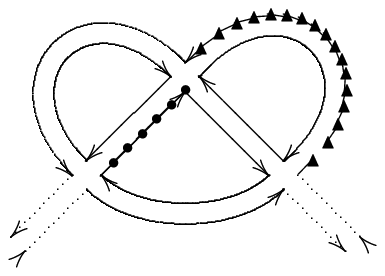,scale=0.9,bb=79 601 182 673}}
       \put(-2,6){\mbox{\scriptsize$\di{m^1+1}{m^2}$}}
       \put(4,-1){\mbox{\scriptsize$\di{n^1+1}{n^2}$}}
       \put(35,4){\mbox{\scriptsize$\di{m^1}{m^2}$}}
       \put(27,1){\mbox{\scriptsize$\di{n^1}{n^2}$}}
       \put(0,13){\mbox{\scriptsize$\di{0}{0}$}}
       \put(13,25.5){\mbox{\scriptsize$\di{0}{0}$}}
       \put(16,25){\mbox{\scriptsize$\di{m^1+1}{m^2}$}}
   \end{picture}}
+ \sqrt{m^1{+}1}~~ \parbox{37\unitlength}{\begin{picture}(35,28)
       \put(0,0){\epsfig{file=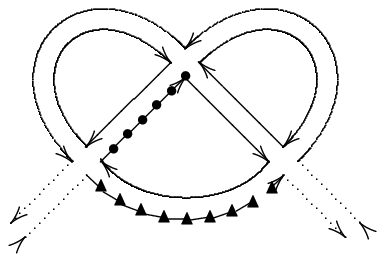,scale=0.9,bb=79 601 182 669}}
       \put(-2,6){\mbox{\scriptsize$\di{m^1+1}{m^2}$}}
       \put(4,-1){\mbox{\scriptsize$\di{n^1+1}{n^2}$}}
       \put(34,13){\mbox{\scriptsize$\di{0}{0}$}}
       \put(35,4){\mbox{\scriptsize$\di{m^1}{m^2}$}}
       \put(27,1){\mbox{\scriptsize$\di{n^1}{n^2}$}}
       \put(0,13){\mbox{\scriptsize$\di{0}{0}$}}
       \put(15,25){\mbox{\scriptsize$\di{0}{0}$}}
       \put(19,25){\mbox{\scriptsize$\di{1}{0}$}}
   \end{picture}}
\label{brezel4b}
\\
& 
+ \left(~~
\parbox{37\unitlength}{\begin{picture}(35,24)
       \put(0,0){\epsfig{file=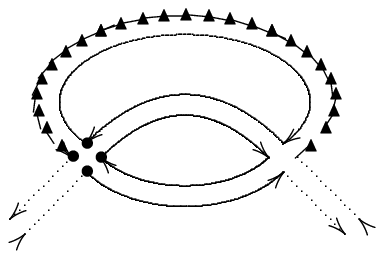,scale=0.9,bb=71 617 174 685}}
       \put(-3,6){\mbox{\scriptsize$\di{m^1+1}{m^2}$}}
       \put(4,-1){\mbox{\scriptsize$\di{n^1+1}{n^2}$}}
       \put(35,4){\mbox{\scriptsize$\di{m^1}{m^2}$}}
       \put(27,1){\mbox{\scriptsize$\di{n^1}{n^2}$}}
   \end{picture}}
+ \sqrt{m^1{+}1}~~ \parbox{37\unitlength}{\begin{picture}(35,24)
       \put(0,0){\epsfig{file=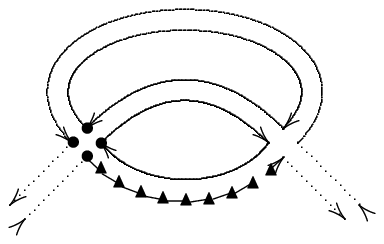,scale=0.9,bb=71 623 174 686}}
       \put(-3,6){\mbox{\scriptsize$\di{m^1+1}{m^2}$}}
       \put(1,13){\mbox{\scriptsize$\di{1}{0}$}}
       \put(4,-1){\mbox{\scriptsize$\di{n^1+1}{n^2}$}}
       \put(33,13){\mbox{\scriptsize$\di{0}{0}$}}
       \put(35,4){\mbox{\scriptsize$\di{m^1}{m^2}$}}
       \put(27,1){\mbox{\scriptsize$\di{n^1}{n^2}$}}
   \end{picture}} ~~\right)
\nonumber
\\*
& \qquad\qquad \times\left(\parbox{28\unitlength}{\begin{picture}(25,28)
       \put(0,0){\epsfig{file=at2f,scale=0.9,bb=79 605 154 680}}
       \put(1,7){\mbox{\scriptsize$\di{0}{0}$}}
       \put(4,-1){\mbox{\scriptsize$\di{0}{0}$}}
       \put(13,7){\mbox{\scriptsize$\di{0}{0}$}}
       \put(17,12){\mbox{\scriptsize$\di{0}{0}$}}
       \put(17,24){\mbox{\scriptsize$\di{0}{0}$}}
       \put(23,17){\mbox{\scriptsize$\di{0}{0}$}}
   \end{picture}}\right)_{\!\!\Lambda_c} .
\label{brezel4c}
\end{align}
\end{subequations}
The line (\ref{brezel4a}) corresponds to the first graph in the braces
$\{~\}$ of (\ref{brezel-a-b}) for both graphs on the lhs of
(\ref{brezel4}). These graphs are already irrelevant\footnote{In the
  right graph (\ref{brezel4a}) the composite propagator is according
  to (\ref{Qpl1}) bounded by $\frac{C}{\Omega\theta\Lambda^2}
  \frac{1}{\theta \Lambda^2}$ so that the combination with the
  prefactor $\sqrt{(m^1{+}1)(n^1{+}1)}$ leads to the ratio
  $\sqrt{(\frac{m^1{+}1}{\theta \Lambda^2}) (\frac{m^1{+}1}{\theta
      \Lambda^2})}$ by which (\ref{brezel4a}) is suppressed over the
  first graph on the lhs of (\ref{brezel4}).} so that no further
decomposition is necessary. The second graph in the braces $\{~\}$ of
(\ref{brezel-a-b}), inserted into the lhs of (\ref{brezel4}), yields
the line (\ref{brezel4b}). Finally, the last part of
(\ref{brezel-a-b}) leads to the line (\ref{brezel4c}).

Let us also look at the relevant contribution
$m_1=k=n_2=\di{m^1}{m^2}$, $n_1=m_2=\di{n^1}{n^2}$ of the graph
(\ref{graph-at2}). The history $a$-$c$-$d$-$e$-$b$ contains irrelevant
subgraphs only:
\begin{subequations}
\begin{align}
&\parbox{45\unitlength}{\begin{picture}(45,28)
       \put(0,0){\epsfig{file=at2,scale=0.9,bb=71 593 191 669}}
       \put(0,8){\mbox{\scriptsize$\di{m^1}{m^2}$}}
       \put(7,2){\mbox{\scriptsize$\di{n^1}{n^2}$}}
       \put(36,8){\mbox{\scriptsize$\di{m^1}{m^2}$}}
       \put(31,2){\mbox{\scriptsize$\di{n^1}{n^2}$}}
       \put(18,26){\mbox{\scriptsize$\di{m^1}{m^2}$}}
   \end{picture}}
+(m^1{+}m^2{+}n^1{+}n^2{-}1) \parbox{45\unitlength}{\begin{picture}(45,28)
       \put(0,0){\epsfig{file=at2,scale=0.9,bb=71 593 191 669}}
       \put(0,8){\mbox{\scriptsize$\di{m^1}{m^2}$}}
       \put(7,2){\mbox{\scriptsize$\di{n^1}{n^2}$}}
       \put(36,8){\mbox{\scriptsize$\di{m^1}{m^2}$}}
       \put(31,2){\mbox{\scriptsize$\di{n^1}{n^2}$}}
       \put(3,15){\mbox{\scriptsize$\di{0}{0}$}}
       \put(14,4){\mbox{\scriptsize$\di{0}{0}$}}
       \put(36,15){\mbox{\scriptsize$\di{0}{0}$}}
       \put(26,4){\mbox{\scriptsize$\di{0}{0}$}}
       \put(20,26){\mbox{\scriptsize$\di{0}{0}$}}
     \end{picture}}
\nonumber
\\*
&- m^1 \parbox{45\unitlength}{\begin{picture}(45,28)
       \put(0,0){\epsfig{file=at2,scale=0.9,bb=71 593 191 669}}
       \put(0,8){\mbox{\scriptsize$\di{m^1}{m^2}$}}
       \put(7,2){\mbox{\scriptsize$\di{n^1}{n^2}$}}
       \put(36,8){\mbox{\scriptsize$\di{m^1}{m^2}$}}
       \put(31,2){\mbox{\scriptsize$\di{n^1}{n^2}$}}
       \put(3,15){\mbox{\scriptsize$\di{1}{0}$}}
       \put(14,4){\mbox{\scriptsize$\di{0}{0}$}}
       \put(36,15){\mbox{\scriptsize$\di{1}{0}$}}
       \put(26,4){\mbox{\scriptsize$\di{0}{0}$}}
       \put(20,26){\mbox{\scriptsize$\di{1}{0}$}}
   \end{picture}}
-m^2 \parbox{45\unitlength}{\begin{picture}(45,28)
       \put(0,0){\epsfig{file=at2,scale=0.9,bb=71 593 191 669}}
       \put(0,8){\mbox{\scriptsize$\di{m^1}{m^2}$}}
       \put(7,2){\mbox{\scriptsize$\di{n^1}{n^2}$}}
       \put(36,8){\mbox{\scriptsize$\di{m^1}{m^2}$}}
       \put(31,2){\mbox{\scriptsize$\di{n^1}{n^2}$}}
       \put(3,15){\mbox{\scriptsize$\di{0}{1}$}}
       \put(14,4){\mbox{\scriptsize$\di{0}{0}$}}
       \put(36,15){\mbox{\scriptsize$\di{0}{1}$}}
       \put(26,4){\mbox{\scriptsize$\di{0}{0}$}}
       \put(20,26){\mbox{\scriptsize$\di{0}{1}$}}
     \end{picture}}
\nonumber
\\*
&- n^1 \parbox{45\unitlength}{\begin{picture}(45,28)
       \put(0,0){\epsfig{file=at2,scale=0.9,bb=71 593 191 669}}
       \put(0,8){\mbox{\scriptsize$\di{m^1}{m^2}$}}
       \put(7,2){\mbox{\scriptsize$\di{n^1}{n^2}$}}
       \put(36,8){\mbox{\scriptsize$\di{m^1}{m^2}$}}
       \put(31,2){\mbox{\scriptsize$\di{n^1}{n^2}$}}
       \put(3,15){\mbox{\scriptsize$\di{0}{0}$}}
       \put(14,4){\mbox{\scriptsize$\di{1}{0}$}}
       \put(36,15){\mbox{\scriptsize$\di{0}{0}$}}
       \put(26,4){\mbox{\scriptsize$\di{1}{0}$}}
       \put(20,26){\mbox{\scriptsize$\di{0}{0}$}}
   \end{picture}}
-n^2 \parbox{45\unitlength}{\begin{picture}(45,28)
       \put(0,0){\epsfig{file=at2,scale=0.9,bb=71 593 191 669}}
       \put(0,8){\mbox{\scriptsize$\di{m^1}{m^2}$}}
       \put(7,2){\mbox{\scriptsize$\di{n^1}{n^2}$}}
       \put(36,8){\mbox{\scriptsize$\di{m^1}{m^2}$}}
       \put(31,2){\mbox{\scriptsize$\di{n^1}{n^2}$}}
       \put(3,15){\mbox{\scriptsize$\di{0}{0}$}}
       \put(14,4){\mbox{\scriptsize$\di{0}{1}$}}
       \put(36,15){\mbox{\scriptsize$\di{0}{0}$}}
       \put(26,4){\mbox{\scriptsize$\di{0}{1}$}}
       \put(20,26){\mbox{\scriptsize$\di{0}{0}$}}
     \end{picture}}
\nonumber
\\
& =
\parbox{40\unitlength}{\begin{picture}(35,28)
       \put(0,0){\epsfig{file=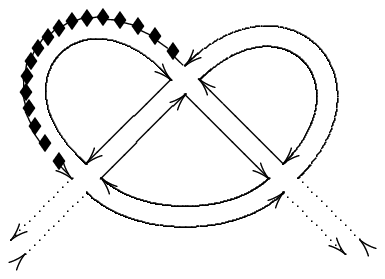,scale=0.9,bb=79 601 182 674}}
       \put(0,6){\mbox{\scriptsize$\di{m^1}{m^2}$}}
       \put(6,1){\mbox{\scriptsize$\di{n^1}{n^2}$}}
       \put(34,6){\mbox{\scriptsize$\di{m^1}{m^2}$}}
       \put(27,1){\mbox{\scriptsize$\di{n^1}{n^2}$}}
       \put(12,1){\mbox{\scriptsize$\di{0}{0}$}}
       \put(34,13){\mbox{\scriptsize$\di{0}{0}$}}
       \put(23,1){\mbox{\scriptsize$\di{0}{0}$}}
       \put(21,25){\mbox{\scriptsize$\di{0}{0}$}}
       \put(15,25){\mbox{\scriptsize$\di{m^1}{m^2}$}}
   \end{picture}}
+ \parbox{40\unitlength}{\begin{picture}(35,28)
       \put(0,0){\epsfig{file=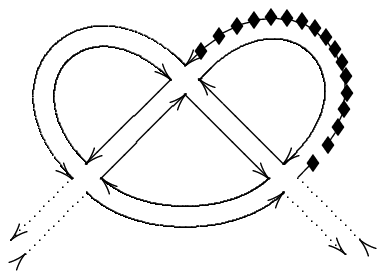,scale=0.9,bb=79 601 182 674}}
       \put(0,6){\mbox{\scriptsize$\di{m^1}{m^2}$}}
       \put(6,1){\mbox{\scriptsize$\di{n^1}{n^2}$}}
       \put(34,6){\mbox{\scriptsize$\di{m^1}{m^2}$}}
       \put(27,1){\mbox{\scriptsize$\di{n^1}{n^2}$}}
       \put(12,1){\mbox{\scriptsize$\di{0}{0}$}}
       \put(1,13){\mbox{\scriptsize$\di{0}{0}$}}
       \put(23,1){\mbox{\scriptsize$\di{0}{0}$}}
       \put(14,25){\mbox{\scriptsize$\di{0}{0}$}}
       \put(18,25){\mbox{\scriptsize$\di{m^1}{m^2}$}}
   \end{picture}}
+ \parbox{40\unitlength}{\begin{picture}(35,28)
       \put(0,0){\epsfig{file=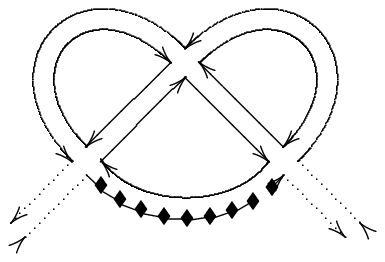,scale=0.9,bb=79 601 182 669}}
       \put(0,6){\mbox{\scriptsize$\di{m^1}{m^2}$}}
       \put(6,1){\mbox{\scriptsize$\di{n^1}{n^2}$}}
       \put(34,6){\mbox{\scriptsize$\di{m^1}{m^2}$}}
       \put(27,1){\mbox{\scriptsize$\di{n^1}{n^2}$}}
       \put(1,13){\mbox{\scriptsize$\di{0}{0}$}}
       \put(34,13){\mbox{\scriptsize$\di{0}{0}$}}
       \put(16,24){\mbox{\scriptsize$\di{0}{0}$}}
   \end{picture}}
\label{brezel0a}
\\
& + \parbox{37\unitlength}{\begin{picture}(35,28)
       \put(0,0){\epsfig{file=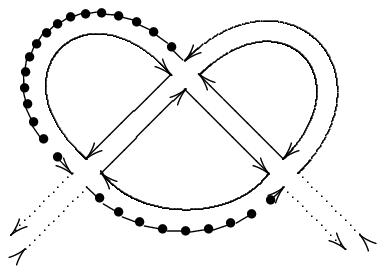,scale=0.9,bb=79 601 182 673}}
        \put(0,6){\mbox{\scriptsize$\di{m^1}{m^2}$}}
       \put(6,1){\mbox{\scriptsize$\di{n^1}{n^2}$}}
       \put(34,6){\mbox{\scriptsize$\di{m^1}{m^2}$}}
       \put(27,1){\mbox{\scriptsize$\di{n^1}{n^2}$}}
       \put(18,25){\mbox{\scriptsize$\di{m^1}{m^2}$}}
   \end{picture}}
+\parbox{37\unitlength}{\begin{picture}(35,28)
       \put(0,0){\epsfig{file=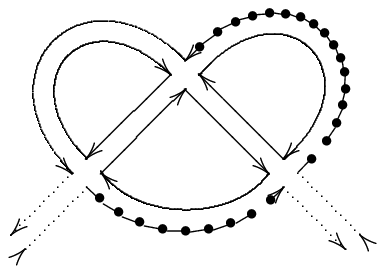,scale=0.9,bb=79 601 182 673}}
        \put(0,6){\mbox{\scriptsize$\di{m^1}{m^2}$}}
       \put(6,1){\mbox{\scriptsize$\di{n^1}{n^2}$}}
       \put(34,6){\mbox{\scriptsize$\di{m^1}{m^2}$}}
       \put(27,1){\mbox{\scriptsize$\di{n^1}{n^2}$}}
       \put(1,13){\mbox{\scriptsize$\di{0}{0}$}}
       \put(15,25){\mbox{\scriptsize$\di{0}{0}$}}
       \put(18,25){\mbox{\scriptsize$\di{m^1}{m^2}$}}
   \end{picture}} 
\label{brezel0b}
\\
&+ \parbox{37\unitlength}{\begin{picture}(35,28)
       \put(0,0){\epsfig{file=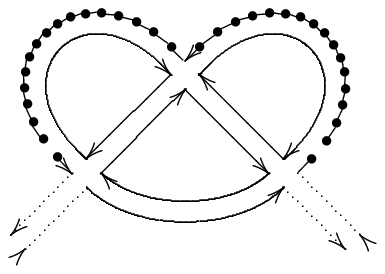,scale=0.9,bb=79 601 182 673}}
       \put(0,6){\mbox{\scriptsize$\di{m^1}{m^2}$}}
       \put(6,1){\mbox{\scriptsize$\di{n^1}{n^2}$}}
       \put(34,6){\mbox{\scriptsize$\di{m^1}{m^2}$}}
       \put(27,1){\mbox{\scriptsize$\di{n^1}{n^2}$}}
       \put(12,1){\mbox{\scriptsize$\di{0}{0}$}}
       \put(23,1){\mbox{\scriptsize$\di{0}{0}$}}
       \put(17,25){\mbox{\scriptsize$\di{m^1}{m^2}$}}
   \end{picture}}
- m^1~  \parbox{37\unitlength}{\begin{picture}(35,28)
       \put(0,0){\epsfig{file=at2q,scale=0.9,bb=79 601 182 673}}
       \put(0,6){\mbox{\scriptsize$\di{m^1}{m^2}$}}
       \put(6,1){\mbox{\scriptsize$\di{n^1}{n^2}$}}
       \put(34,6){\mbox{\scriptsize$\di{m^1}{m^2}$}}
       \put(27,1){\mbox{\scriptsize$\di{n^1}{n^2}$}}
       \put(0,13){\mbox{\scriptsize$\di{1}{0}$}}
       \put(35,13){\mbox{\scriptsize$\di{1}{0}$}}
       \put(12,1){\mbox{\scriptsize$\di{0}{0}$}}
       \put(23,1){\mbox{\scriptsize$\di{0}{0}$}}
       \put(18,24){\mbox{\scriptsize$\di{1}{0}$}}
   \end{picture}}
- m^2~  \parbox{37\unitlength}{\begin{picture}(35,28)
       \put(0,0){\epsfig{file=at2q,scale=0.9,bb=79 601 182 673}}
       \put(0,6){\mbox{\scriptsize$\di{m^1}{m^2}$}}
       \put(6,1){\mbox{\scriptsize$\di{n^1}{n^2}$}}
       \put(34,6){\mbox{\scriptsize$\di{m^1}{m^2}$}}
       \put(27,1){\mbox{\scriptsize$\di{n^1}{n^2}$}}
       \put(0,13){\mbox{\scriptsize$\di{0}{1}$}}
       \put(35,13){\mbox{\scriptsize$\di{0}{1}$}}
       \put(12,1){\mbox{\scriptsize$\di{0}{0}$}}
       \put(23,1){\mbox{\scriptsize$\di{0}{0}$}}
       \put(18,24){\mbox{\scriptsize$\di{0}{1}$}}
   \end{picture}}
\label{brezel0c}
\end{align}
\end{subequations}
The line (\ref{brezel0a}) corresponds to (\ref{decA1-b-a}), 
the line (\ref{brezel0b}) to (\ref{decA1-a}) and the 
line (\ref{brezel0c}) to (\ref{decA1-b-b}).

If the history of contractions contains relevant or marginal
subgraphs, we first have to decompose the subgraphs into the reference
function with vanishing external indices and an irrelevant remainder.
For instance, the decomposition relative to the history
$a$-$b$-$c$-$d$-$e$ would be
\begin{align}
&\parbox{42\unitlength}{\begin{picture}(42,26)
       \put(0,0){\epsfig{file=at2,scale=0.9,bb=71 593 191 669}}
       \put(0,8){\mbox{\scriptsize$\di{m^1}{m^2}$}}
       \put(7,2){\mbox{\scriptsize$\di{n^1}{n^2}$}}
       \put(36,8){\mbox{\scriptsize$\di{m^1}{m^2}$}}
       \put(31,2){\mbox{\scriptsize$\di{n^1}{n^2}$}}
       \put(18,26){\mbox{\scriptsize$\di{m^1}{m^2}$}}
   \end{picture}}
\mapsto~~ \parbox{37\unitlength}{\begin{picture}(35,28)
       \put(0,0){\epsfig{file=at2d,scale=0.9,bb=79 601 182 673}}
       \put(-1,6){\mbox{\scriptsize$\di{m^1}{m^2}$}}
       \put(5,0){\mbox{\scriptsize$\di{n^1}{n^2}$}}
       \put(36,4){\mbox{\scriptsize$\di{m^1}{m^2}$}}
       \put(28,1){\mbox{\scriptsize$\di{n^1}{n^2}$}}
       \put(17,24){\mbox{\scriptsize$\di{m^1}{m^2}$}}
   \end{picture}}
+~~ \parbox{37\unitlength}{\begin{picture}(35,28)
       \put(0,0){\epsfig{file=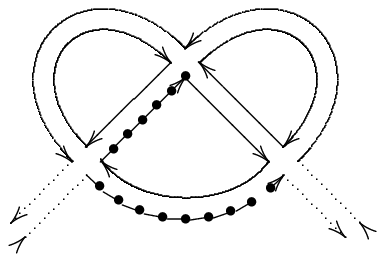,scale=0.9,bb=79 601 182 669}}
       \put(-1,6){\mbox{\scriptsize$\di{m^1}{m^2}$}}
       \put(5,0){\mbox{\scriptsize$\di{n^1}{n^2}$}}
       \put(36,4){\mbox{\scriptsize$\di{m^1}{m^2}$}}
       \put(28,1){\mbox{\scriptsize$\di{n^1}{n^2}$}}
       \put(15,25){\mbox{\scriptsize$\di{0}{0}$}}
       \put(18,25){\mbox{\scriptsize$\di{m^1}{m^2}$}}
       \put(1,13){\mbox{\scriptsize$\di{0}{0}$}}
   \end{picture}}
\nonumber
\\*
& +  \parbox{37\unitlength}{\begin{picture}(35,28)
       \put(0,0){\epsfig{file=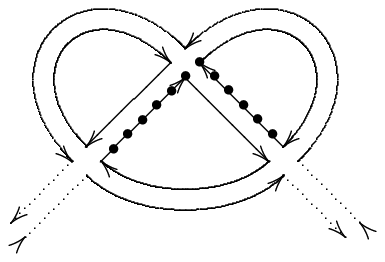,scale=0.9,bb=79 601 182 669}}
       \put(-1,6){\mbox{\scriptsize$\di{m^1}{m^2}$}}
       \put(5,0){\mbox{\scriptsize$\di{n^1}{n^2}$}}
       \put(36,4){\mbox{\scriptsize$\di{m^1}{m^2}$}}
       \put(28,1){\mbox{\scriptsize$\di{n^1}{n^2}$}}
       \put(15,25){\mbox{\scriptsize$\di{0}{0}$}}
       \put(18,25){\mbox{\scriptsize$\di{m^1}{m^2}$}}
       \put(1,13){\mbox{\scriptsize$\di{0}{0}$}}
       \put(13,1){\mbox{\scriptsize$\di{0}{0}$}}
       \put(23,1){\mbox{\scriptsize$\di{0}{0}$}}
   \end{picture}}
+ ~~
\parbox{24\unitlength}{\begin{picture}(22,26)
      \put(0,0){\epsfig{file=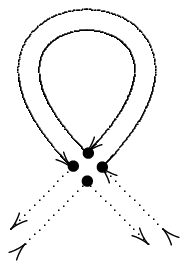,scale=0.9,bb=79 617 126 687}}
      \put(-1,6){\mbox{\scriptsize$\di{m^1}{m^2}$}}
      \put(4,0){\mbox{\scriptsize$\di{n^1}{n^2}$}}
      \put(9,0){\mbox{\scriptsize$\di{n^1}{n^2}$}}
      \put(16,4){\mbox{\scriptsize$\di{m^1}{m^2}$}}
\end{picture}}
\left( \parbox{37\unitlength}{\begin{picture}(35,28)
       \put(0,0){\epsfig{file=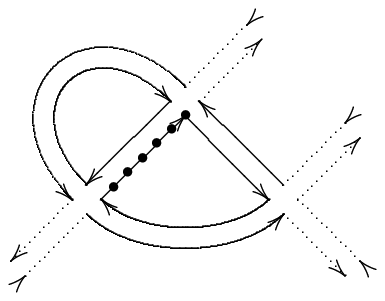,scale=0.9,bb=79 605 182 680}}
       \put(-1,6){\mbox{\scriptsize$\di{0}{0}$}}
       \put(5,0){\mbox{\scriptsize$\di{0}{0}$}}
       \put(33,7){\mbox{\scriptsize$\di{0}{0}$}}
       \put(28,1){\mbox{\scriptsize$\di{0}{0}$}}
       \put(28,14){\mbox{\scriptsize$\di{0}{0}$}}
       \put(23,19){\mbox{\scriptsize$\di{0}{0}$}}
       \put(17,24){\mbox{\scriptsize$\di{0}{0}$}}
   \end{picture}} ~\right)_{\!\!\Lambda_e} 
\nonumber
\\
& 
+ \left(~~
\parbox{37\unitlength}{\begin{picture}(35,25)
       \put(0,0){\epsfig{file=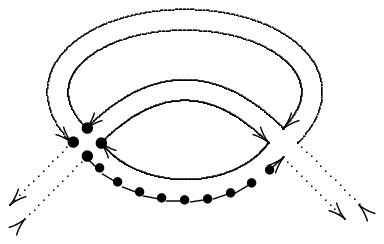,scale=0.9,bb=71 623 174 686}}
       \put(-1,6){\mbox{\scriptsize$\di{m^1}{m^2}$}}
       \put(4,-1){\mbox{\scriptsize$\di{n^1}{n^2}$}}
       \put(35,4){\mbox{\scriptsize$\di{m^1}{m^2}$}}
       \put(27,1){\mbox{\scriptsize$\di{n^1}{n^2}$}}
   \end{picture}}
+~~ \parbox{37\unitlength}{\begin{picture}(35,25)
       \put(0,0){\epsfig{file=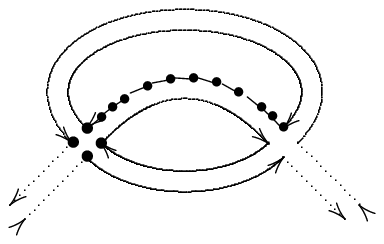,scale=0.9,bb=71 623 174 686}}
       \put(-1,6){\mbox{\scriptsize$\di{m^1}{m^2}$}}
       \put(4,-1){\mbox{\scriptsize$\di{n^1}{n^2}$}}
       \put(35,4){\mbox{\scriptsize$\di{m^1}{m^2}$}}
       \put(27,1){\mbox{\scriptsize$\di{n^1}{n^2}$}}
       \put(12,1){\mbox{\scriptsize$\di{0}{0}$}}
       \put(23,1){\mbox{\scriptsize$\di{0}{0}$}}
   \end{picture}} ~~\right)
\left(\parbox{28\unitlength}{\begin{picture}(25,28)
       \put(0,0){\epsfig{file=at2f,scale=0.9,bb=79 605 154 680}}
       \put(1,7){\mbox{\scriptsize$\di{0}{0}$}}
       \put(4,-1){\mbox{\scriptsize$\di{0}{0}$}}
       \put(13,7){\mbox{\scriptsize$\di{0}{0}$}}
       \put(17,12){\mbox{\scriptsize$\di{0}{0}$}}
       \put(17,24){\mbox{\scriptsize$\di{0}{0}$}}
       \put(23,17){\mbox{\scriptsize$\di{0}{0}$}}
   \end{picture}}\right)_{\!\!\Lambda_c}
\nonumber
\\
& 
+~~ \parbox{24\unitlength}{\begin{picture}(22,26)
      \put(0,0){\epsfig{file=a12a,scale=0.9,bb=79 617 126 687}}
      \put(-1,6){\mbox{\scriptsize$\di{m^1}{m^2}$}}
      \put(4,0){\mbox{\scriptsize$\di{n^1}{n^2}$}}
      \put(9,0){\mbox{\scriptsize$\di{n^1}{n^2}$}}
      \put(16,4){\mbox{\scriptsize$\di{m^1}{m^2}$}}
\end{picture}}
\left(~~\parbox{37\unitlength}{\begin{picture}(35,18)
       \put(0,0){\epsfig{file=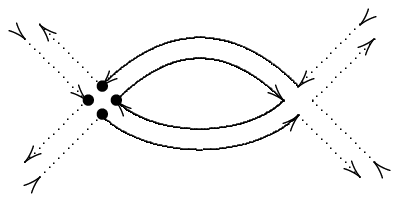,scale=0.9,bb=71 630 174 676}}
       \put(-1,7){\mbox{\scriptsize$\di{0}{0}$}}
       \put(4,-1){\mbox{\scriptsize$\di{0}{0}$}}
       \put(32,7){\mbox{\scriptsize$\di{0}{0}$}}
       \put(27,1){\mbox{\scriptsize$\di{0}{0}$}}
       \put(8,13){\mbox{\scriptsize$\di{0}{0}$}}
       \put(26,13){\mbox{\scriptsize$\di{0}{0}$}}
\end{picture}}~
\right)_{\!\!\Lambda_e}   
\left(\parbox{28\unitlength}{\begin{picture}(25,28)
       \put(0,0){\epsfig{file=at2f,scale=0.9,bb=79 605 154 680}}
       \put(1,7){\mbox{\scriptsize$\di{0}{0}$}}
       \put(4,-1){\mbox{\scriptsize$\di{0}{0}$}}
       \put(13,7){\mbox{\scriptsize$\di{0}{0}$}}
       \put(17,12){\mbox{\scriptsize$\di{0}{0}$}}
       \put(17,24){\mbox{\scriptsize$\di{0}{0}$}}
       \put(23,17){\mbox{\scriptsize$\di{0}{0}$}}
   \end{picture}}\right)_{\!\!\Lambda_c}
\end{align}

\section{Asymptotic behaviour of the propagator}
\label{appB}

\begin{figure}[h!t]
\begin{picture}(130,100)
\put(-27,-120){\epsfig{file=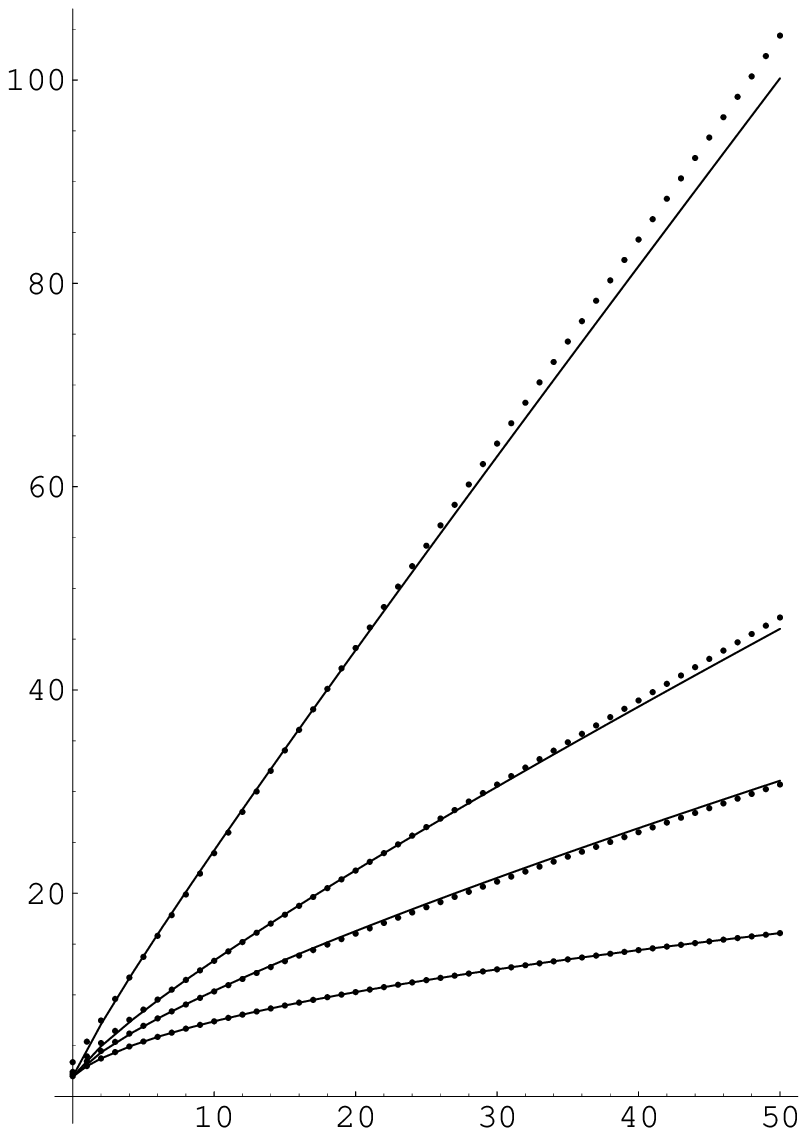,scale=0.72,bb=0 0 598 843}}
\put(54,15){\mbox{\scriptsize$\Omega=0$}}
\put(50,24){\mbox{\scriptsize$\Omega=0.05$}}
\put(50,45){\mbox{\scriptsize$\Omega=0.1$}}
\put(43,83){\mbox{\scriptsize$\Omega=0.3$}}
\put(48,-120){\epsfig{file=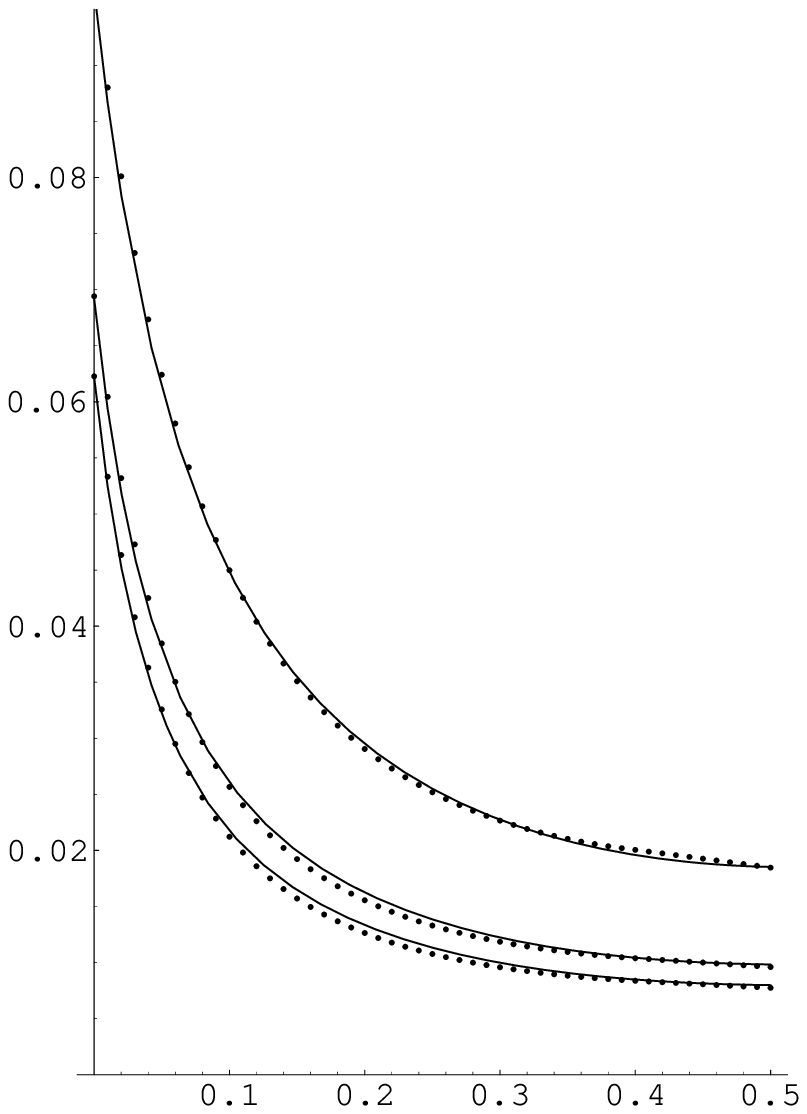,scale=0.72,bb=0 0 598 843}}
\put(127,10.5){\mbox{\scriptsize$\mathcal{C}=50$}}
\put(127,16.5){\mbox{\scriptsize$\mathcal{C}=40$}}
\put(127,25){\mbox{\scriptsize$\mathcal{C}=20$}}
\end{picture}
\caption[Maximum of the propagator]{
  Comparison of $\max \Delta_{mn;kl}^{\mathcal{C}}/\theta$ at
  $\mu_0=0$ (dots) with
  $\big(\sqrt{\frac{1}{\pi}(16\, \mathcal{C}{+}12)} + \frac{6
    \Omega}{1+ 2\Omega^3+2\Omega^4} \mathcal{C}\big)^{-1}$ (solid line). The
  left plot shows the inverses of both the propagator and its
  approximation over $\mathcal{C}$ for various values of $\Omega$. The
  right plot shows the propagator and its approximation 
  over $\Omega$ for various values of $\mathcal{C}$. }
\label{fig-Form4-1}
\end{figure}

For the power-counting theorem we need asymptotic formulae about
the scaling behaviour of the cut-off propagator $\Delta^K_{nm;lk}$ and
certain index summations.  We shall restrict ourselves to the case
$\theta_1=\theta_2=\theta$ and deduce these formulae from the
numerical evaluation of the propagator for a representative class of
parameters and special choices of the parameters where we can compute
the propagator exactly. These formulae involve the cut-off propagator
\begin{align}
  \Delta_{\di{m^1}{m^2}\di{n^1}{n^2};\di{k^1}{k^2}\di{l^1}{l^2}
}^{\mathcal{C}} := \left\{ \begin{array}{ll}
      \Delta_{\di{m^1}{m^2}\di{n^1}{n^2};\di{k^1}{k^2}\di{l^1}{l^2}} 
\quad & \text{for } \mathcal{C} \leq
      \max(m^1,m^2,n^1,n^2,k^1,k^2,l^1,l^2) \leq 2\mathcal{C}\;,
      \\
      0 & \text{otherwise\;,}
\end{array}\right.
\end{align}
which is the restriction of
$\Delta_{\di{m^1}{m^2}\di{n^1}{n^2};\di{k^1}{k^2}\di{l^1}{l^2}}$ to
the support of the cut-off propagator $\Lambda
\frac{\partial}{\partial \Lambda}
\Delta_{\di{m^1}{m^2}\di{n^1}{n^2};\di{k^1}{k^2}\di{l^1}{l^2}}^K(\Lambda)$
appearing in the Polchinski equation, with
$\mathcal{C}=\theta \Lambda^2$.
\\[\bigskipamount]
{\bf Formula 1:}
\begin{align}
 \max_{m^r,n^r,k^r,l^r} 
\Big|  \Delta_{\di{m^1}{m^2}\di{n^1}{n^2};
\di{k^1}{k^2}\di{l^1}{l^2}}^{\mathcal{C}} \Big|_{\mu_0=0} 
 &\approx \frac{\theta\; \delta_{m+k,n+l}}{
\sqrt{\frac{1}{\pi}(16\, \mathcal{C}{+}12)} 
+ \frac{6 \Omega}{1+ 2\Omega^3+2\Omega^4} \mathcal{C}} 
\label{Form4-1}
\end{align}
We demonstrate in Figure~\ref{fig-Form4-1} for selected values of the
parameters $\Omega,\mathcal{C}$ that $\theta/(\max
\Delta_{mn;kl}^{\mathcal{C}})$ at $\mu_0=0$ is
asymptotically reproduced by $\sqrt{\frac{1}{\pi}(16\, \mathcal{C}{+}12)} 
+ \frac{6 \Omega}{1+ 2\Omega^3+2\Omega^4} \mathcal{C}$.% 
\begin{figure}[h!t]
\begin{picture}(130,90)
\put(-26,-190){\epsfig{file=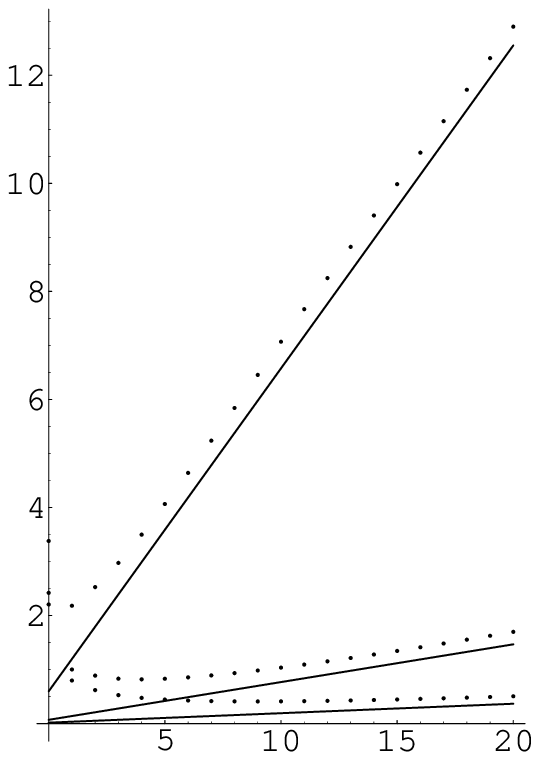,scale=0.9,bb=0 0 598 843}}
\put(47,9){\mbox{\scriptsize$\Omega=0.05$}}
\put(47,16){\mbox{\scriptsize$\Omega=0.1$}}
\put(47,54){\mbox{\scriptsize$\Omega=0.3$}}
\put(48,-190){\epsfig{file=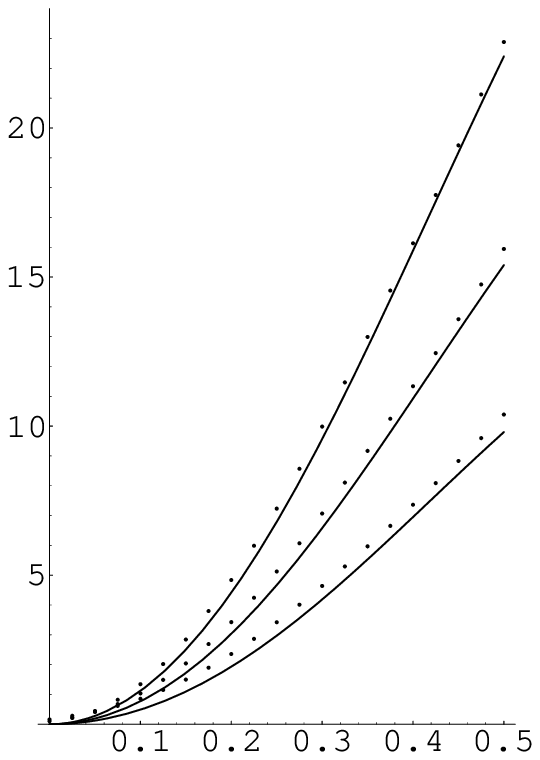,scale=0.9,bb=0 0 598 843}}
\put(127,60){\mbox{\scriptsize$\mathcal{C}=15$}}
\put(125,40){\mbox{\scriptsize$\mathcal{C}=10$}}
\put(122,24){\mbox{\scriptsize$\mathcal{C}=6$}}
\end{picture}
\caption[Sum of the propagator]{Comparison of $\theta/\big(\max_{m} 
\sum_{l} \max_{n,k} |\Delta_{mn;kl}^{\mathcal{C}}| \big)$ at $\mu_0=0$
(dots) with 
$7 \Omega^2 (\mathcal{C}+1)/(1{+}2\Omega^2)$ (solid line). 
The left plot shows the inverse propagator and its approximation 
over $\mathcal{C}$ for three values of $\Omega$, whereas the right plot
shows the inverse propagator and its approximation 
over $\Omega$ for three values of $\mathcal{C}$.}
 \label{fig-Form4-2}
\end{figure}

\bigskip
\noindent
{\bf Formula 2:}
\begin{align}
\max_{m^r} 
\sum_{l^1,l^2 \in \mathbb{N}} 
\max_{k^r,n^r} 
\Big|  \Delta_{\di{m^1}{m^2}\di{n^1}{n^2};
\di{k^1}{k^2}\di{l^1}{l^2}}^{\mathcal{C}} \Big|_{\mu_0=0} 
 &\approx \frac{\theta\,(1{+} 2\Omega^3)}{7 \Omega^2 (\mathcal{C}+1)} \;.
\label{Form4-2}
\end{align}
We demonstrate in Figure~\ref{fig-Form4-2} that $\theta/\big(\max_{m} 
\sum_{l} \max_{n,k} |\Delta_{mn;kl}^{\mathcal{C}}| \big)$
is for $\mu_0=0$ asymptotically given by $7 \Omega^2 (\mathcal{C}+1)/(1{+}2\Omega^3)$.%

\begin{figure}[h!b]
\begin{picture}(130,105)
\put(-23,-127){\epsfig{file=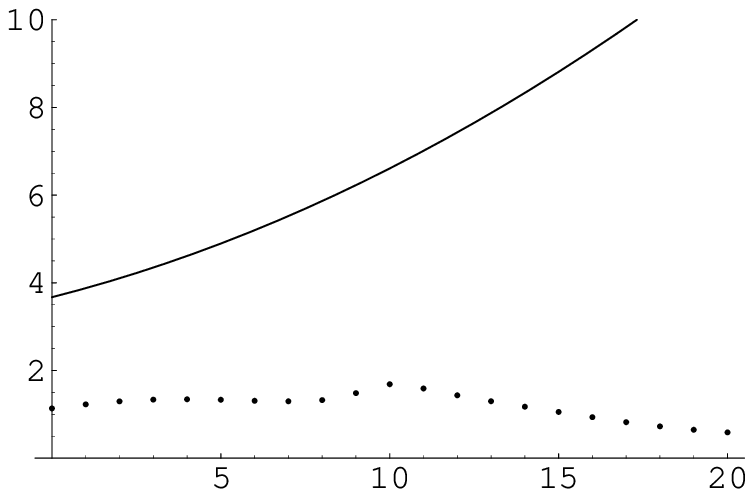,scale=0.74,bb=0 0 598 843}}
\put(45,-127){\epsfig{file=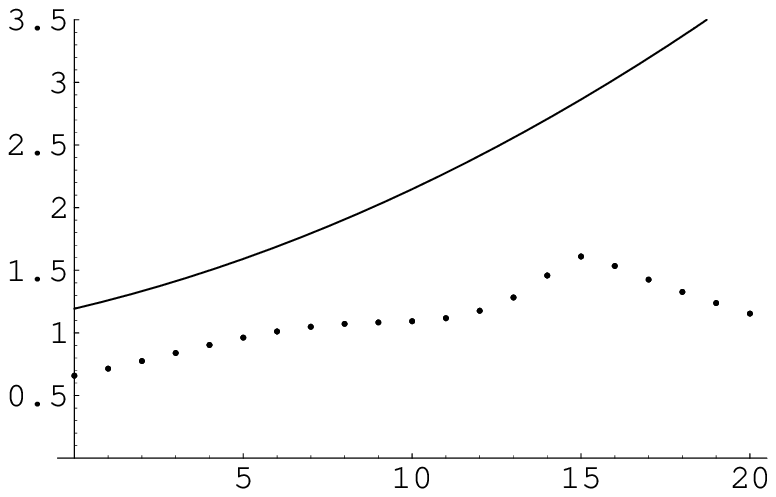,scale=0.74,bb=0 0 598 843}}
\put(-23,-175){\epsfig{file=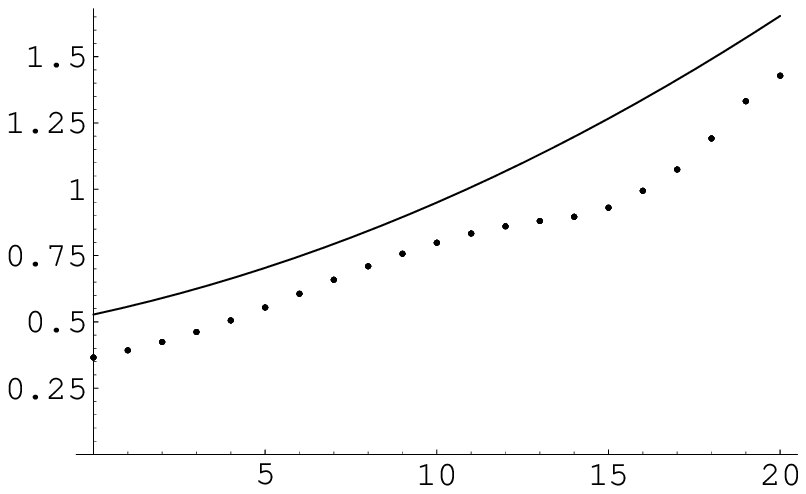,scale=0.74,bb=0 0 598 843}}
\put(47,77){\mbox{\scriptsize$\Omega=0.05$}}
\put(47,72){\mbox{\scriptsize$\mathcal{C}=10$}}
\put(87,87){\mbox{\scriptsize$\Omega=0.05$}}
\put(87,82){\mbox{\scriptsize$\mathcal{C}=15$}}
\put(50,25){\mbox{\scriptsize$\Omega=0.05$}}
\put(50,20){\mbox{\scriptsize$\mathcal{C}=20$}}
\end{picture}
\caption[Sum with gap of the propagator]{
  The index summation $\displaystyle \frac{1}{\theta}
  \Big(\sum_{l\;,~\|m-l\|_1 \geq 5} \max_{k,r} \big|
  \Delta_{mn;kl}^{\mathcal{C}}\big|\Big)$ of the cut-off propagator
 at $\mu_0=0$  (dots) compared with $ \frac{\theta\,(1{-} \Omega)^4 \big(15 +
    \frac{4}{5} \|m\|_\infty + \frac{1}{25}\|m\|_\infty^2\big)}{ \Omega^2
    (\mathcal{C}{+}1)^3}$ (solid line), both plotted over
  $\|m\|_\infty$.}
\label{fig-Form4-3a}
\end{figure}

\begin{figure}[h!t]
\begin{picture}(130,78)
\put(-23,-126){\epsfig{file=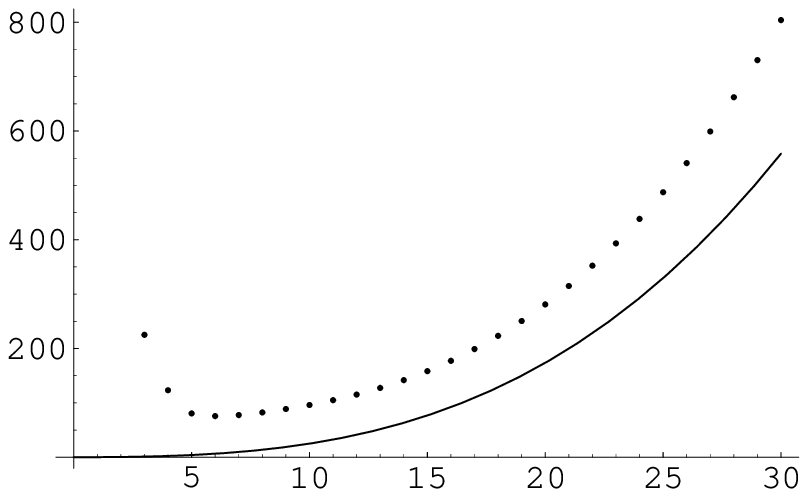,scale=0.67,bb=0 0 598 843}}
\put(48,-126){\epsfig{file=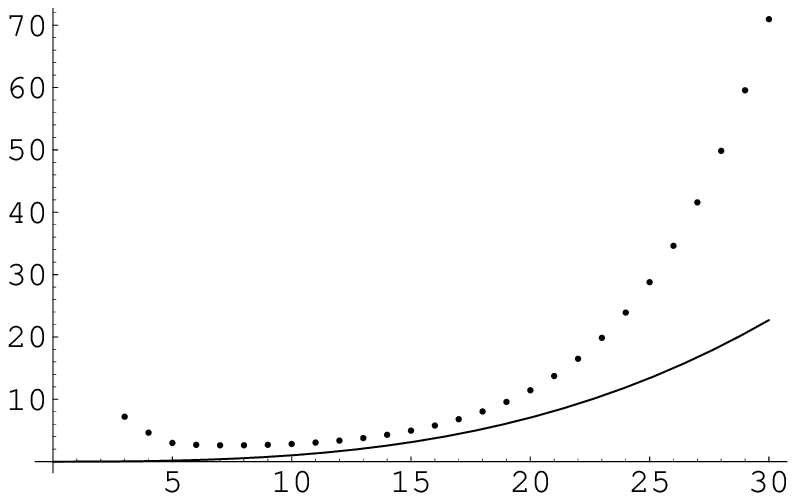,scale=0.67,bb=0 0 598 843}}
\put(-21,-164){\epsfig{file=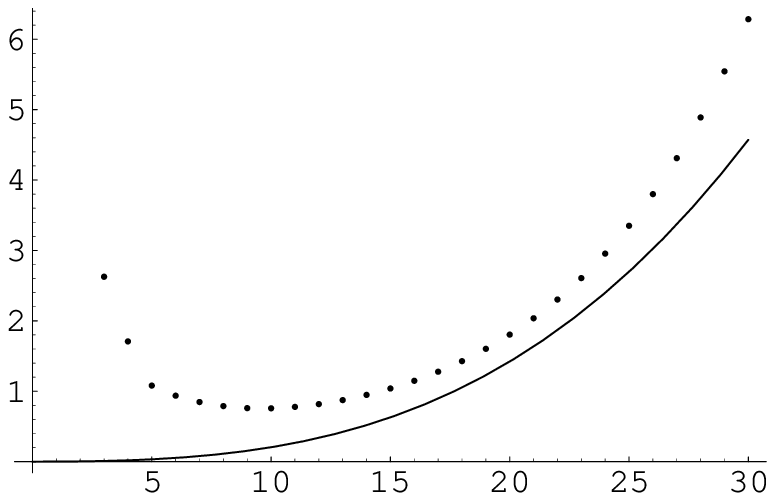,scale=0.67,bb=0 0 598 843}}
\put(18,67){\mbox{\scriptsize$\Omega=0.3$}}
\put(18,62){\mbox{\scriptsize$\|m\|_\infty=5$}}
\put(90,67){\mbox{\scriptsize$\Omega=0.1$}}
\put(90,62){\mbox{\scriptsize$\|m\|_\infty=5$}}
\put(18,28){\mbox{\scriptsize$\Omega=0.05$}}
\put(18,22){\mbox{\scriptsize$\|m\|_\infty=5$}}
\end{picture}
\caption[Sum with gap of the propagator]{The inverse 
  $\displaystyle\theta \Big(\sum_{l\;,~\|m-l\|_1 \geq 5} \max_{k,r}
  \big| \Delta_{mn;kl}^{\mathcal{C}}\big|\Big)^{-1}$ of the summed
  propagator at $\mu_0=0$ (dots) compared with $\frac{\Omega^2
    (\mathcal{C}{+}1)^3}{(1{-} \Omega)^4\big(15 + \frac{4}{5}
    \|m\|_\infty + \frac{1}{25}\|m\|_\infty^2\big)}$ (solid line),
  both plotted over $\mathcal{C}$.}
\label{fig-Form4-3b}
\end{figure}

\begin{figure}[h!b]
\begin{picture}(130,118)
\put(-21,-83){\epsfig{file=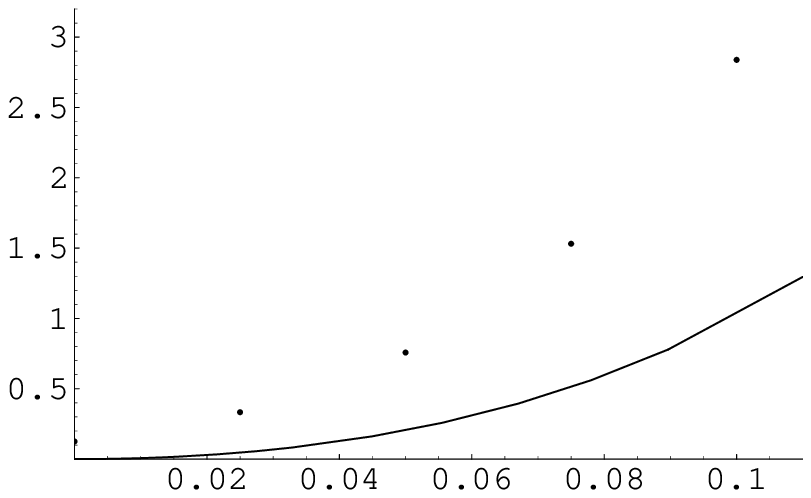,scale=0.65,bb=0 0 598 843}}
\put(48,-83){\epsfig{file=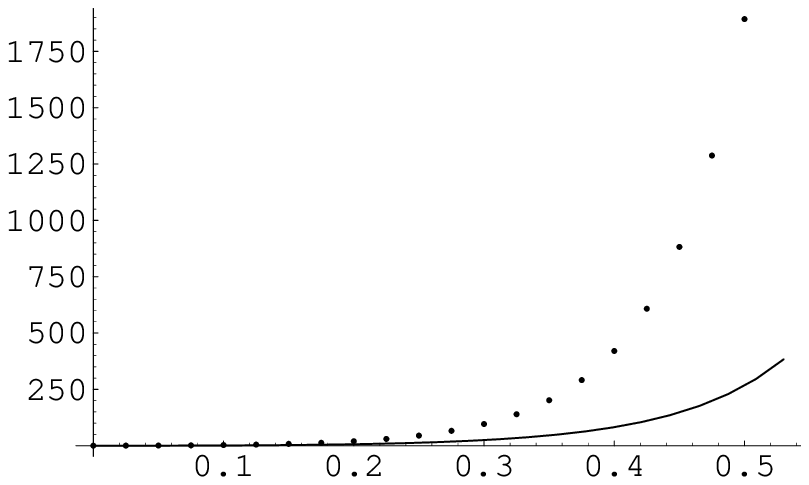,scale=0.65,bb=0 0 598 843}}
\put(-21,-120){\epsfig{file=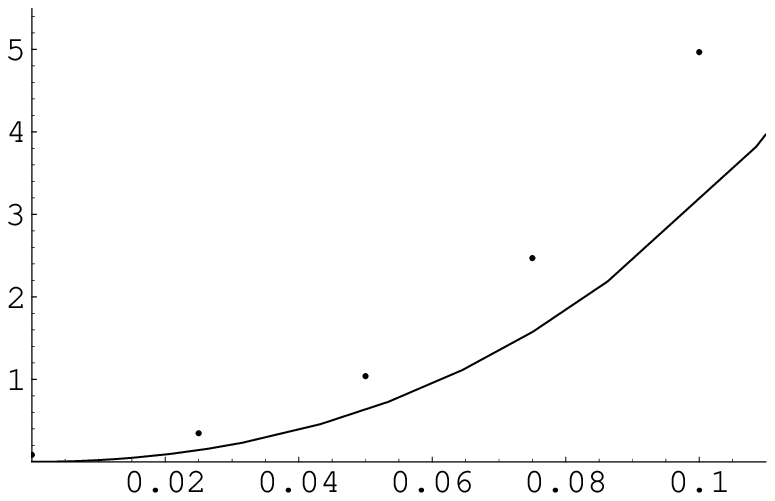,scale=0.65,bb=0 0 598 843}}
\put(48,-120){\epsfig{file=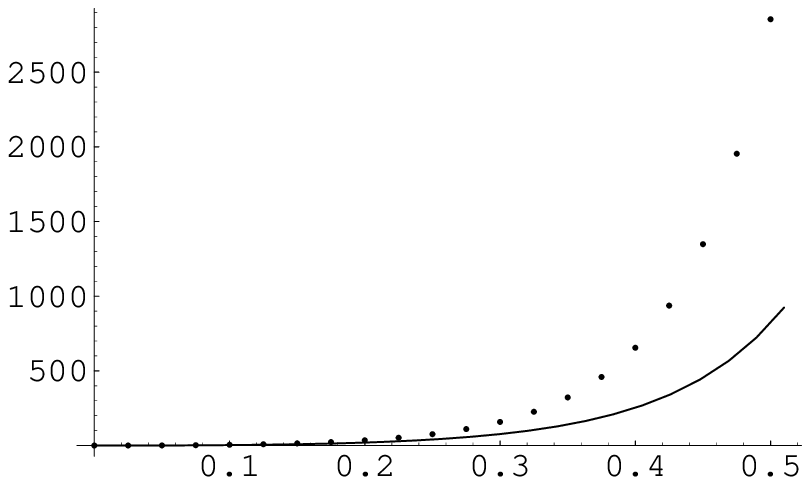,scale=0.65,bb=0 0 598 843}}
\put(-21,-157){\epsfig{file=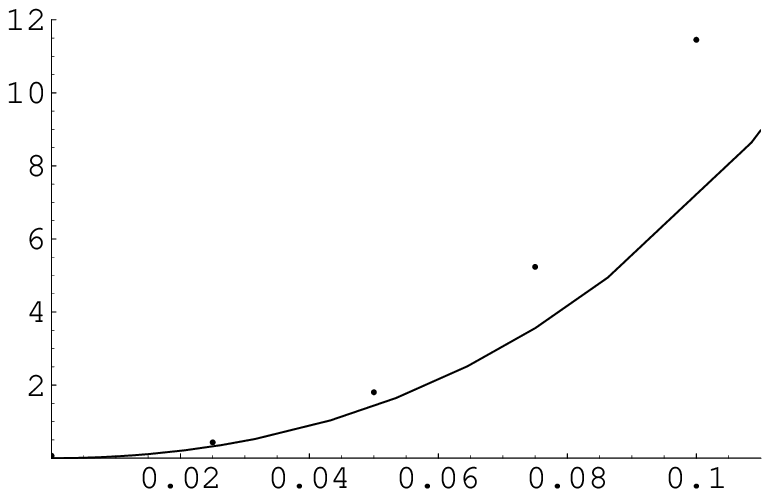,scale=0.65,bb=0 0 598 843}}
\put(48,-157){\epsfig{file=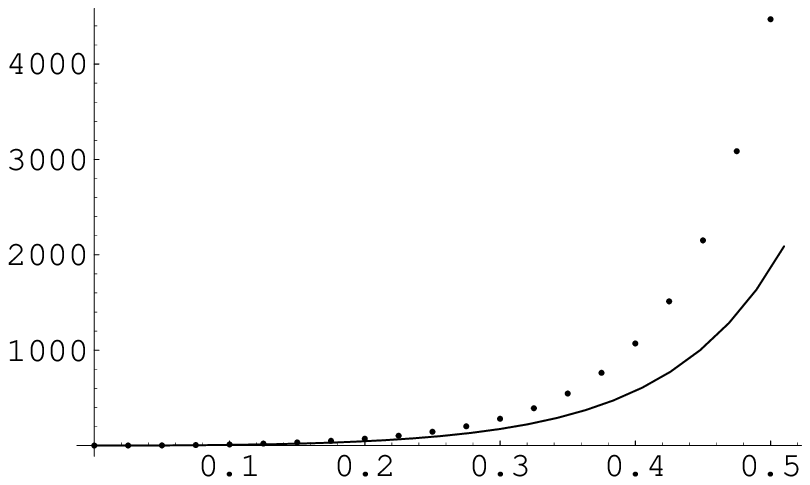,scale=0.65,bb=0 0 598 843}}
\put(20,105){\mbox{\scriptsize$\mathcal{C}=10$}}
\put(20,100){\mbox{\scriptsize$\|m\|_\infty=5$}}
\put(90,105){\mbox{\scriptsize$\mathcal{C}=10$}}
\put(90,100){\mbox{\scriptsize$\|m\|_\infty=5$}}
\put(20,65){\mbox{\scriptsize$\mathcal{C}=15$}}
\put(20,60){\mbox{\scriptsize$\|m\|_\infty=5$}}
\put(90,65){\mbox{\scriptsize$\mathcal{C}=15$}}
\put(90,60){\mbox{\scriptsize$\|m\|_\infty=5$}}
\put(20,25){\mbox{\scriptsize$\mathcal{C}=20$}}
\put(20,20){\mbox{\scriptsize$\|m\|_\infty=5$}}
\put(90,25){\mbox{\scriptsize$\mathcal{C}=20$}}
\put(90,20){\mbox{\scriptsize$\|m\|_\infty=5$}}
\end{picture}
\caption[Sum with gap of the propagator]{The inverse 
  $\displaystyle\theta \Big(\sum_{l\;,~\|m-l\|_1 \geq 5} \max_{k,r}
  \big| \Delta_{mn;kl}^{\mathcal{C}}\big|\Big)^{-1}$ of the summed
  propagator at $\mu_0=0$ (dots) compared with $\frac{\Omega^2
    (\mathcal{C}{+}1)^3}{(1{-} \Omega)^4\big(15 + \frac{4}{5}
    \|m\|_\infty + \frac{1}{25}\|m\|_\infty^2\big)}$ (solid line),
  both plotted over $\Omega$.}
\label{fig-Form4-3c}
\end{figure}

\bigskip
\noindent
{\bf Formula 3:}
\begin{align}
\sum_{\mbox{\scriptsize$\begin{array}{c}l^1,l^2 \in \mathbb{N} \\ 
\|m-l\|_1 \geq 5\end{array}$}} 
\max_{k^r,n^r} 
\Big|  \Delta_{\di{m^1}{m^2}\di{n^1}{n^2};
\di{k^1}{k^2}\di{l^1}{l^2}}^{\mathcal{C}} \Big|_{\mu_0=0}
 &\leq \frac{\theta\,(1{-} \Omega)^4 \big(15 +
    \frac{4}{5} \|m\|_\infty + \frac{1}{25}\|m\|_\infty^2\big)}{
\Omega^2 (\mathcal{C}{+}1)^3} \;.
\label{Form4-3}
\end{align}
We verify (\ref{Form4-3}) for several choices of the parameters 
in Figures~\ref{fig-Form4-3a}, \ref{fig-Form4-3b} and \ref{fig-Form4-3c}.

\end{appendix}

\end{document}